\documentclass[a4paper,11pt]{article}
\usepackage{jheppub}% http://jhep.sissa.it/jhep/help/JHEP/TeXclass/DOCS/jheppub.sty
    \usepackage{etoolbox}% http://ctan.org/pkg/etoolbox
    \makeatletter
    \patchcmd{\maketitle}{\@fpheader}{}{}{}
    \makeatother
    
    \usepackage[normalem]{ulem}
\usepackage{tocloft}
\usepackage{amsfonts}
\usepackage{amsmath}
\usepackage{amssymb}
\usepackage{float,tikz, extarrows, tikz-cd}
\usetikzlibrary{decorations.pathmorphing,calc}
\tikzset{snake it/.style={decorate, decoration=snake}}
\usepackage{array}
\usepackage{bigints}
\usepackage{textcomp}
\usepackage{bm}
\usepackage{booktabs}
\usepackage{color}
\usepackage{dsfont}
\usepackage{float}
\usepackage{framed}
\usepackage{graphicx}
\usepackage{tcolorbox}
\usepackage{indentfirst}
\usepackage{mathrsfs}
\usepackage{multirow}
\usepackage{pdflscape}
\usepackage{setspace}
\usepackage{titlesec}
\usepackage{wrapfig}
\usepackage{mathtools}
\usepackage[all]{xy}
\usepackage{young}
\usepackage[vcentermath]{youngtab}
\usepackage{relsize}
\usepackage{stackengine}
\usepackage{verbatim}
\usepackage{slashed}
\usepackage[compat=1.0.0]{tikz-feynman}
\usepackage{adjustbox}
\usepackage{subcaption}
\usepackage{caption}
\def\tilde{\widetilde}

\def\bar{\overline}

\def\1{{\mathds 1}}

\DeclareMathOperator{\Tr}{\mathrm{Tr}}
\DeclareMathAlphabet{\mathbfsf}{OT1}{cmss}{bx}{n}

\newcommand{\beq}{\begin{equation}\begin{aligned}}
\newcommand{\eeq}{\end{aligned}\end{equation}}

\newcommand{\lam}{\lambda}
\newcommand{\Lam}{\Lambda}

\newcommand{\be}{\beta}

\newcommand{\ov}{\over}

\newcommand{\bea}{\begin{eqnarray}}
\newcommand{\eea}{\end{eqnarray}}

\newcommand{\beqa}{\begin{eqnarray}}
\newcommand{\eeqa}{\end{eqnarray}}
\newcommand{\beqar}{\begin{eqnarray*}}
\newcommand{\eeqar}{\end{eqnarray*}}

\def\({\left(} \def\){\right)}
\def\[{\left[} \def\]{\right]}

\title{Thermal order in large N conformal gauge theories}

\author[a]{Soumyadeep Chaudhuri} 
\author[b, c, d]{\!, Changha Choi}
\author[a]{\!, Eliezer Rabinovici}

\vspace{8cm}

\affiliation[a]{Racah Institute, The Hebrew University. Jerusalem 9190401, Israel}

\affiliation[b]{Simons Center for Geometry and Physics,  SUNY, Stony Brook, NY 11794, USA}

\affiliation[c]{C.N. Yang Institute for Theoretical Physics, SUNY, Stony Brook, NY 11794, USA}

\affiliation[d]{Kavli Institute for Theoretical Physics, University of California, Santa Barbara, CA 93106, USA}

\vspace{8cm}

\emailAdd{chaudhurisoumyadeep@gmail.com} 
\emailAdd{changha.choi@stonybrook.edu}  
\emailAdd{eliezer@mail.huji.ac.il}

\abstract{In this work we explore the possibility of spontaneous breaking of global symmetries at all nonzero temperatures for conformal field theories (CFTs) in $D=4$ space-time dimensions.  We show that such a symmetry-breaking indeed occurs in certain families of non-supersymmetric  large N gauge theories at a planar limit. We also show that this phenomenon is accompanied by the system remaining in a persistent Brout-Englert-Higgs (BEH) phase at any temperature. These analyses are motivated by the work done in  \cite{Chai:2020zgq, Chai:2020onq} where symmetry-breaking was observed in all thermal states for certain CFTs  in fractional dimensions. 

In our case, the theories demonstrating the above features have gauge  groups which are specific products of $SO(N)$ in one family and $SU(N)$ in the other. Working in a perturbative regime at the $N\rightarrow \infty$ limit, we show that the  beta functions in these theories yield circles of fixed points in the space of couplings. We explicitly check this structure up to two loops and then present a  proof of its survival under all loop corrections. We show that under certain conditions, an interval on this circle of fixed points demonstrates both the spontaneous breaking of a global symmetry as well as a persistent BEH phase at all nonzero temperatures.  The  broken global symmetry is $\mathbb Z_2$ in one family of theories and $U(1)$ in the other. The  corresponding order parameters  are expectation values of the determinants of  bifundamental scalar fields in these theories. We characterize these symmetries as baryon-like symmetries in the respective models.}

\begin{document}
\maketitle

\section{Introduction}\label{sec:intro}

The study of the patterns of spontaneous symmetry breaking (SSB) and their potential restoration plays a key role in understanding the phase structures of matter. 
From the onset of such investigations it was noted that SSB is a low temperature property. As the temperature is increased, various symmetries that were spontanously broken get restored \cite{Kirzhnits:1972iw, Kirzhnits:1972ut, Dolan:1973qd, Kirzhnits:1976ts, Klimenko:1988ng, Bimonte:1996cw, Gavela:1998ux}. The generality of this phenomenon was first challenged in \cite{Weinberg:1974hy}, and later followed up by \cite{Orloff:1996yn, Bimonte:1999tw, Pinto:1999pg, Komargodski:2017dmc, Tanizaki:2017qhf, Dunne:2018hog, Wan:2019oax, Hong:2000rk} as well as many explorations of interesting phenomenological implications for the physics of  the early universe \cite{Mohapatra:1979qt, Langacker:1980kd, Salomonson:1984rh,Dodelson:1989ii,Dodelson:1991iv,Dvali:1995cj,Meade:2018saz}. However, these works typically involved either UV-incomplete theories or models with imaginary or random chemical potentials. The former prevents investigating whether the symmetries are restored beyond a UV scale, while the latter requires the identification of concrete unitary theories which reproduce the results.  Some holographic constructions  also explored the issue of symmetry restoration at finite temperatures \cite{Gubser:2008px, Hartnoll:2008kx, Hartnoll:2008vx, Buchel:2009ge,  Donos:2011ut, Alberte:2017oqx, Gursoy:2018umf, Buchel:2018bzp, Buchel:2020thm, Buchel:2020xdk, Buchel:2020jfs}. In such setups, non-restoration of global symmetries in  QFTs translates to the violation of the no-hair theorem for black holes in the dual theories of gravity. However, in all  the examples that were studied, either the relevant symmetries are restored at some critical temperature or a metastable symmetry-broken phase exists at high temperatures. So, the feasibility of a symmetry-broken phase which corresponds to the true thermal vacuum up to arbitrarily high temperatures for a unitary UV-complete theory has remained unresolved for several decades.

Recently, in \cite{Chai:2020zgq, Chai:2020onq} it was shown that there exist UV complete systems consisting of fundamental scalar fields in which global continuous symmetries remain consistently broken at arbitrarily high temperatures. The models examined there are Wilson-Fisher-like conformal field theories (CFTs) in $(4-\epsilon)$ dimensions\footnote{It was shown in \cite{Hogervorst:2015akt} that some CFTs in fractional dimensions with finite number of degrees of freedom  contain operators of complex dimensions. This raises questions about the unitarity of such theories defined in fractional dimensions.} symmetric under the group $O(N_1)\times O(N_2)$. In the regime  $0<\epsilon\ll 1$ and $N_1,N_2\gg 1$ with $N_1\neq N_2$, these models exhibit various patterns of spontaneous breaking of the $O(N_1)$ and $O(N_2)$ symmetries in thermal states. Studying CFTs enables one to extend results obtained for one temperature to all temperatures. The reader is referred to \cite{Chai:2020zgq} for a detailed exposition of the method and the results.

In this paper we venture to study this issue of persistent symmetry breaking in $(3+1)$-dimensional large N gauge theories. This  positions these theories nearer to realistic particle physics systems than those considered in \cite{Chai:2020zgq, Chai:2020onq}. Moreover, introducing gauge symmetries allows for both asymptotically free theories as well as Banks-Zaks-like conformal field theories \cite{Belavin:1974gu, PhysRevLett.33.244, Banks:1981nn}. In either of these cases, the UV completion of the theory is guaranteed.  The addition of the gauge particles also enriches the possible phase structures; in particular, it allows the system to be in the Brout-Englert-Higgs (BEH) phase. We will discuss the persistence of both the BEH phase and the associated spontaneous breaking of a global symmetry in the $N\rightarrow \infty$ (planar/ Veneziano \cite{Veneziano:1976wm}) limit of these models. We will show that these two phenomena go hand in hand in these theories. We recall that in the standard model of
cosmology the electroweak BEH phase is supposed to be modified above a certain temperature \cite{Senaha:2020mop}. It will turn out that for obtaining the persistent phase behavior,
we  need to consider systems which are invariant under a gauge group $G$ of the following form:
\beq
G=\prod_{i=1}^2(G_i\times G_i),
\eeq
where the $G_i$'s are either  $SO(N_{ci})$ or $SU(N_{ci})$ with  two  general ranks $N_{c1}$ and $N_{c2}$. We will call the model where $G_i=SO(N_{ci})$ the {\it real double bifundamental model}, and the one where $G_i=SU(N_{ci})$ the  {\it complex double bifundamental model}. As we will show in section \ref{subsec:planar duality}, there is a perturbative equivalence between these two models in the planar limit \footnote{In this paper, we use the term `planar limit' to denote the strictly $N=\infty$ limit. This limit of QFT has been explored in many other contexts (see e.g. \cite{tHooft:1974pnl,Bardeen:1983rv,Aharony:2011jz}) which enjoy various interesting physical properties that otherwise cannot be continuously obtained from a finite but arbitrarily large $N$.} .

From the structure of the above-mentioned gauge group, we can see that there are two sectors (labeled by $i$) in these models. All the fields belonging to one of these two sectors are invariant under the gauge transformations in the other sector. The matter content in each sector comprises of $N_{fi}$  flavors of massless fermions transforming  in the fundamental representation of  $G_i$, and $N_{ci}^2$ massless scalar fields $\Phi_i$ which transform in the bifundamental representation of $(G_i\times G_i)$. We will work in the Veneziano limit where the ratios $\frac{N_{fi}}{N_{ci}}$ and $\frac{N_{c2}}{N_{c1}}$ approach finite values as $N_{ci},N_{fi}\rightarrow\infty$, and the different couplings in the model also scale with appropriate powers of $\frac{1}{N_{c1}}$ and $\frac{1}{N_{c2}}$. A more detailed description of these models will be provided in section \ref{sec:dbm introduction}.

We will show that  for each sector in these models, there is a global internal symmetry which is  $\mathbb Z_2$ for $G_i=SO(N_{ci})$ and $U(1)$ for $G_i=SU(N_{ci})$. We will call this a {\it baryon symmetry} in the model since the corresponding order parameter is the expectation value of a gauge invariant composite operator made  of $N_{ci}$ number of scalar fields. It is one of these baryon symmetries that will be spontaneously broken at arbitrary nonzero temperatures in some parameter regimes. Along with this symmetry breaking, we will find the system to exist in the BEH phase in any thermal state.

We will analyze this persistent symmetry breaking and the associated BEH phase by first  determining the fixed points in the RG flow of the 't Hooft couplings in the double bifundamental models. For this, we will compute the 2-loop beta functions of the different couplings in the limit  $N_{ci}\rightarrow\infty$. Searching for the corresponding fixed points  will yield the following  result:  In addition to discrete  fixed points, there is a  fixed circle in the space of couplings for these 2-loop beta functions. We will also show that this fixed circle in fact survives  in the planar limit even when all higher loop corrections to the beta functions are taken into account.\footnote{The RG flow of the different couplings will depend on the renormalization scheme\cite{Ryttov:2012ur, Ryttov:2012nt, Shrock:2014zca}. In this paper, we work with dimensional regularization and the modified minimal subtraction ($\overline{\text{MS}}$) scheme. We expect the existence of the conformal manifolds (with the topology of a circle) in the double bifundamental models to persist under all sensible  changes in the renormalization scheme.}

Such fixed circles in large N theories were explored by the authors of \cite{Kiritsis:2008at}. Their proof for the existence of exactly marginal interactions between two CFTs was based on the structure of 1-loop beta functions of  the corresponding couplings. Since the double bifundamental models fall exactly in the class of systems that they considered, it is expected that there would be a fixed circle for  the 1-loop beta functions of the couplings. Here, we take a step further and prove that the fixed circles in these models survive under all loop corrections to the beta functions at the planar limit. To the best of our knowledge, such a result has not been proved earlier for any four dimensional non-supersymmetric large N gauge theory. Moreover, the proof does not rely on any connection to supersymmetric  or holographic setups.\footnote{We refer the reader to the discussion in \cite{Bashmakov:2017rko} for the status of the search for conformal manifolds in non-SUSY theories anchored in SUSY ones.}

To study the possibility of spontaneous breaking of the baryon symmetry and the persistent BEH phase for points lying on this fixed circle, we will consider the thermal effective potential of the scalar fields. We will first show that this effective potential is bounded from below for all points on the fixed circle.  Next, we will determine the parameter regime in which the effective potential can have a minimum away from the origin in the field space for a subset of points on the fixed circle. Such a minimum of the potential at a nonzero field configuration implies both that the system is in a BEH phase and that the thermal expectation value of the baryon operator does not vanish. The latter leads to the spontaneous breaking of the  baryon symmetry. Assuming $N_{c2}<N_{c1}$, we will show that there is an upper bound on the ratio $r\equiv\lim\limits_{N_{c1},N_{c2}\rightarrow\infty}\frac{N_{c2}}{N_{c1}}$ for  spontaneous breaking of the baryon symmetry to be even possible at any point on the fixed circle. For each value of $r$ below this bound, we will identify the fixed points which demonstrate a symmetry-broken phase at nonzero temperatures.

Before moving on, let us pause here to take a short detour. Enlarging the scope of the theories under investigation to include gauge theories indeed adds more structure. It also brings along with it a large number of subtleties and a considerable amount of small print. Moreover, although our work deals with the spontaneous breaking of global symmetries in gauge theories without gravity, there is evidence that a theory of quantum gravity cannot have any global symmetry\cite{Harlow:2018tng}. Thus, it is useful to summarize  the nuances that arise in  the study of phases due to  gauge symmetries. We will now discuss some of these nuances and explain the terminology we will be using in this work. 

Compact gauge symmetries are known to express themselves in several possible phases. The Coulomb phase, the Brout-Englert-Higgs (BEH) phase and the confinement phase manifest themselves in nature and are the building blocks of the present standard model. The oblique confinement phase may have manifested itself  in interesting condensed matter systems and the conformal phase will also hopefully find its place in nature. There are order parameters which distinguish each of them, but the language used to characterize phase transitions corresponding to spontaneous breaking of global symmetries is not strictly appropriate for the phases associated with gauge symmetries. There is no SSB of a gauge symmetry \cite{Elitzur:1975im}. Furthermore, it was shown \cite{Banks:1979fi, Fradkin:1978dv} that in a class of theories containing bosons or fermions in a representation of the gauge group which is not trivial under the center of the gauge group, there are no phase transitions. The quantitative properties of the system may cross over from one appropriate effective description to another as parameters are changed, but there is no phase transition involved in the process.\footnote{Had there been a deconfinement phase transition, the order parameter for that transition would not have been of the zero form type.} 
These cases include the standard model with its bosons and its quarks, each being  in the fundamental representation of the relevant group.
Moreover, it was shown \cite{Banks:1979fi} that this also implies that in the presence of temperature there is no phase transition, only a crossover. So far, indeed, no experimental evidence has been found for such a transition. The predicted crossover behavior could be validated in future experiments\cite{Ding:2015ona}.

We return now to explain how this pertains to the type of models we discuss in this work which contain bifundamental scalars and fundamental fermions. The constraints on the phase diagram imposed by the presence of matter in the fundamental representation do not prevent a priori a transition characterized by the spontaneous breaking of some global symmetry. As we will see, this is indeed the case in the main examples we construct. An additional feature is that  we are considering  fixed points in the RG flow of these models at the planar limit. Thus, at this limit, the theories of our interest are conformal. The lack of any intrinsic scale in such conformal theories guarantees that there will not be any phase transition or crossover in them at any nonzero temperature. However, this does not preclude the possibility of a transition (or a crossover) when one switches on the temperature. Beyond this initial possible phase transition, the system  persistently remains in a single phase as the temperature is increased to arbitrarily high values. 

Before relating the above general feature of CFTs to the large $N$ fixed points of our models, let us contrast this feature with the non-persistent behaviors of theories without conformal symmetry. For instance, in an asymptotically free theory with adjoint scalars, if the system is in a BEH phase at zero temperature, its non-persistent behavior (under increase in temperature) could be to first pass through a phase transition to a confinement phase and then to deconfine. It is also possible to study a non-persistent behavior by  deforming  a CFT. For example, consider $\mathcal{N} = 4$ SUSY Yang Mills theory with a non-abelian compact gauge group of rank $N_c$ in 4 dimensions. At the planar limit, one can infer the structure of this CFT from AdS/CFT correspondence. If the world volume is $\mathbb R^4$, then at $T=0$ there is an exact flat potential. This makes it possible to spontaneously break the scale invariance by choosing an appropriate vacuum. For all the vacua of the theory at $T=0$,  we expect a phase transition from the system having $O(1)$ degrees of freedom to $O(N_c^2)$ degrees of freedom as soon as a temperature is turned on\footnote{We expect this transition to happen even for large but finite $N_c$.}. Now, the above CFT can be deformed by taking the world volume to be $S^3\times S^1$. Then the flat direction is lifted and the extra scale introduces the dimensionless parameter $TR$ where $R$ is the radius of the sphere. From AdS/CFT correspondence, we know that this allows  a transition from $O(1)$ to $O(N_c^2)$ degrees of freedom at a finite nonzero temperature determined by $TR$ \cite{Witten:1998zw}. 

Let us now come back to the discussion of the phases of the large $N$ fixed points in our models. For all these fixed points, we will see that there is no flat direction of the potential at $T=0$. This prevents the possibility of spontaneously breaking the scale invariance at zero temperature. Moreover, the ground state does not demonstrate the breaking of the afore-mentioned baryon symmetries in the model. However,  for a subset of these large $N$ fixed points in certain parameter regimes,  the system undergoes a transition to a phase where one of the baryon symmetries is spontaneously broken as soon as a temperature is switched on. This phase then survives up to arbitrarily high temperatures. This persistent symmetry breaking is accompanied by the Higgsing of a subset of the gauge bosons. Thus, the system can be qualitatively characterized to be in a BEH-like phase at all nonzero temperatures. It will be called in this work ``being in a persistent BEH phase" or simply ``being in the BEH phase". We may lapse into less rigorous terminology but we trust the reader will not be misled.

\paragraph{Organization of the paper:} As a precursor to all the analyses mentioned above, in section \ref{sec:model} we begin by  discussing some familiar examples of conformal QCDs and show explicitly that they fail to exhibit spontaneous breaking  of some global symmetries at nonzero temperatures. These examples help us to introduce the analytic techniques necessary for studying  the possibility of SSB in a thermal state. Moreover, they create the  grounds for introducing the double bifundamental models as their natural extensions later in the same section.

 In section \ref{sec:rdb model}, we first determine the perturbative fixed points of the planar beta functions in the real double bifundamental model. We demonstrate that up to the 2-loop contributions to these beta functions, there is a conformal manifold in the space of couplings which has the topology of a circle. We then show that the thermal effective potential of the scalar fields is stable at all points on this fixed circle. Later, we go on to identify the conditions under which a subset of points on this fixed circle can show spontaneous breaking of a baryon symmetry at any nonzero temperature. Moreover, we demonstrate that such a symmetry breaking is always associated with the Higgsing of a subset of gauge bosons in the model. This implies a persistent BEH phase for these fixed points.
 
 In section \ref{sec: cdb model}, we study the complex double bifundamental model. We demonstrate a planar equivalence between this model and the real double bifundamental model. This allows us to extend most of the results of section \ref{sec:rdb model} to this model. In particular, we find that there is a fixed circle for the 2-loop planar beta functions of this theory. Just as before, under certain conditions, a subset of points on this fixed circle demonstrates both  spontaneous breaking of a baryon symmetry as well as a persistent BEH phase at all nonzero temperatures.
 
 In section \ref{sec: survival of fixed circle}, we present a diagrammatic argument for the survival of the fixed circle under all loop corrections to the beta functions in the planar limit. In the process, we also prove an important feature of these planar beta functions, i.e., they are independent of the ratio  $r=\frac{N_{c2}}{N_{c1}}$.
 
 In section \ref{sec: Conclusion}, we conclude by summarizing our results and commenting  on the possibility of the survival of the fixed points under finite $N$ corrections to the beta functions.

In appendix \ref{app: bifund QCD baryon symmetry}, we show that the Lagrangian of the bifundamental scalar QCD introduced in section \ref{sec:model} is indeed invariant under the $\mathbb Z_2$ transformation that we characterize as the baryon symmetry in the model. We also show that this symmetry can be interpreted as an automorphism of the set of classes of gauge-equivalent field configurations in the theory. The analysis presented in this appendix can be generalized to show similar properties of the baryon symmetries in the double bifundamental models.

In  appendix \ref{appsec:qcd}, we present a detailed review of the two-loop beta functions in a general QCD with special emphasis on the case where the gauge group is semi-simple. We also use these general expressions to compute the beta functions of the theories that serve as precursors to the double bifundamental models.

In appendix \ref{app: rdb beta fns.},  we  derive the two-loop beta functions of the real double bifundamental model using the techniques reviewed in the previous appendix.
 
 In appendix \ref{app: fixed point constraints}, we analyze some constraints on the unitary fixed points of the planar beta functions in the  real double bifundamental model. During this analysis, we restrict our attention to the 2-loop beta functions of the  gauge couplings and the 1-loop beta functions of the quartic scalar couplings.

 In appendix \ref{app: minimum of  potential}, we determine the minima of the thermal effective potential of the scalar fields in the real double bifundamental model for the fixed points where a baryon symmetry is broken.
 
 In appendix \ref{app: scaling of diagrams}, we discuss the scaling of different planar diagrams in the double bifundamental models at the large N limit.
 
  In appendix \ref{app: double trace beta fns.}, we derive some general expressions for the planar beta functions of the double trace couplings in the double bifundamental models. These expressions relate these beta functions to the corresponding wave function and vertex renormlizations. 
 
 In appendix \ref{app: finite N corrections}, we study the finite N corrections to the fixed points in the real double bifundamental model by considering double expansions of the beta functions in powers of $\frac{1}{N}$ and the 't Hooft couplings. 
% At the end of the appendix, we also mention a limitation of this analysis.

\section{Survey of gauge theories}\label{sec:model}
In this section, we will first examine two simple variants of QCDs which contain scalars along with fermions. The gauge groups in these two models are $SO(N_c)$ and  $SO(N_c)\times SO(N_c)$ respectively. The corresponding scalar fields transform in the fundamental and the bifundamental representations of the gauge groups.\footnote{For similar models with $SU(N_c)$ gauge symmetry, the Banks-Zaks fixed points were qualitatively argued to not exhibit symmetry-breaking at nonzero temperatures in \cite{Chai:2020zgq}.} We will study the Banks-Zaks fixed points of these models  in the following subsection. There we will explicitly show that for each of these fixed points, the minimum of the thermal effective potential of the scalar fields lies at the origin of the field space. This implies a lack of thermal order in these CFTs. Experience in working out these models will then guide us on how to extend them to get theories with the dual phenomena of spontaneous breaking of a global symmetry and the persistent BEH phase at all temperatures. We will discuss these extensions, viz., the double bifundamental models, in section \ref{sec:dbm introduction}.
\subsection{Warm-up: QCD with fundamental scalars / Bifundamental scalar QCD} \label{sec:simple}
\subsubsection{QCD with fundamental scalars }
Let us begin by considering a QCD with the  gauge group\footnote{The actual gauge group in this model is $Spin(N_c)$ which is a universal cover of $SO(N_c)$. We neglect this technicality from now on to simplify the presentation.} $SO(N_c)$. The matter content in this theory consists of $N_f$ flavors of  Majorana fermions ($\psi_a^{(q)}$), and $N_s$ flavors of  scalar fields ($\phi_a^{(p)}$),\footnote{The superscripts in $\psi_a^{(q)}$ and $\phi_a^{(p)}$ are the indices corresponding to the flavors of the fermion and the scalar respectively. The subscripts denote the color indices.} each of which  transforms in the fundamental representation of the gauge group. The scalar fields interact with each other via quartic interactions as given later in \eqref{vector QCD:renorm. Lagrangian}.

To tame the UV divergences in the theory, we need to specify a scheme for regularization and renormalization. For all the models we introduce in this paper including the one that we are currently discussing, we choose to work in the dimensional regularization and the $\overline{\text{MS}}$ scheme. In this scheme, we  take the renormalized masses of all the fields to be zero.\footnote{Setting the renormalized masses to zero at any energy scale in the $\overline{\text{MS}}$ scheme ensures that they remain zero at all energy scales.}

The renormalized Lagrangian\footnote{We will  suppress the  gauge-fixing and ghost terms in the Lagrangians of the different models that will be introduced in this paper.} of this model in the above-mentioned scheme is
\begin{equation}
\begin{split}
\mathcal L_{\text{Vector}}=&-\frac{1}{4}F_{\mu\nu}^A F^{\mu\nu A}+\frac{i}{2}\overline{\psi}_a^{(p)} (\slashed{D}\psi^{(p)})_a+\frac{1}{2}\Big(D_{\mu} \phi^{(p)} \Big)_a\Big(D^{\mu} \phi^{(p)} \Big)_a\\
&-\tilde h \phi_a^{(p)}\phi_b^{(p)}\phi_b^{(q)}\phi_a^{(q)}-\tilde f \phi_a^{(p)}\phi_a^{(p)}\phi_b^{(q)}\phi_b^{(q)}.
\end{split}
\label{vector QCD:renorm. Lagrangian}
\end{equation}
Here  $F_{\mu\nu}^A$ is the coefficient of the generator $T_A$ in the expansion of the field strength. We take these generators to be normalized such that the second Dynkin index of the fundamental representation of $SO(N_c)$ is $\frac{1}{2}$, i.e., $\text{Tr}\Big[T_AT_B\Big]=\frac{1}{2}\delta_{AB}$.

The quartic terms in the potential of the above Lagrangian can be expressed in terms of symmetric couplings $\lambda_{abcd}^{(pqrs)}$ as follows:
\begin{equation}
\frac{\lambda_{abcd}^{(pqrs)}}{4!}\phi_a^{(p)}\phi_b^{(q)}\phi_c^{(r)}\phi_d^{(s)}
=\tilde h \phi_a^{(p)}\phi_b^{(p)}\phi_b^{(q)}\phi_a^{(q)}+\tilde f \phi_a^{(p)}\phi_a^{(p)}\phi_b^{(q)}\phi_b^{(q)},\\
\end{equation}
where
\begin{equation}
\begin{split}
\lambda_{abcd}^{(pqrs)}=&4\widetilde{h}\Bigg[\delta^{pq}\delta^{rs}(\delta_{ac}\delta_{b d}+\delta_{ad}\delta_{bc})+\delta^{pr}\delta^{qs}(\delta_{ab}\delta_{cd}+\delta_{ad}\delta_{bc})+\delta^{ps}\delta^{qr}(\delta_{ab}\delta_{cd}+\delta_{ac}\delta_{bd})\Bigg]\\
&+8\widetilde{f}\Bigg[\delta^{pq}\delta^{rs}\delta_{ab}\delta_{cd}+\delta^{pr}\delta^{qs}\delta_{ac}\delta_{bd}+\delta^{ps}\delta^{qr}\delta_{ad}\delta_{bc}\Bigg].
\end{split}
\label{vector qcd: symm. couplings}
\end{equation}

Note that this model has a global $O(N_s)$ symmetry which mixes the different flavors of scalars. We will show that this flavor symmetry is unbroken at nonzero temperatures for the fixed points in the RG flow of the model. We will also see that associated with this is the absence of the BEH phase in all thermal states.
To demonstrate these, let us  work in the Veneziano limit where  $N_{c},N_f,N_s\rightarrow \infty$ , while the ratios  $N_{f,s}/N_c$ are kept fixed at finite values $x_{f,s}$. In this limit, the couplings in the model scale with $N_c$ and $N_s$ as follows:
\begin{equation}
g^2=\frac{16\pi^2\lambda}{N_c},\ \tilde h= \frac{16\pi^2 h}{N_c}, \tilde f= \frac{16\pi^2 f}{N_c N_s}, 
\label{vector QCD: 't Hooft couplings}
\end{equation}
where $g$ is the gauge coupling, and $\lambda$, $h$ and $f$ are the  't Hooft couplings. The beta functions of these couplings are derived in appendix \ref{appsec:simple}. Here we provide the forms of the  two-loop beta function of the gauge coupling and the one-loop  beta functions of the quartic couplings in the Veneziano limit:
\beq 
~&\beta_{\lam}=- {22-4x_f-x_s\ov 6} \lam^2  + {13 x_f+4 x_s-34\ov 6}\lam^3,
\\&\beta_h=8(1+x_s)h^2 -3h \lambda 
   +\frac{3\lambda ^2 }{32},
\\&\beta_f=8f^2 +16(1+x_s)f h -3f \lambda 
  +24 h^2 x_s+\frac{3 \lambda ^2 x_s}{32}.
\eeq
To obtain  the fixed points corresponding to these beta functions, let us first set $\beta_\lambda=0$ which gives us
\beq
 \lambda= \frac{22-4x_f-x_s}{13x_f+4x_s-34}.
\eeq
For the validity of the perturbation theory at the fixed point, one must have 
\beq
x_f=\frac{22-x_s}{4}-\epsilon,
\eeq
where $\epsilon$ is a small positive number.

Solving $\beta_h=\beta_f=0$, we get the following values of the quartic couplings at the fixed points:
\begin{equation}
\begin{split}
&h=\lambda\Big(\frac{3-\sigma_1\sqrt{6-3 x_s}}{16(1+x_s)}\Big),\\
&f=\frac{\sqrt{3}\sigma_1}{16(1+x_s)}\lambda\Bigg[\sqrt{2-x_s}(1+x_s)+\sigma_2\sqrt{2-(13-6\sigma_1\sqrt{6-3x_s})x_s+x_s^2-2 x_s^3}\Bigg],
\end{split}
\end{equation}
where $\sigma_1$ and $\sigma_2$ can take the values $1$ and $-1$.
Note that for each of these fixed points, the coupling  $h$ and $f$ are real \footnote{The reality of the couplings is necessary for the corresponding theory to be unitary.} only if $x_s$ is below a certain value $x_s^\text{max}$. These upper bounds on $x_s$ for the different fixed points are as follows \footnote{Here, let us remark that identical bounds on $x_s$ were already obtained for a model of QCD with fundamental scalar equipped with the $SU(N_c)$ gauge group in \cite{Benini:2019dfy} (see also \cite{Hansen:2017pwe} for similar bounds in models where $N_c$ is finite). The similarity of these bounds in the model considered in \cite{Benini:2019dfy}  and the model that we are presently discussing is a direct consequence of a planar equivalence between them. This perturbative equivalence between the two models in the planar limit is similar to the one we discuss in section \ref{subsec:planar duality}.}:
\beq
x_s^{\text{max}}\approx 
\begin{dcases}
0.84 \text{ when $\sigma_1=1$ },\\
0.073 \text{ when $\sigma_1=-1$ }.
\end{dcases}
\eeq 
Note that these upper bounds on $x_s$ are all less than 1. Therefore, for the unitary fixed points in this theory, $N_s < N_c$. In such a situation, as shown in several papers (see e.g. \cite{Choi:2018ohn}\footnote{The theories considered in this work  were defined in $(2+1)$ dimensions. Since the scalar fields, by definition, are in the trivial representation of the spacetime symmetry, the arguments are equally applicable to $(3+1)$-dimensional theories.}), the flavor symmetry is guaranteed to be preserved  under certain conditions even if the scalar fields get thermal expectation values. The pattern of such expectation values would always result in a BEH phase with a residual $SO(N_c-N_s)$ gauge symmetry and no surviving Nambu-Goldstone bosons. These conditions are obeyed in the cases we discuss and anyhow, as we will show below, the scalar fields actually have zero expectation value in any thermal state. Hence, the flavor symmetry is preserved without the Higgsing of any of the gauge fields.

\begin{figure}[H]
\begin{subfigure}{.5\textwidth}
  \centering
  \scalebox{0.8}{\includegraphics[width=.8\linewidth]{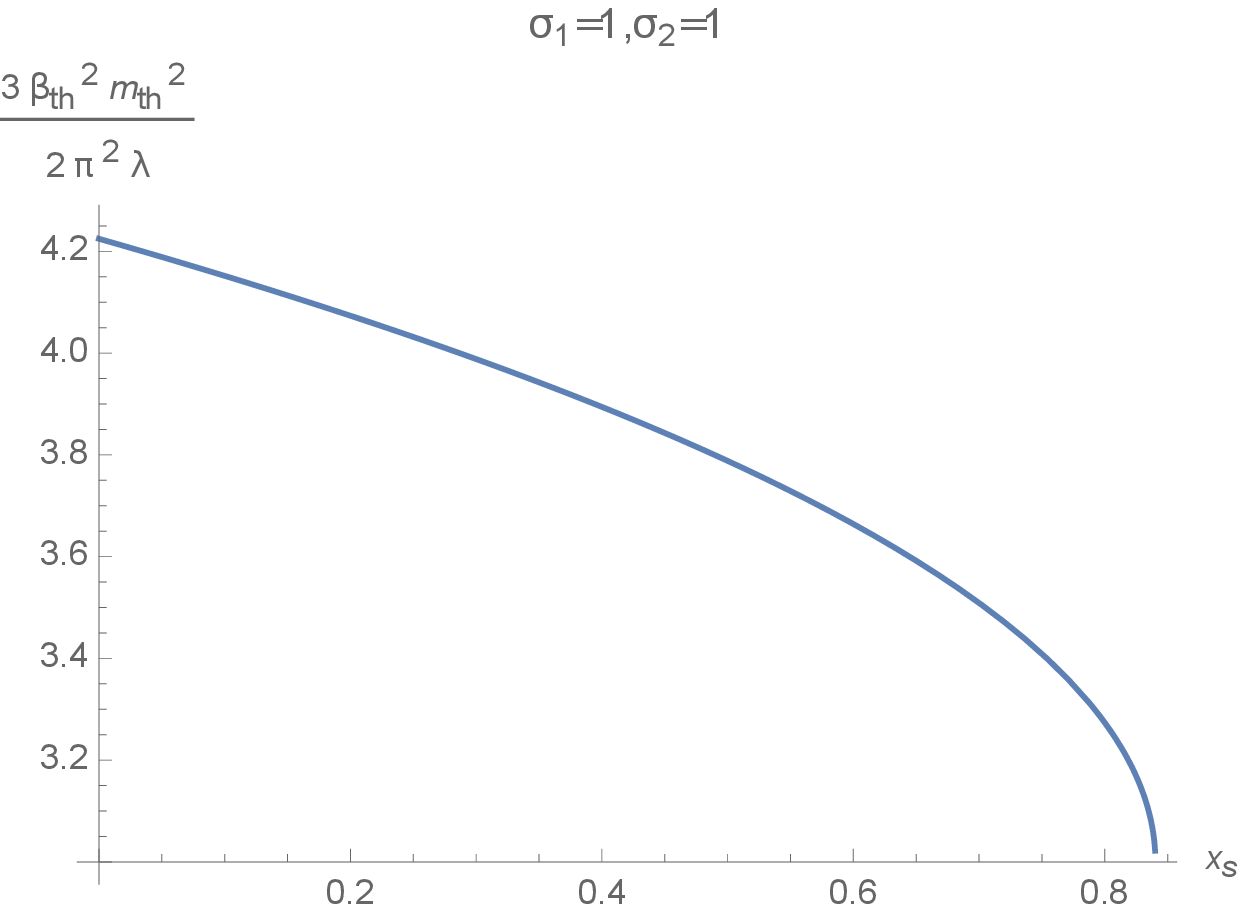}}
\end{subfigure}%
\begin{subfigure}{.5\textwidth}
  \centering
  \scalebox{0.8}{\includegraphics[width=.8\linewidth]{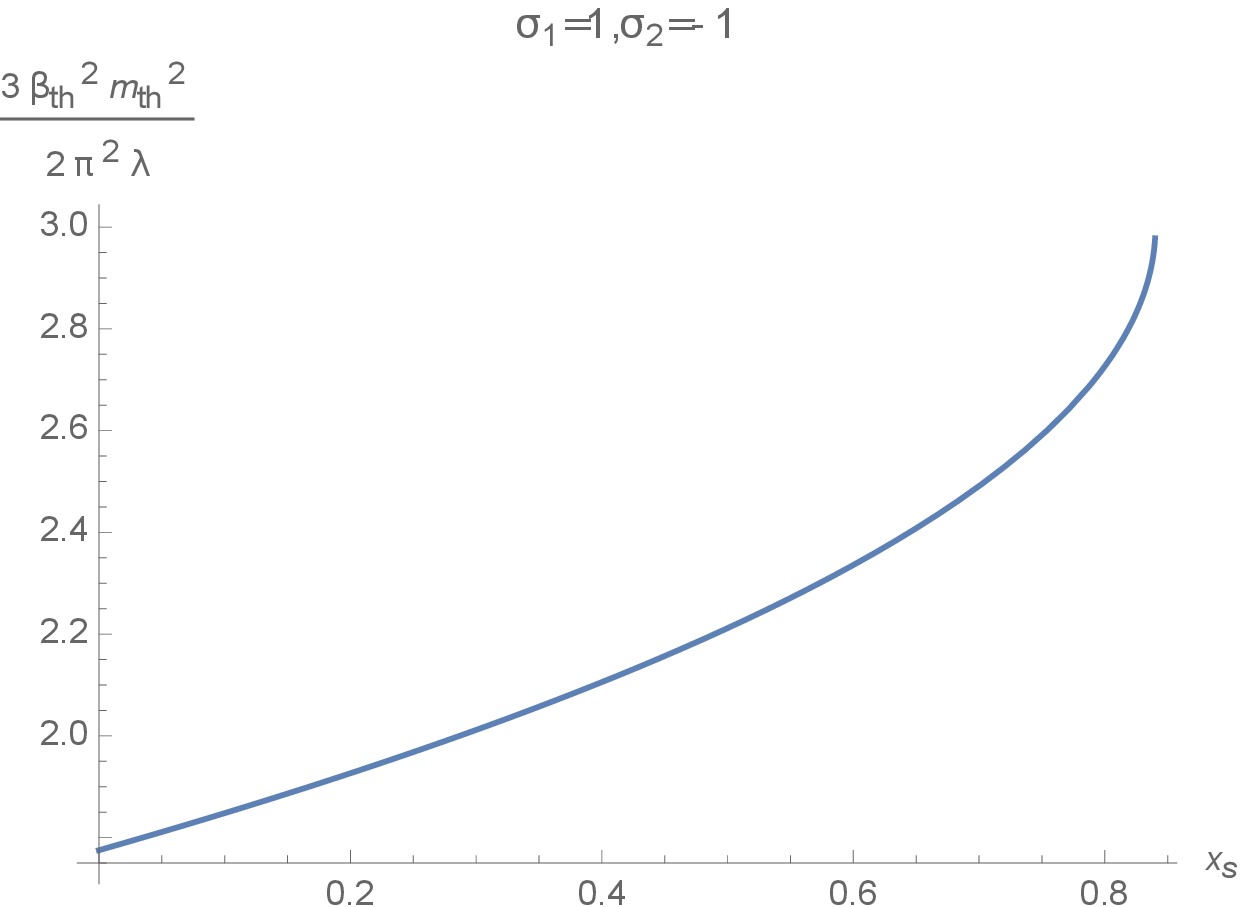}}
\end{subfigure}
\begin{subfigure}{.5\textwidth}
  \centering
  \scalebox{0.8}{\includegraphics[width=.8\linewidth]{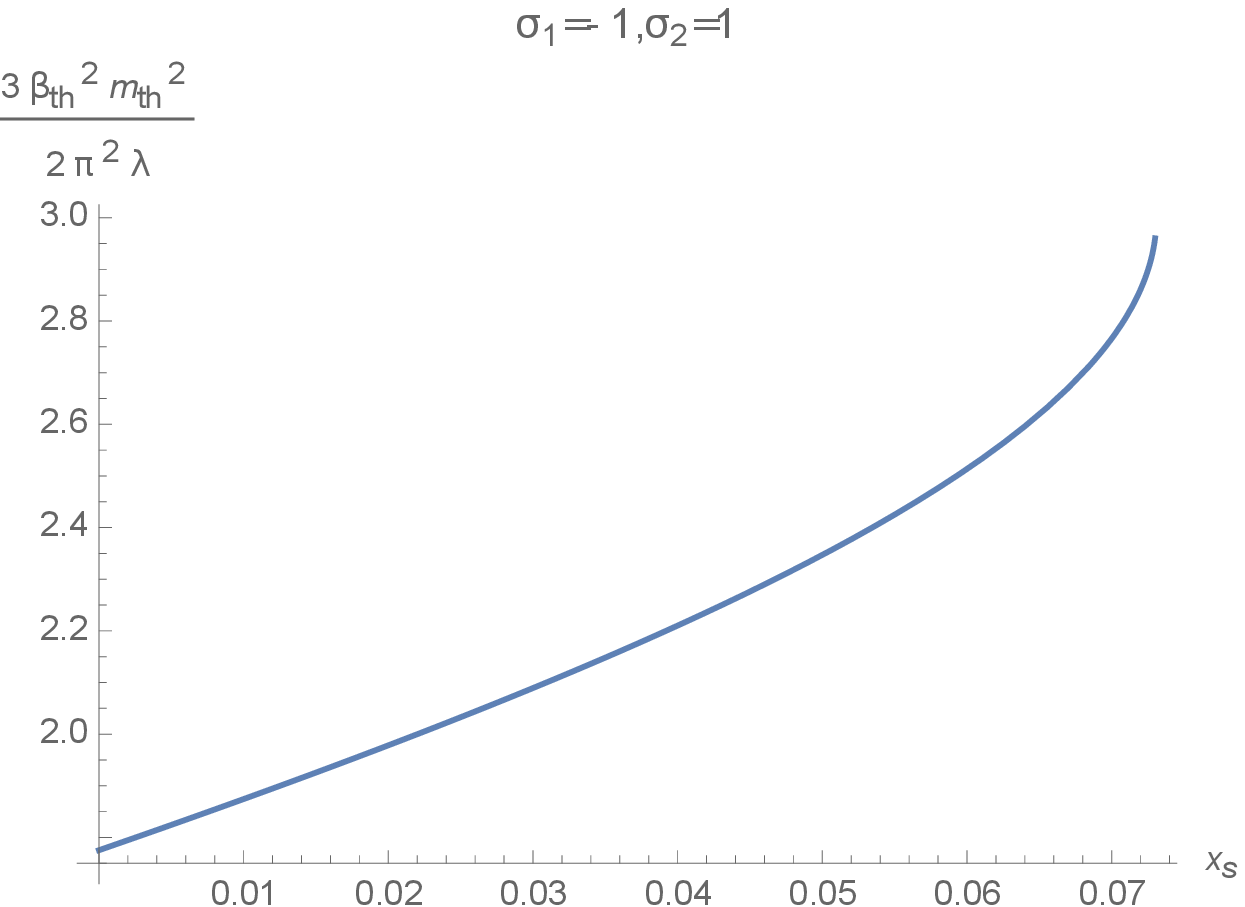}}
\end{subfigure}%
\begin{subfigure}{.5\textwidth}
  \centering
  \scalebox{0.8}{\includegraphics[width=.8\linewidth]{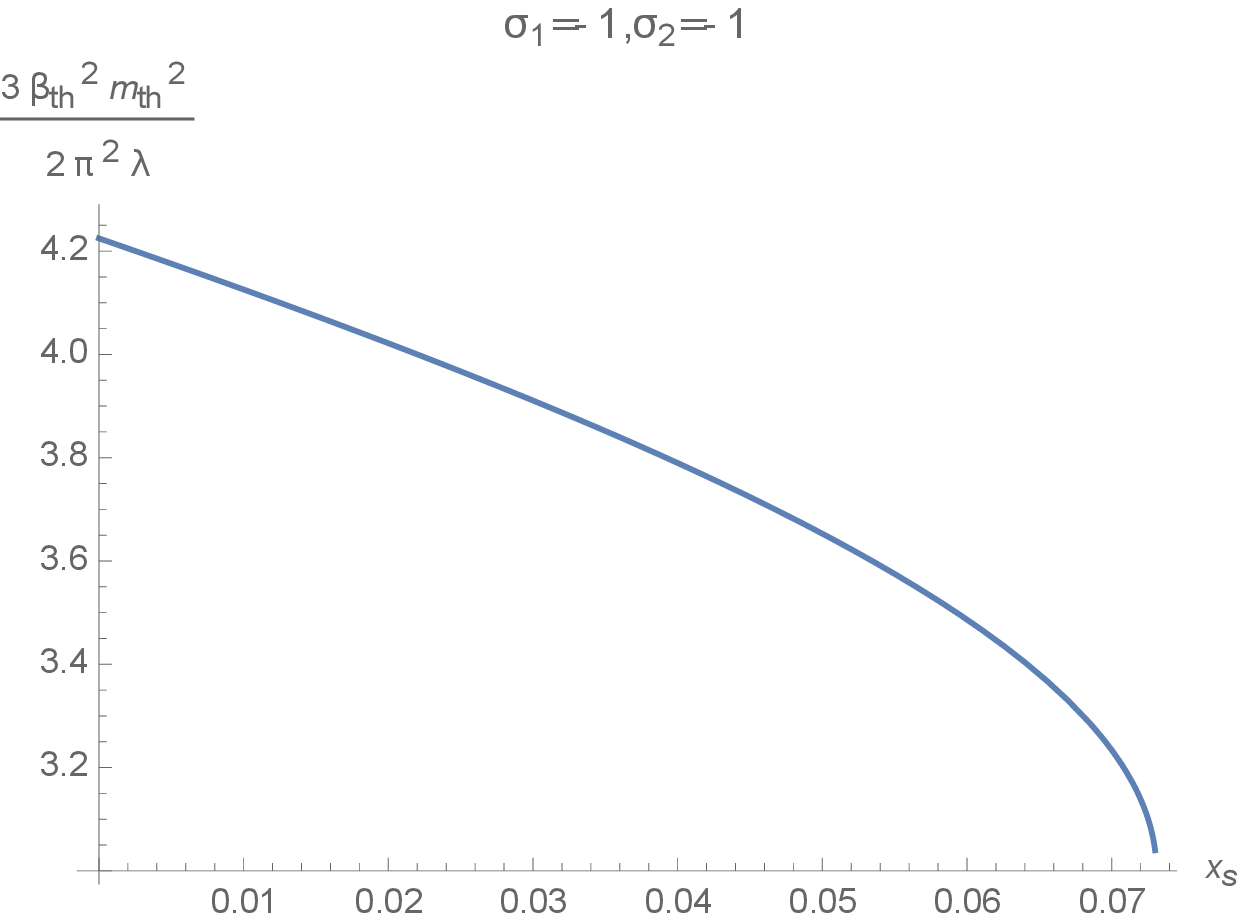}}
\end{subfigure}
\caption{Plots of $\frac{3\beta_{\text{th}}^2 m_{\text{th}}^2}{2\pi^2\lambda}$ vs. $x_s$ for the different fixed points}
\label{fig:vector qcd:thermal mass vs xs}
\end{figure}

To demonstrate this, we will look at the quadratic terms in the thermal effective potential of the scalar fields at the temperature $\frac{1}{\beta_{\text{th}}}$. 
These terms take the following form:
\beq
V_{\text{quadratic}}\Big(\frac{1}{\beta_{\text{th}}}\Big)=\frac{1}{2} \Big(\mathcal{M}^2\Big)_{ab}^{(pq)}\phi_a^{(p)}\phi_b^{(q)},
\eeq
where  $\mathcal{M}^2$ is the thermal mass matrix of the scalar fields at the temperature $\frac{1}{\beta_{\text{th}}}$. The contribution of 1-loop diagrams to this matrix was evaluated for general $(3+1)$-dimensional gauge theories with scalar fields in \cite{Weinberg:1974hy}. For the  model that we are considering currently, the form of this 1-loop thermal mass matrix is 
\begin{equation}
\begin{split}
\Big(\mathcal{M}^2\Big)_{ab}^{(pq)}
&=\frac{\beta_{\text{th}}^{-2}}{24}\Bigg[\lambda_{abcc}^{(pqrr)}+\frac{6g^2 }{2} \Big(T_{cd}^{(r)}(S)\Big)_{ae}^{(p)}\Big(T_{cd}^{(r)}(S)\Big)_{eb}^{(q)}\Bigg],
\end{split}
\end{equation}
where  $T_{cd}^{(r)}(S)$  are the generators of $SO(N_c)$ in the representation of the $r^{\text{th}}$ flavor of scalar fields. The components of these generators are 
\begin{equation}
\begin{split}
\Big(T_{cd}^{(r)}(S)\Big)_{ab}^{(p)}=-\frac{i}{2}\delta^{rp}(\delta_{ac}\delta_{bd}-\delta_{ad}\delta_{bc}).
\end{split}
\label{vector QCD: generators}
\end{equation}
Substituting the expression of the symmetric coupling $\lambda_{abcd}^{(pqrs)}$ given in \eqref{vector qcd: symm. couplings} and the  values of the generators given above, we get the following expression of the thermal mass matrix:
\begin{equation}
\begin{split}
\Big(\mathcal{M}^2\Big)_{ab}^{(pq)}
&=m_{\text{th}}^2\delta^{pq}\delta_{ab},
\end{split}
\end{equation}
where $m_{\text{th}}^2$ is the thermal mass (squared) of the scalar fields, whose value is given by
\begin{equation}
m_{\text{th}}^2=\frac{2\pi^2\beta_{\text{th}}^{-2}}{3}\Bigg[8\Big\{\Big(x_s+1+\frac{1}{N_c}\Big)h+\Big(1 +\frac{2}{N_c N_s}\Big)f\Big\}+\frac{3\lambda }{2} \Big(1-\frac{1}{N_c}\Big)\Bigg].
\end{equation}
In the $N_c, N_s\rightarrow\infty$ limit it reduces to
\begin{equation}
m_{\text{th}}^2=\frac{2\pi^2\beta_{\text{th}}^{-2}}{3}\Bigg[8(x_s+1)h+8 f+\frac{3\lambda }{2}\Bigg].
\end{equation}
If $m_{\text{th}}^2$ is positive, then the effective potential of the scalar fields must have a minimum at the origin of the field space. In that case the expectation of the scalar fields would be zero and the flavor symmetry of the scalar fields would be preserved in a thermal state. This would also mean that the $SO(N_c)$ gauge symmetry remains unbroken in any thermal state, and consequently, the system is never in the BEH phase. To determine whether this is indeed the case, we need to check whether $m_{\text{th}}^2$ is positive at the fixed points for  all $x_s\leq x_s^{\text{max}}$. For this range of $x_s$, the values of $m_{\text{th}}^2$ at the different fixed points are shown in figure \ref{fig:vector qcd:thermal mass vs xs}.

From these graphs, we can see that indeed $m_{\text{th}}^2>0$ at each of the fixed points for any value of $x_s\leq x_s^\text{max}$. Hence, we can conclude that the flavor symmetry is not spontaneously broken and the system is not in a BEH-like phase for these fixed points at nonzero temperatures.

%%%%%%%%%%%%%%%%%%%%%%%%%%%%%%%%%%%%%%%%%%%%%%%%%%%%%%%%%%%%%%%%%%%%%%%%%%%%%%%%%%%%

\subsubsection{Bifundamental scalar QCD}
In the preceding analysis of the model of QCD with fundamental scalars, we found that unitary fixed points could exist only when $N_s<x_s^{\text{max}} N_c$ where $x_s^{\text{max}} <1$. We also mentioned that in such a scenario, even if the scalar fields had nonzero thermal expectation values (which they do not), the flavor symmetry would have been preserved along with the Higgsing of some of the gauge fields.  It is then natural to wonder whether this bound on $N_s$ can be somehow lifted. Here we will show that it is indeed possible to lift the bound by first gauging the $SO(N_s)$ part of the flavor symmetry and then taking $N_s=N_c$.\footnote{We also add a set of Majorana fermions which transform in the fundamental representation of this $SO(N_c)$.} In this case, a $\mathbb Z_2$ transformation remains as a  global symmetry. Unfortunately, we will find that  this  symmetry remains unbroken in a thermal state for the Banks-Zaks-like fixed points of the model. However,  the discussion of this model will serve as a prelude to the double bifundamental scalar QCDs that will be introduced in the following subsection.  As we will see, some of the large N fixed points in such double bifundamental models indeed demonstrate symmetry breaking at nonzero temperatures along with a persistent BEH phase.

With this motivation, let us now introduce the bifundamental scalar QCD model as a generalization of the model of QCD with fundamental scalars that we discussed earlier. When the $SO(N_s)$ part of the flavor symmetry in this model is gauged and $N_s$ is set equal to $N_c$, we end up with a theory whose gauge group is $SO(N_c)\times SO(N_c)$. Now, there is an $N_c\times N_c$ matrix of massless real scalar fields which we will collectively denote by $\Phi$. The components of the matrix $\Phi$ will be denoted by $\phi_{ai}$. These scalar fields transform in the bifundamental representation of the gauge group, thus justifying the name given to the model. In addition to the scalar fields,  for each of the two $SO(N_c)$'s there are $N_f$ flavors of massless Majorana fermions which transform in the fundamental representation of that $SO(N_c)$ and are singlets under the other $SO(N_c)$. We will represent these two sets of Majorana fermions by $\psi^{(p)}_a$ and $\chi^{(p)}_i$.\footnote{As before, the superscripts denote the flavor indices and subscripts denote the color indices.} The renormalized Lagrangian of this model is given by
\begin{equation}
\begin{split}
\mathcal L_{\text{Bifund}}=&-\frac{1}{4}\sum_{\alpha=1}^2(F_{\alpha})_{\mu\nu}^A(F_{\alpha})^{\mu\nu A}+\frac{i}{2}\overline{\psi}_a^{(p)} (\slashed{D}\psi^{(p)})_a+\frac{i}{2}\overline{\chi}_i^{(p)} (\slashed{D}\chi^{(p)})_i+\frac{1}{2}\text{Tr}\Bigg[\Big(D_{\mu} \Phi \Big)^T D^{\mu} \Phi\Bigg]\\
&-\widetilde{h}\text{Tr}\Big[\Phi^T\Phi\Phi^T\Phi\Big]-\widetilde{f}\text{Tr}\Big[\Phi^T\Phi\Big]\text{Tr}\Big[\Phi^T\Phi\Big].
\end{split}
\label{Bifundamental scalar QCD: renorm. Lagrangian}
\end{equation}
Here $(F_{1})_{\mu\nu}^{ab}$ and $(F_{2})_{\mu\nu}^{ij}$ are the field strengths of the gauge fields $(V_{1})_{\mu}^{ab}$ and $(V_2)_{\mu}^{ij}$ which correspond to the two $SO(N_c)$'s. These gauge fields are anti-symmetric under the exchange of color indices, i.e.,  $(V_{1})_{\mu}^{ab}=-(V_{1})_{\mu}^{ba}$ and $(V_2)_{\mu}^{ij}=-(V_2)_{\mu}^{ji}$. Thus, while summing over the components of $(F_{\alpha})_{\mu\nu}^{ab}$, as in the first term of the above Lagrangian, one should introduce a factor of $\frac{1}{2}$ to count only the independent components. We take the gauge couplings for the two gauge fields to be the same (say, $g$)\footnote{The equality of these two couplings is guaranteed to be preserved under RG flow due to the exchange symmetry discussed below.}. The quartic terms involving the scalar fields in the potential can be expressed in terms of symmetric couplings $\lambda_{ai,bj,ck,dl}$ as follows:
\begin{equation}
\frac{\lambda_{ai,bj,ck,dl}}{4!}\phi_{ai}\phi_{bj}\phi_{ck}\phi_{dl}
=\widetilde{h}\text{Tr}\Big[\Phi^T\Phi\Phi^T\Phi\Big]+\widetilde{f}\text{Tr}\Big[\Phi^T\Phi\Big]\text{Tr}\Big[\Phi^T\Phi\Big],\\
\end{equation}
where
\begin{equation}
\begin{split}
\lambda_{a i,b j,c k,d l}\equiv&4\widetilde{h}\Bigg[\delta_{ij}\delta_{kl}(\delta_{ac}\delta_{b d}+\delta_{ad}\delta_{bc})+\delta_{ik}\delta_{jl}(\delta_{ab}\delta_{cd}+\delta_{ad}\delta_{bc})+\delta_{il}\delta_{jk}(\delta_{ab}\delta_{cd}+\delta_{ac}\delta_{bd})\Bigg]\\
&+8\widetilde{f}\Bigg[\delta_{ij}\delta_{kl}\delta_{ab}\delta_{cd}+\delta_{ik}\delta_{jl}\delta_{ac}\delta_{bd}+\delta_{il}\delta_{jk}\delta_{ad}\delta_{bc}\Bigg].
\end{split}
\label{bifundamental qcd: symm. couplings}
\end{equation}

In this model there is a $U(N_f)$ flavor symmetry for each of the two sets of fermions, $\psi$ and $\chi$. There is also a $\mathbb Z_2$ symmetry which exchanges the gauge fields corresponding to the two $SO(N_c)$'s, along with a $\psi^{(p)}\leftrightarrow \chi^{(p)}$ exchange and a transformation $\Phi\rightarrow \Phi^T$ of the scalar fields. In addition, there is another $\mathbb Z_2$ symmetry that transforms the fields as follows\footnote{The form of these transformations as given in \eqref{Z2 symmetry in bifund scalar QCD} would change under a gauge transformation. As shown in appendix \ref{app: bifund QCD baryon symmetry}, the proper way to interpret this $\mathbb Z_2$ symmetry is to think of it as an automorphism of a set of different equivalence classes of field configurations. The configurations in each class are related by gauge transformations, whereas those belonging to different classes are gauge-inequivalent. Note that the above interpretation is consistent with the existence of the gauge-invariant order parameter $\langle[\det \Phi]\rangle$ for this symmetry .
}: 
\begin{equation}
\begin{split}
&\phi_{ai}\rightarrow -\phi_{ai} \ \forall\ i\in\{1,\cdots N_c\},\\
&\psi_a^{(p)}\rightarrow -\psi_a^{(p)}\ \forall\ p\in\{1,\cdots N_f\},\\
& (V_{1})_\mu^{ab}\rightarrow -(V_{1})_\mu^{ab},\ (V_{1})_\mu^{ba}\rightarrow -(V_{1})_\mu^{ba}\ \forall\ b\neq a,\\
\end{split}
\label{Z2 symmetry in bifund scalar QCD}
\end{equation}
for a fixed value of $a$.\footnote{One could equivalently consider reflection of the elements in a column of $\Phi$ along with similar reflections of the components of $\chi^{(p)}$ and $(V_2)_\mu$. But such transformations are related to the reflection of the elements in a row by the afore-mentioned exchange of the two $SO(N_c)$'s.} In appendix \ref{app: bifund QCD baryon symmetry}, we have shown that the Lagrangian in \eqref{Bifundamental scalar QCD: renorm. Lagrangian} is indeed invariant under these transformations. Note that such transformations for the different values of $a$ are all related by global $SO(N_c)$ gauge transformations, and hence only one of them should be considered as an independent global $\mathbb Z_2$ symmetry. For specificity, we can take this symmetry to be the $\mathbb Z_2$ transformation corresponding to $a=1$. It is this $\mathbb Z_2$ symmetry whose spontaneous breaking at nonzero temperatures (or the lack of it) is of interest to us. One can consider the gauge invariant quantity $\langle [\det \Phi]\rangle$ as an order parameter for this symmetry\footnote{Note that $\det \Phi$ is a composite operator, and hence it has to be regulated and renormalized appropriately. One can work with  dimensional regularization and the $\overline{\text{MS}}$ scheme to preserve the gauge invariance of the operator. We put square brackets on the two sides of $\det \Phi$ to indicate that we are talking about such a renormalized operator.}. Since this order parameter is the expectation value of a baryonic operator (the renormalized determinant of $\Phi$), we will call this symmetry the {\it baryon symmetry}. The above-mentioned order parameter for this symmetry  is related to the determinant of the expectation values of the scalar fields as follows:
\beq
\langle [\det \Phi]\rangle=\det\langle \Phi\rangle+\text{ quantum corrections }.
\eeq
These quantum corrections are suppressed compared to the classical result $\det\langle \Phi\rangle$ due to the smallness of the couplings in the  perturbative regime. 
Now, if $\langle \Phi\rangle\neq0$, then it can be brought to the following diagonal form by an appropriate gauge transformation\footnote{The argument for this is analogous to the one presented for the real double bifundamental model in appendix \ref{app: minimum of  potential}.}:
\begin{equation*}
 \langle \Phi\rangle \propto \text{diag}\{\pm1,1,1,\cdots,1\}.
 \end{equation*} 
 As a consequence, $\det \langle\Phi\rangle\neq 0$ when $\langle\Phi\rangle\neq 0$. Hence, a nonzero  expectation value of the field $\Phi$ indicates a spontaneous breaking of the $\mathbb Z_2$ symmetry. Another important consequence of such an expectation value is that the $SO(N_c)\times SO(N_c)$ gauge symmetry would be broken down to a smaller subgroup leading to the Higgsing of some of the gauge bosons and the system existing in a BEH phase.

To explore the possibility of this dual phenomena for the fixed points in the model, let us first study the RG flow of the couplings. For this,  let us work in the $N_c\rightarrow\infty$ limit. In this limit, we take the couplings in the model to scale with $N_c$ as follows:
\begin{equation}
g^2=\frac{16\pi^2\lambda}{N_c},\ \tilde h= \frac{16\pi^2 h}{N_c},\ \tilde f= \frac{16\pi^2 f}{N_c^2}, 
\label{bifundamental QCD: 't Hooft couplings}
\end{equation}
where $\lambda$, $h$ and $f$ are the  't Hooft couplings.

The beta functions of these couplings in the planar limit are given by (see \ref{appsec:simple})
\beq \label{eq:planarbetabifund}
~& \beta_{\lam}=-{21-4x_f\ov 6} \lam^2  + {26 x_f-54\ov12}\lam^3,
\\&\beta_h=16h^2 -6h \lambda 
  +\frac{3\lambda ^2 }{16},\\
&\beta_f=8f^2 +32f h+24 h^2 -6f \lambda 
  +\frac{9 \lambda ^2 }{16}.
\eeq
Repeating the steps that were followed in case of the model of QCD with fundamental scalars, first let us set $\beta_\lambda=0$ which gives us
\beq 
\lam={21-4x_f\ov 13 x_f-27} .
\eeq
To ensure the validity of perturbation theory, we will work in the regime where  $x_f \rightarrow {21\ov 4}-\epsilon$ with $0<\epsilon\ll 1$. There are two unitary fixed points where the quartic couplings are given by 
\beq
h={3-\sqrt{6}\ov 16}\lam,\ f={2\sqrt{6} +\sigma  \sqrt{18\sqrt{6}-39}\ov 16}\lam
\eeq
with $\sigma=\pm1$.

To determine whether the $\mathbb Z_2$ symmetry is broken or not, we can again compute the 1-loop thermal mass matrix of the scalar fields at a temperature $\frac{1}{\beta_{\text{th}}}$. It is given by
\begin{equation}
\begin{split}
\Big(\mathcal{M}^2\Big)_{ai,bj}
&=\frac{\beta_{\text{th}}^{-2}}{24}\Bigg[\lambda_{ai,bj,ck,ck}+6 g^2\sum_{\beta=1}^2 \Big(T_A^{\beta}(S)\Big)_{ai,eu}\Big(T_A^{\beta}(S)\Big)_{eu,bj}\Bigg],
\end{split}
\end{equation}
where $\lambda_{ai,bj,ck,dl}$ are the symmetric couplings introduced in equations \eqref{bifundamental qcd: symm. couplings}, and $T_A^{1}(S)$ and $T_A^{2}(S)$ are the generators of the  two $SO(N_c)$'s in the representations corresponding to the scalar fields. The components of these generators are given by
\begin{equation}
\begin{split}
\Big(T_{cd}^{1}(S)\Big)_{ai,bj}=-\frac{i}{2}\delta_{ij}(\delta_{ac}\delta_{bd}-\delta_{ad}\delta_{bc}),\ \Big(T_{kl}^{2}(S)\Big)_{ai,bj}=-\frac{i}{2}\delta_{ab}(\delta_{ik}\delta_{jl}-\delta_{il}\delta_{jk}).
\end{split}
\label{bifundamental qcd: generators}
\end{equation}
Substituting the values of the symmetric couplings and the generators, we get
\begin{equation}
\begin{split}
\Big(\mathcal{M}^2\Big)_{ai,bj}=m^2_{\text{th}}\delta_{ij}\delta_{ab}\\
\end{split}
\end{equation}
where
\beq
m^2_{\text{th}}=16\pi^2 \beta_{\text{th}}^{-2}\left[{2\ov3} \Big(1+{1 \ov 2 N_c}\Big) h+{1\ov3}\Big(1+{2\ov N_{c}^2}\Big) f+{1\ov 8}\Big(1-{1\ov N_{c} }\Big)\lam \right].
\eeq
In the $N_c\rightarrow\infty$ limit, $m^2_{\text{th}}$ reduces to
\beq
m^2_{\text{th}}=16\pi^2 \beta_{\text{th}}^{-2}\left[{2\ov3} h+{1\ov3}f+{1\ov 8}\lam \right]
=\frac{\pi^2\beta_{\text{th}}^{-2}\lam}{3}\Bigg(12+\sigma\sqrt{18\sqrt{6}-39}\Bigg)\approx \frac{\pi^2\beta_{\text{th}}^{-2}\lam}{3}\Bigg(12+ 2.25 \sigma\Bigg).
\eeq
From the above expression it is clear that $m^2_{\text{th}}>0$. Consequently, the thermal expectation value of the scalar field $\Phi$ is zero for both the fixed points. Therefore,  the baryon  symmetry remains unbroken and the persistent BEH phase is absent at all temperatures for these fixed points.

\subsection{Double bifundamental models} 
\label{sec:dbm introduction}

In the previous subsection we discussed the fixed points of the RG flows of two models, viz., QCDs with additional fundamental and bifundamental scalars, where we saw that the global symmetries remain unbroken and the gauge bosons do not get Higgsed at nonzero temperatures. Now, we will consider two natural extensions\footnote{These extensions are partly motivated by the work  \cite{Aitken:2019mtq}.}  of these models which interestingly lead to fixed points in the large N limit where certain global symmetries are spontaneously broken  and the system is in a persistent BEH phase at all nonzero temperatures. The first extension will involve taking two copies of the bifundamental scalar QCD model and introducing a quartic coupling between the scalar fields in the two copies.  We will call this model the {\it real double bifundamental model}. The second extension will be to modify the  gauge group by promoting the orthogonal groups to unitary groups. Moreover, the Majorana fermions will be promoted to Dirac fermions, and the real scalar fields will be promoted to complex scalar fields. We will call this model the {\it complex double bifundamental model}. Let us now discuss these models in more detail.
\subsubsection{Introduction to the real double bifundamental model}
The bifundamental scalar QCD that we introduced in the previous subsection was symmetric under the gauge group $SO(N_{c})\times SO(N_c)$. The matter content in that model consisted of $N_f$ flavors of two sets of Majorana fermions, each being a singlet under one of the  $SO(N_c)$'s and transforming in the fundamental representation of the other, and real scalar fields which transformed in the  bifundamental representation of the gauge group. Now, let us consider two copies of this model. The  gauge group  is then
\beq
G= \prod\limits_{i=1}^2 \Big(SO(N_{ci})\times SO(N_{ci})\Big)
 \eeq
 where the two ranks $N_{c1}$ and $N_{c2}$ are possibly unequal. For each of the two sectors (labeled by $i$), the matter content is the same as before, viz., two sets of Majorana fermions (each with $N_{fi}$ flavors), $(\psi_i^{(p)})_{a_i}$ and $(\chi_i^{(p)})_{{j_i}}$, transforming in the fundamental representation of one of the $SO(N_{ci})$'s, and an $(N_{ci}\times N_{ci})$ matrix of scalar fields denoted by $\Phi_i$ which transforms in the bifundamental representation of $SO(N_{ci})\times SO(N_{ci})$. All the fields in any one sector are invariant under the gauge transformations in the other sector.  
 
\begin{figure}[!h]
\makebox[\textwidth][c]{ \scalebox{1.5}{
\begin{tikzpicture}
    \draw[thick] (-1.8,1.4) circle (14pt) (-1.8,-1.4) circle (14pt)  (1.8,1.4) circle (14pt) (1.8,-1.4) circle (14pt);
    \filldraw[blue, fill opacity=0.3] (-1.8,1.4) circle (14pt) (-1.8,-1.4) circle (14pt)  (1.8,1.4) circle (14pt) (1.8,-1.4) circle (14pt);
     \draw[thick]  (-4,1.4) circle (14pt) (-4,-1.4) circle  (14pt)  (4,1.4) circle (14pt) (4,-1.4) circle  (14pt) ;
        \filldraw[red, fill opacity=0.3]   (-4,1.4) circle (14pt) (-4,-1.4) circle  (14pt)  (4,1.4) circle (14pt) (4,-1.4) circle  (14pt) ;
    \draw[thick] (-90:14pt) ++(-1.8,1.4)  --($ (90:14pt)+(-1.8,-1.4)$); 
    \draw[thick] (-90:14pt) ++(1.8,1.4)  --($ (90:14pt)+(1.8,-1.4)$); 
 \draw[thick] (180:14pt) ++(-1.8,1.4)  --($ (0:14pt)+(-4,1.4)$); 
\draw[thick] (0:14pt) ++(1.8,1.4)  --($ (180:14pt)+(4,1.4)$); 
\draw[thick] (180:14pt) ++(-1.8,-1.4)  --($ (0:14pt)+(-4,-1.4)$); 
\draw[thick] (0:14pt) ++(1.8,-1.4)  --($ (180:14pt)+(4,-1.4)$); 

\draw[ dashed] (1,0)  --($ (-1,0)$); 

    \node at (0,0.3) {$\mathbf{\zeta}$};
    \node at (-1.8,1.4) {$\mathbf{N_{c1}}$};
    \node at (-1.8,-1.4) {$\mathbf{N_{c1}}$};
    \node at (1.8,1.4) {$\mathbf{N_{c2}}$};
    \node at (1.8,-1.4) {$\mathbf{N_{c2}}$};
     \node at (-4,1.4) {$\mathbf{N_{f1}}$};
    \node at (-4,-1.4) {$\mathbf{N_{f1}}$};
    \node at (4,1.4) {$\mathbf{N_{f2}}$};
    \node at (4,-1.4) {$\mathbf{N_{f2}}$};
    \node at (4,-1.4) {$\mathbf{N_{f2}}$};
     \node at (-1.4,0) {$\mathbf{\Phi_1}$};
      \node at (1.4,0) {$\mathbf{\Phi_2}$};
       \node at (-2.9,1.7) {$\mathbf{\psi_1}$};
        \node at (-2.9,-1.7) {$\mathbf{\chi_1}$};
           \node at (2.9,1.7) {$\mathbf{\psi_2}$};
        \node at (2.9,-1.7) {$\mathbf{\chi_2}$}; 
\end{tikzpicture}}}
\caption{A schematic diagram of the field content of the double bifundamental models.
}  \label{fig:dbf}
\end{figure}
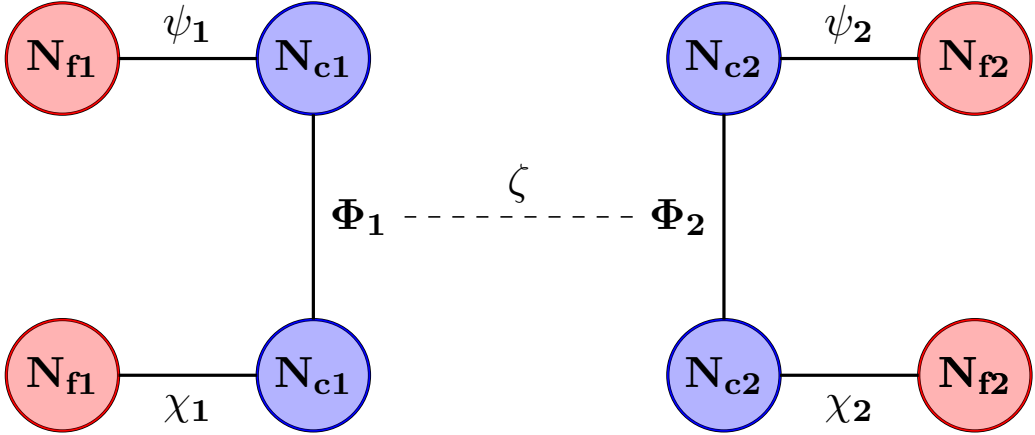

 Apart from the interactions between the matter fields within each sector that were already present in the bifundamental scalar model, let us now introduce a double trace interaction which couples the scalar fields in the two sectors.  A schematic diagram for the matter content of this theory and the interaction between the two sectors is provided in figure \ref{fig:dbf}. The renormalized Lagrangian of the model is given by
 \begin{equation}
\begin{split}
\mathcal{L}_{\text{RDB}}=&-\frac{1}{4}\sum_{i=1}^2\sum_{\alpha=1}^2(F_{i\alpha})_{\mu\nu}^A(F_{i\alpha})^{\mu\nu A}+\frac{i}{2}\sum_{i=1}^2\Big(\overline{\psi}_i^{(p)}\Big)_{a_i} \Big(\slashed{D}(\psi_i^{(p)})\Big)_{a_i}+\frac{i}{2}\sum_{i=1}^2\Big(\overline{\chi}_i^{(p)}\Big)_{j_i} \Big(\slashed{D}(\chi_i^{(p)})\Big)_{j_i}\\
&+\frac{1}{2}\sum_{i=1}^2\text{Tr}\Bigg[\Big(D_{\mu} \Phi_i \Big)^T D^{\mu} \Phi_i\Bigg]-\sum_{i=1}^2\widetilde{h}_i\text{Tr}\Big[\Phi_i^T\Phi_i\Phi_i^T\Phi_i\Big]-\sum_{i=1}^2\widetilde{f}_i\text{Tr}\Big[\Phi_i^T\Phi_i\Big]\text{Tr}\Big[\Phi_i^T\Phi_i\Big]\\
&-2\widetilde{\zeta}\text{Tr}\Big[\Phi_1^T\Phi_1\Big]\text{Tr}\Big[\Phi_2^T\Phi_2\Big].
\end{split}
\label{rdb: lagrangian}
\end{equation}
where $(F_{i\alpha})_{\mu\nu}$ is the field strength corresponding to the gauge field $(V_{i\alpha})_\mu$.

We will be studying this model in the Veneziano limit ($N_{c1},N_{c2}\rightarrow\infty$) where $\frac{N_{fi}}{N_{ci}}\rightarrow x_{fi}$, $\frac{N_{c2}}{N_{c1}}\rightarrow r$, and the different couplings scale as follows:
\begin{equation}
g_i^2=\frac{16\pi^2\lambda_i}{N_{ci}},\ \tilde h_i= \frac{16\pi^2 h_i}{N_{ci}}, \tilde f_i= \frac{16\pi^2 f_i}{N_{ci}^2}, \tilde \zeta= \frac{16\pi^2 \zeta}{N_{c1}N_{c2}}. 
\label{dbm: 't Hooft couplings}
\end{equation}
Here $g_i$ is the gauge coupling of the $i^{\text{th}}$ sector and $\lambda_i, h_i, f_i$ and $\zeta$ are the 't Hooft couplings.

The symmetries in the model are essentially the same as those for the bifundamental model, except now we have one set of symmetries for each sector. We will be interested mainly in the $\mathbb Z_2$ baryon symmetries for the two sectors. The action of the baryon symmetry in the  $i^{\text{th}}$ sector on the fields in that sector is as follows\footnote{Just as in case of the bifundamental scalar QCD, this $\mathbb Z_2$ symmetry can be interpreted as an automorphism of a set of  classes of gauge-equivalent configurations.}: 
\begin{equation}
\begin{split} 
&\Phi_i\rightarrow \mathcal{T}_i\Phi_i,\ \\
&\psi_i^{(p)}\rightarrow \mathcal{T}_i \psi_i^{(p)}\ \forall\ p\in\{1,\cdots, N_{fi}\} ,\\
& (V_{i1})_\mu\rightarrow \mathcal{T}_i (V_{i1})_\mu \mathcal{T}_i^{-1},
\end{split}
\label{Z2 baryon symmetry in RDB model}
\end{equation}
where $\mathcal{T}_i$ is an $N_{ci}\times N_{ci}$ diagonal matrix of the following form:
\beq
\mathcal{T}_i\equiv\text{diag}\{-1,1,\cdots,1\}.
\eeq
Here, the transformation has a nontrivial action only on the first row of the scalar field $\Phi_i$, the first component of the fermion $\psi_i^{(p)}$ for each flavor $p$, and the first row and first column of the gauge field $(V_{i1})_{\mu}$. For each individual sector, this is just the transformation given in \eqref{Z2 symmetry in bifund scalar QCD} for the bifundamental scalar QCD with $a=1$.  As earlier, we can choose $\langle[\det\Phi_i]\rangle$ to be an order parameter for this  symmetry. A nonzero thermal expectation value of $\Phi_i$ would imply that this order parameter is nonzero and consequently, the baryon symmetry is broken in a thermal state. As earlier, this would also mean that some of the gauge bosons in the $i^{\text{th}}$ sector are Higgsed leading to the system being in a persistent BEH phase.

In section \ref{sec:rdb model}, we will undertake a study of the fixed points in this model  to see whether these dual phenomena actually occur in a thermal state. There we will find that this model has two very interesting properties in the Veneziano limit:
\begin{enumerate}
\item Considering up to two loop contributions to the planar beta functions, there is a conformal manifold. It is topologically a circle in the space of the double trace couplings. Later in section \ref{sec: survival of fixed circle}, we will show that this conformal manifold survives even when all higher loop corrections to the planar beta functions are taken into account. 
\item When the ratio of the ranks $N_{c1}$ and $N_{c2}$ is sufficiently away from $1$, the baryon symmetry corresponding to the sector with the smaller rank is spontaneously broken in a thermal state for a subset of points on this conformal manifold. Moreover, at these fixed points, the gauge symmetry in this sector is broken down from $SO(N_{ci})\times SO(N_{ci})$ to just $SO(N_{ci})$ in any thermal state. This leads to Higgsing of half of the gauge bosons in this sector, and thus the system is in a persistent BEH phase.
\end{enumerate}

We will defer further discussion of these features to section \ref{sec:rdb model}, and next introduce a closely related model which shares many of the properties mentioned above.

\subsubsection{Introduction to the complex double bifundamental model}
In the preceding introduction to the real double bifundamental model we saw that the gauge group in that model has the following structure:
\beq
G=\sum_{i=1}^2 G_i\times G_i,
\eeq
where $G_i=SO(N_{ci})$. In addition, there is a discrete ($\mathbb Z_2$) global symmetry for each sector which we claimed to be spontaneously broken in a thermal state at a subset of fixed points. This should naturally lead to the following question: {\it Is there any model where a continuous global symmetry is similary broken at arbitrarily high temperatures?} 

To address this question, we will now extend the above-mentioned model by taking the  group $G_i$ to be the special unitary group $SU(N_{ci})$. We will further promote the Majorana fermions to Dirac fermions transforming in the fundamental representation of $SU(N_{ci})$, and the real scalar fields to complex scalar fields transforming in the bifundamtal representation of $SU(N_{ci})\times SU(N_{ci})$.\footnote{In this representation, the scalar fields tranform as $\Phi_i\rightarrow U_i\Phi_i W_i^\dag$ where $U_i\times W_i\in SU(N_{ci})\times SU(N_{ci})$.} The schematic diagram for the matter content in this model is still the one shown in figure \ref{fig:dbf}. The  renormalized Lagrangian of the model is given by
 \begin{equation}
\begin{split}
\mathcal{L}_{\text{CDB}}=&-\frac{1}{4}\sum_{i=1}^2\sum_{\alpha=1}^2(F_{i\alpha})_{\mu\nu}^A(F_{i\alpha})^{\mu\nu A}+i\sum_{i=1}^2\Big(\overline{\psi}_i^{(p)}\Big)_{a_i} \Big(\slashed{D}(\psi_i^{(p)})\Big)_{a_i}+i\sum_{i=1}^2\Big(\overline{\chi}_i^{(p)}\Big)_{j_i} \Big(\slashed{D}(\chi_i^{(p)})\Big)_{j_i}\\
&+\sum_{i=1}^2\text{Tr}\Bigg[\Big(D_{\mu} \Phi_i \Big)^\dag D^{\mu} \Phi_i\Bigg]-\sum_{i=1}^2\widetilde{h}_i\text{Tr}\Big[\Phi_i^\dag\Phi_i\Phi_i^\dag\Phi_i\Big]-\sum_{i=1}^2\widetilde{f}_i\text{Tr}\Big[\Phi_i^\dag\Phi_i\Big]\text{Tr}\Big[\Phi_i^\dag\Phi_i\Big]\\
&-2\widetilde{\zeta}\text{Tr}\Big[\Phi_1^\dag\Phi_1\Big]\text{Tr}\Big[\Phi_2^\dag\Phi_2\Big].
\end{split}
\label{cdb: lagrangian}
\end{equation}
The components $(F_{i\alpha})_{\mu\nu}^A$ are the coefficients of the  generators of $SU(N_{ci})$ in the expansion of the field strength $(F_{i\alpha})_{\mu\nu}$ corresponding to the gauge fields  $(V_{i\alpha})_{\mu}$. As before, these generators are normalised such that the corresponding second Dynkin index is $\frac{1}{2}$.

Analogous to the previous model, we take the gauge couplings for the two $SU(N_{ci})$'s in the $i^{\text{th}}$ sector to be the same (say, $g_i$). Therefore, there is a $\mathbb Z_2$ symmetry for each sector (labeled by $i$) which exchanges the two sets of gauge fields in that sector, along with the exchange of the $\psi_i^{(p)}$ and $\chi_i^{(p)}$ fields, and the transformation $\Phi_i\rightarrow \Phi_i^\dag$ of the scalar fields. In addition, due to the absence of  mass terms of the fermions and  Yukawa interactions coupling the scalars to the fermions, there is  a chiral symmetry of this Lagrangian. This chiral symmetry leads to an $SU(N_{f_i})_L\times SU(N_{f_i})_R \times U(1)_V$ flavor symmetry for each kind of fermion in the $i^{\text{th}}$ sector. Moreover, there is a symmetry under charge conjugation of each of these fermions. But the symmetries that we will mainly be interested in are two $U(1)$ baryon symmetries, one for each sector, which transform the  fields in the respective sectors as follows \footnote{Let us note here that the transformations given in \eqref{U1 baryon symmetry in CDB model} can be reduced to the following simpler form by an appropriate global gauge transformation:
\begin{equation}
\Phi_i\rightarrow e^{i\theta/N_{ci}}\Phi_i,\
 \psi_i^{(p)}\rightarrow e^{i\theta/N_{ci}} \psi_i^{(p)} \ \forall\ p\in\{1,\cdots, N_{fi}\}.
\end{equation}
Furthermore, the above transformation of the fermionic fields can be undone by a corresponding $U(1)_V$ transformation. This would leave us with a $U(1)/N_{ci}$ transformation acting just on the scalar fields. We choose to work with the transformation given in \eqref{U1 baryon symmetry in CDB model} rather than this simpler transformation to keep the $U(1)$ symmetry manifest in the $N_{ci}\rightarrow\infty$ limit. As earlier, this $U(1)$ symmetry can also be interpreted as an automorphism of the set of classes of gauge-equivalent field configurations.}:
\begin{equation}
\begin{split} 
&\Phi_i\rightarrow (\mathcal{T}_i)_\theta\Phi_i,\ \\
&\psi_i^{(p)}\rightarrow (\mathcal{T}_i)_\theta \psi_i^{(p)}\ \forall\ p\in\{1,\cdots, N_{fi}\} ,\\
& (V_{i1})_\mu\rightarrow(\mathcal{T}_i)_\theta(V_{i1})_\mu (\mathcal{T}_i)_\theta^{-1},
\end{split}
\label{U1 baryon symmetry in CDB model}
\end{equation}
where, for each $\theta\in[0,2\pi)$, $(\mathcal{T}_i)_\theta$ is an $N_{ci}\times N_{ci}$ diagonal matrix of the following form:
\beq
(\mathcal{T}_i)_\theta\equiv\text{diag}\{e^{i\theta},1,\cdots,1\}.
\eeq 
This transformation leaves the remaining fields in the $i^{\text{th}}$ sector, as well as all the fields in the other sector, unchanged. We can again take $\langle[\det\Phi_i]\rangle$ to be an order parameter for the baryon symmetry in the $i^{\text{th}}$ sector. As before, a nonzero  expectation   value of $\Phi_i$ in a thermal states implies a nonzero value of this order parameter resulting in the spontaneous breaking of the corresponding baryon symmetry and a persistent BEH phase. 

To study the possibility of the above phenomena, we will look at the fixed points of this model in the Veneziano limit. As mentioned earlier, in this limit $N_{c1},N_{c2}\rightarrow\infty$, $\frac{N_{fi}}{N_{ci}}\rightarrow x_{fi}$ and $\frac{N_{c2}}{N_{c1}}\rightarrow r$, while the quartic couplings scale exactly as given in \eqref{dbm: 't Hooft couplings}.

The analysis of this model will be simplified by the fact that it is dual to the real double bifundamental model in the Veneziano limit. We will discuss this planar equivalence  between the  two models in section \ref{subsec:planar duality}. As a consequence of this equivalence, there will be a conformal manifold (a fixed circle) for this model in the planar limit. Moreover, when the two ranks $N_{c1}$ and $N_{c2}$ are sufficiently different, we will see that the  baryon symmetry corresponding to the smaller sector is spontaneously broken in a thermal state for some of the fixed points on the manifold. For these fixed points, the gauge symmetry in the smaller sector is broken down from $SU(N_{ci})\times SU(N_{ci})$ to $SU(N_{ci})$ in any thermal state, which leads to the Higgsing of half of the gauge bosons in this sector. Thus the system remains in a persistent BEH phase at all temperatures for these fixed points. In section \ref{sec: cdb model}, we will discuss these features of the complex double bifundamental model in detail.

\section{Real double bifundamental model}
\label{sec:rdb model}
In this section we will restrict our attention to the real double bifundamental model. First, in the following subsection, we will determine the weakly coupled fixed points in this model by employing perturbation theory in the planar (Veneziano) limit that was mentioned in the previous section. We will find that in this limit, the beta functions of the different couplings yield nontrivial fixed points which are analogous to the Banks-Zaks fixed points in more familiar QCDs. 

We will show that the planar 2-loop beta functions of the gauge couplings do not depend on the quartic 't Hooft couplings. Hence, the values of the gauge couplings at the fixed points can be determined independently.  We will work in a regime where these gauge couplings are small. For this, we will see that the ratios $x_{fi}\equiv \frac{N_{fi}}{N_{ci}}$ need to be tuned as follows:
\beq
x_{fi}\rightarrow \frac{21}{4}-\epsilon_i,
\eeq
where $0<\epsilon_i\ll 1$. Choosing the gauge couplings to be small, in turn, will allow us to determine the fixed points of the beta functions of the quartic  couplings perturbatively. By studying the 1-loop contributions to these beta functions, we will find that instead of a discrete set of fixed points, there is a conformal manifold (topologically, a circle) in the space of couplings. Moreover, we will demonstrate that this degeneracy in the fixed points survives even after the contributions of 2-loop diagrams to the beta functions are taken into account. 

Later, in subsection \ref{subsec: rdb symmetry breaking analysis},  we will analyze the thermal effective potential of the scalar fields at the above-mentioned large N fixed points. We will demonstrate that this effective potential is stable at all points on the fixed circle. We will also show that when the ratio $r\equiv\frac{N_{c2}}{N_{c1}}$ is below a certain bound, some of the points on the fixed circle demonstrate spontaneous breaking of the baryon symmetry  in the second sector at all nonzero temperatures. Along with this, we will see that at these fixed points, the gauge symmetry in the second sector gets broken down to $SO(N_{c2})$ in any thermal state, and the system exists in a persistent BEH phase.

\subsection{Fixed points in the planar limit}
\label{subsec rdb: planar fixed points}
The beta functions of the couplings in the real double bifundamental model are determined up to the contributions of 2-loop diagrams in the appendix \ref{app: rdb beta fns.}. Here, we provide their expressions at the leading order in the planar limit. 
\begin{itemize}
\item Beta functions of the gauge couplings (up to 2-loops):
\begin{equation}
\begin{split}
&\beta_{\lambda_i}=-\Big(\frac{21-4 x_{fi}}{6}\Big)\lambda_i^2+\Big(\frac{-27+13 x_{fi}}{6}\Big)\lambda_i^3\  .\\
\end{split}
\label{rdb: gauge coupling planar beta fn}
\end{equation}

\item 1-loop beta functions of the quartic couplings:
\begin{equation}
\begin{split}
&\beta_{h_i}^{\text{1-loop}}=16 h_i^2-6 h_i\lambda_i+\frac{3}{16}\lambda_i^2\ ,\\
&\beta_{f_i}^{\text{1-loop}}=8 f_i^2+32 f_ih_i-6 f_i\lambda_i+24h_i^2+\frac{9}{16}\lambda_i^2+8\zeta^2\ ,\\
&\beta_{\zeta}^{\text{1-loop}}=\zeta\Bigg[8 f_1+8 f_2+16 h_1+16 h_2-3\lambda_1-3\lambda_2\Bigg]\ .
\end{split}
\label{rdb: quartic coupling planar 1-loop beta fn}
\end{equation}

\item 2-loop corrections to the beta functions of the quartic couplings:
\begin{equation}
\begin{split}
&\beta_{h_i}^{\text{2-loop}}=-96 h_i^3+40\lambda_i h_i^2+\Big(\frac{5 x_{fi}}{3}-\frac{13}{2}\Big)\lambda_i^2h_i+\Big(\frac{1}{4}-\frac{x_{fi}}{6}\Big)\lambda_i^3\ ,\\
&\beta_{f_i}^{\text{2-loop}}=-384 h_i^3-160 h_i^2 f_i+96\lambda_i h_i^2+128\lambda_i h_i f_i+32\lambda_i f_i^2+32\lambda_{i^\prime} \zeta^2\\
&\qquad\qquad+9\lambda_i^2 h_i+\Big(\frac{5 x_{fi}}{3}-\frac{7}{2}\Big)\lambda_i^2 f_i+\frac{1}{4}\Big(3-2x_{fi}\Big)\lambda_i^3\ ,\\
&\beta_{\zeta}^{\text{2-loop}}=\zeta\Bigg[-80\Big(h_1^2+h_2^2\Big)+32\lambda_1\Big(f_1+2h_1\Big)+32\lambda_2\Big(f_2+2h_2\Big)\\
&\qquad\qquad\quad+\lambda_1^2\Big(\frac{-21+10x_{f1}}{12}\Big)+\lambda_2^2\Big(\frac{-21+10x_{f2}}{12}\Big)\Bigg]\ .
\end{split}
\label{rdb: quartic coupling planar 2-loop beta fn}
\end{equation}

Note that in the expression of $\beta_{f_i}^{\text{2-loop}}$ given above, one of the terms has the coupling $\lambda_{i^\prime}$. Here, $i^\prime$ denotes the complement of $i$, i.e., $i^\prime=2$ when $i=1$, and $i^\prime=1$ when $i=2$.

\end{itemize}

\subsubsection{Fixed points of the 2-loop beta functions of the gauge couplings and 1-loop beta functions of the quartic couplings}
We will now determine the fixed points of the RG flow of the couplings where the two sectors are coupled to each other, i.e., $\zeta\neq0$. To determine these fixed points, let us first set $\beta_{\lambda_i}=0 $. A nontrivial solution of this equation is 
\begin{equation}
\begin{split}
\lambda_i=\frac{21-4x_{fi}}{-27+13x_{fi}}.
\end{split}
\end{equation}
As mentioned earlier, we can see that gauge couplings at the fixed point become small while remaining positive when $x_{f1}$ and $x_{f2}$ approach  the value $\frac{21}{4}$ from below.

Next, we set $\beta_{h_i}^{\text{1-loop}}=0$ resulting in
\begin{equation}
\begin{split}
& h_i=\Big(\frac{3\pm\sqrt{6}}{16}\Big)\lambda_i.
\end{split}
\end{equation}

Now, for convenience, let us define the following combinations of the quartic couplings: 
\begin{equation}
\begin{split}
f_{p,m}\equiv\frac{f_1\pm f_2}{2},\ h_{p,m}\equiv\frac{h_1\pm h_2}{2}\ .
\end{split}
\end{equation}
In terms of these couplings, the 1-loop beta functions are as follows
\begin{equation}
\begin{split}
\beta_{f_p}^{\text{1-loop}}=&f_m\Big[32 h_m-3(\lambda_1-\lambda_2)\Big]+8 \Big[\zeta^2+(f_p+ h_p)(f_p+ 3h_p)\Big]+ 8f_m^2-3 f_p(\lambda_1+\lambda_2)\\
&+24 h_m^2+\frac{9}{32}(\lambda_1^2+\lambda_2^2),\\
\beta_{f_m}^{\text{1-loop}}=& f_m\Big[16 f_p+32h_p-3(\lambda_1+\lambda_2)\Big] +32 f_p h_m-3 f_p(\lambda_1-\lambda_2)+48 h_p h_m+\frac{9}{32}(\lambda_1^2-\lambda_2^2),\\
\beta_{\zeta}^{\text{1-loop}}=&\zeta\Big[16 f_p+32h_p-3(\lambda_1+\lambda_2)\Big].
\end{split}
\end{equation}
In appendix \ref{app: fixed point constraints}, we show that a unitary fixed point of the above beta functions with $\zeta\neq0$ can exist only if $x_{f1}=x_{f2}\equiv x_f$ which leads to the  equality of the two gauge couplings:
\begin{equation}
\begin{split}
\lambda_1=\lambda_2=\lambda=\frac{21-4x_{f}}{-27+13x_{f}}.
\end{split}
\end{equation}
Moreover, as proved in the same appendix, at such a unitary fixed point, we must have 
\begin{equation}
\begin{split}
h_1=h_2=h=\Big(\frac{3-\sqrt{6}}{16}\Big)\lambda,
\end{split}
\end{equation}
or equivalently,
\begin{equation}
\begin{split}
h_p=\Big(\frac{3-\sqrt{6}}{16}\Big)\lambda,\ h_m=0.
\end{split}
\end{equation}
Plugging these solutions into the beta functions we get the following simpler expressions:
\begin{equation}
\begin{split}
\beta_{f_p}^{\text{1-loop}}=&8 \Big[\zeta^2+(f_p+ h_p)(f_p+ 3h_p)\Big]+ 8f_m^2-6 f_p \lambda+\frac{9}{16}\lambda^2,\\
\beta_{f_m}^{\text{1-loop}}=& f_m\Big[16 f_p+32h_p-6\lambda\Big], \\
\beta_{\zeta}^{\text{1-loop}}=&\zeta\Big[16 f_p+32h_p-6\lambda\Big].
\end{split}
\end{equation}
Now it turns out that for the fixed points with $\zeta\neq0$, $\beta_{f_m}^{\text{1-loop}}=\beta_{\zeta}^{\text{1-loop}}=0$ degenerates into a single equation:
\begin{equation}
\begin{split}
16 f_p+32h_p-6\lambda=0\implies f_p=\frac{\sqrt{6}}{8}\lambda.
\end{split}
\end{equation}
Here we have substituted $h_p$ by its value  given above to obtain $f_p$ in terms of $\lambda$. Substituting these values of $f_p$ and $h_p$ in $\beta_{f_p}^{\text{1-loop}}$, we get 
\begin{equation}
\begin{split}
\beta_{f_p}^{\text{1-loop}}
=& 8(\zeta^2+f_m^2)+ \Big(\frac{39-18\sqrt{6}}{32}\Big)\lambda^2\ .\\
\end{split}
\end{equation}
Finally setting $\beta_{f_p}^{\text{1-loop}}=0$, we get
 \begin{equation}
\begin{split}
 8(\zeta^2+f_m^2)= \Big(\frac{18\sqrt{6}-39}{32}\Big)\lambda^2\ .\\
\end{split}
\end{equation}
Therefore, we see that there is a circle of fixed points for the 1-loop beta functions of the quartic couplings. In the following subsection, we will show that this fixed circle survives even when the 2-loop contributions to the beta functions of the quartic couplings are taken into account.

\subsubsection{2-loop corrections to the fixed points of the quartic couplings}

To analyze the corrections to the fixed points after including the 2-loop contributions to the beta functions of the quartic couplings, we split these couplings as follows:
\begin{equation}
\begin{split}
&h_i=h_0+\delta h_i,\\
&f_i=f_{0i}+\delta f_i,\\
&\zeta=\zeta_0+\delta \zeta,
\end{split}
\end{equation}
where $h_0, f_{0i}$ and $\zeta_0$ are solutions to the fixed point criteria arising from the 1-loop beta functions. These quantites are $O(\lambda)$, and as derived in the previous subsection, they satisfy the following conditions:
\begin{equation}
\begin{split}
h_0=\Big(\frac{3-\sqrt{6}}{16}\Big)\lambda,\ f_{01}+f_{02}=\frac{\sqrt{6}}{4}\lambda,\ 
 8\Big(\zeta_0^2+\frac{1}{4}(f_{01}-f_{02})^2\Big)=\Big(\frac{18\sqrt{6}-39}{32}\Big)\lambda^2.
\end{split}
\end{equation}
The quantities $\delta h_i,\delta f_i$ and $\delta \zeta$ are corrections to these fixed point values due to 2-loop contributions. They are $O(\lambda^2)$. 

Now, retaining terms up to $O(\lambda^3)$ in the 1-loop beta functions, we have 
\begin{equation}
\begin{split}
&\beta_{h_i}^{\text{1-loop}}=\beta_{h_{0}}^{\text{1-loop}}+32 h_{0}\delta h_i-6 \lambda\delta h_i,\\
&\beta_{f_{i}}^{\text{1-loop}}=\beta_{f_{0i}}^{\text{1-loop}}+16 f_{0i}\delta f_i+32 f_{0i}\delta h_i+32 h_0\delta f_i-6 \lambda \delta f_i+48 h_{0}\delta h_i+16\zeta_0 \delta \zeta,\\
&\beta_{\zeta}^{\text{1-loop}}=\beta_{\zeta_0}^{\text{1-loop}}+\Bigg[8 f_{01}+8 f_{02}+32 h_{0}-6\lambda\Bigg]\delta \zeta+\zeta_0\Bigg[8 \delta f_1+8 \delta f_2+16 \delta h_1+16 \delta h_2\Bigg].
\end{split}
\end{equation}
The quantities $\beta_{h_{0}}^{\text{1-loop}}$, $\beta_{f_{0i}}^{\text{1-loop}}$ and $\beta_{\zeta_0}^{\text{1-loop}}$ are the 1-loop beta functions with the couplings  $h_{0}$, $f_{0i}$ and $\zeta_0$ . They vanish because $h_{0}$, $f_{0i}$ and $\zeta_0$ constitute a fixed point of the 1-loop beta functions.

Now, demanding that $\delta h_i$, $\delta f_i$ and $\delta \zeta$ correspond to the fixed point of the overall beta functions (including 2-loop corrections) up to $O(\lambda^3)$, we get the following equations:
\begin{equation}
\begin{split}
&32 h_{0}\delta h_i-6 \lambda\delta h_i=96 h_{0}^3-40\lambda h_{0}^2-\Big(\frac{5 x_{f}}{3}-\frac{13}{2}\Big)\lambda^2h_{0}-\Big(\frac{1}{4}-\frac{x_{f}}{6}\Big)\lambda^3,\\
\end{split}
\label{eqn: delta hi}
\end{equation}
\begin{equation}
\begin{split}
&16 f_{0i}\delta f_i+32 f_{0i}\delta h_i+32 h_{0}\delta f_i-6 \lambda \delta f_i+48 h_{0}\delta h_i+16\zeta_0 \delta \zeta\\
&=384 h_{0}^3+160 h_{0}^2 f_{0i}-96\lambda h_{0}^2-128\lambda h_{0} f_{0i}-32\lambda f_{0i}^2-32\lambda \zeta_0^2\\
&\quad-9\lambda^2 h_{0}-\Big(\frac{5 x_{f}}{3}-\frac{7}{2}\Big)\lambda^2 f_{0i}-\frac{1}{4}\Big(3-2x_{f}\Big)\lambda^3,\\
\end{split}
\label{eqn: delta fi}
\end{equation}
\begin{equation}
\begin{split}
&\Bigg[8 f_{01}+8 f_{02}+32 h_{0}-6\lambda\Bigg]\delta \zeta+\zeta_0\Bigg[8 \delta f_1+8 \delta f_2+16 \delta h_1+16 \delta h_2\Bigg]\\
&=\zeta_0\Bigg[160 h_{0}^2-128\lambda h_0-32\lambda\Big(f_{01}+f_{02}\Big)-\lambda^2\Big(\frac{-21+10x_{f}}{6}\Big)\Bigg].\\
\end{split}
\label{eqn: delta zeta}
\end{equation}
From \eqref{eqn: delta hi}, we get
\begin{equation}
\begin{split}
\delta h_1=\delta h_2=\delta h\equiv \lambda^2\Bigg[\frac{93\sqrt{6}-201+8(7-5\sqrt{6})x_f}{768\sqrt{6}}\Bigg].
\end{split}
\end{equation}

To analyze the solutions of equations \eqref{eqn: delta fi} and \eqref{eqn: delta zeta}, let us introduce the following combinations of the couplings:
\begin{equation}
\begin{split}
&f_{0p}\equiv\frac{1}{2}(f_{01}+f_{02}),\ f_{0m}\equiv\frac{1}{2}(f_{01}-f_{02}),\\
&\delta f_{p}\equiv\frac{1}{2}(\delta f_{1}+\delta f_{2}),\ \delta f_{m}\equiv\frac{1}{2}(\delta f_{1}-\delta f_{2}).\\
\end{split}
\end{equation}

After substituting $h_0,\ f_{0p},\ \delta h$ and $(\zeta_0^2+f_{0m}^2)$ by their values,  equations \eqref{eqn: delta fi} and \eqref{eqn: delta zeta} lead to the following equations:  
\begin{equation}
\begin{split}
&16\Big(\zeta_0\delta \zeta+ f_{0m}\delta f_m\Big)=-\lambda^3\Bigg[\frac{-5784+5607\sqrt{6}+8(51\sqrt{6}-172)x_f}{1536}\Bigg],\\
\end{split}
\label{2-loop correction to radius}
\end{equation}
\begin{equation}
\begin{split}
&8f_{0m }\delta f_p=-\lambda^2f_{0m}\Bigg[\frac{2160+339\sqrt{6}+56\sqrt{6}x_f}{288}\Bigg],\\
\end{split}
\end{equation}
\begin{equation}
\begin{split}
&8\zeta_0\delta f_p=-\lambda^2\zeta_0\Bigg[\frac{2160+339\sqrt{6}+56\sqrt{6}x_f}{288}\Bigg],\\
\end{split}
\end{equation}
Note that the last two equations can be satisfied by setting
\begin{equation}
\begin{split}
\delta f_p=-\lambda^2\Bigg[\frac{2160+339\sqrt{6}+56\sqrt{6}x_f}{2304}\Bigg].\\
\end{split}
\end{equation}
Therefore, we see that the  2-loop contributions to the beta functions do not impose any additional constraint on $\zeta_0$ and $f_{0m}$. From this, we can conclude that the fixed circle survives even after accounting for the 2-loop terms in the  planar beta functions. Moreover, note that the LHS of \eqref{2-loop correction to radius} is proportional to $\delta(\zeta^2+f_m^2)$. This means that this equation gives the 2-loop correction to the radius of the fixed circle. In section \ref{sec: survival of fixed circle}, we will prove that this fixed circle actually survives under all higher loop corrections in the planar limit.

%%%%%%%%%%%%%%%%%%%%%%%%%%%%%%%%%%%%%%%%%%%%%%%%%%%%%%%%%%%%%%%%%%%%%%%%

\subsection{Analysis of symmetry breaking at finite temperature}
\label{subsec: rdb symmetry breaking analysis}
In this subsection, we will analyze the pattern of symmetry breaking at finite temperature for the large N fixed points in our model. We will begin by demonstrating that the effective potential at any finite temperature is stable for all points on the fixed circle (at least up to leading order in $\lambda$). Then we will compute the thermal masses of the fields $\Phi_1$ and $\Phi_2$ and identify the conditions under which some of the fixed points demonstrate the dual phenomena of spontaneous breaking of a baryon symmetry and a persistent BEH phase at nonzero temperatures.
In what follows, we will  truncate the values of the different couplings on the fixed circle to their leading order. 

\subsubsection{Stability of the effective potential}
\label{subsubsec rdb:potential stability}
The leading order terms in the thermal effective potential of the fields $\Phi_1$ and $\Phi_2$ are $O(\lambda)$. At this order, there are only terms which are quadratic and quartic in the fields. The quadratic terms lead to thermal masses of these fields which will be discussed in the following subsection. Here we will consider only the quartic terms which would determine whether the potential is stable (i.e., bounded from below) or not.\footnote{The coefficients of these quartic terms in the classical potential are $O(\lambda)$. The quantum corrections to this classical potential are suppressed by higher powers of $\lambda$.} These  terms are of the following form:
\begin{equation}
\begin{split}
V_{\text{quartic}}=&\widetilde{h}_1\text{Tr}\Big[\rho_1^T\rho_1\Big]+\widetilde{h}_2\text{Tr}\Big[\rho_2^T\rho_2\Big]+\widetilde{f}_1\Big(\text{Tr}\Big[\rho_1\Big]\Big)^2+\widetilde{f}_2\Big(\text{Tr}\Big[\rho_2\Big]\Big)^2+2\widetilde{\zeta}\text{Tr}\Big[\rho_1\Big]\text{Tr}\Big[\rho_2\Big],
\end{split}
\label{potential:quartic terms}
\end{equation}
where $\rho_1$ and $\rho_2$ are symmetric matrices defined as  follows:
\begin{equation}
\begin{split}
\rho_1\equiv \Phi_ 1^T\Phi_ 1,\ \rho_2\equiv \Phi_ 2^T\Phi_ 2.
\end{split}
\end{equation}
The first two terms in \eqref{potential:quartic terms} are manifestly non-negative if $\widetilde{h}_1>0$ and $\widetilde{h}_2>0$. A sufficient condition for the remaining terms to be non-negative is 
\begin{equation}
\begin{split}
\widetilde{f}_1>0,\ \widetilde{f_2}>0,\ \text{ and }\ \widetilde{f}_1\widetilde{f}_2>\widetilde{\zeta}^2. 
\end{split}
\end{equation}
After rescaling the couplings by appropriate powers of $N_{c1}$ and $N_{c2}$, these conditions can be summarized as follows:
\begin{equation}
\begin{split}
h_1> 0,\ h_2> 0,\ f_1> 0,\ f_2> 0, \ (f_1 f_2-\zeta^2)>0.
\end{split}
\label{positivity of different couplings}
\end{equation}
Let us now check whether these conditions are satisfied for the points on the fixed circle. At these points, we have
\begin{equation}
\begin{split}
h_1=h_2=h= \Big(\frac{3-\sqrt{6}}{16}\Big)\lambda,\ f_p\equiv\frac{1}{2}(f_1+f_2)=\frac{\sqrt{6}}{8}\lambda,
\end{split}
\end{equation}
and 
 \begin{equation}
\begin{split}
 \zeta^2+f_m^2= \Big(\frac{18\sqrt{6}-39}{256}\Big)\lambda^2,\\
\end{split}
\label{eqn:fixed circle}
\end{equation}
where $f_m\equiv \frac{1}{2}(f_1-f_2)$.

From the values of $h_1$ and $h_2$ we can  conclude that they are positive since $\Big(\frac{3-\sqrt{6}}{16}\Big)\approx 0.034>0$. To see that the bounds on $f_1$ and $f_2$ are satisfied, note that \eqref{eqn:fixed circle} imposes the following constraint on the value of $f_m$:
 \begin{equation}
 | f_m |\leq \Big(\frac{\sqrt{18\sqrt{6}-39}}{16}\Big)\lambda.\\
\end{equation}
From this we can conclude that
 \begin{equation}
\begin{split}
&f_{1,2}=f_p\pm f_m \geq \frac{\sqrt{6}}{8}\lambda- \Big(\frac{\sqrt{18\sqrt{6}-39}}{16}\Big)\lambda=\frac{2\sqrt{6}-\sqrt{18\sqrt{6}-39}}{16}\lambda\approx 0.165\lambda>0.\\
\end{split}
\end{equation}
Finally, note that 
 \begin{equation}
\begin{split}
&f_1 f_2 -\zeta^2=(f_p+f_m)(f_p-f_m)-\zeta^2=f_p^2-(f_m^2+\zeta^2)=\Big(\frac{63-18\sqrt{6}}{256}\Big)\lambda^2\approx 0.074 \lambda^2>0.
\end{split}
\end{equation}
Therefore, we see that all the conditions sufficient for the non-negativity of the quartic terms in the potential are satisfied at all points on the fixed circle. In fact, we have proved a slightly stronger condition, i.e., these quartic terms are strictly positive-definite for nonzero values of the field configurations.
 This means that as long as the thermal masses are non-negative, the potential increases steadily when one moves away from the origin along any direction in the field space. This is true even when either of the thermal masses is zero ruling out the possibility of there being any flat direction\footnote{We refer the reader to \cite{Bardeen:1983rv, Rabinovici:1987tf, Karananas:2019fox, Chai:2020zgq, Chai:2020onq} for examples of non-supersymmetric theories with such flat directions in the planar limit.} in such a scenario.

%%%%%%%%%%%%%%%%%%%%%%%%%%%%%%%%%%%%%%%%%%%%%%%%%%%%%%%%%%%%%%%%%%%%%%%%

\subsubsection{Thermal masses}
\label{subsec rdb: thermal masses}
Now, let us discuss the quadratic terms in the potential. These terms have the following form:
 \begin{equation}
V_{\text{quadratic}}=\frac{1}{2}\Big(\mathcal{M}^2\Big)_{ai,bj}\phi_{ai}\phi_{bj}.
\end{equation}
As earlier, we can evaluate the contribution of 1-loop Feynman diagrams to the thermal mass matrix $\mathcal{M}^2$ by using the general expression derived in \cite{Weinberg:1974hy}. For the model that we are presently considering, this expression  is given by
\begin{equation}
\begin{split}
\Big(\mathcal{M}^2\Big)_{ai,bj}
&=\frac{\beta_{\text{th}}^{-2}}{24}\Bigg[\sum_{c,k}\lambda_{ai,bj,ck,ck}+6\sum_{l=1}^2\sum_{\beta=1}^2\sum_{e,u} g_l^2\Big(T_A^{l\beta}(S)\Big)_{ai,eu}\Big(T_A^{l\beta}(S)\Big)_{eu,bj}\Bigg].
\end{split}
\end{equation}
Here $\beta_{\text{th}}^{-1}$ is the temperature. 
$\lambda_{ai,bj,ck,dl}$ are the symmetric couplings introduced in equations \eqref{symmetrised couplings def 1},  \eqref{symmetrised couplings def 2} and  \eqref{symmetrised couplings def 3}.  $T_A^{11}(S)$ and $T_A^{12}(S)$ are the generators of the two $SO(N_{c1})$'s in the representations corresponding to the  scalar fields in the model. Their explicit forms are given in the equations \eqref{T11 generators} and \eqref{T12 generators}. Similarly, $T_A^{21}(S)$ and $T_A^{22}(S)$ are the generators of the two $SO(N_{c2})$'s and their forms can be obtained by a $(1\leftrightarrow2)$ exchange of the indices in  \eqref{T11 generators} and \eqref{T12 generators}.

Using the explicit forms of the symmetric couplings and the generators, one can show that the matrix $\mathcal{M}^2$ takes the following form:
\begin{equation}
\begin{split}
\Big(\mathcal{M}^2\Big)=
\begin{pmatrix}
m_{\text{th},1}^2\mathrm{I}_{N_{c1}\times N_{c1}} & \mathrm{0}_{N_{c1}\times N_{c2}}\\
\mathrm{0}_{N_{c2}\times N_{c1}} & m_{\text{th},2}^2\mathrm{I}_{N_{c2}\times N_{c2}} \\
\end{pmatrix}.
\end{split}
\end{equation}
where  $m_{\text{th},1}^2$ and $m_{\text{th},2}^2$ are the 1-loop thermal masses (squared) of the 2 fields with the following values:
\begin{equation}
\begin{split}
&m_{\text{th},1}^2=\frac{16\pi^2\beta_{\text{th}}^{-2}}{3}\Bigg[\Big(2+\frac{1}{N_{c1}}\Big)h_1+\Big(1+\frac{2}{N_{c1}^2}\Big)f_1+\frac{N_{c2}}{N_{c1}}\zeta+\frac{3}{8} \Big(1-\frac{1}{N_{c1}}\Big)\lambda_1\Bigg],\\
&m_{\text{th},2}^2=\frac{16\pi^2\beta_{\text{th}}^{-2}}{3}\Bigg[\Big(2+\frac{1}{N_{c2}}\Big)h_2+\Big(1+\frac{2}{N_{c2}^2}\Big)f_2+\frac{N_{c1}}{N_{c2}}\zeta+\frac{3}{8} \Big(1-\frac{1}{N_{c2}}\Big)\lambda_2\Bigg].\\
\end{split}
\end{equation}
If either $m_{\text{th},1}^2$ or $m_{\text{th},2}^2$ is negative, then the minimum of the potential will be away from the origin in the field space, and hence at least one of the order parameter $\langle[\det\Phi_1]\rangle$ or $\langle[\det\Phi_2]\rangle$ will be nonzero. This would indicate a spontaneous breaking of the baryon symmetry and the Higgsing of some of the gauge bosons in the respective sector.

Thus, to determine whether these dual phenomena actually occur at any of the points on the large N fixed circle, we need to compute the thermal masses at these points. The values of the couplings at these fixed points are as follows:
\begin{equation}
\begin{split}
\lambda_1=\lambda_2=\lambda=\frac{21-4x_{f}}{-27+13x_{f}},\ 
h_1=h_2=h=\Big(\frac{3-\sqrt{6}}{16}\Big)\lambda,\  f_p\equiv\frac{f_1+f_2}{2}=\frac{\sqrt{6}}{8}\lambda.
\end{split}
\end{equation}
In addition, the double trace couplings  $\zeta$ and $f_m\equiv\frac{f_1-f_2}{2}$ are constrained to lie on the circle $ \zeta^2+f_m^2= \Big(\frac{18\sqrt{6}-39}{256}\Big)\lambda^2$. Substituting the values of the above couplings in the expressions of the thermal masses and taking the limit $N_{ci}\rightarrow\infty$, we get
\begin{equation}
\begin{split}
&m_{\text{th},1}^2=\frac{16\pi^2\beta_{\text{th}}^{-2}}{3}\Bigg[\frac{3}{4}\lambda+f_m+r \zeta\Bigg],\ m_{\text{th},2}^2=\frac{16\pi^2\beta_{\text{th}}^{-2}}{3}\Bigg[\frac{3}{4}\lambda-f_m+\frac{\zeta}{r}\Bigg],\\
\end{split}
\label{rdb: fixed point thermal masses}
\end{equation}
where the parameter $r$ is defined as follows:
\begin{equation}
\begin{split}
{r \equiv \frac{N_{c2}}{N_{c1}}.}
\end{split}
\end{equation}
This parameter did not enter in the 2-loop planar beta functions of the couplings\footnote{In section \ref{sec: survival of fixed circle}, we will show that this feature of the planar beta functions holds at all orders of the 't Hooft couplings.}. However, as we can see now, it leaves its imprint in the effective potential of the scalar fields through the thermal masses. There are constraints on the value of this parameter which need to be satisfied to get a symmetry-broken phase at non-zero temperatures. Moreover, even when the constraints on $r$ are satisfied, only a subset of points on the fixed circle demonstrate a symmetry-broken phase.  We will next discuss these conditions on $r$ and the fixed points. In what follows, we will assume that $r<1$, i.e., $N_{c1}>N_{c2}$. The results that we will get can be generalized to the case $r>1$, i.e., $N_{c1}<N_{c2}$, by a $(1\leftrightarrow2)$ exchange of indices everywhere.

\subsubsection{Conditions for symmetry breaking}
To analyze the conditions on the parameter $r$  and the fixed points for breaking of the baryon symmetry and a persistent BEH phase at nonzero temperatures, let us first consider the following relation between the couplings $f_m$ and $\zeta$ at any point on the fixed circle:
 \begin{equation}
\begin{split}
 f_m=\pm \sqrt{\Big(\frac{18\sqrt{6}-39}{256}\Big)\lambda^2-\zeta^2}.
\end{split}
\label{fm expression}
\end{equation}
This relation imposes the following bounds on the values of $\zeta$ at these fixed points:
 \begin{equation}
\begin{split}
 -\frac{\sqrt{18\sqrt{6}-39}}{16}\lambda\leq \zeta \leq \frac{\sqrt{18\sqrt{6}-39}}{16}\lambda.
\end{split}
\label{bounds on zeta}
\end{equation}
To avoid carrying along the factor of $\lambda$ throughout our analysis, let us define a normalized version ($z$) of the coupling $\zeta$ by stripping off the factor of $\lambda$:
\begin{equation}
z\equiv\frac{\zeta}{\lambda}.
\end{equation}
The above-mentioned bounds on $\zeta$ then lead to the following bounds on the normalized coupling $z$:
 \begin{equation}
\begin{split}
 -\frac{\sqrt{18\sqrt{6}-39}}{16}\leq z \leq \frac{\sqrt{18\sqrt{6}-39}}{16}.
\end{split}
\label{bounds on z}
\end{equation}

Now, by using equations \eqref{rdb: fixed point thermal masses} and \eqref{fm expression}, we can express the thermal masses of the scalar fields in terms of the parameter $r$ and the normalised coupling $z$ as follows:
\begin{equation}
\begin{split}
&m_{\text{th},1}^2=\frac{16\pi^2\beta_{\text{th}}^{-2}}{3}\lambda\Bigg[\frac{3}{4}\pm\sqrt{\Big(\frac{18\sqrt{6}-39}{256}\Big)-z^2}+r z\Bigg],\\
&m_{\text{th},2}^2=\frac{16\pi^2\beta_{\text{th}}^{-2}}{3}\lambda\Bigg[ \frac{3}{4}\mp\sqrt{\Big(\frac{18\sqrt{6}-39}{256}\Big)-z^2}+\frac{z}{r}\Bigg].\\
\end{split}
\label{rdb:fixed point thermal masses}
\end{equation}
Note that the above expression of $m_{\text{th},1}^2$ is positive-definite as can be seen from the following inequalities:
\begin{equation}
\begin{split}
\Bigg|\pm\sqrt{\Big(\frac{18\sqrt{6}-39}{256}\Big)-z^2}+r z \Bigg|
&\leq \sqrt{\Big(\frac{18\sqrt{6}-39}{256}\Big)}+r|z|\\
&\leq (1+r)\sqrt{\Big(\frac{18\sqrt{6}-39}{256}\Big)}\\
&< 2\sqrt{\Big(\frac{18\sqrt{6}-39}{256}\Big)}\approx 0.282<\frac{3}{4}.
\end{split}
\end{equation}
Here, in the first line, we have used the relation $|a+b|\leq |a|+|b|$, and the fact that  $\sqrt{\Big(\frac{18\sqrt{6}-39}{256}\Big)-z^2}\leq \sqrt{\Big(\frac{18\sqrt{6}-39}{256}\Big)}$. In the second line, we have imposed the upper bound on $|z|$  given in \eqref{bounds on z}. In the last  line, we have used the fact that $r<1.$

Now, let us see under what conditions $m_{\text{th},2}^2$ can be negative. First note that when $z=0$, $m_{\text{th},2}^2$ is positive:
\begin{equation}
\begin{split}
m_{\text{th},2}^2\Big|_{z=0}=\frac{16\pi^2\beta_{\text{th}}^{-2}}{3}\lambda\Bigg[ \frac{3}{4}\mp\sqrt{\Big(\frac{18\sqrt{6}-39}{256}\Big)}\Bigg]>0.\\
\end{split}
\end{equation}
We can also see that  for any particular value of $r$,  $m_{\text{th},2}^2$ in either of the branches in \eqref{rdb:fixed point thermal masses} is a continuous function of $z$. Therefore, for any  fixed $r$, if  $m_{\text{th},2}^2$ has to be negative at some value of $z$ in the range $( -\frac{\sqrt{18\sqrt{6}-39}}{16},  \frac{\sqrt{18\sqrt{6}-39}}{16})$, then it must also pass through zero at some point  (say, $z_0$) in this interval. At this point we have
\begin{equation}
\begin{split}
 &\frac{3}{4}\mp\sqrt{\Big(\frac{18\sqrt{6}-39}{256}\Big)-z_0^2}+\frac{z_0}{r}=0\\
   \implies &\Big(1+\frac{1}{r^2}\Big) z_0^2+\frac{3}{2r}z_0+\Big(\frac{183-18\sqrt{6}}{256}\Big)=0.\\
\end{split}
\label{eqn for zero mass}
\end{equation}
For the existence of a real solution of this equation, the discriminant of the corresponding quadratic polynomial must be non-negative, i.e.,
\begin{equation}
\begin{split}
&\Big(\frac{3}{2r}\Big)^2-4\Big(1+\frac{1}{r^2}\Big)\Big(\frac{183-18\sqrt{6}}{256}\Big)\geq 0\\
&\implies  r \leq \sqrt{\frac{6\sqrt{6}-13}{61-6\sqrt{6}}} \approx 0.191.\\
\end{split}
\end{equation}
This puts an upper bound on the value of $r$ for which a symmetry-broken phase can exist for any of the points on the fixed circle.

Now, for every value of $r$ lying below this upper bound, there are two solutions of the quadratic equation given above:
\begin{equation}
\begin{split}
z_0 &=\frac{-\frac{3}{2r}\pm\sqrt{\Big(\frac{3}{2r}\Big)^2-4\Big(1+\frac{1}{r^2}\Big)\Big(\frac{183-18\sqrt{6}}{256}\Big)}}{2\Big(1+\frac{1}{r^2}\Big)}\\
&=r\Bigg[\frac{-12\pm \sqrt{\Big(18\sqrt{6}-183\Big)r^2+\Big(18\sqrt{6}-39\Big)}}{16\Big(1+r^2\Big)}\Bigg].\\
\end{split}
\end{equation}
Since $(183-18\sqrt{6})>0$, we have $\sqrt{\Big(\frac{3}{2r}\Big)^2-4\Big(1+\frac{1}{r^2}\Big)\Big(\frac{183-18\sqrt{6}}{256}\Big)}<\frac{3}{2r}$. Therefore, both these solutions are negative. Hence, the value of $z$ (or equivalently, $\zeta$) for which $m_{\text{th},2}^2<0$ is always negative\footnote{This can also be seen more directly from  \eqref{rdb: fixed point thermal masses} since $f_m \leq \frac{\sqrt{18\sqrt{6}-39}}{16}\lambda<{3\ov 4}\lam$  for all points on the fixed circle.}.

Notice that when we went from the first line of equation \eqref{eqn for zero mass} to its second line, we blurred the distinction between the two branches of the linear equation.  These two branches correspond to the two possible signs of the coupling $f_m$ as given in \eqref{fm expression}. For a positive $f_m$, one should choose the $(-)$ sign in the first line of \eqref{eqn for zero mass} . Similarly, for a negative $f_m$, one should choose the $(+)$ sign in the same equation. Let us now see to which of these branches  the two solutions of the quadratic equation belong.

\paragraph{Solution 1: $z_0=r\Bigg[\frac{-12+ \sqrt{\Big(18\sqrt{6}-183\Big)r^2+\Big(18\sqrt{6}-39\Big)}}{16\Big(1+r^2\Big)}\Bigg]$}\mbox{}

In this case, we have
\begin{equation}
\begin{split}
&\Big(\frac{18\sqrt{6}-39}{256}\Big)-z_0^2
=\Bigg[\frac{12 r^2+\sqrt{\Big(18\sqrt{6}-39\Big)-\Big(183-18\sqrt{6}\Big)r^2}}{16(1+r^2)}\Bigg]^2.\\
\end{split}
\end{equation}
Note that both the terms in the numerator of the RHS of the above equation are non-negative. Therefore, taking the positive square roots of both  sides of this equation, we get
\begin{equation}
\begin{split}
\sqrt{\Big(\frac{18\sqrt{6}-39}{256}\Big)-z_0^2}
&=\Bigg[\frac{12 r^2+\sqrt{\Big(18\sqrt{6}-39\Big)-\Big(183-18\sqrt{6}\Big)r^2}}{16(1+r^2)}\Bigg]
=\frac{3}{4}+\frac{z_0}{r}.\\
\end{split}
\end{equation}
From the above equation we get
\begin{equation}
\begin{split}
\frac{3}{4}-\sqrt{\Big(\frac{18\sqrt{6}-39}{256}\Big)-z_0^2}+\frac{z_0}{r}=0.
\end{split}
\end{equation}
Therefore, we find that this solution always belongs to the branch where $f_m>0$.

\paragraph{Solution 2: $z_0=r\Bigg[\frac{-12- \sqrt{\Big(18\sqrt{6}-183\Big)r^2+\Big(18\sqrt{6}-39\Big)}}{16\Big(1+r^2\Big)}\Bigg]$}\mbox{}

In this case, we have
\begin{equation}
\Big(\frac{18\sqrt{6}-39}{256}\Big)-z_0^2
=\Bigg[\frac{12 r^2-\sqrt{\Big(18\sqrt{6}-39\Big)-\Big(183-18\sqrt{6}\Big)r^2}}{16(1+r^2)}\Bigg]^2.
\label{branch check 1: negative fm}
\end{equation}
Now, the second term in the numerator of the RHS of the above equation is negative. So, while taking the square root of both sides of the equation, one has to be careful about the sign of the quantity within the brackets. To check this sign, let us look at the magnitude of the expression within the root in the RHS. This expression can be re-cast as follows:
\begin{equation}
\begin{split}
\Big(18\sqrt{6}-39\Big)-\Big(183-18\sqrt{6}\Big)r^2=(12r^2)^2+\Big(18\sqrt{6}-39-144 r^2\Big)(1+r^2).\\
\end{split}
\end{equation}
Therefore this expression will have a magnitude less than (or equal to) $(12 r^2)^2$ (leading to a positive  sign  of the quantity in  brackets in the RHS of \eqref{branch check 1: negative fm}) only if
\begin{equation}
\begin{split}
&\Big(18\sqrt{6}-39-144 r^2\Big)\leq0 \implies r\geq \frac{\sqrt{18\sqrt{6}-39}}{12}\approx 0.188.\\
\end{split}
\end{equation}
Thus, for $  \frac{\sqrt{18\sqrt{6}-39}}{12}\leq r \leq  \sqrt{\frac{6\sqrt{6}-13}{61-6\sqrt{6}}}$, we have 
\begin{equation}
\begin{split}
\sqrt{\Big(\frac{18\sqrt{6}-39}{256}\Big)-z_0^2}
=&\Bigg[\frac{12 r^2-\sqrt{\Big(18\sqrt{6}-39\Big)-\Big(183-18\sqrt{6}\Big)r^2}}{16(1+r^2)}\Bigg]
=\frac{3}{4}+\frac{z_0}{r},\\
\end{split}
\end{equation}
and we can again conclude that the solution lies in the branch where $f_m>0$.

For $r<\frac{\sqrt{18\sqrt{6}-39}}{12}$, we have 
\begin{equation}
\begin{split}
\sqrt{\Big(\frac{18\sqrt{6}-39}{256}\Big)-z_0^2}
=&-\Bigg[\frac{12 r^2-\sqrt{\Big(18\sqrt{6}-39\Big)-\Big(183-18\sqrt{6}\Big)r^2}}{16(1+r^2)}\Bigg]
=-\Big(\frac{3}{4}+\frac{z_0}{r}\Big),\\
\end{split}
\end{equation}
and the solution lies in the branch where $f_m<0$.

Let us summarize the above results:
\begin{itemize}
\item When $  \frac{\sqrt{18\sqrt{6}-39}}{12}\leq r <  \sqrt{\frac{6\sqrt{6}-13}{61-6\sqrt{6}}}$, there are two fixed points on the same branch ($f_m>0$) for which the thermal mass (squared) $m_{\text{th},2}^2$ vanishes. The values of the normalized coupling $z$ at these two fixed points are:
\begin{equation}
\begin{split}
z_0^{(1)}=r\Bigg[\frac{-12+ \sqrt{\Big(18\sqrt{6}-183\Big)r^2+\Big(18\sqrt{6}-39\Big)}}{16\Big(1+r^2\Big)}\Bigg], \\
z_0^{(2)}=r\Bigg[\frac{-12-\sqrt{\Big(18\sqrt{6}-183\Big)r^2+\Big(18\sqrt{6}-39\Big)}}{16\Big(1+r^2\Big)}\Bigg],
\end{split}
\label{zeros of thermal mass 2}
\end{equation}
where manifestly $z_0^{(2)}<z_0^{(1)}<0$.
\item When $ r< \frac{\sqrt{18\sqrt{6}-39}}{12}$, again we have two fixed points  for which  $m_{\text{th},2}^2$ vanishes. The values of $z$ at these fixed points are again given by the  expressions in \eqref{zeros of thermal mass 2}. However, the fixed point with $\zeta=z_0^{(1)}\lambda$ lies in the branch  $f_m>0$, whereas the one with $\zeta=z_0^{(2)}\lambda$ lies in the branch  $f_m<0$.
\end{itemize}

Now that we have identified the fixed points where $m_{\text{th},2}^2=0$, it is possible to determine the points at which it is negative. To apply the argument of continuity of $m_{\text{th},2}^2$ as we move along the fixed circle, we need to work in a particular branch. If there are two fixed points where $m_{\text{th},2}^2$ vanishes in that branch, then in the interval between these two points it will be negative. On the other hand, if there is only one fixed point (say, at $z=z_0$) in that branch where $m_{\text{th},2}^2$ vanishes, then it will be negative for all values of $z$  below $z_0$.

Before enumerating all the fixed points where $m_{\text{th},2}^2<0$, let us briefly comment on the points where the  afore-mentioned branches meet. At these points, we have $f_m=0$ and $z=\sigma \frac{\sqrt{18\sqrt{6}-39}}{16}$ where $\sigma=\pm 1$. From \eqref{rdb:fixed point thermal masses}, we find that at these points:
\begin{equation}
\begin{split}
&m_{\text{th},2}^2=\frac{16\pi^2\beta_{\text{th}}^{-2}}{3}\lambda\Bigg[ \frac{3}{4}+\sigma\frac{\sqrt{18\sqrt{6}-39}}{16r}\Bigg].\\
\end{split}
\end{equation}
Therefore, $m_{\text{th},2}^2>0$ at all values of $r$ for the fixed point with $\sigma=1$. For the fixed point with $\sigma=-1$, $m_{\text{th},2}^2\geq0$ when $r\geq\frac{\sqrt{18\sqrt{6}-39}}{12}$, and  $m_{\text{th},2}^2<0$ when $r<\frac{\sqrt{18\sqrt{6}-39}}{12}$.

\begin{figure}[H]
\begin{subfigure}{.5\textwidth}
  \centering
  \scalebox{0.8}{\includegraphics[width=0.8\linewidth]{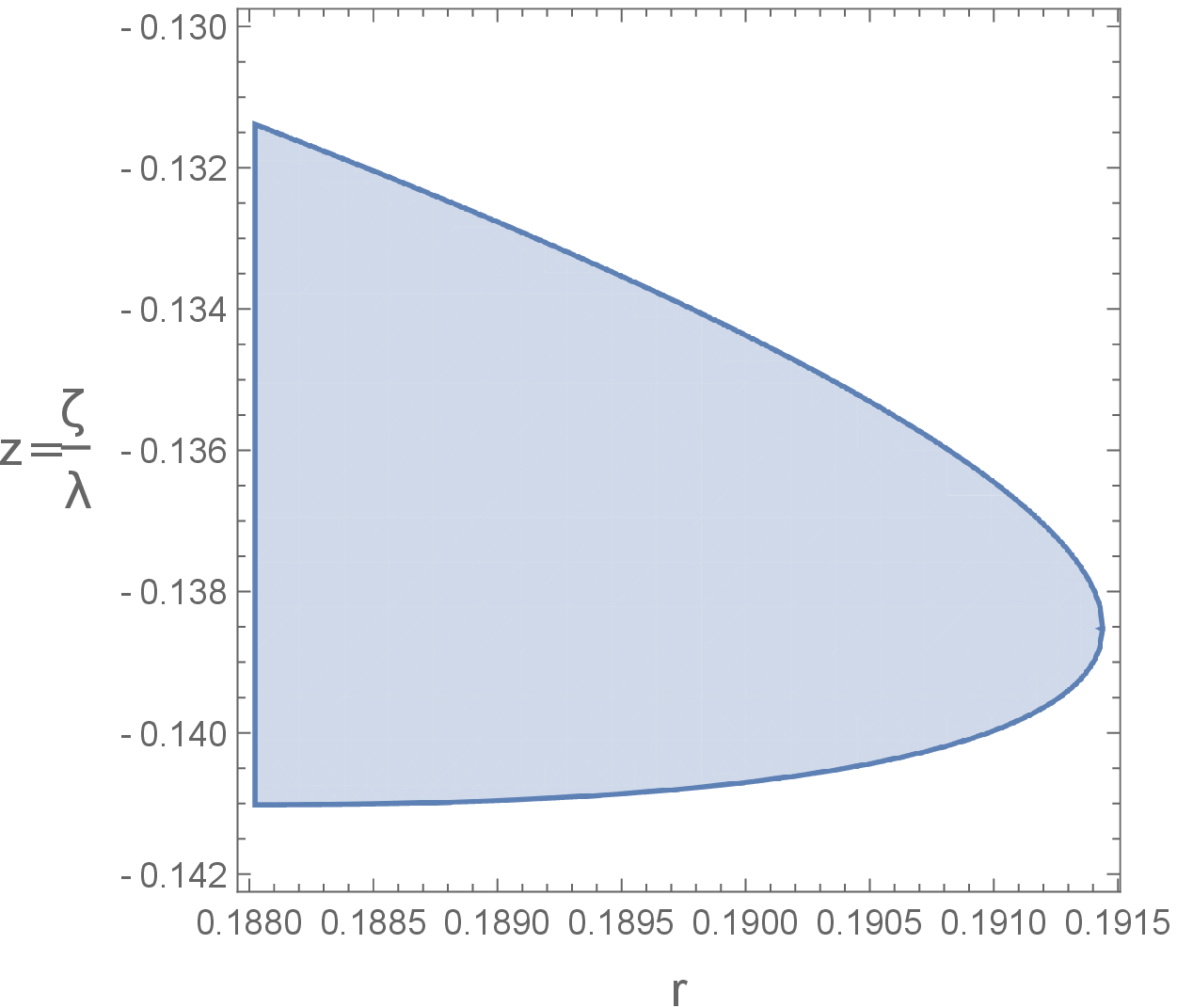}}
  \caption{$ r\in  \Big[\frac{\sqrt{18\sqrt{6}-39}}{12}, \sqrt{\frac{6\sqrt{6}-13}{61-6\sqrt{6}}}\ \Big),\ f_m> 0$}
\end{subfigure}%
\begin{subfigure}{.5\textwidth}
  \centering
  \scalebox{0.8}{\includegraphics[width=0.8\linewidth]{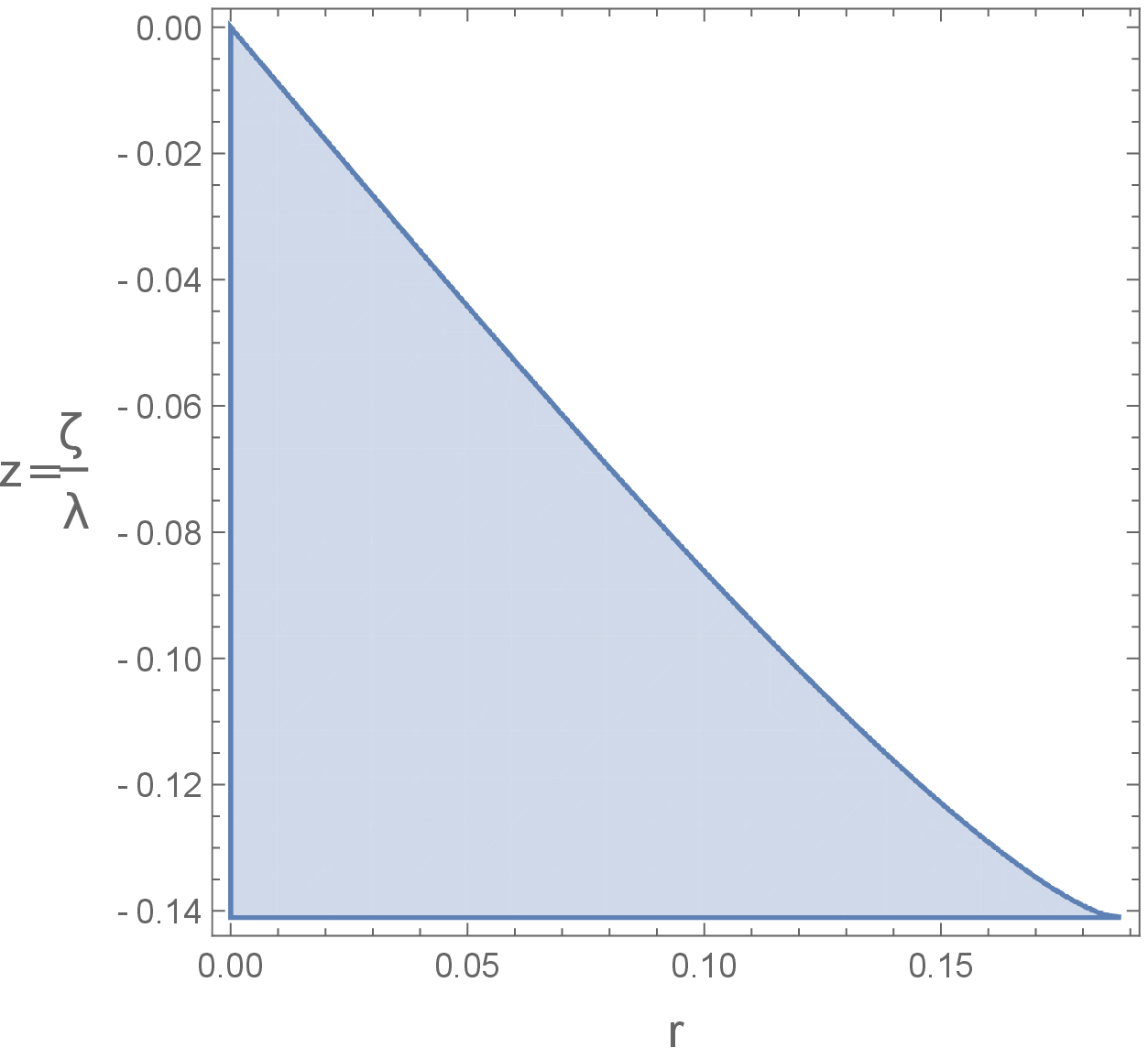}}
  \caption{$r\in(0, \frac{\sqrt{18\sqrt{6}-39}}{12}),\ f_m< 0$}
\end{subfigure}
\begin{subfigure}{.5\textwidth}
  \centering
  \scalebox{0.8}{\includegraphics[width=0.8\linewidth]{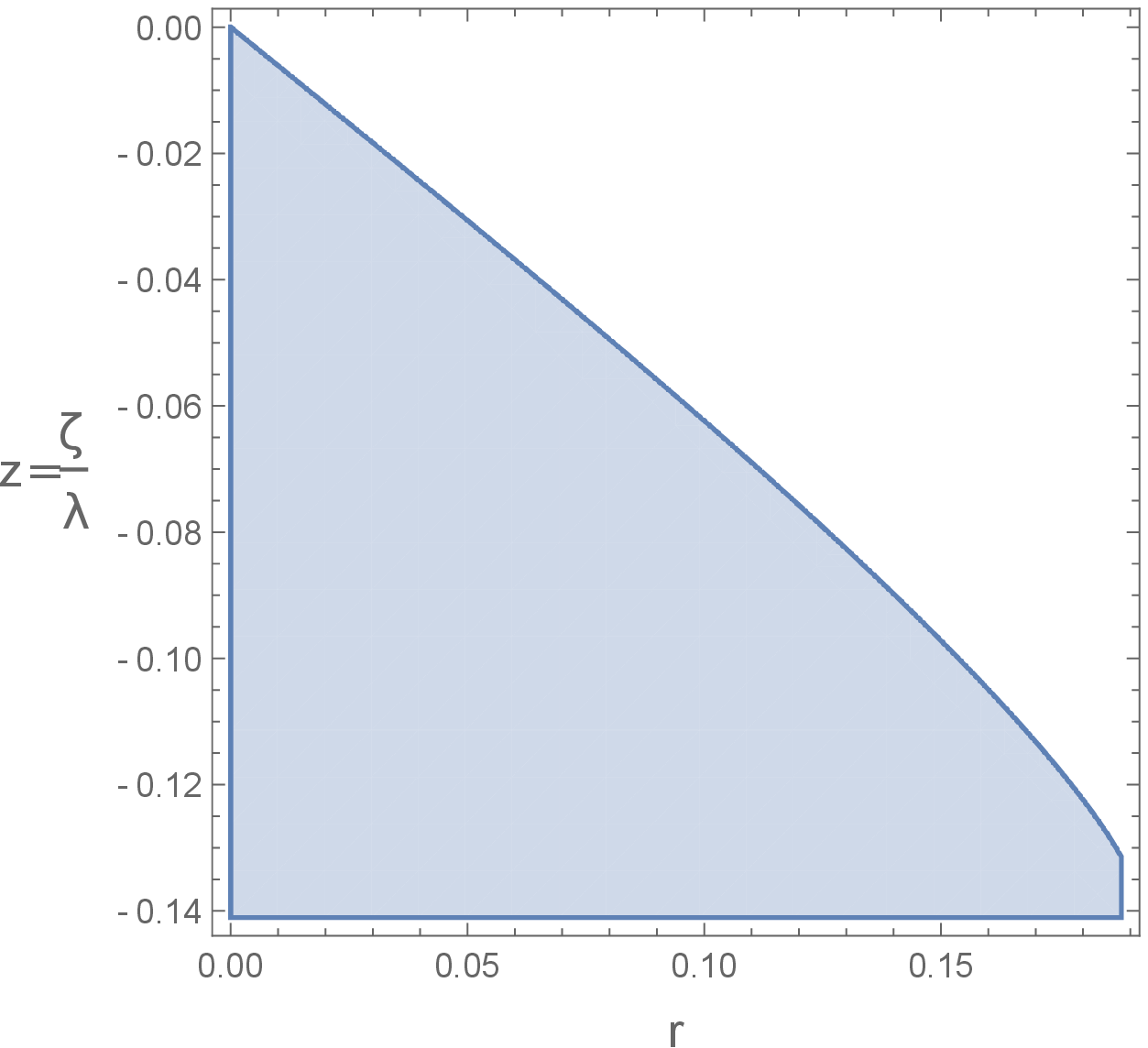}}
    \caption{$r\in(0, \frac{\sqrt{18\sqrt{6}-39}}{12}),\ f_m\geq  0$}
\end{subfigure}%
\caption{Fixed points in the real double bifundamental model for which $m_{\text{th},2}^2<0$.}
\label{fig:RDB fixed points with thermal order}
\end{figure}

Thus, from the above analysis we can finally conclude that $m_{\text{th},2}^2<0$ 
%and the baryon symmetry in the second sector is broken at nonzero temperatures 
for the fixed points satisfying any one of the following conditions: 
\begin{enumerate}
\item  $ r\in  \Big[\frac{\sqrt{18\sqrt{6}-39}}{12}, \sqrt{\frac{6\sqrt{6}-13}{61-6\sqrt{6}}}\ \Big)$,  $\zeta\in (z_0^{(2)}\lambda,z_0^{(1)}\lambda)$, and $f_m >0$,
\item  $ r\in(0, \frac{\sqrt{18\sqrt{6}-39}}{12})$, $\zeta\in [z_{\text{min}}\lambda,z_0^{(1)}\lambda)$, and $f_m\geq 0$,
\item  $ r\in(0, \frac{\sqrt{18\sqrt{6}-39}}{12})$, $\zeta\in (z_{\text{min}}\lambda,z_0^{(2)}\lambda)$, and  $f_m<0$.
\end{enumerate}
Here $z_0^{(1)}$ and $z_0^{(2)}$ are the values of the normalized coupling $z$ defined in \eqref{zeros of thermal mass 2}, and $z_{\text{min}}$ is the lower bound on the value of $z$ which was given in \eqref{bounds on z}, i.e.,
\beq
z_{\text{min}}= -\frac{\sqrt{18\sqrt{6}-39}}{16}.
\label{value of zmin}
\eeq
We provide graphical plots of these fixed points in figure \ref{fig:RDB fixed points with thermal order}.

In appendix \ref{app: minimum of  potential} we have shown that for all these fixed points, the thermal expectation values of the scalar fields (up to gauge transformations) have the following forms:
\beq
\langle\Phi_1\rangle=0,\ \langle\Phi_2\rangle=\Bigg(-\frac{N_{c2}}{64\pi^2}\frac{m_{\text{th},2}^2}{h_2+f_2}\Bigg)^{\frac{1}{2}}\ \text{diag}\{\pm1 ,1 ,1 ,\cdots ,1\},
\label{rdb: thermal expectation values}
\eeq
where `diag' stands for a diagonal matrix with the corresponding diagonal entries given in the brackets. Note that in the zero temperature limit, $\langle \Phi_2\rangle=0$  due to the vanishing of  $m_{\text{th},2}^2$. This means that both the baryon symmetries are unbroken in the ground state.

For any nonzero temperature, the above forms of the  thermal expectation values of the scalar fields lead to the following consequences:
\begin{enumerate}
\item There are two gauge inequivalent vacua with the order parameter $\langle[\det \Phi_2]\rangle\neq 0$. This leads to the spontaneous breaking of the baryon symmetry in the second sector at all nonzero temperatures.
\item For either of the two vacua, the $SO(N_{c2})\times SO(N_{c2})$ gauge symmetry in the second sector is  broken down to  $SO(N_{c2})$. To see this, let us consider a gauge transformation $O_1\times O_2$  belonging to $SO(N_{c2})\times SO(N_{c2})$ which leaves the thermal expectation values given in \eqref{rdb: thermal expectation values} invariant, i.e., 
\beq
O_1 \langle\Phi_2\rangle O_2^T=\langle\Phi_2\rangle.
\eeq
This relates $O_2$ to $O_1$ as follows:
\beq
O_2=\langle\Phi_2\rangle^{-1} O_1 \langle\Phi_2\rangle.
\eeq
Thus, we see that the residual gauge symmetry is just $SO(N_{c2})$. Consequently, half of the generators in the second sector are broken. This, in turn, leads to half of the gauge bosons in this sector getting Higgsed, and the system being in the Brout-Englert-Higgs (BEH) phase at all nonzero temperatures.
\end{enumerate}

Therefore, we find that the spontaneous breaking of the baryon symmetry and the persistent BEH phase always occur together in this model as mentioned in section \ref{sec:intro}.

\section{Complex double bifundamental model}
\label{sec: cdb model}
In this section we will turn our attention to the complex double bifundamental model. We will begin by demonstrating that this model is actually dual to the real double bifundamental model in the Veneziano limit. Therefore, in this limit, all the conclusions that we arrived at in the last section are equally valid for the complex double bifundamental model. In particular, we will see that for the planar beta functions in this model, there is a fixed circle in the space of couplings. Just as before, by looking at the thermal masses of the scalar fields we will determine the conditions under which a subset of these fixed points demonstrate spontaneous breaking of the baryon symmetry and a persistent BEH phase at nonzero temperatures.
  
\subsection{Planar equivalence between the two double bifundamental models} \label{subsec:planar duality}

Let us now demonstrate that the  real and the complex double bifundamental models are perturbatively equivalent in the Veneziano limt. This equivalence between the two models rests on the procedure of orbifolding a parent theory to obtain an equivalent daughter theory.\footnote{This is a counterpart of string orbifold equivalence discussed in \cite{Kachru:1998ys,Lawrence:1998ja}.} In this case, a real double bifundamental model symmetric under the gauge group $G^{(R)}\equiv\prod\limits_{i=1}^2 SO(2 N_{ci})\times SO(2 N_{ci})$ and with $2N_{fi}$ flavors of Majorana fermions in the $i^{\text{th}}$ sector serves as the parent theory. On the other hand, the daughter theory is a complex double bifundamental model symmetric under the gauge group $G^{(C)}\equiv\prod\limits_{i=1}^2 SU(N_{ci})\times SU(N_{ci})$ and with $N_{fi}$ flavors of Dirac fermions in the $i^{\text{th}}$ sector.

 The idea of orbifolding a parent theory is to first identify a discrete global symmetry in it. One can then project all its fields to the components that are invariant under this symmetry group.  Operators composed  of these projected fields are also invariant under the discrete global symmetry  and hence, are known as {\it neutral operators}.  One can next project the  Lagrangian of the parent theory to its part containing such neutral operators by simply removing  the  terms that are not invariant under the discrete symmetry.  The form of this truncated Lagrangian of the parent theory should be similar to the Lagrangian of the daughter theory. Based on this similarity between the two Lagrangians, it can be shown that in the planar limit there is a mapping between the couplings in the two theories under which the correlators of  neutral operators in the parent theory are equal to the corresponding correlators in the daughter theory. In this precise sense, the two theories are dual to each other in the planar limit. We will next show how this procedure of orbifolding of the real double bifundamental model leads to its planar equivalence with the complex double bifundamental model. Our arguments for this equivalence will closely follow those given for similar dualities in  \cite{Bershadsky:1998cb, Schmaltz:1998bg}.\footnote{See also \cite{Dymarsky:2005nc,Dymarsky:2005uh,Pomoni:2008de, Cherman:2010jj,Hanada:2011ju,Dunne:2016nmc,Aitken:2019shs,Jepsen:2020czw} for discussions on the various aspects of the planar equivalence.}

To demonstrate the planar equivalence between the two models, let us first express the Majorana fermions in the real double bifundamental model in terms of Weyl spinors as follows \footnote{From here onwards, we shall be working in the Weyl representation of the Clifford algebra. }:  
 \begin{equation}
\begin{split}
 \psi_{i}^{(p)}
 =\begin{dcases}
\begin{pmatrix}\psi_{i,L}^{(p)}\\ i\sigma^2\Big(\psi_{i,L}^{(p)}\Big)^*\end{pmatrix}\ \text{for}\ p\in \{1,\cdots,N_{fi}\},\\
\begin{pmatrix}-i\sigma^2\Big(\psi_{i,R}^{(p-N_{fi})}\Big)^*\\ \psi_{i,R}^{(p-N_{fi})}\end{pmatrix}\ \text{for}\ p\in \{N_{fi}+1,\cdots,2N_{fi}\},
 \end{dcases}
\end{split}
\end{equation}
\begin{equation}
\begin{split}
 \chi_{i}^{(p)}
 =\begin{dcases}
\begin{pmatrix}\chi_{i,L}^{(p)}\\ i\sigma^2\Big(\chi_{i,L}^{(p)}\Big)^*\end{pmatrix}\ \text{for}\ p\in \{1,\cdots,N_{fi}\},\\
\begin{pmatrix}-i\sigma^2\Big(\chi_{i,R}^{(p-N_{fi})}\Big)^*\\ \chi_{i,R}^{(p-N_{fi})}\end{pmatrix}\ \text{for}\ p\in \{N_{fi}+1,\cdots,2N_{fi}\},
 \end{dcases}
\end{split}
\end{equation}
where the spinors with the subscript `L' are left-handed, while those with the subscript `R' are right-handed. Substituting the above expressions of the Majorana fermions in equation \eqref{rdb: lagrangian}, we get the following form of the Lagrangian (after removing some total derivatives):
 \begin{equation}
\begin{split}
&\mathcal{L}_{\text{RDB}}\\
=&-\frac{1}{4}\sum_{i=1}^2\sum_{\alpha=1}^2(F_{i\alpha})_{\mu\nu}^A(F_{i\alpha})^{\mu\nu A}+i\sum_{i=1}^2\Big(\psi_{i,L}^{(q)}\Big)_{a_i}^\dag \overline{\sigma}^\mu   \Big(D_\mu\psi_{i,L}^{(q)}\Big)_{a_i}+i\sum_{i=1}^2\Big(\chi_{i,L}^{(q)}\Big)_{j_i}^\dag \overline{\sigma}^\mu   \Big(D_\mu\chi_{i,L}^{(q)}\Big)_{j_i}\\
&+i\sum_{i=1}^2\Big(\psi_{i,R}^{(q)}\Big)_{a_i}^\dag\sigma^\mu   \Big(D_\mu\psi_{i,R}^{(q)}\Big)_{a_i}+i\sum_{i=1}^2\Big(\chi_{i,R}^{(q)}\Big)_{j_i}^\dag \sigma^\mu   \Big(D_\mu\chi_{i,R}^{(q)}\Big)_{j_i}+\frac{1}{2}\sum_{i=1}^2\text{Tr}\Bigg[\Big(D_{\mu} \Phi_i \Big)^T D^{\mu} \Phi_i\Bigg]\\
&-\sum_{i=1}^2\widetilde{h}_i^{R}\text{Tr}\Big[\Phi_i^T\Phi_i\Phi_i^T\Phi_i\Big]-\sum_{i=1}^2\widetilde{f}_i^{R}\text{Tr}\Big[\Phi_i^T\Phi_i\Big]\text{Tr}\Big[\Phi_i^T\Phi_i\Big]-2\widetilde{\zeta}^{R}\text{Tr}\Big[\Phi_1^T\Phi_1\Big]\text{Tr}\Big[\Phi_2^T\Phi_2\Big],
\end{split}
\end{equation}
where $\overline{\sigma}^\mu=(\mathbb I, -\overrightarrow{\sigma})$, and the dummy index $q$ runs over the values $\{1,\cdots, N_{fi}\}$ for the fermions in the $i^{\text{th}}$ sector. We have put a superscript $R$ on the couplings to distinguish them from the couplings in the dual complex double bifundamental model.

Now, note that the above Lagrangian is invariant under the following $\mathbb Z_2$ transformation of the fields\footnote{Here, the fields in both the sectors are transformed  simultaneously.}:
\beq 
(V_{i\alpha})_\mu ~&\rightarrow~ J_i (V_{i\alpha})_\mu  J_i^T,
\\ 
 \Phi_i ~&\rightarrow~ J_i \Phi_i J_i^T,\\
(\psi_{i,L}^{(q)} ,\ \psi_{i,R}^{(q)}) ~&\rightarrow~ (- i J_i \psi_{i,L}^{(q)},\ -i J_i\psi_{i,R}^{(q)})\ ,\\
(\chi_{i,L}^{(q)} ,\ \chi_{i,R}^{(q)}) ~&\rightarrow~ (- i J_i \chi_{i,L}^{(q)},\ -i J_i\chi_{i,R}^{(q)})\ ,
\label{Z2 symmetry in rdb}
\eeq
where 
\beq
J_i=i\sigma_2 \otimes \mathbb I_{N_{ci}\times N_{ci}}=\begin{pmatrix}  0_{N_{ci}\times N_{ci}} &  \mathbb I_{N_{ci}\times N_{ci}}\\
 -\mathbb I_{N_{ci}\times N_{ci}} &  0_{N_{ci}\times N_{ci}}\end{pmatrix}.
 \eeq
To project the fields on to their neutral components, we  demand the invariance of the fields under the above transformations. This yields the following structure of the projected fields:
 \begin{equation}
\begin{split}  
&(V_{i\alpha})_{\mu} 
=\begin{pmatrix}
(B^A_{i\alpha})_{\mu} & -(C^S_{i\alpha})_{\mu}\\
(C^S_{i\alpha})_{\mu} & (B^A_{i\alpha})_{\mu} 
\end{pmatrix},\ 
\Phi_i
=\begin{pmatrix}
\Phi_{i1} & \Phi_{i2}\\
-\Phi_{i2}&\Phi_{i1}
\end{pmatrix},\\
&\Big(\psi_{i,L}^{(p)}\Big)_{a_i+N_{ci}}=i\Big(\psi_{i,L}^{(p)}\Big)_{a_i},\ \Big(\psi_{i,R}^{(p)}\Big)_{a_i+N_{ci}}=i\Big(\psi_{i,R}^{(p)}\Big)_{a_i}\ \forall\ a_i\in\{1,\cdots,N_{ci}\},\\
&\Big(\chi_{i,L}^{(q)}\Big)_{j_i+N_{ci}}=i\Big(\chi_{i,L}^{(q)}\Big)_{j_i},\ \Big(\chi_{i,R}^{(q)}\Big)_{j_i+N_{ci}}=i\Big(\chi_{i,R}^{(q)}\Big)_{j_i}\ \forall\ j_i\in\{1,\cdots,N_{ci}\}.
\end{split} 
\end{equation}
Here $(B_{i\alpha}^A)_\mu$ and $(C_{i\alpha}^S)_\mu$ are imaginary $(N_{ci}\times N_{ci})$-matrices which are anti-symmetric and symmetric respectively. Similarly, $\Phi_{i1}$ and $\Phi_{i2}$ are real $(N_{ci}\times N_{ci})$-matrices, but they do not have any symmetry property under transposition.

Let us now combine these projected fields to define a new set of fields:
\begin{equation}
\begin{split}
&(\widetilde{\mathcal{V}}_{i\alpha})_{\mu}\equiv \sqrt{2}\Big[(B^A_{i\alpha})_{\mu}-i(C^S_{i\alpha})_{\mu}\Big],\ \tilde \Phi_i\equiv \Phi_{i1}+i\Phi_{i2},\\
&\Big(\tilde \Psi_{i,L}^{(q)}\Big)_{a_i}\equiv \sqrt{2}\Big(\psi_{i,L}^{(q)}\Big)_{a_i}, \Big(\tilde \Psi_{i,R}^{(q)}\Big)_{a_i}\equiv\sqrt{2}\Big(\psi_{i,R}^{(q)}\Big)_{a_i} \  \forall\ a_i\in\{1,\cdots,N_{ci}\},\\
&\Big(\tilde \chi_{i,L}^{(q)}\Big)_{j_i}\equiv \sqrt{2}\Big(\chi_{i,L}^{(q)}\Big)_{j_i}, \Big(\tilde \chi_{i,R}^{(q)}\Big)_{j_i}\equiv\sqrt{2}\Big(\chi_{i,R}^{(q)}\Big)_{j_i}\  \forall\ j_i\in\{1,\cdots,N_{ci}\}.
 \end{split}
\end{equation}

Here, $(\widetilde{\mathcal{V}}_{i\alpha})_{\mu}$ is an $N_{ci}\times N_{ci}$ Hermitian matrix. Therefore it can serve as a generator of $U(N_{ci})$. Note that this is distinct from a generator of the group $SU(N_{ci})$ in the complex double bifundamental model (the daughter theory). However, the difference in the number of generators between the two cases is suppressed by a factor of $\frac{1}{N_{ci}^2}$ compared to the total number of generators in either of them. Hence, this difference can be ignored while computing correlators in the planar limit \cite{Cherman:2010jj}.  With this caveat, let us go on treating $(\widetilde{\mathcal{V}}_{i\alpha})_{\mu}$  as a gauge field in the projected theory.
The field strength corresponding to this gauge field is given by
\begin{equation}
\begin{split} 
(\widetilde{\mathcal{F}}_{i\alpha})_{\mu\nu}
&=\partial_\mu(\widetilde{\mathcal{V}}_{i\alpha})_\nu-\partial_\nu(\widetilde{\mathcal{V}}_{i\alpha})_\mu -\frac{i g_i^{R}}{\sqrt{2}}\Big[(\widetilde{\mathcal{V}}_{i\alpha})_\mu, (\widetilde{\mathcal{V}}_{i\alpha})_\nu \Big],\\
\end{split} 
\end{equation}
where $g_{i}^{R}$ is the gauge coupling in the $i^{\text{th}}$ sector of the parent theory.

The Weyl fermions introduced above can be combined together to form Dirac fermions as shown below:
\begin{equation}
\begin{split}
&\Big(\tilde \Psi_{i}^{(q)}\Big)_{a_i} \equiv\begin{pmatrix}\Big(\tilde \Psi_{i,L}^{(q)}\Big)_{a_i} \\ \Big(\tilde \Psi_{i,R}^{(q)}\Big)_{a_i}\end{pmatrix},\ 
\Big(\tilde \chi_{i}^{(q)}\Big)_{j_i} \equiv\begin{pmatrix}\Big(\tilde \chi_{i,L}^{(q)}\Big)_{j_i} \\ \Big(\tilde \chi_{i,R}^{(q)}\Big)_{j_i}\end{pmatrix}.
 \end{split}
\end{equation}

Now writing the Lagrangian of the parent theory in terms of the gauge fields $(\widetilde{\mathcal{V}}_{i\alpha})_{\mu}$, the complex scalar fields $\tilde \Phi_i$, and the  Dirac fermions $(\tilde \Psi_{i}^{(q)},\tilde \chi_{i}^{(q)})$, we get
 \begin{equation}
\begin{split}
\mathcal{L}_{\text{RDB}}^{(\text{proj.})}=&-\frac{1}{2}\text{Tr}\Bigg[(\widetilde{\mathcal{F}}_{i\alpha})_{\mu\nu}(\widetilde{\mathcal{F}}_{i\alpha})^{\mu\nu}\Bigg]+i\sum_{i=1}^2 \Big(\overline{\tilde \Psi}_{i}^{(q)}\Big)_{a_i}\Big(\slashed{D}\tilde \Psi_{i}^{(q)}\Big)_{a_i}+i\sum_{i=1}^2 \Big(\overline{\tilde \chi}_{i}^{(q)}\Big)_{j_i}\Big(\slashed{D}\tilde \chi_{i}^{(q)}\Big)_{j_i}\\
&+\sum_{i=1}^2\text{Tr}\Bigg[(D_\mu \tilde \Phi_i)^\dag  D_\mu \tilde \Phi_i\Bigg]-2\sum_{i=1}^2\widetilde{h}_i^{R}\text{Tr}\Bigg[\tilde\Phi_i^\dag\tilde\Phi_i\tilde\Phi_i^\dag\tilde\Phi_i\Bigg]-4\sum_{i=1}^2\widetilde{f}_i^{R}\text{Tr}\Big[\tilde\Phi_i^\dag\tilde\Phi_i\Big]\text{Tr}\Big[\tilde\Phi_i^\dag\tilde\Phi_i\Big]\\
&-8\widetilde{\zeta}^{R}\text{Tr}\Big[\tilde\Phi_1^\dag\tilde\Phi_1\Big]\text{Tr}\Big[\tilde\Phi_2^\dag\tilde\Phi_2\Big],
\end{split}
\label{rdb:projected Lagrangian}
\end{equation}
where
\begin{equation}
\begin{split}
D_\mu \tilde \Phi_i &\equiv\partial_\mu\tilde\Phi_{i}-\frac{i g_i^{R}}{\sqrt{2}}(\widetilde{\mathcal{V}}_{i1})_{\mu}\tilde\Phi_{i}+\frac{ig_i^{R}}{\sqrt{2}}\tilde\Phi_{i} (\widetilde{\mathcal{V}}_{i2})_{\mu}^\dag,\\
D_\mu\tilde \Psi_{i}^{(q)} &\equiv \partial_\mu\tilde \Psi_{i}^{(q)} - \frac{ig_i^{R}}{\sqrt{2}}(\widetilde{\mathcal{V}}_{i1})_{\mu}\tilde \Psi_{i}^{(q)},\\
D_\mu\tilde \chi_{i}^{(q)} &\equiv \partial_\mu\tilde \chi_{i}^{(q)} - \frac{ig_i^{R}}{\sqrt{2}}(\widetilde{\mathcal{V}}_{i2})_{\mu}\tilde \chi_{i}^{(q)}.\\
\end{split}
\end{equation}

Finally,  to obtain  the Lagrangian of the daughter theory, we need to rescale the above Lagrangian by a factor of $\Gamma^{-1}$, where $\Gamma$ is the order of the discrete global symmetry group in the parent theory. This is necessary to account for contribution of projectors on the fields to the correlators of neutral operators in the parent theory\cite{Bershadsky:1998cb, Schmaltz:1998bg}. In our case, $\Gamma=2$ since the global symmetry group is $\mathbb Z_2$. To express this rescaled Lagrangian in terms of canonically normalized fields, let us rescale the fields appearing in \eqref{rdb:projected Lagrangian} as follows:
 \begin{equation}
\begin{split}
&(\widetilde{\mathcal{V}}_{i\alpha}^\prime)_\mu\equiv\frac{1}{\sqrt{2}}(\widetilde{\mathcal{V}}_{i\alpha})_\mu,\ 
\tilde \Psi_i^{\prime(q)}\equiv\frac{1}{\sqrt{2}}\tilde \Psi_i^{(q)},\ 
\tilde\chi_i^{\prime(q)}\equiv\frac{1}{\sqrt{2}}\tilde \chi_i^{(q)},\ 
\tilde \Phi_i^\prime\equiv\frac{1}{\sqrt{2}}\tilde \Phi_i.
\end{split}
\end{equation}
In terms of these rescaled fields, the Lagrangian of the daughter theory takes the following form:

 \begin{equation}
\begin{split}
\mathcal{L}_{\text{daughter}}=&-\frac{1}{2}\text{Tr}\Bigg[(\widetilde{\mathcal{F}}^\prime_{i\alpha})_{\mu\nu}(\widetilde{\mathcal{F}}^\prime_{i\alpha})^{\mu\nu}\Bigg]+i\sum_{i=1}^2 \Big(\overline{\tilde \Psi^\prime}_{i}^{(q)}\Big)_{a_i}\Big(\slashed{D}\tilde \Psi_{i}^{\prime(q)}\Big)_{a_i}+i\sum_{i=1}^2 \Big(\overline{\tilde \chi^\prime}_{i}^{(q)}\Big)_{j_i}\Big(\slashed{D}\tilde\chi_{i}^{\prime(q)}\Big)_{j_i}\\
&+\sum_{i=1}^2\text{Tr}\Bigg[(D_\mu \tilde \Phi_i^\prime)^\dag  D_\mu \tilde \Phi_i^\prime\Bigg]-4\sum_{i=1}^2\widetilde{h}_i^R\text{Tr}\Bigg[\tilde\Phi_i^{\prime\dag}\tilde\Phi_i^\prime\tilde\Phi_i^{\prime\dag}\tilde\Phi_i^\prime\Bigg]\\
&-8\sum_{i=1}^2\widetilde{f}_i^R\text{Tr}\Big[\tilde\Phi_i^{\prime\dag}\tilde\Phi_i^\prime\Big]\text{Tr}\Big[\tilde\Phi_i^{^\prime\dag}\tilde\Phi_i^\prime\Big]-16\widetilde{\zeta}^R\text{Tr}\Big[\tilde\Phi_1^{\prime\dag}\tilde\Phi_1^\prime\Big]\text{Tr}\Big[\tilde\Phi_2^{\prime\dag}\tilde\Phi_2^\prime\Big],
\end{split}
\end{equation}
where
\begin{equation}
\begin{split} 
(\widetilde{\mathcal{F}}^\prime_{i\alpha})_{\mu\nu}
&\equiv\partial_\mu(\widetilde{\mathcal{V}^\prime}_{i\alpha})_\nu-\partial_\nu(\widetilde{\mathcal{V}^\prime}_{i\alpha})_\mu -i g_i^R\Big[(\widetilde{\mathcal{V}^\prime}_{i\alpha})_\mu, (\widetilde{\mathcal{V}^\prime}_{i\alpha})_\nu \Big],\\
D_\mu \tilde \Phi_i^\prime &\equiv\partial_\mu\tilde\Phi_{i}^\prime - i g_i^R(\widetilde{\mathcal{V}^\prime}_{i1})_{\mu}\tilde\Phi_{i}^\prime + i g_i^R\tilde\Phi_{i}^\prime (\widetilde{\mathcal{V}^\prime}_{i2})_{\mu}^\dag,\\
D_\mu\Big(\tilde \Psi_{i}^{\prime(q)}\Big) &\equiv \partial_\mu\tilde \Psi_{i}^{\prime(q)} - i  g_i^R(\widetilde{\mathcal{V}^\prime}_{i1})_{\mu}\tilde \Psi_{i}^{\prime(q)},\\
D_\mu\Big(\tilde \chi_{i}^{\prime(q)}\Big) &\equiv \partial_\mu\tilde \chi_{i}^{\prime(q)} - i g_i^R(\widetilde{\mathcal{V}^\prime}_{i2})_{\mu}\tilde \chi_{i}^{\prime(q)}.\\
\end{split}
\end{equation}
Note that if we expand the field strengths $(\widetilde{\mathcal{F}}^\prime_{i\alpha})_{\mu\nu}$ in terms of generators $T_{i\alpha}^A$ normalized so that the corresponding second Dynkin index is $\frac{1}{2}$, then the Lagrangian given above has  the same structure as the Lagrangian of the complex double bifundamental model  in \eqref{cdb: lagrangian}. In fact, the two Lagrangians are exactly identical if we impose the following  relations between the couplings:
\begin{equation}
\begin{split}
g_i^{C}=g_i^{R},\ \tilde h_i^{C}= 4 \tilde h_i^{R},\ \tilde f_i^{C} =8 \tilde f_i^{R},\ \tilde \zeta^{C} =8 \tilde\zeta^{R} .
\end{split}
\label{planar equiv: coupling relations}
\end{equation}
where we have now introduced the superscript `C' to indicate that the couplings correspond to the complex double bifundamental model. 
From \eqref{planar equiv: coupling relations}, one can derive the relations between the 't Hooft couplings\footnote{Note that while defining the 't Hooft couplings in the parent theory, one has to take $N_{ci}\rightarrow 2 N_{ci}$ in the relations given in \eqref{dbm: 't Hooft couplings}.} in the two dual theories  to be
 \begin{equation}
\begin{split}
\lambda_i^{C}=\frac{\lambda_i^{R}}{2},\ h_i^{C}=2h_i^{R},\ f_i^{C}=2f_i^{R},\ \zeta^{C}=2\zeta^{R}.
  \end{split}
  \label{duality relations}
\end{equation}

As a consistency check, we provide the forms of the 2-loop beta functions of the gauge couplings and the 1-loop beta functions of the quartic couplings in the complex double bifundamental model at the Veneziano limit:
\beq \label{eq:su1}
\beta_{\lam_i^C}^{\text{2-loop}}&=-\left({21-4x_{fi} \ov 3}\right)(\lam_i^C)^2+\left({-54+26 x_{fi} \ov 3}\right) (\lam_i^C)^3,
\\ \beta_{h_i^C}^{\text{1-loop}}&=8(h_i^C)^2-12\lam_i^C h_i^C+{3\ov 2}(\lam_i^C)^2,
\\ \beta_{f_i^C}^{\text{1-loop}}&=4(f_i^C)^2+16 f_i^C h_i^C+12 (h_i^C)^2+4(\zeta^C)^2-12\lam_i^C f_i^C+{9\ov 2}(\lam_i^C)^2,
\\ \beta_{\zeta^C}^{\text{1-loop}}&=\zeta^C\sum_{i=1}^2\left( 4f_i^C+  8h_i^C-6\lam_i^C\right).
\eeq
These beta functions have been computed by  methods analogous to those illustrated for the real double bifundamental model in appendix \ref{app: rdb beta fns.}. One can compare these beta functions to those of the real double bifundamental model given in \eqref{rdb: gauge coupling planar beta fn} and \eqref{rdb: quartic coupling planar 1-loop beta fn}, and see that they are consistent with the relations given in \eqref{duality relations}.

From the planar equivalence between the two double bifundamental models, we can  obtain the unitary fixed points (with $\zeta^C\neq 0$) corresponding to the beta functions given above. As earlier, for the existence of such unitary fixed points, the ratios $x_{fi}\equiv \frac{N_{fi}}{N_{ci}}$ must satisfy the condition that $x_{f1}=x_{f2}\equiv x_f$. When this condition is satisfied, the unitary fixed points are all the points lying on the following manifold in the coupling space:
\beq
\lam_1^C=\lam_2^C=\lam^C\equiv \frac{21-4x_f}{-54+26x_f},\ h_1^C=h_2^C=h^C\equiv\Big(\frac{3-\sqrt{6}}{4}\Big)\lambda^C,\\
f_p^C\equiv\frac{f_1^C+f_2^C}{2}=\sqrt{\frac{3}{2}}\lambda^C,\ 
 (\zeta^C)^2+(f_m^C)^2= \Big(\frac{18\sqrt{6}-39}{16}\Big)(\lambda^C)^2\ ,
 \label{cdb: fixed points}
\eeq
where $f_m^C\equiv\frac{f_1^C-f_2^C}{2}$. As before, this conformal manifold has the topology of a circle. The orbifold equivalence with the real double bifundamental model ensures that this fixed circle would survive even after taking the 2-loop contributions to the planar beta functions into account. In fact, in section \ref{sec: survival of fixed circle}, we will prove that this fixed circle survives under all loop corrections at the planar limit.

\subsection{Analysis of thermal effective potential and symmetry breaking}
Let us now discuss the thermal effective potential of the scalar fields in the complex double bifundamental model for the fixed points given in \eqref{cdb: fixed points}. From here onwards, we will drop the superscript `C' over the couplings.

First we will check the stability of the potential at these fixed points. The quartic terms in the potential (up to leading order in $\lambda$)  are
\beq
V_{\text{quartic}}=\sum_{i=1}^2\widetilde{h}_i\text{Tr}\Big[\Phi_i^\dag\Phi_i\Phi_i^\dag\Phi_i\Big]+\sum_{i=1}^2\widetilde{f}_i\text{Tr}\Big[\Phi_i^\dag\Phi_i\Big]\text{Tr}\Big[\Phi_i^\dag\Phi_i\Big]+2\widetilde{\zeta}\text{Tr}\Big[\Phi_1^\dag\Phi_1\Big]\text{Tr}\Big[\Phi_2^\dag\Phi_2\Big].
\eeq
As in the case of the real double bifundamental model, the single trace interactions are manifestly positive because 
\beq
\tilde h_i=\frac{16\pi^2}{N_{ci}}h=\frac{16\pi^2}{N_{ci}}\Big(\frac{3-\sqrt{6}}{4}\Big)\lambda>0. 
\eeq
To ensure that the other terms are positive, we need to  check the positivity of $\tilde f_1$, $\tilde f_2$ and  $(\tilde f_1\tilde f_2-\tilde \zeta^2)$. But this is guaranteed by the relations in equation \eqref{duality relations}, and the positivity of the corresponding quantities in the dual real double bifundamental model.\footnote{See the proof of this in section \ref{subsubsec rdb:potential stability}.} Therefore, we can  conclude that the potential is stable for all points on the fixed circle.

Next, consider the quadratic terms in the potential which have the following form:
\beq
V_{\text{quadratic}}=\frac{1}{2}\sum_{i=1}^2 m_{\text{th},i}^2 \text{Tr}\Big[\Phi_i^\dag\Phi_i\Big],
\eeq
where the thermal masses (squared) are\footnote{These thermal masses are computed by methods analogous to those discussed in section \ref{subsec rdb: thermal masses}.}
\beq
m_{\text{th},1}^2=&16\pi^2 \beta_{\text{th}}^{-2}\Bigg[{1\ov3} h_1+{1\ov6}\Big(1+{1\ov N_{c1}^2}\Big) f_1+{1\ov6}{N_{c2}\ov N_{c1}} \zeta+{1\ov 4}\Big(1-{1\ov N_{c1}^2 }\Big)\lam_1 \Bigg],
\\
m_{\text{th},2}^2=&16\pi^2 \beta_{\text{th}}^{-2}\Bigg[{1\ov3} h_2+{1\ov6}\Big(1+{1\ov N_{c2}^2}\Big) f_2+{1\ov6}{N_{c1}\ov N_{c2}} \zeta+{1\ov 4}\Big(1-{1\ov N_{c2}^2 }\Big)\lam_2 \Bigg].
\eeq
In the Veneziano limit ($N_{ci}\rightarrow\infty$), we can drop all the subleading terms in $\frac{1}{N_{ci}}$ and substitute the couplings by their values at the fixed points (given in \eqref{cdb: fixed points}) to obtain
\beq
m_{\text{th},1}^2&
=\frac{16\pi^2}{3} \beta_{\text{th}}^{-2}\Bigg[{3\lam\ov2}+{f_m\ov2} +{r\zeta\ov2} \Bigg],
\ 
m_{\text{th},2}^2
=\frac{16\pi^2}{3} \beta_{\text{th}}^{-2}\Bigg[{3\lam\ov2}-{f_m\ov2} +{\zeta\ov2r} \Bigg],
\eeq
where $r\equiv\lim\limits_{N_{c1},N_{c2}\rightarrow \infty}\frac{N_{c2}}{N_{c1}}$. Note that  these thermal masses are equal to the corresponding thermal masses in the dual real double bifundamental model (see \eqref{rdb: fixed point thermal masses}) under the mapping between the couplings in the two models given in \eqref{duality relations}. This is actually a consequence of the perturbative orbifold equivalence between the  two models being valid in thermal states.  Since the proof of the perturbative planar (orbifold) equivalence \cite{Bershadsky:1998cb} between the parent  and the daughter theories  relies only on some combinatorics arising from the projection under a discrete automorphism, it can be extended to the finite temperature case without any major modification.\footnote{It would be interesting to see if the equivalence holds non-perturbatively for the double bifundamental models. See \cite{Kovtun:2003hr,Kovtun:2004bz, Armoni:2004uu, Unsal:2006pj} for the criteria for  such non-perturbative planar equivalences between various models.} This ensures   the equality of the  $1$-loop planar  diagrams \cite{Weinberg:1974hy,Dolan:1973qd}  contributing to the thermal masses in the two theories.

The equality of the thermal masses in the two dual theories allows us to just use the results obtained in section \ref{subsec: rdb symmetry breaking analysis} to investigate the conditions under which one of the thermal masses (squared) is negative. We quote these conditions below. 

First, without any loss of generality, we assume that $N_{c2}<N_{c1}$, i.e., $r<1$. Under this assumption, on one hand, $m_{\text{th},1}^2>0$ for all points on the fixed circle. On the other hand, $m_{\text{th},2}^2<0$ for the fixed points satisfying any one of the following conditions: 
\begin{enumerate}
\item  $ r\in  \Big[\frac{\sqrt{18\sqrt{6}-39}}{12}, \sqrt{\frac{6\sqrt{6}-13}{61-6\sqrt{6}}}\ \Big]$,  $\zeta\in (4z_0^{(2)}\lambda,4z_0^{(1)}\lambda)$, and $f_m >0$,
\item  $ r\in(0, \frac{\sqrt{18\sqrt{6}-39}}{12})$, $\zeta\in [4z_{\text{min}}\lambda,4z_0^{(1)}\lambda)$, and $f_m\geq 0$,
\item  $ r\in(0, \frac{\sqrt{18\sqrt{6}-39}}{12})$, $\zeta\in (4z_{\text{min}}\lambda,4z_0^{(2)}\lambda)$, and  $f_m<0$.
\end{enumerate}
Here $z_0^{(1)}$ and $z_0^{(2)}$ are the values defined in \eqref{zeros of thermal mass 2}, and $z_{\text{min}}$ is given in \eqref{value of zmin}. For the reader's convenience, we provide these values once more below:
\beq
&z_0^{(1)}=r\Bigg[\frac{-12+ \sqrt{\Big(18\sqrt{6}-183\Big)r^2+\Big(18\sqrt{6}-39\Big)}}{16\Big(1+r^2\Big)}\Bigg], \\
&z_0^{(2)}=r\Bigg[\frac{-12-\sqrt{\Big(18\sqrt{6}-183\Big)r^2+\Big(18\sqrt{6}-39\Big)}}{16\Big(1+r^2\Big)}\Bigg],\\
&z_{\text{min}}= -\frac{\sqrt{18\sqrt{6}-39}}{16}.
\eeq
The graphical plots of these fixed points are similar to those given in figure \ref{fig:RDB fixed points with thermal order}. The only difference is that the quantity $z$ in those figures has to be now defined as $z=\frac{\zeta}{4\lambda}$.

For all these fixed points, the thermal expectation values of the scalar fields (up to  gauge transformations) are as follows\footnote{The arguments for arriving at these expectation values are exactly analogous to those in appendix \ref{app: minimum of  potential} for the real double bifundamental model.}:
\beq
\langle \Phi_1\rangle=0,\ \langle \Phi_2\rangle=\Bigg(-\frac{N_{c2}}{64\pi^2}\frac{m_{\text{th},2}^2}{h_2+f_2}\Bigg)^{\frac{1}{2}}\ \text{diag}\{e^{i\theta},1, 1, \cdots,1\},
\label{cdb: thermal expectation values}
\eeq
where $\theta\in [0,2\pi)$. Note that in the zero temperature limit, we again have $\langle \Phi_2\rangle=0$ and both the baryon symmetries are unbroken. For any nonzero temperature, we have a circle of gauge inequivalent vacua each of which develops a nonzero thermal expectation value $\langle[\det \Phi_2]\rangle\neq 0$. This leads to the spontaneous breaking of the baryon symmetry in the second sector as in case of the real double bifundamental model. Moreover, the gauge symmetry in this sector is broken down to $SU(N_{c2})$. This can be seen by checking that a gauge transformation $U_1\times U_2$  belonging to $SU(N_{c2})\times SU(N_{c2})$ can leave the thermal expectation values $\langle \Phi_2\rangle$ invariant only if 
\beq
U_2=\langle\Phi_2\rangle^{-1} U_1 \langle\Phi_2\rangle.
\eeq
Thus, only one of the two unitary transformations is independent and the residual gauge symmetry is  $SU(N_{c2})$. As a result of this,  half of the generators in the second sector are broken, and accordingly, half of the gauge bosons in this sector get Higgsed. Therefore, we find that just as in case of the real double bifundamental model, the spontaneous breaking of the baryon symmetry and the persistent BEH phase are linked to each other in the complex double bifundamental model.

\section{Survival of the fixed circle at all orders in the planar limit}
\label{sec: survival of fixed circle}
In the previous two sections, we saw that the 2-loop planar beta functions in the real double bifundamental models yield conformal manifolds which have the topology of a circle. This intriguing result raises the question of whether such conformal manifolds survive under higher order contributions in the 't Hooft couplings to the planar beta functions. In this section we will show that it is indeed the case. This means that the double bifundamental models belong to the list of interesting non-supersymmetric theories which have conformal manifolds in the planar limit \cite{Bardeen:1983rv, Rabinovici:1987tf, Chai:2020zgq, Chai:2020onq}.

We will first prove a lemma in the following subsection that the planar beta functions in these models are independent of the ratio $r=\frac{N_{c2}}{N_{c1}}$. This ratio is the only quantity that could have led to an asymmetry in the forms of the planar beta functions of the couplings in the two sectors.\footnote{A difference in the values of $x_{f1}\equiv \frac{N_{f1}}{N_{c1}}$ and $x_{f2}\equiv \frac{N_{f2}}{N_{c2}}$ could also have led to an asymmetry between the two sectors. But we have already shown that in such a scenario, there is no unitary fixed point with $\zeta\neq 0$. Therefore, we keep assuming that $x_{f1}=x_{f2}$ in the $N_{c1}, N_{c2}\rightarrow\infty$ limit.}  The above-mentioned lemma ensures the absence of such an asymmetry in the planar limit.

While proving this lemma, we will illustrate the planar diagrams that contribute to connected correlators in the ground state of these models. To extract the planar beta  functions from such correlators we  work in  dimensional regularization\footnote{In dimensional regularization, we take the theories to be defined in $(4-\varepsilon)$ dimensions.}  and the $\overline{\text{MS}}$ scheme. In this scheme, the contributions to the beta functions come from the coefficients of the $O(\frac{1}{\varepsilon})$ terms in such connected correlators, i.e., the residues of these correlators corresponding to the pole at  $\varepsilon=0$.

One advantage of working with dimensional regularization is that all  diagrams with a massless tadpole\footnote{Here, by tadpole, we mean a subdiagram which has only one external vertex. It doesn't matter how this vertex is connected to the rest of the diagram. The loop integrals in such a subdiagram vanish in dimensional regularization when all the propagators in such loops are massless.} vanish  in this scheme\cite{kleinert2001critical}. The vanishing of such diagrams is not essential to the arguments that we will propose as such tadpoles can at most contribute to the renormalization of the masses \footnote{In fact, these tadpoles are responsible for the generation of the thermal masses of the scalar fields. In a  thermal state, the vanishing of such tadpoles is spoiled by the contributions of the Matsubara modes.} and they do not affect the beta functions in any mass-independent renormalization scheme. However, it simplifies the analysis slightly since we have taken the renormalized masses of all the fields in the double bifundamental models to be zero.

Before proceeding further, let us make a brief comment about our convention for representing different diagrams in this section. The fields in the double bifundamental models  are matrices, and hence an appropriate convention for drawing the corresponding Feynman diagrams would be 't Hooft's double line notation. Indeed, we  make use of this notation while discussing the large $N$ scaling of different diagrams in appendix \ref{app: scaling of diagrams}. However, in this section, we will represent the propagators of the scalar fields by single lines  to avoid unnecessary clutter. We hope the reader will not be confused by this slightly unconventional notation.

Now, without further ado, let us delve into the proof of the afore-mentioned lemma. As we will see in section \ref{sec:proof}, the insights gained while proving this lemma will be essential to the proof of the survival of the fixed circle in the planar limit.

\subsection{Lemma: planar beta functions are independent of $r=\frac{N_{c2}}{N_{c1}}$}
\label{subsec: r independence}

 In sections \ref{sec:rdb model} and \ref{sec: cdb model}, we found that the two-loop planar beta functions in the double bifundamental models are independent of the ratio $r=\frac{N_{c2}}{N_{c1}}$. Here, we will show that this feature of the planar beta functions actually persists at all orders in the 't Hooft couplings. 

Before proceeding with the argument, we note that the interactions in the double bifundamental models can be divided into two classes: single trace and double trace interactions. The couplings corresponding to the former are the gauge couplings ($g_{1}$ and $g_{2}$)\footnote{Note that each gauge coupling also covers a corresponding ghost coupling. Furthermore, gauge fixing terms do not affect the large $N$ counting since they only affect gauge propagators which are normalized to be $O(1)$.} and the quartic couplings $\tilde h_1$ and $\tilde h_2$, whereas those corresponding to the latter are the quartic couplings $\tilde f_1,\tilde f_2$ and $\tilde \zeta$. 

Now, let us consider the double bifundamental model as two initially decoupled bifundamental scalar QCDs coupled  through the double trace coupling $\tilde \zeta$. 
Note that $\tilde \zeta$ is the only coupling that mediates an interaction between the two bifundamental scalar QCDs. 
Hence, if the theory starts with $\tilde \zeta=0$ at any energy scale, it will be preserved along the RG flow. In this case, as we saw already, each bifundamental scalar QCD's large $N$ planar beta function would be determined solely by the corresponding 't Hooft couplings ($\lambda_i\propto N_{ci} g_i^2$, $h_i\propto N_{ci} \tilde h_i$, $f_i\propto N_{ci}^2\tilde f_i$).

Here, a question naturally arises: what happens to the structure of the beta functions after a non-zero double trace interaction $\tilde \zeta$ between the two sectors is turned on? Since each decoupled sector could have an independent large $N$ limit (i.e. $N_{c1}\neq N_{c2}$), one might wonder whether the planar beta functions of the  couplings would depend on the additional parameter $r=\frac{N_{c2}}{N_{c1}}$. We will now argue that this is not the case when the 't Hooft coupling $\zeta$ is defined with an appropriate normalization as $\zeta\propto N_{c1}N_{c2}\tilde \zeta$.

To prove the $r$-independence of the planar beta functions, we will use  the following feature of the connected planar diagrams that contribute to  such beta functions: 
{\it Any  double trace vertex in  such a diagram links two otherwise disconnected planar subdiagrams.}\footnote{See e.g. \cite{OBrien:1984hvc} which discusses general large $N$ scaling involving multi-trace vertices.} We have provided an argument for this in appendix \ref{app: scaling of diagrams} by identifying the connected diagrams that have dominant contributions in the large N limit. Here let us just explain what is meant by the above statement. Consider  a connected planar diagram with a double trace vertex. If we remove this vertex from the diagram and join the  pairs of legs with identical colors that were initially attached to it, then we are still left with a planar diagram. But according to the above statement, it is no longer a connected diagram. Rather, it has two disconnected pieces each of which is a connected planar subdiagram.

\begin{figure}[H]
\begin{subfigure}{.5\textwidth}
\centering
\begin{tikzpicture}
\draw[thick,color=red] (-1,0) -- (1,0);
\draw[red, pattern=north east lines, pattern color=red] (0, 0.5) circle (0.5);
\end{tikzpicture}
\caption{}
\label{subfig:2-point tadpole diagram a}
\end{subfigure}
\begin{subfigure}{.5\textwidth}
\centering
\begin{tikzpicture}
\draw[thick,color=red] (-1,0) -- (1,0);
\draw[blue, pattern=north east lines, pattern color=blue] (0, 0.5) circle (0.5);
\end{tikzpicture}
\caption{}
\label{subfig:2-point tadpole diagram b}
\end{subfigure}
\caption{Examples of planar diagrams with 2 external legs and a massless tadpole: Here we use red color for components in the first sector, and  blue color for the same in the second sector.  In both the diagrams, the color indices of the two external propagators  are the same.  The blobs comprise of  single trace vertices and  propagators belonging to the respective sector. These blobs are massless tadpoles.}
\label{fig:2-point tadpole diagrams}
\end{figure}
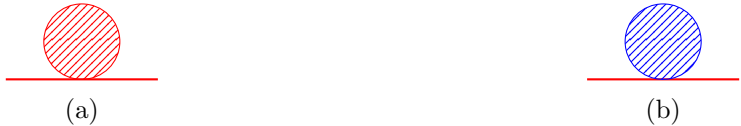

Now, let us see how this feature of connected planar diagrams constrains the form of the planar beta functions. Note that these beta functions receive contributions from both vertex renormalizations and wave function renormalizations.\footnote{We note that there is a scheme-independent way to understand the reason why tadpole diagrams do not contribute to the wave function renormalization from their external momentum independence.} Let us first discuss why the wave function renormalizations do not lead to a dependence of the planar beta functions on the ratio $r$. The contributions to such wave function renormalizations come from connected planar diagrams with two external legs. If such a planar diagram contains a double trace vertex, then as we just argued, it can be decomposed into two disconnected pieces. But one of these disconnected piece would always be a tadpole (eg. see the diagrams in figure \ref{fig:2-point tadpole diagrams}). As we mentioned earlier, such tadpoles contribute only to the renormalization of masses of the attached propagator. But since all the propagators here are massless, these diagrams simply vanish. Therefore, there is no planar diagram with a double trace vertex which contributes to the wave function renormalizations. This means that in the planar limit, these  wave function renormalizations  do not depend on the coupling $\zeta$ which connects the two sectors and they are just functions of the single trace couplings in the respective sectors. Hence, their contributions to the beta functions do not generate any $r$-dependence.

Now, let us turn our attention to the vertex renormalizations of the different couplings. First, consider the single trace couplings $\lambda_i$ and $h_i$. We will show that, just like the wave function renormalizations, the vertex renormalizations of these single trace couplings are independent of the double trace couplings in the planar limit. This will, in turn, imply that the planar beta functions of the single trace couplings are totally independent of the double trace couplings\footnote{This statement is a well-known fact for such large N gauge theories (see for example, \cite{Dymarsky:2005uh,Kiritsis:2008at}). It can be generalized to  cases involving multi-trace couplings.}. To prove this, let us work with a specific example, viz., the planar vertex renormalization of the coupling $h_1$. The structure of this vertex renormalization can be understood from connected diagrams with 4 external legs of the scalar field $\Phi_1$. We take the color indices corresponding to these external legs to be $(ai,bj,aj,bi)$ with $a\neq b$ and $i\neq j$. Now, for these external legs, if there is a connected planar diagram containing a double trace vertex, then by the arguments given above we can split it into two disconnected pieces by removing this vertex and joining the residual legs with identical colors. The only way to get such a diagram without generating a vanishing tadpole  is to let each of the disconnected pieces have a pair of external legs. But then the two external legs for either of the two disconnected pieces would have different color indices which is impossible given the structure of the interaction vertices. This means that there cannot be any connected planar diagram with double trace vertices for the external legs that we specified above. Consequently, the planar vertex renormalization of $h_1$ does not receive any contribution from the double trace couplings. Similar arguments can be given to show that the above property is shared by the vertex renormalizations of the other single trace couplings. Together with the same feature of the wave function renormalizations, this conclusively demonstrates that the planar beta functions of the single trace couplings are independent of the double trace couplings. As a byproduct of this, we see that for a fixed $i$, $\be_{\lambda_{i}}$ and $\be_{h_{i}}$ depend only on $\lambda_{i}$ and $h_i$ in the planar limit because there is no connected planar graph involving the coupling $\zeta$ which couples the two sectors. In other words, the planar beta functions of the single trace couplings are only generated from the single trace couplings in the same sector, which proves their $r$-independence.

\begin{figure}[H]
\centering
\scalebox{0.8}{\begin{tikzpicture}
\draw[thick,color=red] (-2,1) -- (-1,0);
\draw[thick,color=red] (-2,-1) -- (-1,0);
\draw[thick,color=red] (2,1) -- (1,0);
\draw[thick,color=red] (2,-1) -- (1,0);
\draw[thick, red, snake it] (0,0) circle (1);
\node[] at (-2.5,1.2) {$(ai)$};
\node[] at (-2.5,-1.2) {$(ai)$};
\node[] at (2.5,1.2) {$(bj)$};
\node[] at (2.5,-1.2) {$(bj)$};
\node[] at (0,1.4) {$(ij)$};
\node[] at (0,-1.4) {$(ij)$};
\end{tikzpicture}}
\caption{Example of a diagram containing only single trace vertices contributing to the vertex renormalization of $f_1$: The solid lines represent propagators of the scalar field $\Phi_1$, whereas the wavy lines represent propagators of the gauge field $V_{12}$. The color indices for the different propagators are indicated in brackets adjacent to the respective propagators. we take the color indices on the two sides to be distinct, i.e., $a\neq b$,\ $i\neq j$. Each of the two vertices contributes a factor of $g_1^2\propto \frac{\lambda_1}{N_{c1}}$. Thus, the diagram scales as $O(\frac{1}{N_{c1}^2})$ in the large $N_{c1}$ limit.}
\label{fig: diagram with single trace couplings contributing to Zf1}
\end{figure}
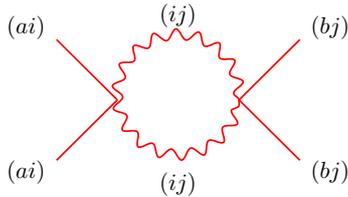

\begin{figure}[H]
\begin{subfigure}{.5\textwidth}
\centering
\begin{tikzpicture}
\draw[thick,color=red] (-1.5,-1) -- (-0,0);
\draw[thick,color=red] (-1.5, 1) -- (-0,0);
\draw[thick,color=red] (1.5,-1) -- (0,0);
\draw[thick,color=red] (1.5,1) -- (0,0);
\draw[red, fill=red,] (0,0) circle (0.5);
\end{tikzpicture}
\caption{Diagrams in the first sector}
\end{subfigure}
\begin{subfigure}{.5\textwidth}
\centering
\begin{tikzpicture}
\draw[thick,color=blue] (-1.5,-1) -- (-0,0);
\draw[thick,color=blue] (-1.5, 1) -- (-0,0);
\draw[thick,color=blue] (1.5,-1) -- (0,0);
\draw[thick,color=blue] (1.5,1) -- (0,0);
\draw[blue, fill=blue,] (0,0) circle (0.5);
\end{tikzpicture}
\caption{Diagrams in the second sector}
\end{subfigure}
\caption{Diagrams without double trace vertices contributing to the vertex renormalizations of $f_1$ and $f_2$: In both the diagrams, the pair of external legs on the left have the same color indices. Similarly,  the pair of external legs on the right have the same color indices. The blobs contain only single trace vertices and propagators belonging to a particular sector.}
\label{fig: general form of diagrams with single trace vertices contributing to Zfi}
\end{figure}
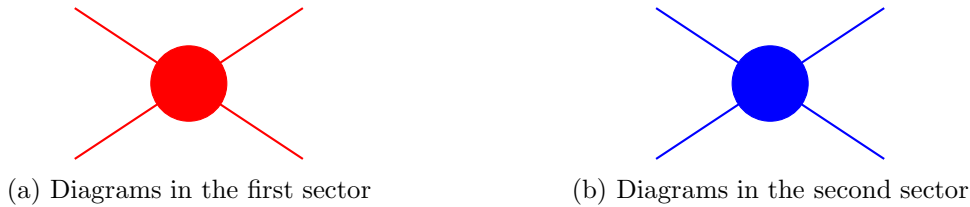

Hence, what remains to be proven is the $r$-independence of the vertex renormalizations for the double trace  couplings $f_i$ and $\zeta$. 
To demonstrate this, let us first consider the planar diagrams contributing to the vertex renormalizations of $f_1$ and $f_2$ which do not contain any double trace vertex. An example of such a diagram is shown in figure \ref{fig: diagram with single trace couplings contributing to Zf1}. 
Such a diagram can be generally expressed as a blob which contains only single trace vertices and propagators belonging to a particular sector\footnote{The blob can also contain counterterm vertices corresponding to the renormalization of the single trace vertices and the wave function renormalization of the fields in the same sector.} and is connected to the external legs as shown in figure \ref{fig: general form of diagrams with single trace vertices contributing to Zfi}. 
We will denote the set of all such blobs in the $i^{\text{th}}$ sector by $B_i^0$. All  diagrams in $B_i^0$ scale as $O(\frac{1}{N_{ci}^2})$ in the large $N_{ci}$ limit.\footnote{We refer the reader to appendix \ref{app: scaling of diagrams} for an analysis of the large N scaling of such diagrams.} Note that these diagrams do not contain the double trace coupling $\zeta$ which mixes the two sectors. Hence, as earlier, their contributions do not lead to an $r$-dependence in the planar vertex renormalizations of $f_1$ and $f_2$.

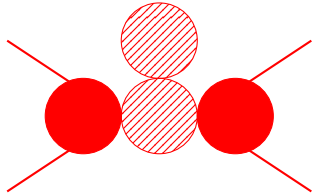
\begin{figure}[H]
\centering
\begin{tikzpicture}
\draw[thick,color=red] (-2,1) -- (-0.5,0);
\draw[thick,color=red] (-2,-1) -- (-0.5,0);
\draw[thick,color=red] (2,1) -- (0.5,0);
\draw[thick,color=red] (2,-1) -- (0.5,0);
\draw[red, pattern=north east lines, pattern color=red] (0,0) circle (0.5);
\draw[red, pattern=north east lines, pattern color=red] (0,1.0) circle (0.5);
\draw[red, pattern=north east lines,fill=red,] (-1,0) circle (0.5);
\draw[red, pattern=north east lines, fill=red] (1,0) circle (0.5);
\end{tikzpicture}
\caption{Example of a diagram with 4 external legs and a massless tadpole: The external and internal blobs are characterized by filled or patterned color respectively. The color indices in the two external propagators on each side are the same. The blobs stand for connected diagrams with propagators and single trace vertices belonging to the first sector. The vertices at which the external blobs connect to the lower internal blob and the vertex connecting the two internal blobs correspond to the double trace coupling $\tilde f_1$. The internal blob on the top is a massless tadpole.}
\label{fig:wave fn. renorm.}
\end{figure}

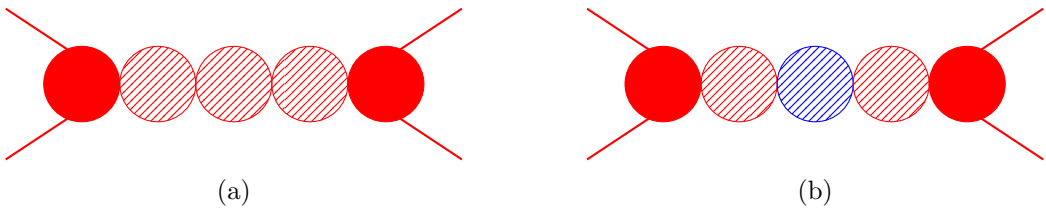
\begin{figure}[H]
\begin{subfigure}{.5\textwidth}
\centering
\begin{tikzpicture}
\draw[thick,color=red] (-3,1) -- (-1.5,0);
\draw[thick,color=red] (-3,-1) -- (-1.5,0);
\draw[thick,color=red] (3,1) -- (1.5,0);
\draw[thick,color=red] (3,-1) -- (1.5,0);
\draw[red, pattern=north east lines, pattern color=red] (-1, 0) circle (0.5);
\draw[red, pattern=north east lines, pattern color=red] (0, 0) circle (0.5);
\draw[red, pattern=north east lines, pattern color=red] (1, 0) circle (0.5);

\draw[red, pattern=north east lines, fill=red] (2, 0) circle (0.5);
\draw[red, pattern=north east lines, fill=red] (-2, 0) circle (0.5);
\end{tikzpicture}
\caption{}
\label{subfig:linear chain of blobs a}
\end{subfigure}
\begin{subfigure}{.5\textwidth}
\centering
\begin{tikzpicture}
\draw[thick,color=red] (-3,1) -- (-1.5,0);
\draw[thick,color=red] (-3,-1) -- (-1.5,0);
\draw[thick,color=red] (3,1) -- (1.5,0);
\draw[thick,color=red] (3,-1) -- (1.5,0);
\draw[red, pattern=north east lines, pattern color=red] (-1, 0) circle (0.5);
\draw[blue, pattern=north east lines, pattern color=blue] (0, 0) circle (0.5);
\draw[red, pattern=north east lines, pattern color=red] (1, 0) circle (0.5);
\draw[red, pattern=north east lines, fill=red] (2, 0) circle (0.5);
\draw[red, pattern=north east lines, fill=red] (-2, 0) circle (0.5);
\end{tikzpicture}
\caption{}
\label{subfig:linear chain of blobs b}
\end{subfigure}
\caption{Examples of linear chains of blobs contributing to the vertex renormalization of a double trace coupling (here $f_1$).}
\label{fig:linear chain of blobs}
\end{figure}

Now, let us determine the planar diagrams containing double trace vertices that contribute to the vertex renormalizations of the double trace couplings. For this, recall our  observation that each double trace vertex in a connected planar diagram links two otherwise disconnected planar subdiagrams. Note that if such a disconnected piece is not attached to any of the external legs, then it is a massless tadpole which vanishes (eg. see figure \ref{fig:wave fn. renorm.}). Therefore, the only connected planar diagrams contributing to the vertex renormalizations of the double trace couplings are linear chains of planar subdiagrams  connected through double trace vertices as shown in figure \ref{fig:linear chain of blobs}. Each of these subdiagrams contains only single trace vertices and propagators belonging to a particular sector. It may also include counterterm vertices corresponding to the wave function renormalization of the  fields and the  renormalizations of the single trace vertices in the same sector.\footnote{As we have already argued, such wave function and vertex renormalizations can depend only on the single trace couplings in the respective sector.} The vertices connecting such subdiagrams can also be counterterm vertices corresponding to the double trace couplings. A lower order term in the  perturbative expansion of such counterterm vertices can contribute to a higher order correction to the vertex renormalization. 

Here, let us make an important observation: there are two classes of planar subdiagrams in such a chain. One of them consists of the planar subdiagrams at the two ends, each of which is attached to two external legs as well as a double trace vertex which connects it to the rest of the diagram. It includes the case where the two external legs directly connect to the double trace vertex. The other class consists of the planar subdiagrams in the middle, each of which is connected to two double trace vertices on the two sides. From now on, we will call these two classes of planar subdiagrams `external blobs' and `internal blobs' respectively.  So there are four distinct types of blobs which we denote by $B^E_1,B^I_1,B^E_2, B^I_2$. Here the superscript indicates whether the blob is external or internal,  and the subscript indicates the sector to which the corresponding propagators and vertices belong. An important point to note here is that each internal blob in $B_i^I$ contributes  a factor of $O(N_{ci}^2)$ to the planar graph, whereas the contribution of each external blob is $O(1)$.\footnote{See appendix \ref{app: scaling of diagrams} for a derivation of the ways in which different diagrams scale in the large $N$ limit.}  From these scalings of the different blobs in the $N_{ci}\rightarrow\infty$ limit along with the property that $f_i$ only connects $B_i^*$ and $B_i^*$ while $\zeta$ only connects $B_1^*$ and $B_2^*$, we can  infer that any planar graph contributing to the vertex renormalization of a double trace quartic coupling does not generate an $r$-dependence. To illustrate this, let us consider the two diagrams in figure \ref{fig:linear chain of blobs}. Both these diagrams contribute to the vertex renormalization of the coupling $f_1$. The diagram in figure \ref{subfig:linear chain of blobs a} consists only of vertices belonging to the first sector. In figure \ref{subfig:linear chain of blobs b}, two of the vertices with the double trace coupling $\tilde f_1$ are replaced by vertices with the coupling $\tilde \zeta$. In addition, the $B_1^I$ blob in the middle is replaced by a $B_2^I$ blob\footnote{Here, let us remark that the external blobs belonging to $B_1^E$ in such a diagram cannot be replaced by  blobs belonging to $B_2^E$ without changing the external legs as well.}. Since the double trace couplings $\tilde f_1$ and $\tilde \zeta$ scale as $\tilde f_1\propto \frac{f_1}{N_{c1}^2}$ and $\tilde \zeta\propto \frac{\zeta}{N_{c1} N_{c2}}$, the replacement of the two vertices leads to an additional factor of $(\frac{N_{c1}}{N_{c2}})^2$ in the diagram shown in figure \ref{subfig:linear chain of blobs b} compared to the one in figure \ref{subfig:linear chain of blobs a}. However, this factor is exactly cancelled by the replacement of the $B_1^I$ blob  by the $B_2^I$ blob. Therefore, we can conclude that the scaling of the two diagrams in the limit $N_{c1}, N_{c2}\rightarrow\infty$ limit is exactly similar, and there is no relative factor of $r=\frac{N_{c2}}{N_{c1}}$ between them. Similar arguments can be applied to other planar diagrams contributing to the vertex renormalizations of the double trace quartic couplings. Consequently, all these planar vertex renormalizations are independent of $r$. Coupled with the same feature of the wave function renormalizations, this proves the $r$-independence of the planar beta functions of all the double trace quartic couplings. This completes our proof of the lemma.

Finally, we remark that one can  extend this lemma to  the case of a general weakly coupled QFT which consists of two originally decoupled 3+1 dimensional QFTs, $T_1$ and $T_2$, deformed by an arbitrary number of quartic double trace scalar operators $\Tr[\Phi_{1}^\dag\Phi_1]\Tr[\Phi_{2}^\dag\Phi_2]$ where the scalar $\Phi_{i}$ belongs to the sector $T_i$.

\subsection{Proof of the survival of the fixed circle in the planar limit} \label{sec:proof}

Let us now prove that the fixed circle survives at all orders of the 't Hooft couplings in the planar limit. The fact that the planar beta functions are independent of the ratio  $r=\frac{N_{c2}}{N_{c1}}$  ensures that  the  beta functions of the  couplings $\lambda_i,\ h_i$ and $f_i$ in the two sectors are related to each other in the planar limit by the following  exchanges:
\beq
\lambda_1\leftrightarrow\lambda_2,\ h_1\leftrightarrow h_2, \ f_1\leftrightarrow f_2.
\eeq
Moreover, as we argued in the previous subsection, the planar beta functions of the single trace couplings ($\lambda_i$ and $h_i$) do not receive contributions from the double trace couplings ($f_i$ and $\zeta$). This guarantees that the RG flows of the single trace couplings in one sector do not depend on the couplings in the other sector.\footnote{The mixing of the two sectors can happen only when the double trace coupling $\zeta$ contributes to the beta functions.} Therefore, the planar beta functions of these single trace couplings take the following forms:
\beq
&\beta_{\lam_i}=\beta_\lam(\lam_i, h_i),\ \beta_{h_i}=\beta_h(h_i,\lam_i).
\label{planar beta functions of single trace couplings}
\eeq
As we have already seen from the 1-loop and 2-loop beta functions, there is a discrete set of solutions of the equations $\beta_\lam(\lam_i, h_i)=\beta_h(h_i,\lam_i)=0$, and only one of these solutions corresponds to unitary fixed points where the two sectors are coupled. At this solution, the single trace couplings in the two sectors are equal, i.e., we have\footnote{$\lambda_0$ and $h_0$  receive corrections from higher loop diagrams to their magnitudes determined from the 2-loop beta functions in the previous sections.} 
\beq
\lam_1=\lam_2=\lam_0, \ h_1=h_2=h_0.
\label{fixed point of single trace couplings at all orders}
\eeq 
We assume that within the domain of validity of perturbation theory, this solution survives at all orders in the 't Hooft couplings. Note that this is consistent with the identical forms of the beta functions of the single trace couplings in the two sectors as shown in \eqref{planar beta functions of single trace couplings}.  

 Next, we will determine the planar beta functions of the double trace couplings in the subspace where $\lambda_i$ and $h_i$ are fixed at their values given in \eqref{fixed point of single trace couplings at all orders}. In appendix \ref{app: double trace beta fns.}, we have shown how these planar beta functions can be expressed in terms of the corresponding wave function and vertex renormalizations. In particular, we have argued there that these beta functions depend only on the wave function and vertex renormalizations in a region where  $\lambda_1=\lambda_2=\lambda$ and $h_1=h_2=h$ with $(\lambda, h)$ lying in the neighborhood of $(\lambda_0,h_0)$. So henceforth, all our discussions will be restricted to such a region in the space of couplings. Let us now look at the forms of the relevant planar vertex and wave function renormalizations in the above-mentioned region. 

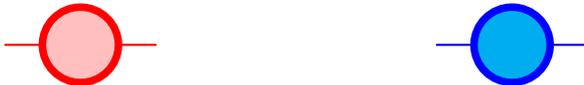
\begin{figure}[H]
\begin{subfigure}{.6\textwidth}
\centering
\begin{tikzpicture}
\draw[thick,color=red] (-1,0) -- (-0.5,0);
\draw[thick,color=red] (0.5,0) -- (1,0);
\draw[red,line width=1mm, fill=pink] (0,0) circle (0.5);
\end{tikzpicture}
\end{subfigure}
\begin{subfigure}{.1\textwidth}
\centering
\begin{tikzpicture}
\draw[thick,color=blue] (-1,0) -- (-0.5,0);
\draw[thick,color=blue] (0.5,0) -- (1,0);
\draw[blue,line width=1mm, fill= cyan] (0,0) circle (0.5);
\end{tikzpicture}
\end{subfigure}
\caption{Blobs contributing to wave function renormalization for $\Phi_1$ and $\Phi_2$.  Each blob is made only of single trace couplings and propagators belonging to the corresponding sector.}
\label{fig:wfn}
\end{figure}

First let us discuss the  wave function renormalizations $Z_{\Phi_1}$ and $Z_{\Phi_2}$ of the scalar fields $\Phi_1$ and $\Phi_2$.  These are determined by demanding that the sums over the planar diagrams of the form shown in figure \ref{fig:wfn} are free of divergences in the $\varepsilon\rightarrow 0$ limit. As we discussed in the previous subsection, these diagrams contain only propagators, single trace vertices and the corresponding counterterm vertices. Since we are probing a region in the coupling space where the single trace couplings in the two sectors are identical, the sums over such diagrams are also the same for the two sectors. Thus, the wave function renormalizations of the scalar fields in the two sectors are equal, i.e., we have
\beq
Z_{\Phi_1}=Z_{\Phi_2}\equiv Z_\Phi,
\eeq
where $Z_\Phi$ is a function of $\lambda$ and $h$.

Next, let us turn our attention to the vertex renormalizations of the double trace couplings. To determine these vertex renormalizations, we will consider the  contributions of different planar diagrams with two external legs on each side as shown in figures \ref{fig: general form of diagrams with single trace vertices contributing to Zfi} and \ref{fig:linear chain of blobs}.\footnote{While considering the sum over such diagrams, we will ignore the contributions of the external legs.} The diagrams in figure \ref{fig: general form of diagrams with single trace vertices contributing to Zfi}  contribute only to the vertex renormalizations of $f_1$ and $f_2$. They contain blobs comprising only of  single  trace vertices and propagators belonging to a particular sector. In the previous subsection we denoted the set of all such blobs in the $i^{\text{th}}$ sector by $B_i^0$. All the diagrams in $B_i^0$ are $O(\frac{1}{N_{ci}^2})$ in the large $N_{ci}$ limit. Let us denote the sum over the diagrams in $B_i^0$  by $C_i^0$. Since the quantity $C_i^0$ receives contributions only from the single trace couplings in the $i^{\text{th}}$ sector and we are considering a region where these single trace couplings are the same for the two sectors, the dependence of $C_i^0$ on the respective sector is solely determined by its overall scaling with $N_{ci}$, i.e., we have
\beq
C_i^0=\frac{C^0}{N_{ci}^2},
\eeq  
where $C^0$ is a function of $\lambda$ and $h$.\footnote{It also depends on the momenta of the external legs.} 

Now let us consider the  class of diagrams shown in figure \ref{fig:linear chain of blobs}. These diagrams are  linear chains of blobs where the two adjacent blobs are connected by double trace vertices or the corresponding counterterm vertices. Each of these blobs is a connected diagram made of propagators and single trace vertices belonging to a single sector. It can also contain counterterm vertices corresponding to the wave function renormalization of the scalar fields and the renormalizations of the single trace vertices belonging to the same sector. There are four distinct classes of such blobs: $B_1^{I}$, $B_1^{E}$, $B_2^{I}$ and $B_2^{E}$. 
Since each blob's loop integrals are independent of the other blobs, the total contribution of a given chain of blobs would be the product of the contributions from the individual blobs. This factorization  implies that a building block of the planar vertex renormalizations is an infinite sum over all possible planar diagrams within a given blob $B_i^*$. Let us denote this sum by $C_i^*$. Note that all the components appearing in the diagrams contributing to $C_i^*$ depend only on the single trace couplings in the $i^\text{th}$ sector. Since we are considering a region where the single trace 't Hooft couplings $\lambda_i$ and $h_i$ are the same in the two sectors, the dependence of the quantity $C_i^*$ on the sector to which it belongs comes only from its overall scaling with $N_{ci}$, i.e., we have\footnote{Here $C^E$ and $C^I$ depend on the values of $\lambda$ and $h$. $C^I$ also depends on the momentum flowing through the linear chain, while $C^E$  depends on the momenta of the two external legs to which the external blob is attached.}
\beq \label{eq:sumblob}
\quad C_i^E=C^E, \quad C_i^I=N_{ci}^2 C^I.
\eeq
Here we have used the fact that the external blob $B_i^E$ consists of connected planar diagrams with two external legs which scale like $O(1)$ while the internal blob $B_i^I$ consists of closed connected planar diagrams which scale like $O(N_{ci}^2)$ as $N_{ci}\rightarrow\infty$. We will call $C_i^{E,I}$  `building blobs' for obvious reasons.

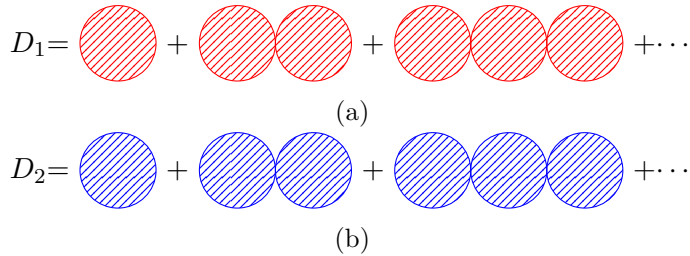
\begin{figure}[H]
\begin{subfigure}{.8\textwidth}
\centering
$D_1$=
\raisebox{-0.4cm}{\begin{tikzpicture}
\draw[red, pattern=north east lines, pattern color=red] (0, 0) circle (0.5);
\end{tikzpicture}}
+
\raisebox{-0.4cm}{\begin{tikzpicture}
\draw[red, pattern=north east lines, pattern color=red] (-0.5, 0) circle (0.5);
\draw[red, pattern=north east lines, pattern color=red] (0.5, 0) circle (0.5);
\end{tikzpicture}}
+
\raisebox{-0.4cm}{\begin{tikzpicture}
\draw[red, pattern=north east lines, pattern color=red] (-1, 0) circle (0.5);
\draw[red, pattern=north east lines, pattern color=red] (0, 0) circle (0.5);
\draw[red, pattern=north east lines, pattern color=red] (1, 0) circle (0.5);
\end{tikzpicture}}
+$\cdots$
\caption{}
\end{subfigure}
\\
\begin{subfigure}{.8\textwidth}
\centering
$D_2$=
\raisebox{-0.4cm}{\begin{tikzpicture}
\draw[blue, pattern=north east lines, pattern color=blue] (0, 0) circle (0.5);
\end{tikzpicture}}
+
\raisebox{-0.4cm}{\begin{tikzpicture}
\draw[blue, pattern=north east lines, pattern color=blue] (-0.5, 0) circle (0.5);
\draw[blue, pattern=north east lines, pattern color=blue] (0.5, 0) circle (0.5);
\end{tikzpicture}}
+
\raisebox{-0.4cm}{\begin{tikzpicture}
\draw[blue, pattern=north east lines, pattern color=blue] (-1, 0) circle (0.5);
\draw[blue, pattern=north east lines, pattern color=blue] (0, 0) circle (0.5);
\draw[blue, pattern=north east lines, pattern color=blue] (1, 0) circle (0.5);
\end{tikzpicture}}
+$\cdots$
\caption{}
\end{subfigure}
\caption{Diagrammatic expansions of $D_1$ and $D_2$: The red blobs correspond to sector 1, while the blue blobs correspond to sector 2.}
\label{fig:D1 and D2 diagrams}
\end{figure}

Now, consider a geometric sum of linear chains of internal building blobs in a given sector connected by the double trace coupling $\tilde f_i$ in that sector (see figure \ref{fig:D1 and D2 diagrams}). We denote these sums for the two sectors by $D_1$ and $D_2$. For convenience, we  absorb the contributions of the counterterms in the vertices corresponding to the double trace couplings. Therefore, the coupling $\tilde f_1$ and $\tilde f_2$ in these sums are multiplied by the corresponding vertex renormalizations $Z_{f_1}$ and $Z_{f_2}$.
Based on this diagrammatic expansion, we can  see that the quantity $D_i$ has the following value\footnote{Here let us make a brief comment about the symmetry factors of the different diagrams in this expansion which will also hold for the expansions given in \eqref{4-point correlators for double trace vertex}. The symmetry factors in these diagrams mostly arise due to the invariance under exchanges of vertices and propagators within individual blobs. Such symmetry factors are absorbed in the definition of the quantity $C^I$ and $C^E$, and we need not worry about them while connecting the different blobs via double trace vertices. The  exceptions to the above statement are diagrams where all the four legs emanating from a double trace vertex are identical. But the contributions of such diagrams are suppressed by powers of  $\frac{1}{N_{ci}}$, and hence they can be ignored in the planar limit.}:
\beq
D_i=C_i^I\sum_{n=0}^\infty\Big[\alpha Z_{f_i}\tilde f_i C_i^I\Big]^n=\frac{C_i^I}{1-\alpha Z_{f_i}\tilde f_i   C_i^I}=\frac{N_{ci}^2 C^I}{1-\alpha Z_{f_i}\tilde f_i N_{ci}^2  C^I}=\frac{N_{ci}^2 C^I}{1-16\pi^2\alpha  Z_{f_i} f_i  C^I},
\label{expression of D_i}
\eeq
where $\alpha$ is a normalization associated with the vertex factor of the couplings $\tilde f_i$ and $\tilde \zeta$. Its exact value is not essential to the arguments that will follow and hence we will leave it unspecified. The only thing that we need to keep in mind is that this normalization is the same for all the double trace couplings which follows from the  way they appear in the Lagrangians of the two double bifundamental models. We  absorb the dimensionful factor $\tilde \mu^\varepsilon$ in the definition of $\alpha$. The energy scale $\tilde \mu$ is related to the renormalization scale $\mu$ in the $\overline{\text{MS}}$ scheme by $\tilde \mu^2=\mu^2 \frac{e^{\gamma_E}}{4\pi}$ with $\gamma_E$ being the Euler-Mascheroni constant. The quantity $C^I$ has mass-dimension $(-\epsilon)$ which can be verified by checking the dimension of the Feynman diagrams appearing in the internal blobs. This leads to $(\alpha C^I)$ being dimensionless.

Now that we have obtained the quantities $D_1$ and $D_2$, we can evaluate the contributions of all planar diagrams with a fixed pair of external legs on the two sides.\footnote{The color indices for the two legs on each side are identical.} Such diagrams are of two types. One class of diagrams do not involve double trace vertices as shown in figure \ref{fig: general form of diagrams with single trace vertices contributing to Zfi} and contribute to  correlators where all the external legs correspond to fields belonging to the same sector. The other class  comprises of diagrams where alternating linear sequences  of $D_1$ and $D_2$ are coupled by the double trace vertex $\tilde \zeta$ (or the corresponding counterterm vertex) internally and connected to an appropriate external building blob $C_i^E$ at each end. Just as in case of the vertices with the coupling $\tilde f_i$, we  absorb the counterterm vertices for $\tilde \zeta$ by multiplying the vertex renormalization $Z_\zeta$ to the vertices connecting $D_1$ and $D_2$.
Now, summing over all such diagrams, we get
\beq
&\Gamma_{11}^{(4)}=\frac{C^0}{N_{c1}^2}+(C^E)^2 \Bigg\{\alpha Z_{f_1}\tilde f_1+\alpha^2 Z_{f_1}^2\tilde f_1^2 D_1\\
&\qquad\qquad\qquad\qquad\quad+Z_\zeta^2 (1+\alpha Z_{f_1}\tilde f_1 D_1)^2\alpha\tilde \zeta D_2 \alpha\tilde \zeta\Bigg(\sum_{n=0}^\infty (Z_\zeta^2 D_1\alpha\tilde \zeta D_2 \alpha\tilde \zeta)^n  \Bigg)\Bigg\},\\
&\Gamma_{22}^{(4)}=\frac{C^0}{N_{c2}^2}+(C^E)^2 \Bigg\{ \alpha Z_{f_2}\tilde f_2+\alpha^2 Z_{f_2}^2\tilde f_2^2 D_2\\
&\qquad\qquad\qquad\qquad\quad+Z_\zeta^2 (1+\alpha Z_{f_2}\tilde f_2 D_2)^2\alpha \tilde \zeta D_1 \alpha\tilde \zeta\Bigg(\sum_{n=0}^\infty (Z_\zeta^2 D_2\alpha\tilde \zeta D_1 \alpha\tilde \zeta)^n  \Bigg)\Bigg\},\\
&\Gamma_{12}^{(4)}=(C^E)^2 \left\{(1+\alpha Z_{f_1} \tilde f_1 D_1)(1+\alpha Z_{f_2} \tilde f_2  D_2)\alpha Z_\zeta\tilde\zeta \Bigg(\sum_{n=0}^\infty(Z_\zeta^2 D_2 \alpha\tilde\zeta D_1 \alpha\tilde\zeta )^n\Bigg) \right\},
\label{4-point correlators for double trace vertex}
\eeq
where $\Gamma_{ij}^{(4)}$ is the sum over planar diagrams when the pairs of external legs on the two sides correspond to the fields $\Phi_i$ and $\Phi_j$. The first term in the expression of $\Gamma_{11}^{(4)}$ or $\Gamma_{22}^{(4)}$ is due to the contributions of diagrams without double trace vertices. In the remaining terms, the overall factor of $(C^E)^2$ comes from the two external building blobs at the two ends\footnote{The $C^E$ at each end depends on the momenta of the external legs to which the corresponding external blob is attached. Since the momenta at the two ends can be distinct, the values of $C^E$ at the two ends can also be different. Nevertheless, to avoid unnecessary clutter, we  use the shorthand notation $(C^E)^2$ to denote the product of the two $C^E$'s. Note that this product can be taken to be the same while evaluating the vertex renormalizations of all the double trace couplings by choosing the same set of momenta for the external legs in all these cases. The vertex renormalizations are determined by demanding that the expressions in  \eqref{4-point correlators for double trace vertex} are free of divergences in the $\varepsilon\rightarrow 0$ limit. The momentum-dependent pieces in these expressions do not contribute to these vertex renormalizations.}. The first two terms within the curly braces in the expression of $\Gamma_{11}^{(4)}$ or $\Gamma_{22}^{(4)}$ correspond to diagrams without any $\tilde \zeta$ vertex. The other terms correspond to diagrams with different numbers of insertions of the $\tilde \zeta$ vertex. Note that the total number of such insertions in these diagrams must be even as the components on the two sides of such a vertex belong to different sectors while the external legs at the two ends of these  diagrams belong to the same sector.  The factor $(1+\alpha Z_{f_i}\tilde f_i D_i)^2$ arises as a coefficient of these terms to include  diagrams where there is a chain of blobs appearing in the expansion of $D_i$ at either end, as well  as diagrams where the first double trace vertex on either end corresponds to $\tilde \zeta$. The expansion in the expression of $\Gamma_{12}^{(4)}$ can be understood similarly. 

Performing the sum over the geometric series appearing in \eqref{4-point correlators for double trace vertex} and using the expressions of $D_i$ given in \eqref{expression of D_i}, we get the following re-summed expressions:
\beq
&\frac{N_{c1}^2}{16\pi^2}\frac{\Gamma_{11}^{(4)}}{\alpha}\\
&=\frac{C^0}{16\pi^2\alpha}+(C^E)^2\Bigg\{\frac{Z_{f_1} f_1-16 \pi^2\alpha C^I(Z_{f_1} Z_{f_2} f_1 f_2-Z_{\zeta}^2\zeta^2)}{1-16\pi^2\alpha C^I(Z_{f_1}f_1+Z_{f_2}f_2)+(16\pi^2\alpha C^I)^2(Z_{f_1} Z_{f_2} f_1 f_2-Z_{\zeta}^2\zeta^2) }\Bigg\},\\
&\frac{N_{c2}^2}{16\pi^2}\frac{\Gamma_{22}^{(4)}}{\alpha}\\
&=\frac{C^0}{16\pi^2\alpha}+(C^E)^2\Bigg\{\frac{Z_{f_2} f_2-16 \pi^2\alpha C^I (Z_{f_1} Z_{f_2} f_1 f_2-Z_{\zeta}^2\zeta^2)}{1-16\pi^2\alpha C^I(Z_{f_1} f_1+Z_{f_2} f_2)+(16\pi^2\alpha C^I)^2(Z_{f_1}Z_{f_2}f_1 f_2-Z_{\zeta}^2\zeta^2) }\Bigg\} ,\\
&\frac{N_{c1}N_{c2}}{16\pi^2}\frac{\Gamma_{12}^{(4)}}{\alpha}=(C^E)^2\Bigg\{\frac{Z_{\zeta}\zeta}{1-16\pi^2\alpha C^I(Z_{f_1}f_1+Z_{f_2}f_2)+(16\pi^2\alpha C^I)^2(Z_{f_1}Z_{f_2}f_1 f_2-Z_{\zeta}^2\zeta^2) }\Bigg\},
\label{4-point correlators for double trace vertex: resummed expressions}
\eeq
where we have now switched to the 't Hooft couplings $f_i=\frac{N_{ci}^2}{16\pi^2}\tilde f_i$ and $\zeta=\frac{N_{c1}N_{c2}}{16\pi^2}\tilde \zeta$. 
To extract the vertex renormalizations in the $\overline{\text{MS}}$ scheme from the above expressions, we need to demand that all  $O(\frac{1}{\varepsilon^n})$ terms in these expressions vanish for $n>0$. This would render these quantities finite in the  $\varepsilon\rightarrow0$ limit. To simplify the analysis of these vertex renormalizations, let us introduce the following quantities:
\beq
\overline{\Gamma}_{ij}^{(4)}\equiv \frac{N_{ci}N_{cj}}{16\pi^2}\Gamma_{ij}^{(4)},\ \overline{f}_i\equiv Z_{f_i} f_i,\ ,\ \overline{\zeta}\equiv Z_{\zeta} \zeta,\ \overline{C^0}\equiv \frac{C^0}{16\pi^2\alpha}.
\eeq
Then the  expressions in \eqref{4-point correlators for double trace vertex: resummed expressions} can be rewritten as
\beq
&\frac{\overline{\Gamma}_{11}^{(4)}}{\alpha}=\overline{C^0}+(C^E)^2\Bigg\{\frac{\overline{f}_1-16 \pi^2\alpha C^I(\overline{f}_1 \overline{f}_2-\overline{\zeta}^2)}{1-16\pi^2\alpha C^I(\overline{f}_1+\overline{f}_2)+(16\pi^2\alpha C^I)^2(\overline{f}_1 \overline{f}_2-\overline{\zeta}^2) }\Bigg\},\\
&\frac{\overline{\Gamma}_{22}^{(4)}}{\alpha}=\overline{C^0}+(C^E)^2\Bigg\{\frac{\overline{f}_2-16 \pi^2\alpha C^I (\overline{f}_1 \overline{f}_2-\overline{\zeta}^2)}{1-16\pi^2\alpha C^I(\overline{f}_1+\overline{f}_2)+(16\pi^2\alpha C^I)^2(\overline{f}_1 \overline{f}_2-\overline{\zeta}^2) }\Bigg\} ,\\
&\frac{\overline{\Gamma}_{12}^{(4)}}{\alpha}=(C^E)^2\Bigg\{\frac{\overline{\zeta}}{1-16\pi^2\alpha C^I(\overline{f}_1+\overline{f}_2)+(16\pi^2\alpha C^I)^2(\overline{f}_1 \overline{f}_2-\overline{\zeta}^2) }\Bigg\}.
\label{simplified expressions of 4-point correlators}
\eeq
We can make a further simplification by introducing 
\beq
\overline{f}_p\equiv\frac{\overline{f}_1+\overline{f}_2}{2}\equiv Z_{f_p}f_{p},\ \overline{f}_m\equiv\frac{\overline{f}_1-\overline{f}_2}{2}\equiv Z_{f_m}f_{m},
\eeq
and taking the sum and difference of the first two expressions in \eqref{simplified expressions of 4-point correlators} as shown below:

\beq
&\frac{\overline{\Gamma}_{11}^{(4)}+\overline{\Gamma}_{22}^{(4)}}{2\alpha}=\overline{C^0}+(C^E)^2\Bigg\{\frac{\overline{f}_p-16 \pi^2\alpha C^I(\overline{f}_p^2 -\overline{f}_m^2-\overline{\zeta}^2)}{1-32\pi^2\alpha C^I\overline{f}_p+(16\pi^2\alpha C^I)^2(\overline{f}_p^2- \overline{f}_m^2-\overline{\zeta}^2) }\Bigg\},\\
&\frac{\overline{\Gamma}_{11}^{(4)}-\overline{\Gamma}_{22}^{(4)}}{2\alpha}=(C^E)^2\Bigg\{\frac{\overline{f}_m}{1-32\pi^2\alpha C^I\overline{f}_p+(16\pi^2\alpha C^I)^2(\overline{f}_p^2- \overline{f}_m^2-\overline{\zeta}^2) }\Bigg\},\\
&\frac{\overline{\Gamma}_{12}^{(4)}}{\alpha}=(C^E)^2\Bigg\{\frac{\overline{\zeta}}{1-32\pi^2\alpha C^I \overline{f}_p+(16\pi^2\alpha C^I)^2(\overline{f}_p^2- \overline{f}_m^2-\overline{\zeta}^2) }\Bigg\}.
\label{simplified expression of the 4 pt. correlators}
\eeq
Notice that demanding  the third expression in \eqref{simplified expression of the 4 pt. correlators} to be free of divergences in the $\varepsilon\rightarrow 0$ limit automatically ensures that the  second expression is also divergence-free if we take
\beq
Z_{f_m}=Z_{\zeta},
\eeq
or equivalently, $\overline{f}_m=\frac{f_m}{\zeta}\overline{\zeta}$.
Assuming this equality, we can rewrite the first and third expressions in \eqref{simplified expression of the 4 pt. correlators} as given below: 
\beq
\frac{\overline{\Gamma}_{11}^{(4)}+\overline{\Gamma}_{22}^{(4)}}{2\alpha}&=\Bigg[\overline{C^0}+(C^E)^2\Bigg\{\frac{Z_{f_p}f_p-16 \pi^2\alpha C^I\Big(Z_{f_p}^2 f_p^2 -Z_{\zeta}^2(f_m^2+\zeta^2)\Big)}{1-32\pi^2\alpha C^I Z_{f_p} f_p+(16\pi^2\alpha C^I)^2\Big(Z_{f_p}^2 f_p^2- Z_{\zeta}^2(f_m^2+\zeta^2)\Big) }\Bigg\}\Bigg],\\
\frac{\overline{\Gamma}_{12}^{(4)}}{\alpha}&=\zeta\Bigg[(C^E)^2\Bigg\{\frac{Z_{\zeta}}{1-32\pi^2\alpha C^I Z_{f_p} f_p+(16\pi^2\alpha C^I)^2\Big(Z_{f_p}^2 f_p^2- Z_{\zeta}^2(f_m^2+\zeta^2)\Big) }\Bigg\}\Bigg].
\eeq
$Z_{f_p}$ and $Z_{\zeta}$ are determined by demanding that the quantities within brackets in the above expressions are free of divergences in the $\varepsilon\rightarrow 0$ limit.  Apart from  depending on $Z_{f_p}$ and $Z_\zeta$, these quantities are functions of  $\lambda, h, f_p$ and  $(f_m^2+\zeta^2)$. Since  $Z_{f_p}$ and $Z_\zeta$ are chosen in the $\overline{\text{MS}}$ scheme such that they just cancel the overall divergences in these quantities, they must also be functions of $\lambda, h, f_p$ and  $(f_m^2+\zeta^2)$.  
So, to summarize, the planar vertex renormalizations $Z_{f_p}, Z_{f_m}$ and $Z_{\zeta}$ depend on the different couplings as shown below:
\beq
Z_{f_p}=Z_{f_p}(\lambda,h, f_p, f_m^2+\zeta^2),\ Z_{f_m}=Z_{\zeta}=Z_{\zeta}(\lambda,h, f_p, f_m^2+\zeta^2).
\eeq

From the planar vertex and wave function renormalizations, we can now evaluate the planar beta functions of the different double trace couplings. As derived in appendix \ref{app: double trace beta fns.}, these beta functions are given by
\beq
&\beta_{ f_1}= f_1 \sum_{A}\kappa_A\frac{\partial}{\partial \kappa_A}\text{Res}\Bigg[Z_{ f_1}-2 Z_{\Phi_1}\Bigg]= f_1 \sum_{A}\kappa_A\frac{\partial}{\partial \kappa_A}\text{Res}\Bigg[Z_{ f_1}-2 Z_{\Phi}\Bigg],\\
&\beta_{ f_2}= f_2 \sum_{A}\kappa_A\frac{\partial}{\partial \kappa_A}\text{Res}\Bigg[Z_{ f_2}-2 Z_{\Phi_2}\Bigg]= f_2 \sum_{A}\kappa_A\frac{\partial}{\partial \kappa_A}\text{Res}\Bigg[Z_{ f_2}-2 Z_{\Phi}\Bigg],\\
&\beta_{ \zeta}= \zeta \sum_{A}\kappa_A\frac{\partial}{\partial \kappa_A}\text{Res}\Bigg[Z_{ \zeta}- Z_{\Phi_1}- Z_{\Phi_2}\Bigg]=\zeta \sum_{A}\kappa_A\frac{\partial}{\partial \kappa_A}\text{Res}\Bigg[Z_{ \zeta}- 2 Z_{\Phi}\Bigg].
\label{beta fns. of double trace couplings}
\eeq
Here, $\text{Res}[\cdots]$ indicates the residue of the quantity within the brackets at the $\varepsilon=0$ pole.  $\kappa_A$ runs over the couplings $\lambda, h, f_1, f_2$  and $\zeta$. The derivatives in the above expressions have to be computed on the subspace $h=h_0$ and $\lambda=\lambda_0$. The second equality in each line of \eqref{beta fns. of double trace couplings} relies on the fact that $Z_{\Phi_1}=Z_{\Phi_2}=Z_\Phi$ when $\lambda_1=\lambda_2=\lambda$ and $h_1=h_2=h$.

Now, let us use differentiation by parts to rewrite the above expressions of $\beta_{f_1}$ and $\beta_{f_2}$ as
\beq
\beta_{ f_1}=&  \Big(\sum_{A}\kappa_A\frac{\partial}{\partial \kappa_A}-1\Big)\text{Res}\Bigg[f_1 Z_{ f_1}-2 f_1 Z_{\Phi}\Bigg],\\
\beta_{ f_2}=&  \Big(\sum_{A}\kappa_A\frac{\partial}{\partial \kappa_A}-1\Big)\text{Res}\Bigg[f_2 Z_{ f_2}-2 f_2 Z_{\Phi}\Bigg],\\
\eeq
Here we have also used the fact that the action of the operator $\sum\limits_A\kappa_A\frac{\partial}{\partial \kappa_A}$ on  $f_i$ is trivial.
From these expressions, we can obtain the planar beta functions of  $f_p=\frac{f_1+f_2}{2}$ and $f_m=\frac{f_1-f_2}{2}$ as given below:
\beq
\beta_{ f_p}=&  \Big(\sum_{A}\kappa_A\frac{\partial}{\partial \kappa_A}-1\Big)\text{Res}\Bigg[f_p Z_{ f_p}-2 f_p Z_{\Phi}\Bigg],\\
\beta_{ f_m}=&  \Big(\sum_{A}\kappa_A\frac{\partial}{\partial \kappa_A}-1\Big)\text{Res}\Bigg[f_m Z_{ f_m}-2 f_m Z_{\Phi}\Bigg].\\
\eeq
Now, using differentiation by parts once more, we can rewrite the above expressions as 
\beq
&\beta_{f_p}=f_p\sum_A \kappa_A\frac{\partial}{\partial \kappa_A}\text{Res}\Bigg[ Z_{ f_p}-2  Z_{\Phi}\Bigg],\\
&\beta_{f_m}=f_m\sum_A \kappa_A\frac{\partial}{\partial \kappa_A}\text{Res}\Bigg[ Z_{ f_m}-2  Z_{\Phi}\Bigg].
\label{beta fns. of fp and fm}
\eeq
By comparing the expressions of $\beta_{ f_m}$ and  $\beta_{ \zeta}$ in \eqref{beta fns. of fp and fm} and \eqref{beta fns. of double trace couplings} respectively, and by using the fact that $Z_{f_m}=Z_\zeta$, we get
\beq
\beta_{f_m}=\frac{f_m }{\zeta }\beta_{\zeta}.
\eeq
This means that the equations obtained by setting all the planar beta functions to zero (with the condition $\zeta\neq 0$) are degenerate. The corresponding degenerate fixed points satisfy the following two equations:
\beq
f_p\sum_A \kappa_A\frac{\partial}{\partial \kappa_A}\text{Res}\Bigg[ Z_{ f_p}-2  Z_{\Phi}\Bigg]=0,\\
\sum_{A}\kappa_A\frac{\partial}{\partial \kappa_A}\text{Res}\Bigg[Z_{ \zeta}- 2 Z_{\Phi}\Bigg]=0.
\label{independent eqns. for fixed circle}
\eeq
Notice that the residues in the above equations are functions of  $\lambda, h, f_p$ and $(f_m^2+\zeta^2)$. Moreover, the operator $\sum_A\kappa_{A}\frac{\partial}{\partial \kappa_A}$ acting on these residues takes the following form:
\beq
\sum_A\kappa_{A}\frac{\partial}{\partial \kappa_A}
&=\lambda_0\frac{\partial}{\partial \lambda}+h_0\frac{\partial}{\partial h}+f_p\frac{\partial}{\partial f_p}+2\Big(f_m^2+\zeta^2\Big)\frac{\partial}{\partial (f_m^2+\zeta^2)}.
\eeq
Therefore, we can see that the two equations in \eqref{independent eqns. for fixed circle} are essentially two functions of $f_p$ and $(f_m^2+\zeta^2)$ set equal to zero.\footnote{The values of $\lambda$ and $h$ in these equations are set to $\lambda_0$ and $h_0$.} From the analysis of 2-loop beta functions, we have already seen that there are perturbatively reliable solutions of these equations. These solutions correspond to the fixed circles in the double bifundamental models. Here we can see that the entire effect of the higher loop corrections is to modify the location of the plane on which the fixed circle lies\footnote{This plane is determined by  $\lambda_0$, $h_0$ and the value of $f_p$ obtained by solving the equations in \eqref{independent eqns. for fixed circle}.} and the radius of the circle\footnote{The radius of the fixed circle is determined by the value of $(f_m^2+\zeta^2)$ obtained by solving the equations in \eqref{independent eqns. for fixed circle}.}.

Thus, we have proved the survival of the fixed circles in the double bifundamental models under all loop corrections in the planar limit. As a result, we can conclude that  all the points on this circle which demonstrate thermal order are genuine CFTs in this limit. Therefore, the respective baryon symmetries in these theories remain broken up to arbitrarily high temperatures and the systems exist in a persistent BEH phase.

\section{Conclusion and discussion}
\label{sec: Conclusion}
In this paper we studied the possibility of spontaneous breaking of global symmetries  at nonzero temperatures for large N conformal gauge theories in $D=4$ space-time dimensions. We started with some familiar QCD models, viz., QCDs with $SO(N_c)$ and $SO(N_c)\times SO(N_c)$ gauge symmetries. 
The matter contents of these theories consist of Majorana fermions in the fundamental representation of the $SO(N_c)$'s, and real scalar fields transforming in the fundamental/bifundamental representation of the gauge group. 
By analyzing the beta functions of the different couplings in the $N_c\rightarrow \infty$ limit, we determined the Banks-Zaks-like fixed points in these models. We showed that the conformal theories at these fixed points fail to demonstrate spontaneous breaking of certain global symmetries (a flavor symmetry for the vector model and a $\mathbb Z_2$ baryon symmetry for the bifundamental model) at nonzero temperatures. 

We then extended these models by taking two copies of the bifundamental scalar QCD (with ranks $N_{c1}$ and $N_{c2}$) and coupling them via a double trace interaction. We called this the  `real double bifundamental model'. Both by direct perturbative computations up to 2-loops as well as general diagrammatic arguments at all orders in the 't Hooft couplings, we showed that this model has the following interesting property: {\it Within the domain of validity of perturbation theory, the planar  beta functions of the different couplings yield a conformal manifold with the topology of a circle.} 

Let us  emphasize here that the proof of the existence of the conformal manifolds in the double bifundamental models at the planar limit in section \ref{sec: survival of fixed circle} is largely model-independent. So it can be used to generate  similar families of models with conformal manifolds. One such family would be models where there are two similar large $N$ QFTs $T_1$ and $T_2$, each having a single matrix-valued scalar $\Phi_i$, with only a double trace interaction between them\footnote{Here, we assume that no other interaction between $T_1$ and $T_2$ is generated along the RG flow. We thank Ofer Aharony for suggestions on this issue.}. For example, when the gauge groups in each theory are products of  $SU(N)$'s or $SO(N)$'s, and the scalar fields are in either the adjoint, the symmetric, the anti-symmetric or the bifundamental representations of the gauge groups, then $\Tr(\Phi_1^\dagger \Phi_1)\Tr(\Phi_2^\dagger \Phi_2)$ can be such a double trace interaction coupling the two theories. It would be interesting to investigate other variants as well, for example, models with multiple double trace interactions between  two similar large N QFTs.

 Let us now return back to the real double bifundamental model. For the fixed points lying on the above-mentioned manifold, we explored the possibility of spontaneous breaking of global symmetries at nonzero temperatures. We found that such a symmetry breaking indeed occurs in certain parameter regimes. The relevant symmetries here are two $\mathbb Z_2$ baryon symmetries, one for each sector. We found that in the zero temperature limit, i.e., in the ground state, both these symmetries remain unbroken. Moreover,  in this limit, the effective potential of the scalar fields steadily increases as one moves away from the origin in the field space along any direction. This is in contrast with the models discussed in \cite{Chai:2020zgq, Chai:2020onq} where, in the planar limit, a flat direction of the potential in the field space allowed for nonzero vacuum expectation values of the fields which led to the spontaneous breaking of  scale invariance. This also made it possible to spontaneously break global  $O(N)$ symmetries in these models even at zero temperature. Here nothing of that sort happens due to the absence of such flat directions of the effective potential.

When a temperature is turned on, the scalar fields in this model pick up thermal masses. If the square of any of these  masses is negative, then the minimum of the thermal effective potential\footnote{In computing the thermal effective potential we have included the microscopic fields. Due to the matter content which does not permit a deconfining phase transition, we have no evidence for the existence of light non-perturbative degrees of freedom to be included in the calculation of the potential.} lies away from the origin in the field space which leads to  nonzero  expectation values of the scalar fields. This, in turn, means that at least one of the $\mathbb Z_2$ baryon symmetries is broken in such a thermal state. Thus, to determine whether these baryon symmetries are spontaneously broken at nonzero temperatures for the points lying on the conformal manifold, one just needs to evaluate the thermal masses at these points. We analysed these thermal masses and made the following observations which constitute the main results of the paper:
{\it When  $N_{c2}<N_{c1}$, the baryon symmetry in the first sector is always unbroken. On top of this, when the ratio $r\equiv\frac{N_{c2}}{N_{c1}}\leq\sqrt{\frac{6\sqrt{6}-13}{61-6\sqrt{6}}}$, then the baryon symmetry in the second sector is spontaneously broken at all temperatures for a certain subset of points on the conformal manifold. Along with this symmetry-breaking, half of the gauge bosons in the second sector are Higgsed.\footnote{It was pointed out in \cite{Linde:1980ts, Gross:1980br} that in certain cases thermal perturbation theory is challenged by IR problems. What we can add here is that for those gauge particles which gain a perturbative mass, the mass is of order $\sqrt{\lambda} T$ which for weak coupling is larger than $\lambda T$. This value shields perturbation theory from the problems which could have been posed by those particles.} Thus, the system exists in a persistent Brout-Englert-Higgs (BEH) phase at all temperatures for such fixed points. Exactly analogous features hold when $N_{c1}<N_{c2}$.}

In addition to observing the above features in the real double bifundamental model, we also studied a closely related model which we called the `complex double bifundamental model'. This model again has two sectors each of which is symmetric under the gauge group $SU(N_{ci})\times SU(N_{ci})$. The matter content in each sector consists of Dirac fermions transforming in the fundamental representation of the individual $SU(N_{ci})$'s in that sector, and a set of complex scalar fields transforming in the bifundamental representation of $SU(N_{ci})\times SU(N_{ci})$. Here, the global symmetries of our interest are two $U(1)$ baryon symmetries, one for each sector. The analysis of the spontaneous breaking of these  symmetries was considerably simplified by a perturbative planar equivalence between this model and the real double bifundamental model with  ranks $2N_{c1}$ and $2 N_{c2}$ in the two sectors. This equivalence allowed us to extend all the features of the real double bifundamental model discussed above to the complex double bifundamental model. Thus, this model demonstrates both a conformal manifold in the planar limit as well as the spontaneous breaking of one of the baryon symmetries at nonzero temperatures for a subset of points on this manifold when the ranks of the two sectors are sufficiently different. Just like before, the breaking of the baryon symmetry is accompanied by a persistent BEH phase at all temperatures.

Here, let us briefly comment on the fact that the persistence of thermal order in our models\footnote{We have not yet determined whether these models  have holographic duals.} does not contradict  the standard results in the AdS/CMT literature  predicting symmetry-restoration in holographic CFTs \cite{Gubser:2008px, Hartnoll:2008kx, Hartnoll:2008vx}. In these works, symmetry breaking was examined by analyzing charged scalar hairs on the AdS side. These hairs correspond to  order parameters which are expectation values of gauge invariant operators with $O(1)$ scaling dimensions in the dual CFT. In our case, the order parameter for the $\mathbb Z_2$ or the $U(1)$ symmetry is the expectation value of the determinant of the scalar fields which has an $O(N)$ scaling dimension. Hence, these symmetries are baryon-like. In  AdS/CFT correspondence, the dual of such a baryon-like operator is interpreted as a wrapped Euclidean D-brane in the AdS bulk which has a point-like intersection with the boundary \cite{Witten:1998xy,Gukov:1998kn,Berenstein:2002ke}. In a given supergravity background, the expectation value of such an operator can be obtained from a regularized partition function evaluated by summing over all possible configurations of the D-brane satisfying appropriate boundary conditions \cite{Klebanov:2007us,Martelli:2007mk,Martelli:2008cm}.  It would be interesting to see whether it is possible to violate the no-hair theorem for black holes by obtaining baryon hairs in the bulk via the above prescription. Such baryon hairs would then correspond to spontaneously broken  baryonic symmetries in thermal states of the dual CFT.

Let us now end with a discussion on how the above-mentioned features of the double bifundamental models may be affected when corrections due to finiteness of $N_{c1}$ and $N_{c2}$ are taken into account. To analyze this, we have studied how the fixed points for the real double bifundamental model are modified by finite $N_{ci}$ corrections in appendix \ref{app: finite N corrections}. There we have taken the two ranks to be of comparable magnitudes, say $O(N)$ with $N$ being a large number. We have worked in a regime where $\frac{1}{N}\ll$'t Hooft couplings. In this regime, for each order in  $\frac{1}{N}$, one can expand the  corresponding terms in the beta functions about the fixed points at the planar limit in powers of the 't Hooft couplings. Since we  have the explicit expressions of these beta functions only up to 2-loops, we have been able to study the $O\Big(\frac{1}{N}\Big)$ corrections to the beta functions only up to the first subleading terms in such an expansion. This analysis shows that the degeneracy in the fixed points is not lifted if the ratios $x_{f1}\equiv\frac{N_{f1}}{N_{c1}}$ and $x_{f2}\equiv\frac{N_{f2}}{N_{c2}}$ are tuned appropriately. Under such a fine-tuning, the closed curve of fixed points remains on a plane in the space of couplings. However, when the ranks of the two sectors are unequal, its shape is deformed away from the circular form that we found in the planar limit. In addition to these observations, we have also looked at the leading order terms in the expansion of the $O\Big(\frac{1}{N^2}\Big)$ terms. For these terms, we have found  the degeneracy in the fixed points to again survive under appropriate fine-tuning of $x_{f1}$ and $x_{f2}$. It would be interesting to see if the above-mentioned features persist up to all orders under suitable constraints on $x_{f1}$ and $x_{f2}$. One way to check this may be to re-sum the series expansions of the beta functions in powers of the 't Hooft couplings (as we have done in section \ref{sec: survival of fixed circle} at the planar limit), and then systematically consider the subleading terms in the $\frac{1}{N}$-expansion.\footnote{See \cite{Chai:2020hnu} for a recent work which discusses how  $1/N$ corrections lift the degeneracy in fixed points of certain large $N$ models.} Even if the degeneracy in the fixed points is lifted at higher orders, there may be isolated fixed points which survive under finite N corrections after such re-summations. It would be interesting to see if any of the large $N$ fixed points  that demonstrate thermal order survive under such finite N corrections. We would like to resolve this issue in the future.

\acknowledgments
We thank Noam Chai, Zohar Komargodski and Michael Smolkin for collaboration in the initial phase of this project. We are grateful to Ofer Aharony, Zohar Komargodski and Michael Smolkin for their  comments after reading a preliminary draft of the paper. We also thank Timothy Jones and Robert Shrock for their comments on a part of the work. We thank George Sterman for useful discussions. E. Rabinovici would like to thank the Institut des Hautes \'Etudes Scientifiques in Bures sur Yvette, the New High Energy Theory Center at Rutgers Physics Department and Center for Cosmology and Particle Physics at New York University for hospitality and support.  S. Chaudhuri and E. Rabinovici are supported by the Israel Science Foundation Center of Excellence (Grant No. 2289/ 18).
C. Choi is supported in part by the Simons Foundation grant 488657 (Simons Collaboration on the Non-Perturbative Bootstrap) and also by the National Science Foundation under Grant No. NSF PHY-1748958 and the Heising-Simons Foundation under the KITP Graduate Fellowship.  

\appendix

\section{Baryon symmetry in the bifundamental scalar QCD}
\label{app: bifund QCD baryon symmetry}
In this appendix, we will show that the $\mathbb{Z}_2$ transformation given in \eqref{Z2 symmetry in bifund scalar QCD} is indeed a global symmetry of the Lagrangian of the bifundamental scalar QCD.  We will also show that this symmetry can be interpreted as an automorphism of the set of equivalence classes of field configurations, where  each class comprises of gauge-equivalent configurations. The analysis presented here can be  extended in a straightforward way to derive similar results for the baryon symmetries in the double bifundamental models.

Let us first show the invariance of the Lagrangian given in \eqref{Bifundamental scalar QCD: renorm. Lagrangian} under the afore-mentioned $\mathbb Z_2$ transformation. For the convenience of the reader, we provide the form of this Lagrangian below\footnote{Here $(F_\alpha)_{\mu\nu}$ is the field strength corresponding to the gauge field $(V_\alpha)_\mu$ which is an imaginary-valued anti-symmetric $N_c\times N_c$ matrix. Just like the scalar field $\Phi$ and the gauge field $(V_\alpha)_\mu$, this field strength is also an $N_c\times N_c$ matrix. We take the fermionic fields ($\psi^{(p)}$ and $\chi^{(p)}$) of each flavor to be column vectors with $N_c$ components.}:
\begin{equation}
\begin{split}
\mathcal L_{\text{Bifund}}=&-\frac{1}{2}\sum_{\alpha=1}^2 \text{Tr}\Big[(F_{\alpha})_{\mu\nu}(F_{\alpha})^{\mu\nu }\Big]+\frac{i}{2}\overline{\psi}^{(p)} (\slashed{D}\psi^{(p)})+\frac{i}{2}\overline{\chi}^{(p)} (\slashed{D}\chi^{(p)})+\frac{1}{2}\text{Tr}\Bigg[\Big(D_{\mu} \Phi \Big)^T D^{\mu} \Phi\Bigg]\\
&-\widetilde{h}\text{Tr}\Big[\Phi^T\Phi\Phi^T\Phi\Big]-\widetilde{f}\text{Tr}\Big[\Phi^T\Phi\Big]\text{Tr}\Big[\Phi^T\Phi\Big].
\end{split}
\end{equation}
Now, let us see how each term in the above Lagrangian transforms under the transformations given in \eqref{Z2 symmetry in bifund scalar QCD}. To be specific, we will set $a=1$ in these transformations. These transformations can be written more compactly as
\begin{equation}
\begin{split} 
&\Phi\rightarrow \mathcal{T}\Phi,\ \\
&\psi^{(p)}\rightarrow \mathcal{T} \psi^{(p)}\ \forall\ p\in\{1,\cdots, N_{f}\} ,\\
& (V_{1})_\mu\rightarrow\mathcal{T}(V_{1})_\mu \mathcal{T}^{-1},
\end{split}
\end{equation}
where $\mathcal{T}$ is an $N_c\times N_c$ diagonal matrix of the following form:
\beq
\mathcal{T}\equiv\text{diag}\{-1,1,\cdots,1\}.
\eeq
The terms  $-\frac{1}{2} \text{Tr}\Big[(F_{2})_{\mu\nu}(F_{2})^{\mu\nu }\Big]$ and $\frac{i}{2}\overline{\chi}^{(p)} (\slashed{D}\chi^{(p)})$ are clearly invariant as the $\mathbb Z_2$ transformation acts trivially on the fields appearing in these terms. The quartic interaction terms are also manifestly invariant as  $(\Phi^T \Phi)$ remains unchanged under the transformation. So we just need to check the invariance of the terms $-\frac{1}{2} \text{Tr}\Big[(F_{1})_{\mu\nu}(F_{1})^{\mu\nu }\Big]$, $\frac{i}{2}\overline{\psi}^{(p)} (\slashed{D}\psi^{(p)})$ and $\frac{1}{2}\text{Tr}\Bigg[\Big(D_{\mu} \Phi \Big)^T D^{\mu} \Phi\Bigg]$. For this, we can first obtain the transformations of the field strength $(F_{1})_{\mu\nu}$ and the covariant derivatives of $\psi^{(p)}$ and $\Phi$ as follows:
\beq
& (F_1)_{\mu\nu}
\rightarrow \mathcal{T} (F_1)_{\mu\nu}\mathcal{T}^{-1},\ 
D_\mu \psi^{(p)}
\rightarrow \mathcal{T} D_\mu \psi^{(p)},\ 
D_\mu \Phi
\rightarrow \mathcal{T} D_\mu \Phi.\\
\eeq
From  these transformations   we can then derive the invariance of the above-mentioned terms as shown below:
\beq
-\frac{1}{2}\text{Tr}\Big[(F_1)_{\mu\nu}(F_1)^{\mu\nu}\Big]&\rightarrow -\frac{1}{2}\text{Tr}\Big[\mathcal{T}(F_1)_{\mu\nu}(F_1)^{\mu\nu}\mathcal{T}^{-1}\Big]=-\frac{1}{2}\text{Tr}\Big[(F_1)_{\mu\nu}(F_1)^{\mu\nu}\Big],\\
\frac{i}{2}\overline{\psi}^{(p)} (\slashed{D}\psi^{(p)})& \rightarrow \frac{i}{2}\overline{\psi}^{(p)} \mathcal{T}^\dag \mathcal{T}(\slashed{D}\psi^{(p)})=\frac{i}{2}\overline{\psi}^{(p)} (\slashed{D}\psi^{(p)}),\\
\frac{1}{2}\text{Tr}\Bigg[\Big(D_{\mu} \Phi \Big)^T D^{\mu} \Phi\Bigg]&\rightarrow \frac{1}{2}\text{Tr}\Bigg[\Big(D_{\mu} \Phi \Big)^T \mathcal{T}^T \mathcal{T} D^{\mu} \Phi\Bigg]=\frac{1}{2}\text{Tr}\Bigg[\Big(D_{\mu} \Phi \Big)^T D^{\mu} \Phi\Bigg].
\eeq
Here, we have used the fact that $\mathcal{T}^\dag=\mathcal{T}^T=\mathcal{T}^{-1}$. This conclusively demonstrates the invariance of the Lagrangian under the  $\mathbb Z_2$ transformation.

Now, let us show that this $\mathbb Z_2$ transformation is an automporhism of a set of  classes of gauge-equivalent field configurations. For this, it is necessary and sufficient to demonstrate that performing the $\mathbb Z_2$ transformation over a gauge transformation is equivalent to performing another gauge transformation over the $\mathbb Z_2$ transformation. To prove this statement, let us consider a gauge transformation $O_1\times O_2\in SO(N_c)\times SO(N_c)$ which transforms the fields as follows:
\beq
&\Phi^\prime=O_1\Phi O_2^T,\\
&(\psi^{(p)})^\prime=O_1\psi^{(p)},\ (\chi^{(p)})^\prime=O_2\chi^{(p)}  \text{ for $p\in\{1,\cdots N_f\}$},\\
&(V_\alpha)_\mu^\prime=O_\alpha (V_\alpha)_\mu O_\alpha^T-\frac{i}{g}(\partial_\mu O_\alpha) O_\alpha^T \text{ for $\alpha\in\{1,2\}.$}
\eeq
Now, if we act $\mathcal{T}$ on the above configurations, we get
\beq
&\mathcal{T}\Phi^\prime=\mathcal{T} O_1\Phi O_2^T=(\mathcal{T} O_1 \mathcal{T} ^{-1})(\mathcal{T} \Phi )O_2^T ,\\
&\mathcal{T} (\psi^{(p)})^\prime=\mathcal{T}  O_1\psi^{(p)}=(\mathcal{T}  O_1\mathcal{T} ^{-1})(\mathcal{T} \Psi^{(p)}),\\
&\mathcal{T} (V_1)_\mu^\prime \mathcal{T}^{-1}=\mathcal{T}O_1 (V_1)_\mu O_1^T\mathcal{T}^{-1}-\frac{i}{g}\mathcal{T}(\partial_\mu O_1) O_1^T \mathcal{T}^{-1}\\
&\qquad\qquad\quad=(\mathcal{T}O_1 \mathcal{T}^{-1})(\mathcal{T} (V_1)_\mu \mathcal{T}^{-1}) (\mathcal{T} O_1^T\mathcal{T}^{-1})-\frac{i}{g}\Big(\partial_\mu (\mathcal{T} O_1 \mathcal{T}^{-1})\Big) \mathcal{T} O_1^T \mathcal{T}^{-1}\\
&\qquad\qquad\quad=(\mathcal{T}O_1 \mathcal{T}^{-1})(\mathcal{T} (V_1)_\mu \mathcal{T}^{-1}) (\mathcal{T} O_1\mathcal{T}^{-1})^T-\frac{i}{g}\Big(\partial_\mu (\mathcal{T} O_1 \mathcal{T}^{-1})\Big) (\mathcal{T} O_1 \mathcal{T}^{-1})^T,
\eeq
while $(\chi^{(q)})^{\prime}$ and $(V_2)_\mu^\prime$ are left invariant. In deriving the last line of the above  equations we have used the fact that $\mathcal{T}^{T}=\mathcal{T}^{-1}$ and $O_1^T=O_1^{-1}$.

From the above expressions we can clearly see that the overall transformation can be obtained by first acting the $\mathbb Z_2$ global transformation on the original fields and then performing a gauge transformation by $(\mathcal{T}O_1 \mathcal{T}^{-1})\times O_2$.\footnote{The fact that this is a gauge transformation can be verifed by checking that $\mathcal{T}O_1 \mathcal{T}^{-1}$ is an orthogonal matrix and $\det\Big[\mathcal{T}O_1 \mathcal{T}^{-1}\Big]=1$.} This means that the $\mathbb Z_2$ transformation maps two configurations related by a gauge transformation to two other configurations which are also related by a gauge transformation. This completes the proof of the statement that the $\mathbb Z_2$ symmetry is an automorphism of the set of classes of gauge-equivalent configurations in the model.

\section{Two-loop beta functions of QCD} \label{appsec:qcd}

In this appendix we will review the two-loop beta functions in a general QCD without any Yukawa interaction. We will mainly follow the presentation of \cite{Machacek:1983tz,Machacek:1984zw} where the authors employed  dimensional regularization and the modified minimal subtraction $(\overline{\text{MS}})$ scheme. We  refer the reader also to \cite{Luo:2002ti} for the detailed forms of such two-loop beta functions.

Let us consider a general QCD Lagrangian (without Yukawa interaction) of the following form\footnote{Just as in the main text, we suppress the gauge-fixing and ghost terms here as well.  At the two-loop level, ghost terms affect the beta functions of quartic couplings through wave function renormalization of the gauge and scalar propagators. On the other hand, the beta functions are independent of the gauge fixing term as they should be.}:
\beq \label{eq:qcd}
\mathcal L^{\text{QCD}}=&-{1\ov 4} 	F^A_{\mu\nu}F^A_{\mu\nu}+{1\ov 2}D_\mu \phi_a D_\mu \phi_a +i\kappa\psi_j^\dagger \slashed D\psi_j-{1\ov 4!}\lam_{abcd}\phi_a \phi_b\phi_c \phi_d+~\text{mass terms},
\eeq
where $\kappa=\frac{1}{2}$ for Majorana fermions and $\kappa=1$ for Dirac fermions.
$F_{\mu\nu}^A$ is the field strength corresponding to a gauge field $V_{\mu}^A$  associated with an arbitrary compact semi-simple Lie group $G$.  $\{\phi_a\}$ and $\{\psi_j\}$ are sets of scalars and fermions respectively, and we denote the representation in which they transform under the gauge group by $S$ and $F$ respectively. We choose the scalars to be real and take the couplings $\lam_{abcd}$  to be fully symmetric under permutation of indices. Note that the indices $a$ and $j$ may contain both gauge and flavor indicies. The matter and the gauge sectors are coupled through the covariant derivatives 
\beq
D_\mu \phi_a\equiv\partial_\mu\phi_a-igV^A_\mu \Big(T_A(S)\Big)_{ab}\phi_b, \\
D_\mu \psi_j\equiv\partial_\mu\psi_j-igV^A_\mu \Big(T_A(F)\Big)_{jk}\psi_k,
\eeq
where $T_A(S)$ and $T_A(F)$ are the generators of the gauge group in the representations $S$ and $F$. Calculation of the beta functions in such a theory largely reduces to a determination of various group theoretical quantities associated with the representations $S$, $F$ and the $G$, and different combinations of the quartic couplings $\lam_{abcd}$. When $G$ is not simple, there are several subtleties in the evaluation of the group theoretical invariants because of  diagrammatic reasons. We will  discuss these subtleties in the following subsections.

\subsection{Gauge couplings}\label{appsec:gauge}
The two-loop beta function for the gauge coupling  $g$ in the general QCD Lagrangian \ref{eq:qcd} is as follows:

\beq \label{eq:betag}
\beta_g \equiv & {dg\ov d \ln \mu}\\
=&-{g^3 \ov (4\pi)^2}\left \{{11\ov 3}C_2(G)-{4\ov 3}\kappa S_2(F) -{1\ov 6} S_2 (S) \right \}
\\&-{g^5 \ov (4\pi)^4} \left\{{34\ov 3} C_2(G)^2-\kappa [4C_2(F)+{20\ov 3}C_2(G) ] S_2(F)-[2C_2(S)+{1\ov 3}C_2(G)]S_2(S) \right\}.
\eeq
The first and second line in the above expression correspond to the one-loop and two-loop contributions. The quantities $C_2(R)$ and $S_2(R)$ are the quadratic Casimir and the second Dynkin index respectively of the representation $R$  ($G$ corresponds to the adjoint representation). Note that there is no contribution from the quartic couplings at this order. Equation \eqref{eq:betag} is valid for any simple gauge group. For a general semi-simple gauge group $G=\prod_iG_i$ with independent gauge coupling $g_i$ for each simple $G_i$, one has to  modify this equation to obtain the beta function of $g_i$ by the following substitutions:
\beq
&g^3 C_2(G)\rightarrow g_i^3 C_2^i(G),\\
&g^3 S_2(R)\rightarrow g_i^3 S_2^i(R)\\
&g^5[C_2(G)]^2\rightarrow g_i^5 [C_2^i(G)]^2\\
&g^5 C_2(G)S_2(R)\rightarrow g_i^5 C_2^i(G)S_2^i(R)\\
&g^5 C_2(R)S_2(R)\rightarrow \sum_j g_i^3g_j^2 C_2^j(R)S_2^i(R),
\eeq
where $C_2^i(R)$ and $S_2^i(R)$ denote the quadratic Casimir and the Dynkin index of the representation R corresponding to the simple Lie group $G_i$. While employing these rules, one should consider only the matter and the gauge fields that transform nontrivially under the group $G_i$.

\begin{figure}[!h]
\centering
\begin{subfigure}[b]{0.23\linewidth}
\centering
\begin{tikzpicture}
  \begin{feynman}
    \vertex (a);
    \vertex [right=1cm of a] (e);
    \vertex [above right=1cm of e] (c);
    \vertex [below right=1cm of e] (d);
    \vertex [below right =1cm of c] (f);
    \vertex [right=1cm of f] (b);
    \diagram* {   (a) -- [boson] (e)--[scalar, quarter left] (c)--[scalar,quarter left](f) -- [boson] (b), (c)--[boson] (d), (e)--[scalar, quarter right] (d)--[scalar, quarter right](f),};
  \end{feynman}
\end{tikzpicture}
\end{subfigure}
\begin{subfigure}[b]{0.23\linewidth}
\centering
\begin{tikzpicture}
  \begin{feynman}
    \vertex (a);
    \vertex [right=1cm of a] (c);
    \vertex [right =1 cm of c] (d);
    \vertex [right =1 cm of d] (b);
        \vertex [above =1 cm of d] (e);
            \vertex [  left =1 cm of e] (f);
    \diagram* {  (b) -- [boson] (d)--[scalar, quarter left] (c)--[boson](a), (c)--[scalar, quarter left]  (f) --[scalar, quarter left](e) --[scalar, quarter left] (d), (e)--[boson, quarter left] (f)};
  \end{feynman}
\end{tikzpicture}
\end{subfigure}
\begin{subfigure}[b]{0.23\linewidth}
\centering
\begin{tikzpicture}
  \begin{feynman}
    \vertex (a);
    \vertex [right=1cm of a] (e);
    \vertex [above right=1cm of e] (c);
    \vertex [below right=1cm of e] (d);
    \vertex [below right =1cm of c] (f);
    \vertex [right=1cm of f] (b);
    \diagram* {   (a) -- [boson] (e)--[fermion, quarter left] (c)--[fermion,quarter left](f) -- [boson] (b), (c)--[boson] (d), (f)--[fermion, quarter left] (d)--[fermion, quarter left](e),};
  \end{feynman}
\end{tikzpicture}
\end{subfigure}
\begin{subfigure}[b]{0.23\linewidth}
\centering
\begin{tikzpicture}
  \begin{feynman}
    \vertex (a);
    \vertex [right=1cm of a] (c);
    \vertex [right =1 cm of c] (d);
    \vertex [right =1 cm of d] (b);
        \vertex [above =1 cm of d] (e);
            \vertex [  left =1 cm of e] (f);
    \diagram* {  (b) -- [boson] (d)--[fermion, quarter left] (c)--[boson](a), (c)--[fermion, quarter left]  (f) --[fermion, quarter left](e) --[fermion, quarter left] (d), (e)--[boson, quarter left] (f)};
  \end{feynman}
\end{tikzpicture}
\end{subfigure}
\caption{Set of two-loop diagrams contributing to $\beta_g$ which is affected by substitution rules for the semi-simple gauge group. We follow a standard convention where the solid line, the dotted line and the wavy line correspond to the fermions, the scalars and the gluons respectively. } \label{fig:gauge2}
\end{figure}
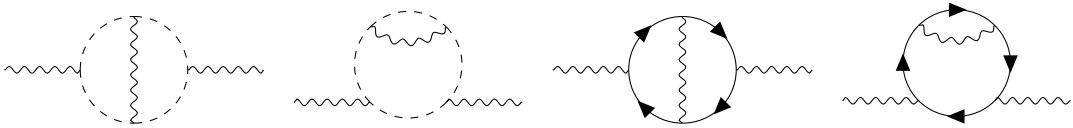

The above substitution rules can be understood diagrammatically (see figure \ref{fig:gauge2}) when we decompose the semi-simple gauge field into the direct sum of simple gauge fields as $V_{\mu}^A T_A(S)\rightarrow (V_i)_\mu^AT_A^i(S)$, $V_{\mu}^A T_A(F)\rightarrow (V_i)_\mu^AT_A^i(F)$.

\subsection{Quartic couplings}\label{appsec:quartics}

\medskip
\noindent
$\bullet$ {\it \textbf{One-loop}}
\medskip

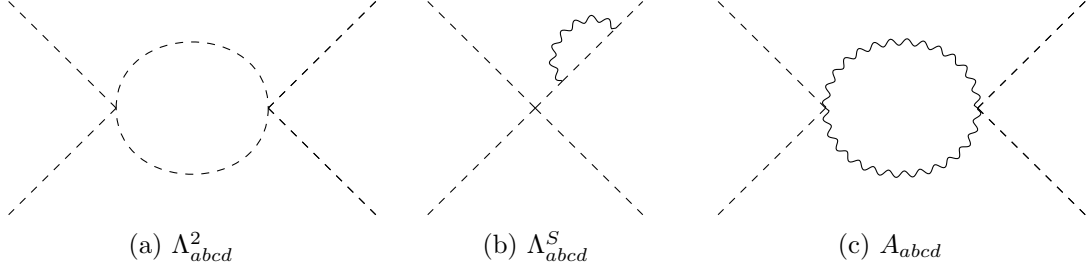
\begin{figure}[!h]
\centering
\begin{subfigure}[b]{0.3\linewidth}
\centering
\begin{tikzpicture}
  \begin{feynman}
    \vertex (e);
    \vertex [right=2cm of e] (f);
    \vertex [above right=2cm of f] (c);
    \vertex [below right=2cm of f] (d);
       \vertex [above left=2cm of e] (a);
    \vertex [below left=2cm of e] (b);
    \diagram* {
      (a) -- [scalar] (e)--[scalar] (b) ,
            (c) -- [scalar] (f)--[scalar] (d),
             (c) -- [scalar] (f)--[scalar] (d),
             (e)--[scalar,half left] (f)--[scalar,half left] (e),
    };
  \end{feynman}
\end{tikzpicture}
\caption{$\Lam^2_{abcd}$} \label{fig:quartic1a}
\end{subfigure}
\begin{subfigure}[b]{0.3\linewidth}
\centering
\begin{tikzpicture}
  \begin{feynman}
    \vertex (e);
    \vertex [above left=2cm of e] (a);
     \vertex [below left=2cm of e] (b);
    \vertex [above right=0.5cm of e] (f);    
    \vertex [above right=1cm of f] (g);
    \vertex [above right=0.5cm of g] (c);
    \vertex [below right=2cm of e] (d);
    \diagram* {
      (e)--[scalar] (a) , (e)--[scalar] (b) , (e)--[scalar] (f) , (e)--[scalar] (d) , (f)--[scalar] (g),(g)--[scalar] (c), (f)--[boson,half left] (g),
    };
  \end{feynman}
\end{tikzpicture}
\caption{$\Lam^S_{abcd}$} \label{fig:quartic1b}
\end{subfigure}
\begin{subfigure}[b]{0.3\linewidth}
\centering
\begin{tikzpicture}
  \begin{feynman}
    \vertex (e);
    \vertex [right=2cm of e] (f);
    \vertex [above right=2cm of f] (c);
    \vertex [below right=2cm of f] (d);
       \vertex [above left=2cm of e] (a);
    \vertex [below left=2cm of e] (b);
    \diagram* {
      (a) -- [scalar] (e)--[scalar] (b) ,
            (c) -- [scalar] (f)--[scalar] (d),
             (c) -- [scalar] (f)--[scalar] (d),
             (e)--[boson ,half left] (f)--[boson,half left] (e),
    };
  \end{feynman}
\end{tikzpicture}
\caption{$A_{abcd}$} \label{fig:quartic1c}
\end{subfigure} 
\caption{One-loop contributions to the running of quartic couplings with corresponding group factors e.g. in the Landau gauge.} \label{fig:quartic1}
\end{figure}

The contribution of one-loop diagrams to the beta function of the quartic coupling $\lambda_{abcd}$ is 

\beq \label{eq:betaquartic}
\beta^{\text{1-loop}}_{abcd}\equiv {d\lambda_{abcd}\ov d\ln \mu} ={1\ov (4\pi)^2}\left \{ \Lambda_{abcd}^2-3g^2 \Lambda_{abcd}^S +3g^4 A_{abcd} \right \},
\eeq
where the group invariants $\Lambda^2_{abcd},\Lam^{S}_{abcd},A_{abcd}$ are defined as follows:
\beq
~&\Lam_{abcd}^2={1\ov 8}\sum_{\text{perm}} \lam_{abef}\lam_{efcd},~~ \Lam^S_{abcd}=\sum_{k=a,b,c,d} C_2(k)\lam_{abcd},
\\&  A_{abcd}={1\ov 4}\sum_{\text{perm}} \{\Lam_{ac,ef}\Lam_{ef,bd}+\Lam_{ae,fd}\Lam_{eb,cf}\}~\text{ with } \Lam_{ab,cd}\equiv\Big(T_A(S)\Big)_{ac}\Big(T_A(S)\Big)_{bd}.
\eeq
Here, `perm' denotes all possible permutations of the indicies $\{a,b,c,d\}$. $C_2(k)$ is the quadratic Casimir of the representation in which the scalar $k$ transforms under the gauge group.  The diagrammatic interpretation of each term is straightforward: $\Lam_{abcd}^2$ and $A_{abcd}$  come from one-particle irreducible vertices made of two quartic and two gauge interaction vertices respectively, while $\Lam^S_{abcd}$  comes from the one-loop anomalous dimension of each external scalar propagators (see the diagrams in figure \ref{fig:quartic1}). When the gauge group is semi-simple, we need to do the following two substitutions in the expressions of the 1-loop beta functions:
\beq
g^2\Lam^S_{abcd}\rightarrow \sum_ig_i^2 (\Lam^S)_{abcd}^i, \\
 g^2\Lam_{ab,cd}\rightarrow \sum_i  g_i^2 (\Lam)_{ab,cd}^i,
 \label{app:1-loop substitution ruels}
\eeq
where $(\Lam^S)_{abcd}^i$ and $(\Lam)_{ab,cd}^i$ are given by
\beq
&(\Lam^S)_{abcd}^i\equiv\sum_{k=a,b,c,d} C_2^i(k)\lam_{abcd},~~
(\Lam)_{ab,cd}^i\equiv\Big(T_A^i(S)\Big)_{ac}\Big(T_A^i(S)\Big)_{bd}.
\eeq 
Based on the above substitution rules, we find it convenient to define 
\beq
(A)_{abcd}^{ij}\equiv {1\ov 4}\sum_{\text{perm}} \{(\Lam)_{ac,ef}^i(\Lam)_{ef,bd}^j+(\Lam)_{ae,fd}^i(\Lam)_{eb,cf}^j\}.
\eeq
Then the substitution rule for the last term in \eqref{eq:betaquartic} is as follows:
\beq
g^4 A_{abcd}\rightarrow \sum_{i,j}g_i^2 g_j^2 (A)_{abcd}^{ij}.
\eeq

\medskip
\noindent
$\bullet$ {\it \textbf{Two-loop}}
\medskip

More complicated structures arise at the level of two-loop diagrams where 11 different types of combinations  contribute to the beta functions:
\beq \label{eq:qb2}
(4\pi)^4\beta_{abcd}^{\text{2-loop}}=&{1\ov2} \sum_k \Lam^2(k)\lam_{abcd}-{\bar \Lam}^3_{abcd}+g^2 (2{\bar\Lam}^{2S}_{abcd}-6\Lam^{2g}_{abcd})
\\&-g^4\left \{\left[{35\ov 3} C_2(G)-{10\ov 3}\kappa S_2(F)-{11\ov 12}S_2(S)\right]\Lam^S_{abcd} -{3\ov 2}\Lam^{SS}_{abcd}-{5\ov 2}A^\lam_{abcd}-{1\ov 2}{\bar A}^\lam_{abcd}\right \}
\\&+g^6\left\{ \left[ {161\ov 6}C_2(G) -{32\ov 3}\kappa  S_2(F) -{7\ov 3}S_2(S) \right ]A_{abcd}-{15\ov 2}A^S_{abcd}+27A^g_{abcd} \right\}.
\eeq
Here the new group factors are defined as follows:
\beq \label{eq:two-loop quartic group}
~&\Lam^2(k)={1\ov 6}\lam_{kcde}\lam_{kcde},~~{\bar{\Lam}}^3_{abcd}={1\ov 4}\sum_{\text{perms}}\lam_{abef}\lam_{cegh}\lam_{dfgh},\\&
{\bar{\Lam}}^{2S}_{abcd}={1\ov 8}\sum_{\text{perm}}\Big(C_2(S)\Big)_{fg}\lam_{abef}\lam_{cdeg},~~
\Lam^{2g}_{abcd}={1\ov 8}\sum_{\text{perms}}\lam_{abef}\lam_{cdgh}\Big(T_A(S)\Big)_{eg}\Big(T_A(S)\Big)_{fh},\\
& \Lam^{SS}_{abcd}=\sum_k C_2(k)^2\lam_{abcd}, ~~  A^{\lam}_{abcd}={1\ov 4}\sum_{\text{perms}}\lam_{abef}\{T_A(S),T_B(S)\}_{ef}\{T_A(S),T_B(S)\}_{cd},\\
& {\bar A}^{\lam}_{abcd}={1\ov 4}\sum_{\text{perms}}\lam_{abef}\{ T_A(S),T_B(S)\}_{ce}\{T_A(S),T_B(S)\}_{df}, ~~ A^S_{abcd}=\sum_k C_2(k) A_{abcd}, \\
&A^g_{abcd}={1\ov 8} f^{ACE}f^{BDE}\sum_{\text{perms}} \{T_A(S),T_B(S)\} _{ab} \{T_C(S),T_D(S)\}_{cd}.
\eeq
Here, $\Big(C_2(S)\Big)_{fg}\equiv \Big(T_A(S)T_A(S)\Big)_{fg}$,  and $f^{ABC}$'s denote the structure constants of the gauge group. 

For the semi-simple gauge group $G=\prod_iG_i$, the following additional substitution rules are necessary because of the internal gluon loops:
\beq
~~g^2C_2(R)~&\rightarrow~ \sum_i g_i^2 C_2^i(R),
\\~~g^4C_2(G)C_2(R)~&\rightarrow~ \sum_i g_i^4C_2^i(G)C_2^i(R),
\\~~g^4C_2(R_1)S_2(R_2)~&\rightarrow~  \sum_i g_i^4 C_2^i(R_1) S_2^i(R_2),
\\~~g^4 C_2(R_1)C_2(R_2)~&\rightarrow ~\sum_{i,j} g_i^2g_j^2 C_2^i(R_1)C_2^j(R_2),
\\~~g^6 S_2(R)A_{abcd}~&\rightarrow ~\sum_{i,j}g_i^4g_j^2S_2^i(R) (A)_{abcd}^{ij},\\
~~g^6 C_2(G)A_{abcd}~&\rightarrow ~\sum_{i,j}g_i^4g_j^2 C_2^i(G) (A)_{abcd}^{ij},
\\
~~g^6 C_2(R)A_{abcd}~&\rightarrow ~\sum_{i,j,k}g_i^2 g_j^2 g_k^2 C_2^k(R) (A)_{abcd}^{ij},
\\~~ g^6 A^g_{abcd}~&\rightarrow ~\sum_i g_i^6 (A^g)_{abcd}^i,
\eeq 
where $(A^g)_{abcd}^i$ refers to the invariant $A^g_{abcd}$ defined in \eqref{eq:two-loop quartic group} for the simple factor $G_i$. Note that the substitution rules for $g^2 \Lam^{2g}_{abcd}$, $g^4 A^\lam_{abcd}$ and $g^4 \overline{A}^\lam_{abcd}$ can be obtained from the rule already given in the second line of \eqref{app:1-loop substitution ruels}.

Next we outline the diagrammatic origin of the different group theoretical contributions to the running of two-loop quartic couplings. For  simplicity, we work in the Landau gauge.

\medskip
\noindent
$\bullet$ {\it ${1\ov2} \sum_k \Lam^2(k)\lam_{abcd}$, $-{\bar \Lam}^3_{abcd}$} 
\medskip

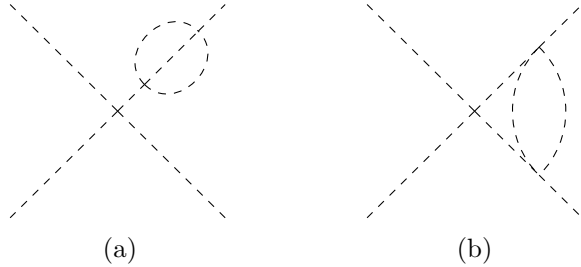
\begin{figure}[!h]
\centering
\begin{subfigure}[b]{0.3\linewidth}
\centering
\begin{tikzpicture}
  \begin{feynman}
    \vertex (e);
    \vertex [above left=2cm of e] (a);
     \vertex [below left=2cm of e] (b);
    \vertex [above right=0.5cm of e] (f);    
    \vertex [above right=1cm of f] (g);
    \vertex [above right=0.5cm of g] (c);
    \vertex [below right=2cm of e] (d);
    \diagram* {
      (e)--[scalar] (a) , (e)--[scalar] (b) , (e)--[scalar] (f) , (e)--[scalar] (d) , (f)--[scalar] (g),(g)--[scalar] (c), (f)--[scalar,half left] (g),  (f)--[scalar,half right] (g),
    };
  \end{feynman}
\end{tikzpicture}
\caption{} \label{fig:2-1}
\end{subfigure}
\begin{subfigure}[b]{0.3\linewidth}
\centering
\begin{tikzpicture}
  \begin{feynman}
    \vertex (e);
    \vertex [above left=2cm of e] (a);
     \vertex [below left=2cm of e] (b);
    \vertex [above right=1.2cm of e] (f);    
    \vertex [below right=1.2cm of e] (g);
    \vertex [above right=.8cm of f] (c);
    \vertex [below right=.8cm of g] (d);
    \diagram* {
      (e)--[scalar] (a) , (e)--[scalar] (b) , (e)--[scalar] (f) , (e)--[scalar] (g) , (f)--[scalar] (c),(g)--[scalar] (d), (f)--[scalar,quarter left] (g),  (f)--[scalar,quarter right] (g),
    };
  \end{feynman}
\end{tikzpicture}
\caption{} \label{fig:2-2}
\end{subfigure}
\caption{$(\text{quartic}^3)$ contributions to the two-loop beta function of quartic couplings.}
\end{figure}

This first contribution comes from the two-loop anomalous dimension of the scalar propagators as shown in the figure \ref{fig:2-1}. The second contribution is a genuine 1-PI contribution to the quartic vertex as shown in the figure \ref{fig:2-2}. No substitution rule is required since there is no gluon propagator.

\medskip
\noindent
$\bullet$ {\it $2 {\bar\Lam}^{2S}_{abcd}g^2$, $-6\Lam^{2g}_{abcd}g^2$}
\medskip

These terms came from all possible insertions of a single gluon propagator to the one-loop quartic 1-PI diagram in the figure \ref{fig:quartic1a} as shown in the figure \ref{fig:2-3,4}. The required substitution rules are $g^2\Lam_{ab,cd}\rightarrow \sum_i  g_i^2 (\Lam)_{ab,cd}^i$ and $g^2C_2(R)\rightarrow\sum_i g_i^2 C_2^i(R)$ to take account all possible gluon propagators. 
 
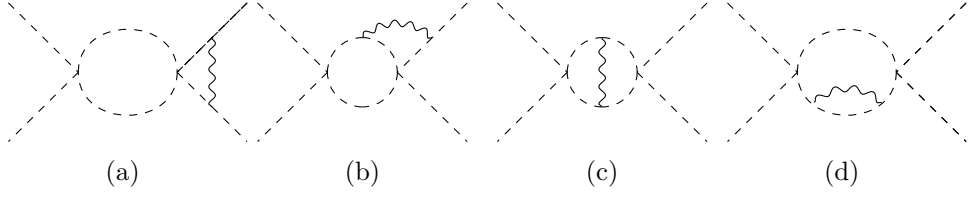
\begin{figure}[!h]
\centering
\begin{subfigure}[b]{0.2\linewidth}
\centering
\begin{tikzpicture}
  \begin{feynman}
    \vertex (e);
    \vertex [right=1.3cm of e] (f);
    \vertex [above right=0.65cm of f] (g);
     \vertex [above right=0.65cm of g] (c);
    \vertex [below right=0.65cm of f] (h);
    \vertex [below right=0.65cm of h] (d);
       \vertex [above left=1.3cm of e] (a);
    \vertex [below left=1.3cm of e] (b);
    \diagram* {
      (a) -- [scalar] (e)--[scalar] (b) ,
            (c) -- [scalar] (f)--[scalar] (g)--[scalar](c),
             (c) -- [scalar] (f)--[scalar] (h)--[scalar](d), (g)--[boson](h),
             (e)--[scalar,half left] (f)--[scalar,half left] (e),
    };
  \end{feynman}
\end{tikzpicture}
\caption{}
\end{subfigure}
\begin{subfigure}[b]{0.2\linewidth}
\centering
\begin{tikzpicture}
  \begin{feynman}
     \vertex (e);
     \vertex [below right=0.65cm of e] (i);
    \vertex [above right=0.65cm of e] (h);
        \vertex [below right=0.65cm of h] (f);
        \vertex [above right=0.65cm of f] (g);
     \vertex [above right=0.65cm of g] (c);
    \vertex [below right=1.3cm of f] (d);
       \vertex [above left=1.3cm of e] (a);
    \vertex [below left=1.3cm of e] (b);
    \diagram* {
      (a) -- [scalar] (e)--[scalar] (b) ,
            (c) -- [scalar](g)-- [scalar] (f)--[scalar] (d),
             (e)--[scalar,quarter left] (h)--[scalar,quarter left] (f)--[scalar,quarter left] (i) --[scalar,quarter left] (e), (h)--[quarter left, boson] (g),};
  \end{feynman}
\end{tikzpicture}
\caption{}
\end{subfigure}
\begin{subfigure}[b]{0.2\linewidth}
\centering
\begin{tikzpicture}
  \begin{feynman}
     \vertex (e);
     \vertex [below right=0.65cm of e] (i);
    \vertex [above right=0.65cm of e] (h);
        \vertex [below right=0.65cm of h] (f);
        \vertex [above right=0.65cm of f] (g);
     \vertex [above right=0.65cm of g] (c);
    \vertex [below right=1.3cm of f] (d);
       \vertex [above left=1.3cm of e] (a);
    \vertex [below left=1.3cm of e] (b);
    \diagram* {
      (a) -- [scalar] (e)--[scalar] (b) ,
            (c) -- [scalar](g)-- [scalar] (f)--[scalar] (d),
             (e)--[scalar,quarter left] (h)--[scalar,quarter left] (f)--[scalar,quarter left] (i) --[scalar,quarter left] (e), (h)--[ boson] (i),};
  \end{feynman}
\end{tikzpicture}
\caption{}
\end{subfigure}
\begin{subfigure}[b]{0.2\linewidth}
\centering
\begin{tikzpicture}
  \begin{feynman}
        \vertex (e);
    \vertex [right=1.3cm of e] (f);
             \vertex [below=0.4cm of e] (m);
            \vertex [below=0.4cm of f] (n);
             \vertex [right=0.22cm of m] (k);
            \vertex [left=0.19cm of n] (l);
    \vertex [above right=1.3cm of f] (c);
    \vertex [below right=1.3cm of f] (d);
       \vertex [above left=1.3cm of e] (a);
    \vertex [below left=1.3cm of e] (b);
    \diagram* {
      (a) -- [scalar] (e)--[scalar] (b) ,
            (c) -- [scalar] (f)--[scalar] (d),
             (c) -- [scalar] (f)--[scalar] (d),
             (e)--[scalar,half left] (f)--[scalar,half left] (e),
             (k)--[boson,quarter left](l),
    };
  \end{feynman}
\end{tikzpicture}
\caption{}
\end{subfigure}
\caption{$(\text{quartic}^2 \cdot \text{gauge}^2)$ contributions to the two-loop beta function of quartic couplings. } \label{fig:2-3,4}
\end{figure}

\medskip
\noindent
$\bullet$ {\it $-g^4\left \{\left[{35\ov 3} C_2(G)-{10\ov 3}\kappa S_2(F)-{11\ov 12}S_2(S)\right]\Lam^S_{abcd} -{3\ov 2}\Lam^{SS}_{abcd}\right \}$}
\medskip

This factor comes from the $g^4$ contribution to the two-loop anomalous dimension of the scalar propagator where the corresponding 8 feynman diagrams are given in the figure 1 of \cite{Machacek:1983tz}. The required substitution rules are
\begin{equation*}
\begin{split}
g^4C_2(G)C_2(R)&\rightarrow \sum_i g_i^4C_2^i(G)C_2^i(R),\\
g^4C_2(R_1)S_2(R_2)&\rightarrow \sum_i g_i^4 C_2^i(R_1) S_2^i(R_2), \\
g^4 C_2(R_1)C_2(R_2)&\rightarrow \sum_{i,j} g_i^2g_j^2 C_2^i(R_1)C_2^j(R_2)
\end{split}
\end{equation*}
to cover all possible gluons inside a scalar propagator together with all possible matter contribution to the gluon propagator.

\medskip
\noindent
$\bullet$ {\it ${5\ov 2}A^\lam_{abcd}g^4$, ${1\ov 2}{\bar A}^\lam_{abcd}g^4$}
\medskip

\begin{figure}[!h]
\centering
\begin{subfigure}[b]{0.3\linewidth}
\centering
\begin{tikzpicture}
  \begin{feynman}
    \vertex (e);
       \vertex [above left=1.3cm of e] (a);
    \vertex [below left=1.3cm of e] (b);
    \vertex [above right=1.3cm of e] (f);
     \vertex [below right=1.3cm of e] (g);
    \vertex [right=1cm of f] (h);
        \vertex [right=1cm of h] (c);
      \vertex [right=1cm of g] (i);
        \vertex [right=1cm of i] (d);
    \diagram* {
      (a) -- [scalar] (e)--[scalar] (b) ,
            (e) -- [scalar] (f)--[scalar] (g)--[scalar](e),
             (f) -- [boson] (h)--[scalar] (c), (g) -- [boson] (i)--[scalar] (d),
             (h)--[scalar] (i),
    };
  \end{feynman}
\end{tikzpicture}
\caption{}\label{fig:2-7-1}
\end{subfigure}
\begin{subfigure}[b]{0.3\linewidth}
\centering
\begin{tikzpicture}
  \begin{feynman}
       \vertex (e);
       \vertex [above left=1.3cm of e] (a);
    \vertex [below left=1.3cm of e] (b);
    \vertex [above right=1.3cm of e] (f);
     \vertex [below right=1.3cm of e] (g);
    \vertex [below right=1.3cm of f] (h);
        \vertex [above right=1.3cm of h] (c);
      \vertex [below right=1.3cm of h] (d);
    \diagram* {
      (a) -- [scalar] (e)--[scalar] (b) ,
            (e) -- [scalar] (f)--[scalar] (g)--[scalar](e),
             (f) -- [boson] (h)--[boson] (g), (c) -- [scalar] (h)--[scalar] (d),          
    };
  \end{feynman}
\end{tikzpicture}
\caption{} \label{fig:2-7-2}
\end{subfigure}
\begin{subfigure}[b]{0.3\linewidth}
\centering
\begin{tikzpicture}
  \begin{feynman}
     \vertex (e);
     \vertex [below right=1cm of e] (i);
    \vertex [above right=1cm of e] (h);
        \vertex [below right=1cm of h] (f);
     \vertex [above right=1.3cm of f] (x);
    \vertex [below right=1.3cm of f] (y);
      \vertex [right=1cm of x] (c);
    \vertex [right=1cm of y] (d);
       \vertex [above left=1.3cm of e] (a);
    \vertex [below left=1.3cm of e] (b);
    \diagram* {
      (a) -- [scalar] (e)--[scalar] (b) ,
            (x) -- [boson](f)-- [boson] (y),
             (e)--[scalar,quarter left] (h)--[scalar,quarter left] (f)--[scalar,quarter left] (i) --[scalar,quarter left] (e), (c)--[scalar](x)--[scalar] (y)--[scalar](d)};
  \end{feynman}
\end{tikzpicture}
\caption{}\label{fig:2-7-3}
\end{subfigure}

\begin{subfigure}[b]{0.2\linewidth}
\centering
\begin{tikzpicture}
  \begin{feynman}
   \vertex (e);
     \vertex [right=1cm of e] (f);
      \vertex [right=1cm of f] (g);
      \vertex [above right=1.3cm of g] (c);
    \vertex [below right=1.3cm of g] (d);
       \vertex [above left=1.3cm of e] (a);
    \vertex [below left=1.3cm of e] (b);
    \diagram* {
      (a) -- [scalar] (e)--[scalar] (b) ,
            (c) -- [scalar](g)-- [scalar] (d),
             (e)--[scalar,half left] (f)--[scalar,half left] (e), (f)--[boson,half left] (g)--[boson,half left] (f)};
  \end{feynman}
\end{tikzpicture}
\caption{} \label{fig:2-7-4}
\end{subfigure}
\begin{subfigure}[b]{0.3\linewidth}
\centering
\begin{tikzpicture}
  \begin{feynman}
    \vertex (e);
    \vertex [above left=1.5cm of e] (a);
     \vertex [below left=1.5cm of e] (b);
    \vertex [above right=1cm of e] (f);    
    \vertex [below right=1cm of e] (g);
    \vertex [above right=.5cm of f] (c);
    \vertex [below right=.5cm of g] (d);
    \diagram* {
      (e)--[scalar] (a) , (e)--[scalar] (b) , (e)--[scalar] (f) , (e)--[scalar] (g) , (f)--[scalar] (c),(g)--[scalar] (d), (f)--[boson,quarter left] (g),  (f)--[boson,quarter right] (g),
    };
  \end{feynman}
\end{tikzpicture}
\caption{} \label{fig:2-8}
\end{subfigure}
\caption{$(\text{quartic} \cdot \text{gauge}^4)$ contributions to the two-loop beta function of quartic couplings. } \label{fig:2-7,8}
\end{figure}
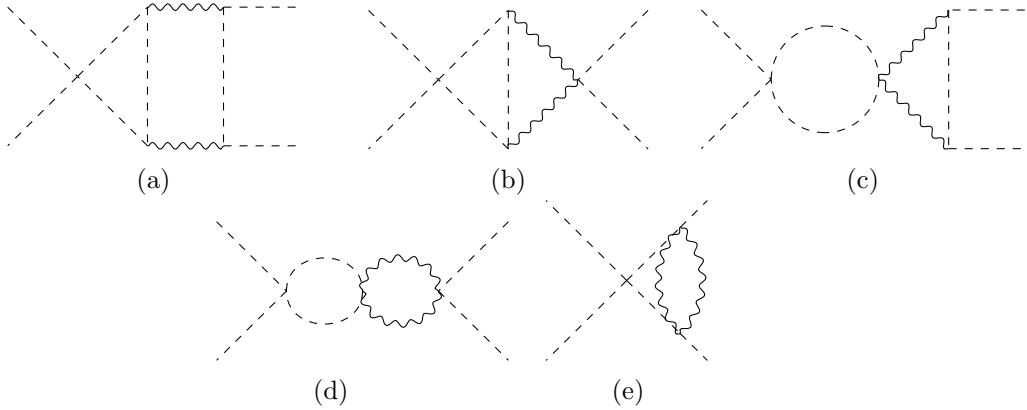

The first term comes from the diagrams \ref{fig:2-7-1} $\sim$ \ref{fig:2-7-4} in figure \ref{fig:2-7,8}, and the second term originates from \ref{fig:2-8}. The required substitution rules are to reflect the fact that each gluon propagator has semi-simple  indices corresponding to the simple factors $\{G_i\}$. 

\medskip
\noindent
$\bullet$ {\it $\left[ {161\ov 6}C_2(G) -{32\ov 3} \kappa S_2(F) -{7\ov 3}S_2(S) \right ]A_{abcd}g^6$, $-{15\ov 2}A^S_{abcd}g^6$, $27A^g_{abcd}g^6$}
\medskip

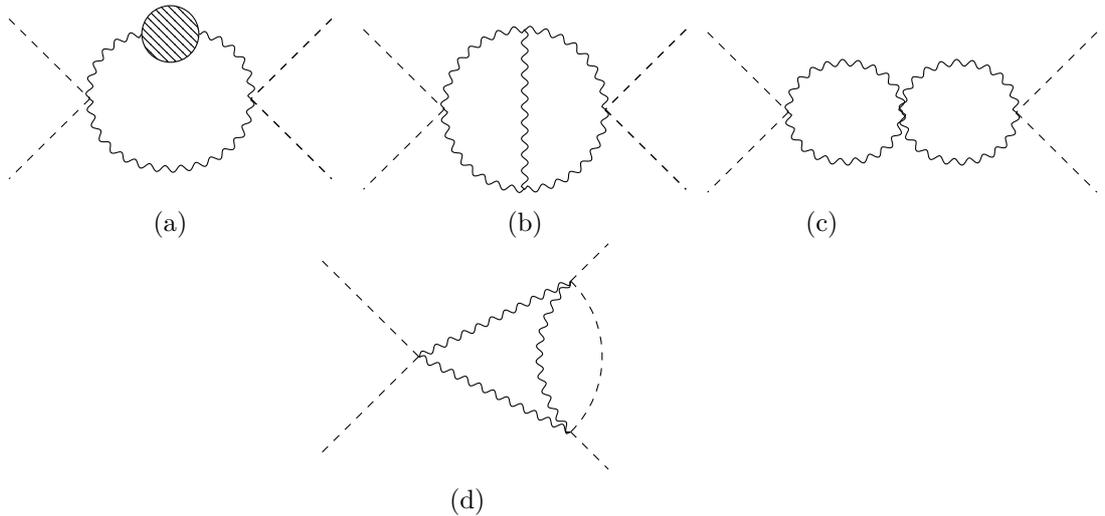
\begin{figure}[!h]
\centering
\begin{subfigure}[b]{0.3\linewidth}
\centering
\begin{tikzpicture}
  \begin{feynman}
    \vertex (e);
    \vertex [above right=1.5cm of e] (g);
     \vertex [below=0.2cm of g] (h);
     \vertex [left =0.38 cm of h] (i); \vertex [right =0.38cm of h](j);
    \vertex[blob] at (h) {};
    \vertex [below right=1.5cm of g] (f);
    \vertex [above right=1.5cm of f] (c);
    \vertex [below right=1.5cm of f] (d);
       \vertex [above left=1.5cm of e] (a);
    \vertex [below left=1.5cm of e] (b);
    \diagram* {
      (a) -- [scalar] (e)--[scalar] (b) ,
            (c) -- [scalar] (f)--[scalar] (d),
             (c) -- [scalar] (f)--[scalar] (d),
 (f)--[boson,half left] (e),
 (e)--[boson,quarter left](i) , (j)--[boson,quarter left](f)
    };
  \end{feynman}
\end{tikzpicture}
\caption{} \label{fig:2f-1}
\end{subfigure} 
\begin{subfigure}[b]{0.3\linewidth}
\centering
\begin{tikzpicture}
  \begin{feynman}
    \vertex (e);
    \vertex [above right=1.5cm of e] (g);
     \vertex [below right=1.5cm of e] (h);
    \vertex [below right=1.5cm of g] (f);
    \vertex [above right=1.5cm of f] (c);
    \vertex [below right=1.5cm of f] (d);
       \vertex [above left=1.5cm of e] (a);
    \vertex [below left=1.5cm of e] (b);
    \diagram* {
      (a) -- [scalar] (e)--[scalar] (b) ,
            (c) -- [scalar] (f)--[scalar] (d),
             (c) -- [scalar] (f)--[scalar] (d),
 (f)--[boson,quarter left] (h)--[boson,quarter left] (e),
 (e)--[boson,quarter left](g)--[boson,quarter left](f), (g)--[boson](h)
    };
  \end{feynman}
\end{tikzpicture}
\caption{} \label{fig:2f-2}
\end{subfigure} 
\begin{subfigure}[b]{0.2\linewidth}
\centering
\begin{tikzpicture}
  \begin{feynman}
   \vertex (e);
     \vertex [right=1.5cm of e] (f);
      \vertex [right=1.5cm of f] (g);
      \vertex [above right=1.5cm of g] (c);
    \vertex [below right=1.5cm of g] (d);
       \vertex [above left=1.5cm of e] (a);
    \vertex [below left=1.5cm of e] (b);
    \diagram* {
      (a) -- [scalar] (e)--[scalar] (b) ,
            (c) -- [scalar](g)-- [scalar] (d),
             (e)--[boson,half left] (f)--[boson,half left] (e), (f)--[boson,half left] (g)--[boson,half left] (f)};
  \end{feynman}
\end{tikzpicture}
\caption{} \label{fig:2f-3}
\end{subfigure}

\centering
\begin{subfigure}[b]{0.3\linewidth}
\centering
\begin{tikzpicture}
  \begin{feynman}
    \vertex (e);
    \vertex [above left=1.8cm of e] (a);
     \vertex [below left=1.8cm of e] (b);
     \vertex[right=2 cm of e] (k);
    \vertex [above =1cm of k] (f);    
    \vertex [below =1cm of k] (g);
    \vertex [above right=.7cm of f] (c);
    \vertex [below right=.7cm of g] (d);
    \diagram* {
      (e)--[scalar] (a) , (e)--[scalar] (b) , (e)--[boson] (f) , (e)--[boson] (g) , (f)--[scalar] (c),(g)--[scalar] (d), (f)--[scalar,quarter left] (g),  (f)--[boson,quarter right] (g),
    };
  \end{feynman}
\end{tikzpicture}
\caption{} \label{fig:2f-4}
\end{subfigure}
\caption{ $(\text{gauge}^6)$ contributions to the two-loop beta function of quartic couplings. } \label{fig:2f}
\end{figure}

These three terms come solely from the cubic or quartic gauge interaction vertices. The four Feynman diagrams shown in  figure \ref{fig:2f} generate these terms. The first diagram \ref{fig:2f-1} generates  $A_{abcd}$ and one needs to use the substitution rules 
\begin{equation*}
g^6 S_2(R)A_{abcd}~\rightarrow ~\sum_{i,j}g_i^4g_j^2S_2^i(R) (A)_{abcd}^{ij}
\end{equation*}
because of the semi-simple index for the gluon lines and the corresponding one-loop insertions. The second and the third diagrams  (\ref{fig:2f-2} and \ref{fig:2f-3}) generate $A^g_{abcd}$ and correspond to addition of the gluon propagator in the figure \ref{fig:quartic1c}. Hence, all the gluons should belong to the same node $G_i$, which is reflected in the substitution rule $g^6 A^g_{abcd}\rightarrow  \sum_i g_i^6 (A^g)_{abcd}^i$. Finally, the fourth diagram \ref{fig:2f-4} generates all three group invariants. While extending $A^S_{abcd}$ to the corresponding invariant for a semi-simple gauge group, one should use the substitution rules for  $A_{abcd}$ and $C_2(R)$ introduced earlier. 

\subsection{QCD with fundamental scalars / Bifundamental scalar QCD}\label{appsec:simple}

Let us now apply the formalism of \ref{appsec:gauge}, \ref{appsec:quartics} to evaluate the two-loop beta function of the gauge coupling and the one-loop beta functions of the quartic couplings for the two models described in the section \ref{sec:simple}.

\medskip
\noindent
$\bullet$ {\it QCD with fundamental scalars}
\medskip

The model of QCD with fundamental scalars is made of $SO(N_c)$ Yang-Mills gauge fields with $N_f$ flavors of massless Majorana fermions and $N_s$ flavors of massless real scalars transforming in the fundamental representation of the gauge group. Its renormalized  Lagrangian is
\begin{equation}
\begin{split}
\mathcal L_{\text{Vector}}=&-\frac{1}{4}F_{\mu\nu}^A F^{\mu\nu A}+\frac{i}{2}\overline{\psi}_a^{(p)} (\slashed{D}\psi^{(p)})_a+\frac{1}{2}\Big(D_{\mu} \phi^{(p)} \Big)_a\Big(D^{\mu} \phi^{(p)} \Big)_a\\
&-\tilde h \phi_a^{(p)}\phi_b^{(p)}\phi_b^{(q)}\phi_a^{(q)}-\tilde f \phi_a^{(p)}\phi_a^{(p)}\phi_b^{(q)}\phi_b^{(q)}.
\end{split}
\end{equation}
To evaluate the beta functions of the different couplings, we need to identify the symmetric couplings and the generators of the gauge group in the representation of the scalar fields. These are given in \eqref{vector qcd: symm. couplings} and  \eqref{vector QCD: generators} respectively. In addition, we need the quadratic Casimirs and the Dynkin indices of the different representations which are given below: 
\begin{equation}
\begin{split}
&C_2(F)
= \frac{N_{c}-1}{4},\ C_2(S)= \frac{N_{c}-1}{4},\ 
C_2(G)= \frac{N_{c}-2}{2},\\
& S_2(F)=  \frac{N_f}{2},\ S_2(S)=  \frac{N_{s}}{2},\ S_2(G)=\ \frac{N_{c}-2}{2}.
\end{split}
\end{equation}
Now, using these, we can evaluate the 2-loop beta function of the gauge coupling  from \eqref{eq:betag}, and the 1-loop beta function of the quartic couplings from \eqref{eq:betaquartic}. We  provide the forms of these beta functions below in terms of the 't Hooft couplings ($\lambda$, $h$ and $f$) defined in  \eqref{vector QCD: 't Hooft couplings}:
\beq \label{eq:betavector}
~& \beta_{\lam}=-\left( {22-4x_f-x_s\ov 6} -{22\ov 3N_c}\right)\lam^2  +\left ( {13 x_f+4 x_s-34\ov 6}-\frac{23 x_f+5 x_s-136}{6 N_c}-\frac{68}{3 N_c^2}\right)\lam^3,
\\&\beta_h^{\text{1-loop}}=\frac{96 f h}{N_c^2 x_s}+h^2 \left(\frac{32}{N_c}+8 x_s+8\right)+h \lambda 
   \left(\frac{3}{N_c}-3\right)+\lambda ^2 \left(\frac{3}{32}-\frac{3}{16 N_c}\right),
\\&\beta_f^{\text{1-loop}}=f^2 \left(\frac{64}{N_c^2 x_s}+8\right)+f h \left(\frac{16}{N_c}+16 x_s+16\right)+24 h^2x_s+f \lambda 
   \left(\frac{3}{N_c}-3\right)+\frac{3 \lambda ^2 x_s}{32},
\eeq
where $x_{f,s}\equiv{N_{f,s}/ N_c}$.

\medskip
\noindent
$\bullet$ {\it Bifundamental scalar QCD}
\medskip

The bifundamental scalar QCD is made of $SO(N_c)\times SO(N_c)$ semi-simple gauge group with two sets of Majorana fermions, $\psi^{(p)}$ and $\chi^{(p)}$, each of which consists of $N_f$ flavors and transforms in the fundamental representation of one of the $SO(N_c)$'s while remaining a singlet under the other $SO(N_c)$. In addition, there is an $N_c\times N_c$ matrix of scalar fields denoted by $\Phi$ which transforms in the bifundamental representation of the gauge group. The renormalized Lagrangian of this model is 
\begin{equation}
\begin{split}
\mathcal L_{\text{Bifund}}=&-\frac{1}{4}\sum_{\alpha=1}^2(F_{\alpha})_{\mu\nu}^A(F_{\alpha})^{\mu\nu A}+\frac{i}{2}\overline{\psi}_a^{(p)} (\slashed{D}\psi^{(p)})_a+\frac{i}{2}\overline{\chi}_i^{(p)} (\slashed{D}\chi^{(p)})_i+\frac{1}{2}\text{Tr}\Bigg[\Big(D_{\mu} \Phi \Big)^T D^{\mu} \Phi\Bigg]\\
&-\widetilde{h}\text{Tr}\Big[\Phi^T\Phi\Phi^T\Phi\Big]-\widetilde{f}\text{Tr}\Big[\Phi^T\Phi\Big]\text{Tr}\Big[\Phi^T\Phi\Big].
\end{split}
\end{equation}
As explained in \ref{appsec:gauge} and \ref{appsec:quartics}, for computing the beta functions of the different couplings, we need to use \eqref{eq:betag}, \eqref{eq:betaquartic} with appropriate substitution rules to take into account the different internal gluon lines corresponds to the each simple gauge group. The diagrams responsible for these rules are given in figure \ref{fig:gauge2} for the two-loop $\beta_\lam$ and  figure \ref{fig:quartic1} for the one-loop $\beta_{f,h}$. The necessary ingredients to compute these beta functions are the symmetric couplings and the generators given in \eqref{bifundamental qcd: symm. couplings} and \eqref{bifundamental qcd: generators} respectively, and the quadratic Casimirs and the second Dynkin indices of the different representations. We provide these quadratic Casimirs and Dynkin indices below:
\begin{equation}
\begin{split}
&C_2^{\alpha}(F)
= \frac{N_{c}-1}{4},\ C_2^{\alpha}(S)= \frac{N_{c}-1}{4},\ 
C_2^{\alpha}(G)= \frac{N_{c}-2}{2},\\
& S_2^{\alpha}(F)=  \frac{N_{f}}{2},\ S_2^{\alpha}(S)=  \frac{N_{c}}{2},\ S_2^{\alpha}(G)=\ \frac{N_{c}-2}{2},
\end{split}
\end{equation}
where the superscript $\alpha$ distinguishes the two $SO(N_c)$'s. Using these ingredients, we can compute the beta functions for the case where the gauge couplings $g_1$ and $g_2$ corresponding to the two $SO(N_c)$'s are equal (say, $g$). We  provide the forms of these beta functions below in terms of the 't Hooft couplings ($\lambda$, $h$ and $f$) defined in  \eqref{bifundamental QCD: 't Hooft couplings}:
\beq \label{eq:betabifund}
~& \beta_{\lam}=-\left( {21-4x_f\ov 6} -{22\ov 3N_c}\right)\lam^2  +\left ( {13 x_f-27\ov6}-\frac{23 x_f-128}{6 N_c}-\frac{68}{3 N_c^2}\right)\lam^3,
\\&\beta_h^{\text{1-loop}}=\frac{96 f h}{N_c^2 }+h^2 \left(\frac{32}{N_c}+16\right)+h \lambda 
   \left(\frac{6}{N_c}-6\right)+\lambda ^2 \left(\frac{3}{16}-\frac{3}{4 N_c}\right),
\\&\beta_f^{\text{1-loop}}=f^2 \left(\frac{64}{N_c^2}+8\right)+f h \left(\frac{16}{N_c}+32\right)+24 h^2 +f \lambda 
   \left(\frac{6}{N_c}-6\right)+\frac{9 \lambda ^2 }{16},
\eeq
where $x_f=\frac{N_f}{N_c}$.

\section{Beta functions in the real double bifundamental model}
\label{app: rdb beta fns.}
In this appendix, we will derive the beta functions of the different couplings in the real double bifundamental model up to the contributions of  2-loop diagrams. For this, we will use the formalism discussed in appendix \ref{appsec:qcd}.

\subsection{Beta functions of the gauge couplings (up to 2-loops)}

In this model the gauge group has the following structure:
\begin{equation}
\begin{split}
G=(G_{11}\times G_{12})\times (G_{21}\times G_{22})
\end{split}
\end{equation}
where each $G_{i\gamma}$ is an orthogonal group with rank $N_{ci}$. We will denote all the scalars and fermions in the $i^{th}$ sector transforming nontrivially under the  group $G_i\equiv (G_{i1}\times G_{i2})$ by $S_i$ and $F_i$ respectively. We will also find it useful to denote all the scalars and the fermions in the model collectively by $S$ and $F$ respectively . 

The beta functions (up to 2-loops) of the gauge couplings are given by
\begin{equation}
\begin{split}
\beta_{g_i}\equiv& \mu\frac{dg_i}{d\mu}\\
=&-\frac{g_i^3}{(4\pi)^2}\Bigg\{\frac{11}{3}C_2^{i\gamma}(G_i)-\frac{4}{6}S_2^{i\gamma}(F_i)-\frac{1}{6}S_2^{i\gamma}(S_i)\Bigg\}\\
&-\frac{g_i^5}{(4\pi)^4}\Bigg\{\frac{34}{3}C_2^{i\gamma}(G_i)^2-\Big(2C_2^{i\gamma}(F_i)+\frac{10}{3}C_2^{i\gamma}(G_i)\Big)S_2^{i\gamma}(F_i)-\frac{1}{3}C_2^{i\gamma}(G_i)S_2^{i\gamma}(S_i)\Bigg\}\\
& +\frac{2g_i^3}{(4\pi)^4}\sum_{j=1}^2\sum_{\rho=1}^2 g_{j}^2C_2^{j\rho}(S_i)S_2^{i\gamma}(S_i).
\end{split}
\end{equation}
In the above expression, $C_2$ and $S_2$ denote the quadratic Casimir and the second Dynkin index of the corresponding representation. The super scripts $(i\gamma)$ indicate the simple Lie group $G_{i\gamma}$ corresponding to which these quantities are computed. The values of these quadratic Casimirs and Dynkin indices are given by $C_2^{j\beta}(R_i)=C_2^{i\beta}(R)\delta^j_i$ and $S_2^{j\beta}(R_i)=S_2^{i\beta}(R)\delta^j_i$ where
\begin{equation}
\begin{split}
&C_2^{i\beta}(F)
= \frac{N_{ci}-1}{4},\ C_2^{i\beta}(S)= \frac{N_{ci}-1}{4},\ 
C_2^{i\beta}(G)= \frac{N_{ci}-2}{2},\\
& S_2^{i\beta}(F)=  \frac{N_{fi}}{2},\ S_2^{i\beta}(S)=  \frac{N_{ci}}{2},\ S_2^{i\beta}(G)=\ \frac{N_{ci}-2}{2}.
\end{split}
\end{equation}
Substituting these values in the expressions of the beta functions of the gauge couplings, we get
\begin{equation}
\begin{split}
\beta_{g_i}=&-\frac{g_i^3}{2(4\pi)^2}\Bigg\{\frac{21N_{ci}-44-4N_{fi}}{6}\Bigg\}\\
&-\frac{ g_i^5}{2(4\pi)^4}\Bigg\{\frac{\Big(27 N_{ci}^2-128 N_{ci}+136\Big)-\Big(13N_{ci}-23\Big) N_{fi}}{6} \Bigg\}.
\end{split}
\end{equation}
From the above expression, one can obtain the beta functions of the rescaled couplings $\lambda_i\equiv \frac{N_{ci}g_i^2}{(4\pi)^2}$ which are given below: 
\begin{equation}
\begin{split}
\beta_{\lambda_i}=&-\lambda_i^2\Bigg\{\frac{21-4x_{fi}}{6}-\frac{22}{3N_{ci}}\Bigg\}-\lambda_i^3\Bigg\{\frac{27-13x_{fi}}{6} +\frac{23 x_{fi}-128}{6N_{ci}} +\frac{68}{3 N_{ci}^2}\Bigg\}.
\end{split}
\end{equation}
Here $x_{fi}\equiv \frac{N_{fi}}{N_{ci}}$.

\subsection{1-loop beta functions of the quartic couplings}
Let us now evaluate the beta functions of the quartic couplings. To use the results worked out for the 1-loop beta functions of these couplings in \cite{Machacek:1984zw}, we will introduce the following couplings which are symmetric under permutation of indices:
\begin{equation}
\begin{split}
\lambda_{a_1 i_1,b_1 j_1,c_1 k_1,d_1 l_1}=&4\widetilde{h}_1\Bigg[\delta_{i_1j_1}\delta_{k_1l_1}(\delta_{a_1c_1}\delta_{b_1 d_1}+\delta_{a_1d_1}\delta_{b_1c_1})+\delta_{i_1k_1}\delta_{j_1l_1}(\delta_{a_1b_1}\delta_{c_1d_1}+\delta_{a_1d_1}\delta_{b_1c_1})\\
&\quad+\delta_{i_1l_1}\delta_{j_1k_1}(\delta_{a_1b_1}\delta_{c_1d_1}+\delta_{a_1c_1}\delta_{b_1d_1})\Bigg]\\
&+8\widetilde{f}_1\Bigg[\delta_{i_1j_1}\delta_{k_1l_1}\delta_{a_1b_1}\delta_{c_1d_1}+\delta_{i_1k_1}\delta_{j_1l_1}\delta_{a_1c_1}\delta_{b_1d_1}+\delta_{i_1l_1}\delta_{j_1k_1}\delta_{a_1d_1}\delta_{b_1c_1}\Bigg],
\end{split}
\label{symmetrised couplings def 1}
\end{equation}

\begin{equation}
\begin{split}
\lambda_{a_2 i_2,b_2 j_2,c_2 k_2,d_2 l_2}=&4\widetilde{h}_2\Bigg[\delta_{i_2j_2}\delta_{k_2l_2}(\delta_{a_2c_2}\delta_{b_2 d_2}+\delta_{a_2d_2}\delta_{b_2c_2})+\delta_{i_2k_2}\delta_{j_2l_2}(\delta_{a_2b_2}\delta_{c_2d_2}+\delta_{a_2d_2}\delta_{b_2c_2})\\
&\quad+\delta_{i_2l_2}\delta_{j_2k_2}(\delta_{a_2b_2}\delta_{c_2d_2}+\delta_{a_2c_2}\delta_{b_2d_2})\Bigg]\\
&+8\widetilde{f}_2\Bigg[\delta_{i_2j_2}\delta_{k_2l_2}\delta_{a_2b_2}\delta_{c_2d_2}+\delta_{i_2k_2}\delta_{j_2l_2}\delta_{a_2c_2}\delta_{b_2d_2}+\delta_{i_2l_2}\delta_{j_2k_2}\delta_{a_2d_2}\delta_{b_2c_2}\Bigg],
\end{split}
\label{symmetrised couplings def 2}
\end{equation}

\begin{equation}
\begin{split}
&\lambda_{a_1 i_1,b_1 j_1, c_2 k_2,d_2 l_2}=\lambda_{a_1 i_1, c_2 k_2,b_1 j_1,d_2 l_2}=\lambda_{a_1 i_1, c_2 k_2,d_2 l_2,b_1 j_1}\\
&=\lambda_{ c_2 k_2, a_1 i_1,b_1 j_1,d_2 l_2}=\lambda_{ c_2 k_2, a_1 i_1,d_2 l_2,b_1 j_1}=\lambda_{ c_2 k_2, d_2 l_2,a_1 i_1,b_1 j_1}\\
&=8\widetilde{\zeta}\delta_{a_1b_1}\delta_{i_1j_1}\delta_{c_2d_2}\delta_{k_2l_2}.
\end{split}
\label{symmetrised couplings def 3}
\end{equation}
In terms of these couplings, the quartic potential of the scalar fields takes the following form:
\begin{equation}
\begin{split}
V_{\text{quartic}}=\frac{1}{4!}\Bigg[&\lambda_{a_1i_1,b_1j_1,c_1k_1,d_1l_1}\phi_{a_1i_1}\phi_{b_1j_1}\phi_{c_1k_1}\phi_{d_1l_1}+\lambda_{a_2i_2,b_2j_2,c_2k_2,d_2l_2}\phi_{a_2i_2}\phi_{b_2j_2}\phi_{c_2k_2}\phi_{d_2l_2}\\
&+\lambda_{a_1i_1,b_1j_1,c_2k_2,d_2l_2}\phi_{a_1i_1}\phi_{b_1j_1}\phi_{c_2k_2}\phi_{d_2l_2}+\lambda_{a_1i_1,b_2j_2,c_1k_1,d_2l_2}\phi_{a_1i_1}\phi_{b_2j_2}\phi_{c_1k_1}\phi_{d_2l_2}\\
&+\lambda_{a_1i_1,b_2j_2,c_2k_2,d_1l_1}\phi_{a_1i_1}\phi_{b_2j_2}\phi_{c_2k_2}\phi_{d_1l_1}+\lambda_{a_2i_2,b_1j_1,c_1k_1,d_2l_2}\phi_{a_2i_2}\phi_{b_1j_1}\phi_{c_1k_1}\phi_{d_2l_2}\\
&+\lambda_{a_2i_2,b_1j_1,c_2k_2,d_1l_1}\phi_{a_2i_2}\phi_{b_1j_1}\phi_{c_2k_2}\phi_{d_1l_1}+\lambda_{a_2i_2,b_2j_2,c_1k_1,d_1l_1}\phi_{a_2i_2}\phi_{b_2j_2}\phi_{c_1k_1}\phi_{d_1l_1}\Bigg]\ .
\end{split}
\end{equation}
where the summation over repeated indices is implicitly assumed. 

The couplings $\widetilde{h}_1$ and $\widetilde{f}_1$, or equivalently the rescaled couplings $h_1=\frac{ N_{c1} \widetilde{h}_1}{16\pi^2}$ and $f_1=\frac{ N_{c1}^2 \widetilde{f}_1}{16\pi^2}$, appear in the expression of $\lambda_{a_1 i_1,b_1 j_1,c_1 k_1,d_1 l_1}$. Therefore, to determine the beta functions of $h_1$ and $f_1$, we will first evaluate the beta function of $\lambda_{a_1 i_1,b_1 j_1,c_1 k_1,d_1 l_1}$. The beta functions of $h_2$ and $f_2$ can be then obtained by a $(1\leftrightarrow 2)$ exchange in the indices. Similarly, we will also evaluate the beta function of the rescaled coupling $\zeta=\frac{ N_{c1} N_{c2} \widetilde{\zeta}}{16\pi^2}$ by evaluating the same for $\lambda_{a_1 i_1,b_1 j_1, c_2 k_2,d_2 l_2}$.

\subsubsection{Evaluation of  $\beta_{h_i}^{\text{1-loop}}$ and $\beta_{f_i}^{\text{1-loop}}$}
The 1-loop beta function of the coupling $\lambda_{a_1i_1,b_1j_1,c_1k_1,d_1l_1}$, as derived in  \cite{Machacek:1984zw}, is  given by
\begin{equation}
\begin{split}
(4\pi)^2\beta_{a_1i_1,b_1j_1,c_1k_1,d_1l_1}^{\text{1-loop}}=&(\Lambda^2)_{a_1 i_1,b_1 j_1,c_1 k_1,d_1 l_1}-3\sum_{i=1}^2\sum_{\beta=1}^2\ g_i^2(\Lambda^S)_{a_1 i_1,b_1 j_1,c_1 k_1,d_1 l_1}^{i\beta}\\
&+3\sum_{i,j=1}^2\sum_{\beta,\gamma=1}^2g_i^2 g_j^2(A)_{a_1 i_1,b_1 j_1,c_1 k_1,d_1 l_1}^{i\beta,j\gamma}
\end{split}
\end{equation}
where
\begin{equation}
\begin{split}
(\Lambda^2)_{a_1i_1,b_1j_1,c_1k_1,d_1l_1}\equiv &\frac{1}{8}\sum_{\text{perms}}\lambda_{a_1i_1,b_1j_1,e u,f v}\lambda_{e u,f v,c_1k_1,d_1l_1},\\
(\Lambda^S)_{a_1i_1,b_1j_1,c_1k_1,d_1l_1}^{i\beta}\equiv & 4C_2^{i\beta}(S_1)\lambda_{a_1i_1,b_1j_1,c_1k_1,d_1l_1},\\
(A)_{a_1i_1,b_1j_1,c_1k_1,d_1l_1}^{i\beta,j\gamma}\equiv & \frac{1}{4}\sum_{\text{perms}}\Big((\Lambda)^{i\beta}_{a_1i_1,c_1k_1;eu,fv}(\Lambda)^{j\gamma}_{eu,fv;b_1j_1,d_1 l_1}\\
&\qquad\quad+(\Lambda)^{i\beta}_{a_1i_1,eu;fv,d_1l_1}(\Lambda)^{j\gamma}_{eu,b_1j_1;c_1k_1,fv}\Big),
\end{split}
\label{h1, f1 1-loop:ingredients}
\end{equation}
with the quantity $(\Lambda)^{i\beta}_{a_1i_1,c_1k_1;eu,fv}$ defined as follows:
\begin{equation}
\begin{split}
(\Lambda)^{i\beta}_{a_1i_1,c_1k_1;eu,fv}\equiv& \Big(T_A^{i\beta}(S)\Big)_{a_1i_1,eu}\Big(T_A^{i\beta}(S)\Big)_{c_1k_1,fv}.\\
\end{split}
\label{h1, f1 1-loop: Lambda def}
\end{equation}
Let us now briefly explain the notations used in the definition of the  above objects. In the first and third lines of \eqref{h1, f1 1-loop:ingredients}, the sums are over all permutations of the indices $(a_1i_1,b_1j_1,c_1k_1,d_1l_1)$. In the second line, $S_1$ denotes the scalar fields in the first sector. They transform in the bifundemental representation of $SO(N_{c1})\times SO(N_{c1})$, and are invariant under the orthogonal transformations in the other sector. Therefore, the quadratic Casimir $C_2^{i\beta}(S_1)$ has the following value:
\begin{equation}
\begin{split}
C_2^{i\beta}(S_1)=\delta_{i1}\Big(\frac{N_{ci}-1}{4}\Big).
\end{split}
\end{equation}
The quantities $T_A^{i\beta}(S)$ in \eqref{h1, f1 1-loop: Lambda def} are the generators of the representation in which the scalar fields transform under the group $G_{i\beta}$.  For example, the generators $T^{1\beta}$ can be chosen to take the following values:
\begin{equation}
\begin{split}
&\Big(T^{11}_{c_1d_1}(S)\Big)_{a_1i_1,b_1j_1}=-\frac{i}{2}\Big[\delta_{c_1a_1}\delta_{d_1b_1}-\delta_{c_1b_1}\delta_{d_1a_1}\Big]\delta_{i_1j_1},\\
& \Big(T^{11}_{c_1d_1}(S)\Big)_{a_1i_1,b_2j_2}=\Big(T^{11}_{c_1d_1}(S)\Big)_{a_2i_2,b_1j_1}=\Big(T^{11}_{c_1d_1}(S)\Big)_{a_2i_2,b_2j_2}=0,\\
\end{split}
\label{T11 generators}
\end{equation}
\begin{equation}
\begin{split}
&\Big(T^{12}_{k_1l_1}(S)\Big)_{a_1i_1,b_1j_1}=-\frac{i}{2}\Big[\delta_{k_1i_1}\delta_{l_1j_1}-\delta_{k_1j_1}\delta_{l_1i_1}\Big]\delta_{a_1b_1},\\
& \Big(T^{12}_{k_1l_1}(S)\Big)_{a_1i_1,b_2j_2}=\Big(T^{12}_{k_1l_1}(S)\Big)_{a_2i_2,b_1j_1}=\Big(T^{12}_{k_1l_1}(S)\Big)_{a_2i_2,b_2j_2}=0.\\
\end{split}
\label{T12 generators}
\end{equation}
Similarly, the generators $T^{2\beta}$ can be chosen by $(1\leftrightarrow 2)$ exchange in the indices of the above expressions. 

Note that the generators $T_{c_1d_1}^{1\beta}(S)$ are antisymmetric under the exchange of $c_1$ and $d_1$.  Generators which are related by such exchanges of indices, therefore, should not be counted as independent generators. Hence, while summing over the generators $T^{1\beta}_{c_1d_1}(S)$ in \eqref{h1, f1 1-loop: Lambda def}, one should introduce a factor of $\frac{1}{2}$ to count only the independent ones.

Now that we have introduced all the ingredients that go into computation of the 1-loop beta function of $\lambda_{a_1i_1,b_1j_1,c_1k_1,d_1l_1}$, we can evaluate the contributions of these terms to the beta functions of $h_1$ and $f_1$. These contributions are enumerated in table \ref{tab:h1,f1-1loop}.

\begin{table}[H]
\centering
\caption{Contribution of different terms to $\beta_{h_1}^{\text{1-loop}}$ and $\beta_{f_1}^{\text{1-loop}}$.} 
\scalebox{0.8}{
\begin{tabular}{|c|c|c|}
\hline
 Contributing term & Contribution to $\beta_{h_1}^{\text{1-loop}}$ & Contribution to $\beta_{f_1}^{\text{1-loop}}$ \\
  \hline 
$(\Lambda^2)_{a_1i_1,b_1j_1,c_1k_1,d_1l_1}$ & $16 h_1^2\Big(1+\frac{2}{N_{c1}}\Big)+\frac{96}{N_{c1}^2}h_1 f_1$ & $24 h_1^2+8 f_1^2\Big(1+\frac{8}{N_{c1}^2}\Big)+16 h_1 f_1\Big(2+\frac{1}{N_{c1}}\Big)+8\zeta^2$\\
  \hline
$-3\sum\limits_{i=1}^2\sum\limits_{\beta=1}^2 g_i^2(\Lambda^S)_{a_1i_1,b_1j_1,c_1k_1,d_1l_1}^{i\beta}$ & $-6\lambda_1 h_1\Big(1-\frac{1}{N_{c1}}\Big) $ & $-6\lambda_1 f_1\Big(1-\frac{1}{N_{c1}}\Big)$\\
\hline
$3\sum\limits_{i,j=1}^2\sum\limits_{\beta,\gamma=1}^2 g_i^2g_j^2(A)_{a_1i_1,b_1j_1,c_1k_1,d_1l_1}^{i\beta,j\gamma}$ & $\frac{3 \lambda_1^2}{16}\Big(1-\frac{4}{N_{c1}}\Big)$ & $\frac{9 \lambda_1^2}{16}$\\
\hline
\end{tabular}
}
\label{tab:h1,f1-1loop}
\end{table}

Adding all the contributions we get
\begin{equation}
\begin{split}
\beta_{h_1}^{\text{1-loop}}=\frac{96h_1 f_1}{N_{c1}^2}+\Big(16+\frac{32}{N_{c1}}\Big)h_1^2-\Big(6-\frac{6}{N_{c1}}\Big)h_1 \lambda_1+\Big(\frac{3}{16}-\frac{3}{4N_{c1}}\Big)\lambda_1^2,
\end{split}
\end{equation}
\begin{equation}
\begin{split}
\beta_{f_1}^{\text{1-loop}}=& 8 \Big(1+\frac{8}{N_{c1}^2}\Big)f_1^2+32 \Big(1+\frac{1}{2N_{c1}}\Big)f_1 h_1-\Big(6-\frac{6}{N_{c1}}\Big)f_1\lambda_1+24 h_1^2+\frac{9\lambda_1^2}{16} +8\zeta^2.
 \end{split}
\end{equation}
Similarly, we can obtain the 1-loop beta functions of $h_2$ and $f_2$ which are given below:
\begin{equation}
\begin{split}
\beta_{h_2}^{\text{1-loop}}=\frac{96h_2 f_2}{N_{c2}^2}+\Big(16+\frac{32}{N_{c2}}\Big)h_2^2-\Big(6-\frac{6}{N_{c2}}\Big)h_2 \lambda_2+\Big(\frac{3}{16}-\frac{3}{4N_{c2}}\Big)\lambda_2^2,
\end{split}
\end{equation}
\begin{equation}
\begin{split}
\beta_{f_2}^{\text{1-loop}}=& 8 \Big(1+\frac{8}{N_{c2}^2}\Big)f_2^2+32 \Big(1+\frac{1}{2N_{c2}}\Big)f_2 h_2-\Big(6-\frac{6}{N_{c2}}\Big)f_2\lambda_2+24 h_2^2+\frac{9\lambda_2^2}{16} +8\zeta^2.
 \end{split}
\end{equation}

\subsubsection{Evaluation of $\beta_{\zeta}^{\text{1-loop}}$}
The 1-loop beta function of the coupling $\lambda_{a_1i_1,b_1j_1,c_2k_2,d_2l_2}$, as derived in  \cite{Machacek:1984zw}, is given by
\begin{equation}
\begin{split}
(4\pi)^2(\beta)_{a_1i_1,b_1j_1,c_2k_2,d_2l_2}^{\text{1-loop}}=&(\Lambda^2)_{a_1 i_1,b_1 j_1,c_2 k_2,d_2 l_2}-3\sum_{i=1}^2\sum_{\beta=1}^2\ g_i^2(\Lambda^S)_{a_1 i_1,b_1 j_1,c_2 k_2,d_2 l_2}^{i\beta}\\
&+3\sum_{i,j=1}^2\sum_{\beta,\gamma=1}^2g_i^2 g_j^2(A)_{a_1 i_1,b_1 j_1,c_2 k_2,d_2 l_2}^{i\beta,j\gamma}
\end{split}
\end{equation}
where
\begin{equation}
\begin{split}
(\Lambda^2)_{a_1i_1,b_1j_1,c_2k_2,d_2l_2}\equiv &\frac{1}{8}\sum_{\text{perms}}\lambda_{a_1i_1,b_1j_1,e u,f v}\lambda_{e u,f v,c_2k_2,d_2l_2},\\
(\Lambda^S)_{a_1i_1,b_1j_1,c_2k_2,d_2l_2}^{i\beta}\equiv & 2\Big(C_2^{i\beta}(S_1)+C_2^{i\beta}(S_2)\Big)\lambda_{a_1i_1,b_1j_1,c_2k_2,d_2l_2},\\
(A)_{a_1i_1,b_1j_1,c_2k_2,d_2l_2}^{i\beta,j\gamma}\equiv & \frac{1}{4}\sum_{\text{perms}}\Big((\Lambda)^{i\beta}_{a_1i_1,c_2k_2;eu,fv}(\Lambda)^{j\gamma}_{eu,fv;b_1j_1,d_2 l_2}\\
&\qquad\quad+(\Lambda)^{i\beta}_{a_1i_1,eu;fv,d_2l_2}(\Lambda)^{j\gamma}_{eu,b_1j_1;c_2k_2,fv}\Big).
\end{split}
\end{equation}
with the quantity $(\Lambda)^{i\beta}_{a_1i_1,c_2k_2;eu,fv}$ defined as follows:
\begin{equation}
\begin{split}
(\Lambda)^{i\beta}_{a_1i_1,c_2k_2;eu,fv}\equiv& \Big(T_A^{i\beta}(S)\Big)_{a_1i_1,eu}\Big(T_A^{i\beta}(S)\Big)_{c_2k_2,fv}.\\
\end{split}
\end{equation}
Here  $S_2$ denotes the scalar fields in the second sector. The rest of the notations are similar to the ones introduced earlier. 

The contributions of these terms to the 1-loop beta function of $\zeta$ are given below in table \ref{tab:zeta-1loop}.

\begin{table}[H]
\centering
\caption{Contribution of different terms to $\beta_{\zeta}^{\text{1-loop}}$ .}
\scalebox{0.8}{
\begin{tabular}{|c|c|}
\hline
 Contributing term & Contribution to $\beta_{\zeta}^{\text{1-loop}}$  \\
  \hline 
$(\Lambda^2)_{a_1i_1,b_1j_1,c_2k_2,d_2l_2}$ & $8\zeta\Bigg[\sum\limits_{i=1}^2\Bigg\{ h_i\Big(2+\frac{1}{N_{ci}}\Big)+f_i\Big(1+\frac{2}{N_{ci}^2}\Big)\Bigg\}+\frac{4\zeta}{N_{c1}N_{c2}}\Bigg]$\\
\hline
$-3\sum\limits_{i=1}^2\sum\limits_{\beta=1}^2 g_i^2(\Lambda^S)_{a_1 i_1,b_1 j_1,c_2 k_2,d_2 l_2}^{i\beta}$ & $-3\zeta\sum\limits_{i=1}^2 \lambda_i \Big(1-\frac{1}{N_{ci}}\Big)$\\
\hline
$3\sum\limits_{i,j=1}^2\sum\limits_{\beta,\gamma=1}^2g_i^2 g_j^2(A)_{a_1 i_1,b_1 j_1,c_2 k_2,d_2 l_2}^{i\beta,j\gamma}$ & $0$\\
\hline
\end{tabular}
}
\label{tab:zeta-1loop}
\end{table}

Adding all the contributions we get
\begin{equation}
\begin{split}
\beta_{\zeta}^{\text{1-loop}}=& \zeta\Bigg[8\Big(1+\frac{2}{N_{c1}^2}\Big)f_1+8\Big(1+\frac{2}{N_{c2}^2}\Big)f_2+16\Big(1+\frac{1}{2N_{c1}}\Big)h_1+16\Big(1+\frac{1}{2N_{c2}}\Big)h_2\\
&\qquad+\frac{32}{N_{c1}N_{c2}}\zeta- \Big(3-\frac{3}{N_{c1}}\Big)\lambda_1- \Big(3-\frac{3}{N_{c2}}\Big)\lambda_2\Bigg].\\
\end{split}
\end{equation}

%%%%%%%%%%%%%%%%%%%%%%%%%%%%%%%%%%%%%%%%%%%%%%%%%%%%%%%%%%%%%%%%%%%%%%%%%%%%%%%%%%%%

\subsection{2-loop beta functions of the quartic couplings}
Let us now turn to the evaluation of 2-loop corrections to the beta functions of the quartic couplings. The  strategy is analogous to the one employed to determine the 1-loop beta functions. We will determine the 2-loop corrections to the beta functions of $h_1$ and $f_1$ by using the expressions of similar corrections to the beta function of $\lambda_{a_1i_1,b_1j_1,c_1k_1,d_1l_1}$. As earlier, the 2-loop beta functions of $h_2$ and $f_2$ can then be obtained by $(1\leftrightarrow 2)$ exchange of the indices. Similarly, to determine the 2-loop beta function of $\zeta$, we will use the expression of the corrections  to the beta function of $\lambda_{a_1i_1,b_1j_1,c_2k_2,d_2l_2}$.

\subsubsection{Evaluation of $\beta_{h_i}^{\text{2-loop}}$ and $\beta_{f_i}^{\text{2-loop}}$}

The expression of the two loop beta function of the coupling $\lambda_{a_1i_1,b_1j_1,c_1k_1,d_1l_1}$ is as follows: 
\begin{equation}
\begin{split}
&(4\pi)^4\beta_{a_1i_1,b_1j_1,c_1k_1,d_1l_1}^{\text{2-loop}}\\
=&\frac{1}{2}\sum_k\Lambda^2(k)\lambda_{a_1i_1,b_1j_1,c_1k_1,d_1l_1}-\overline{\Lambda}^3_{a_1i_1,b_1j_1,c_1k_1,d_1l_1}\\
&+\sum_{i=1}^2\sum_{\beta=1}^2 g_i^2\Big(2(\overline{\Lambda}^{2S})_{a_1i_1,b_1j_1,c_1k_1,d_1l_1}^{i\beta}-6(\Lambda^{2g})_{a_1i_1,b_1j_1,c_1k_1,d_1l_1}^{i\beta}\Big)\\
&-\sum_{i=1}^2\sum_{\beta=1}^2 g_i^4\Big[\frac{35}{3}C_2^{i\beta}(G)-\frac{5}{3}S_2^{i\beta}(F)-\frac{11}{12}S_2^{i\beta}(S)\Big](\Lambda^{S})_{a_1i_1,b_1j_1,c_1k_1,d_1l_1}^{i\beta}\\
&+\sum_{i,j=1}^2\sum_{\beta,\gamma=1}^2g_{i}^2g_{j}^2\Bigg\{\frac{3}{2}(\Lambda^{SS})_{a_1i_1,b_1j_1,c_1k_1,d_1l_1}^{i\beta,j\gamma}+\frac{5}{2}(A^{\lambda})_{a_1i_1,b_1j_1,c_1k_1,d_1l_1}^{i\beta,j\gamma}+\frac{1}{2}(\overline{A}^{\lambda})_{a_1i_1,b_1j_1,c_1k_1,d_1l_1}^{i\beta,j\gamma}\Bigg\}\\
&+\sum_{i,j=1}^2\sum_{\beta,\gamma=1}^2 g_i^4g_j^2\Bigg\{\Big[\frac{161}{6}C_2^{i\beta}(G)-\frac{16}{3}S_2^{i\beta}(F)-\frac{7}{3}S_2^{i\beta}(S)\Big](A)_{a_1i_1,b_1j_1,c_1k_1,d_1l_1}^{i\beta,j\gamma}\Bigg\}\\
&-\frac{15}{2}\sum_{i,j,r=1}^2\sum_{\beta,\gamma,\rho=1}^2 g_i^2g_j^2g_r^2(A^S)_{a_1i_1,b_1j_1,c_1k_1,d_1l_1}^{i\beta,j\gamma,r\rho}+27 \sum_{i=1}^2\sum_{\beta=1}^2 g_i^6(A^g)_{a_1i_1,b_1j_1,c_1k_1,d_1l_1}^{i\beta}.
\end{split}
\label{beta zeta:2-loop expression}
\end{equation}
The sum in the first term of the above expression runs over the values $\{a_1i_1,b_1j_1,c_1k_1,d_1l_1\}$. The different quantities appearing in this expression are defined below:
\begin{equation}
\begin{split}
&\Lambda^2(k)=\frac{1}{6}\lambda_{k,eu,fv,gw}\lambda_{k,eu,fv,gw},\\
&\overline{\Lambda}^3_{a_1i_1,b_1j_1,c_1k_1,d_1l_1}=\frac{1}{4}\sum_{\text{perms}}\lambda_{a_!i_1,b_1j_1,eu,fv}\lambda_{c_1k_1,eu,gw,ht}\lambda_{d_1l_1,fv,gw,ht},\\
&(\overline{\Lambda}^{2S})_{a_1i_1,b_1j_1,c_1k_1,d_1l_1}^{i\beta}=\frac{1}{8}\sum_{\text{perms}}\Big(C_2^{i\beta}(S)\Big)_{fv,gw}\lambda_{a_1i_1,b_1j_1,eu,fv}\lambda_{c_1k_1,d_1l_1,eu,gw},\\
&(\Lambda^{2g})_{a_1i_1,b_1j_1,c_1k_1,d_1l_1}^{i\beta}=\frac{1}{8}\sum_{\text{perms}}\lambda_{a_1i_1,b_1j_1,eu,fv}\lambda_{c_1k_1,d_1l_1,gw,ht}\Big(T^{i\beta}_A(S)\Big)_{eu,gw}\Big(T^{i\beta}_A(S)\Big)_{fv,ht},\\
&(\Lambda^{S})_{a_1i_1,b_1j_1,c_1k_1,d_1l_1}^{i\beta}=4 C_2^{i\beta}(S_1)\lambda_{a_1i_1,b_1j_1,c_1k_1,d_1l_1},\\
&(\Lambda^{SS})_{a_1i_1,b_1j_1,c_1k_1,d_1l_1}^{i\beta,j\gamma}=4 C_2^{i\beta}(S_1)C_2^{j\gamma}(S_1)\lambda_{a_1i_1,b_1j_1,c_1k_1,d_1l_1},\\
&(A^{\lambda})_{a_1i_1,b_1j_1,c_1k_1,d_1l_1}^{i\beta,j\gamma}=\frac{1}{4}\sum_{\text{perms}}\lambda_{a_1i_1,b_1j_1,eu,fv}\{T_A^{i\beta}(S),T_B^{j\gamma}(S)\}_{eu,fv}\{T_A^{i\beta}(S),T_B^{j\gamma}(S)\}_{c_1k_1,d_1l_1},\\
&(\overline{A}^{\lambda})_{a_1i_1,b_1j_1,c_1k_1,d_1l_1}^{i\beta,j\gamma}=\frac{1}{4}\sum_{\text{perms}}\lambda_{a_1i_1,b_1j_1,eu,fv}\{T_A^{i\beta}(S),T_B^{j\gamma}(S)\}_{c_1k_1,eu}\{T_A^{i\beta}(S),T_B^{j\gamma}(S)\}_{d_1l_1,fv},\\
&(A^S)_{a_1i_1,b_1j_1,c_1k_1,d_1l_1}^{i\beta,j\gamma,r\rho}=4 C_2^{r\rho}(S_1)(A)_{a_1i_1,b_1j_1,c_1k_1,d_1l_1}^{i\beta,j\gamma},\\
&(A^g)_{a_1i_1,b_1j_1,c_1k_1,d_1l_1}^{i\beta}=\frac{1}{8}(f_{i\beta})^{ACE}(f_{i\beta})^{BDE}\sum_{\text{perms}}\{T_A^{i\beta}(S),T_B^{i\beta}(S)\}_{a_1i_1,b_1j_1}\{T_C^{i\beta}(S),T_D^{i\beta}(S)\}_{c_1k_1,d_1l_1}.
\end{split}
\end{equation}
Most of the notations in the above expressions have already been introduced in the evaluation of the 1-loop beta functions. The only new elements are the quantitiies $(f_{i\beta})^{ACE}$. These are the structure constants of the group $G^{i\beta}$. For example, the structure constants $f_{11}$ are as follows:
\begin{equation}
\begin{split}
(f_{11})^{m_1n_1,e_1f_1,u_1v_1}=\frac{1}{2}\Bigg[&\delta_{m_1e_1}\Big\{\delta_{n_1u_1}\delta_{f_1v_1}-\delta_{f_1u_1}\delta_{n_1v_1}\Big\}+\delta_{m_1f_1}\Big\{\delta_{e_1u_1}\delta_{n_1v_1}-\delta_{n_1u_1}\delta_{e_1v_1}\Big\}\\
&-\delta_{n_1e_1}\Big\{\delta_{m_1u_1}\delta_{f_1v_1}-\delta_{f_1u_1}\delta_{m_1v_1}\Big\}-\delta_{n_1f_1}\Big\{\delta_{e_1u_1}\delta_{m_1v_1}-\delta_{m_1u_1}\delta_{e_1v_1}\Big\}\Bigg].\\
\end{split}
\end{equation}
The other structure constants $f_{i\beta}$ also have analogous forms. 

With these ingredients, we can evaluate the contribution of each of the terms in  \eqref{beta zeta:2-loop expression} to the 2-loop corrections  ($\beta_{h_1}^{\text{2-loop}}$ and $\beta_{f_1}^{\text{2-loop}}$) in the beta functions of $h_1$ and $f_1$. We provide the forms of these contributions in tables \ref{tab:h1-2loop} and \ref{tab:f1-2loop} below.
\begin{table}[H]
\centering
\caption{Contribution of different terms to $\beta_{h_1}^{\text{2-loop}}$.} 
\scalebox{0.8}{
\begin{tabular}{|c|c|}
\hline
 Contributing term & Contribution to $\beta_{h_1}^{\text{2-loop}}$ \\
  \hline 
  $\frac{1}{2}\sum \limits_k \Lambda^2(k)\lambda_{a_1i_1,b_1j_1,c_1k_1,d_1l_1}$ & $32\Big(1+\frac{2}{N_{c1}}+\frac{3}{N_{c1}^2}\Big)h_1^3+64\Big(\frac{1}{N_{c1}^2}+\frac{2}{N_{c1}^4}\Big)h_1f_1^2$\\
  & $+128\Big(\frac{2}{N_{c1}^2}+\frac{1}{N_{c1}^3}\Big)h_1^2 f_1+ \frac{64}{N_{c1}^2}  h_1\zeta^2$\\
  \hline
  $-\overline{\Lambda}^3_{a_1i_1,b_1j_1,c_1k_1,d_1l_1}$ & $-64\Big(2+\frac{10}{N_{c1}}+\frac{27}{N_{c1}^2}\Big)h_1^3-768\Big(\frac{4}{N_{c1}^2}+\frac{5}{N_{c1}^3}\Big)h_1^2 f_1$\\
  & $-384\Big(\frac{1}{N_{c1}^2}+\frac{14}{N_{c1}^4}\Big)h_1 f_1^2 -\frac{384}{N_{c1}^2}h_1\zeta^2$\\
  \hline
  $2\sum \limits_{i=1}^2\sum \limits_{\beta=1}^2 g_i^2(\overline{\Lambda}^{2S})_{a_1i_1,b_1j_1,c_1k_1d_1l_1}^{i\beta}$ & $\lambda_1\Big(1-\frac{1}{N_{c1}}\Big)\Bigg\{16\Big(1+\frac{2}{N_{c1}}\Big)h_1^2+\frac{96}{N_{c1}^2} h_1 f_1\Bigg\}$\\
  \hline
  $-6\sum \limits_{i=1}^2\sum \limits_{\beta=1}^2 g_i^2(\Lambda^{2g})_{a_1i_1,b_1j_1,c_1k_1d_1l_1}^{i\beta}$ & $\lambda_1\Bigg\{24\Big(1-\frac{4}{N_{c1}^2}\Big)h_1^2+96\Big(\frac{1}{N_{c1}^2}-\frac{2}{N_{c1}^3}\Big)h_1 f_1\Bigg\}$\\
  \hline
  $-\sum \limits_{i=1}^2\sum \limits_{\beta=1}^2 g_i^4\Big[\frac{35}{3}C_2^{i\beta}(G)-\frac{5}{3}S_2^{i\beta}(F)-\frac{11}{12}S_2^{i\beta}(S)\Big]$ & $-\frac{\lambda_1^2h_1}{12} \Big(129-20 x_{f1}-\frac{280}{N_{c1}}\Big) \Big(1-\frac{1}{N_{c1}}\Big)$ \\
  $(\Lambda^{S})_{a_1i_1,b_1j_1,c_1k_1d_1l_1}^{i\beta}$ & \\
  \hline
  $\frac{3}{2}\sum \limits_{i,j=1}^2\sum \limits_{\beta,\gamma=1}^2g_{i}^2g_{j}^2(\Lambda^{SS})_{a_1i_1,b_1j_1,c_1k_1,d_1l_1}^{i\beta,j\gamma}$ & $\frac{3}{2}\Big(1-\frac{1}{N_{c1}}\Big)^2\lambda_1^2h_1$\\
  \hline
    $\frac{5}{2}\sum \limits_{i,j=1}^2\sum \limits_{\beta,\gamma=1}^2g_{i}^2g_{j}^2 (A^{\lambda})_{a_1i_1,b_1j_1,c_1k_1,d_1l_1}^{i\beta,j\gamma}$ &  $\lambda_1^2\Bigg\{\frac{5}{2}\Big(1-\frac{1}{N_{c1}}\Big)h_1+5\Big(\frac{1}{N_{c1}^2}-\frac{4}{N_{c1}^3}\Big)f_1\Bigg\}$\\
  \hline
    $\frac{1}{2}\sum \limits_{i,j=1}^2\sum \limits_{\beta,\gamma=1}^2g_{i}^2g_{j}^2 (\overline{A}^{\lambda})_{a_1i_1,b_1j_1,c_1k_1,d_1l_1}^{i\beta,j\gamma}$ & $\lambda_1^2\Bigg\{\frac{1}{4}\Big(1-\frac{3}{N_{c1}}+\frac{2}{N_{c1}^2}\Big)h_1+\Big(\frac{1}{N_{c1}^2}-\frac{4}{N_{c1}^3}\Big)f_1\Bigg\}$\\
  \hline
  $\sum \limits_{i,j=1}^2\sum \limits_{\beta,\gamma=1}^2 g_i^4g_j^2\Big[\frac{161}{6}C_2^{i\beta}(G)-\frac{16}{3}S_2^{i\beta}(F)-\frac{7}{3}S_2^{i\beta}(S)\Big]$ & $\frac{\lambda_1^3}{192}\Big(1-\frac{4}{N_{c1}}\Big)\Big(147-32x_{f1}-\frac{322}{N_{c1}}\Big)$\\
  $(A)_{a_1i_1,b_1j_1,c_1k_1,d_1l_1}^{i\beta,j\gamma}$ & \\
  \hline
  $-\frac{15}{2}\sum \limits_{i,j,r=1}^2\sum_{\beta,\gamma,\rho=1}^2 g_i^2g_j^2g_r^2(A^S)_{a_1i_1,b_1j_1,c_1k_1,d_1l_1}^{i\beta,j\gamma,r\rho}$ & $-\frac{15}{16}\lambda_1^3\Big(1-\frac{4}{N_{c1}}\Big)\Big(1-\frac{1}{N_{c1}}\Big)$\\
  \hline
  $27 \sum \limits_{i=1}^2\sum_{\beta=1}^2 g_i^6(A^g)_{a_1i_1,b_1j_1,c_1k_1,d_1l_1}^{i\beta}$ & $\frac{27\lambda_1^3}{64}\Big(1-\frac{6}{N_{c1}}+\frac{8}{N_{c1}^2}\Big)$\\
  \hline
\end{tabular}
}
\label{tab:h1-2loop}
\end{table}

%%%%%%%%%%%%%%%%%%%%%%%%%%%%%%%%%%%%%%%%%%

\begin{table}[H]
\centering
\caption{Contribution of different terms to $\beta_{f_1}^{\text{2-loop}}$.} 
\scalebox{0.8}{
\begin{tabular}{|c|c|}
\hline
 Contributing term & Contribution to $\beta_{f_1}^{\text{2-loop}}$ \\
  \hline 
  $\frac{1}{2}\sum \limits_k \Lambda^2(k)\lambda_{a_1i_1,b_1j_1,c_1k_1,d_1l_1}$ & $32\Big(1+\frac{2}{N_{c1}}+\frac{3}{N_{c1}^2}\Big)h_1^2f_1+64\Big(\frac{1}{N_{c1}^2}+\frac{2}{N_{c1}^4}\Big)f_1^3$\\
  & $+128\Big(\frac{2}{N_{c1}^2}+\frac{1}{N_{c1}^3}\Big)h_1 f_1^2+ \frac{64}{N_{c1}^2}  f_1\zeta^2$\\
  \hline
  & $-192\Big(2+\frac{3}{N_{c1}}\Big)h_1^3-192\Big(1+\frac{2}{N_{c1}}+\frac{15}{N_{c1}^2}\Big)h_1^2 f_1$\\
  $-\overline{\Lambda}^3_{a_1i_1,b_1j_1,c_1k_1,d_1l_1}$  & $-1536\Big(\frac{2}{N_{c1}^2}+\frac{1}{N_{c1}^3}\Big)h_1 f_1^2-128\Big(\frac{5}{N_{c1}^2}+\frac{22}{N_{c1}^4}\Big)f_1^3$\\
  & $-256\Big(\frac{2}{N_{c2}^2}+\frac{1}{N_{c1}N_{c2}^2}\Big)h_1\zeta^2-128\Big(\frac{3}{N_{c1}^2}+\frac{2}{N_{c2}^2}+\frac{4}{N_{c1}^2N_{c2}^2}\Big)f_1\zeta^2$\\
  \hline
  $2\sum \limits_{i=1}^2\sum \limits_{\beta=1}^2 g_i^2(\overline{\Lambda}^{2S})_{a_1i_1,b_1j_1,c_1k_1d_1l_1}^{i\beta}$ & $\lambda_1\Big(1-\frac{1}{N_{c1}}\Big)\Bigg\{24 h_1^2+8\Big(1+\frac{8}{N_{c1}^2}\Big)f_1^2$\\
  & $+16\Big(2+\frac{1}{N_{c1}}\Big)h_1 f_1\Bigg\}+8\Big(1-\frac{1}{N_{c2}}\Big)\lambda_2 \zeta^2$\\
  \hline
  $-6\sum \limits_{i=1}^2\sum \limits_{\beta=1}^2 g_i^2(\Lambda^{2g})_{a_1i_1,b_1j_1,c_1k_1d_1l_1}^{i\beta}$ & $24\lambda_1\Bigg\{3 h_1^2+\Big(1-\frac{1}{N_{c1}}+\frac{4}{N_{c1}^2}-\frac{4}{N_{c1}^3}\Big)f_1^2$\\
&$+\Big(4-\frac{2}{N_{c1}}+\frac{2}{N_{c1}^2}\Big)h_1 f_1\Bigg\}+24\lambda_2 \Big(1-\frac{1}{N_{c2}}\Big)\zeta^2$\\
  \hline
  $-\sum \limits_{i=1}^2\sum \limits_{\beta=1}^2 g_i^4\Big[\frac{35}{3}C_2^{i\beta}(G)-\frac{5}{3}S_2^{i\beta}(F)-\frac{11}{12}S_2^{i\beta}(S)\Big]$ & $-\frac{\lambda_1^2 f_1}{12} \Big(129 -20 x_{f1}-\frac{280}{N_{c1}}\Big) \Big(1-\frac{1}{N_{c1}}\Big)$ \\
  $(\Lambda^{S})_{a_1i_1,b_1j_1,c_1k_1d_1l_1}^{i\beta}$ & \\
  \hline
  $\frac{3}{2}\sum \limits_{i,j=1}^2\sum \limits_{\beta,\gamma=1}^2g_{i}^2g_{j}^2(\Lambda^{SS})_{a_1i_1,b_1j_1,c_1k_1,d_1l_1}^{i\beta,j\gamma}$ & $\frac{3}{2}\Big(1-\frac{1}{N_{c1}}\Big)^2\lambda_1^2 f_1$\\
  \hline
    $\frac{5}{2}\sum \limits_{i,j=1}^2\sum \limits_{\beta,\gamma=1}^2g_{i}^2g_{j}^2 (A^{\lambda})_{a_1i_1,b_1j_1,c_1k_1,d_1l_1}^{i\beta,j\gamma}$ &  $\lambda_1^2\Bigg\{\frac{15}{2}\Big(1-\frac{1}{N_{c1}}\Big)h_1+5\Big(1- \frac{1}{N_{c1}}+\frac{3}{N_{c1}^2}\Big)f_1\Bigg\}$\\
  \hline
    $\frac{1}{2}\sum \limits_{i,j=1}^2\sum \limits_{\beta,\gamma=1}^2g_{i}^2g_{j}^2 (\overline{A}^{\lambda})_{a_1i_1,b_1j_1,c_1k_1,d_1l_1}^{i\beta,j\gamma}$ & $\lambda_1^2\Bigg\{\frac{3}{2}\Big(1-\frac{1}{N_{c1}}\Big)h_1+\frac{1}{4}\Big(3-\frac{5}{N_{c1}}+\frac{14}{N_{c1}^2}\Big)f_1\Bigg\}$\\
  \hline
  $\sum \limits_{i,j=1}^2\sum \limits_{\beta,\gamma=1}^2 g_i^4g_j^2\Big[\frac{161}{6}C_2^{i\beta}(G)-\frac{16}{3}S_2^{i\beta}(F)-\frac{7}{3}S_2^{i\beta}(S)\Big]$ & $\frac{\lambda_1^3 }{64}\Big(147-32x_{f1}-\frac{322}{N_{c1}}\Big)$\\
  $(A)_{a_1i_1,b_1j_1,c_1k_1,d_1l_1}^{i\beta,j\gamma}$ & \\
  \hline
  $-\frac{15}{2}\sum \limits_{i,j,r=1}^2\sum_{\beta,\gamma,\rho=1}^2 g_i^2g_j^2g_r^2(A^S)_{a_1i_1,b_1j_1,c_1k_1,d_1l_1}^{i\beta,j\gamma,r\rho}$ & $-\frac{45}{16}\lambda_1^3\Big(1-\frac{1}{N_{c1}}\Big)$\\
  \hline
  $27 \sum \limits_{i=1}^2\sum_{\beta=1}^2 g_i^6(A^g)_{a_1i_1,b_1j_1,c_1k_1,d_1l_1}^{i\beta}$ & $\frac{81\lambda_1^3}{64}\Big(1-\frac{2}{N_{c1}}\Big)$\\
  \hline
\end{tabular}
}
\label{tab:f1-2loop}
\end{table}

Summing up these contributions, we get

\begin{equation}
\begin{split}
\beta_{h_1}^{\text{2-loop}}=&-96\Big(1+\frac{6}{N_{c1}}+\frac{17}{N_{c1}^2}\Big)h_1^3-128\Big(\frac{22}{N_{c1}^2}+\frac{29}{N_{c1}^3}\Big)h_1^2 f_1-64\Big(\frac{5}{N_{c1}^2}+\frac{82}{N_{c1}^4}\Big)h_1 f_1^2\\
&-\frac{320}{N_{c1}^2}h_1\zeta^2+8\Big(5+\frac{2}{N_{c1}}-\frac{16}{N_{c1}^2}\Big)h_1^2\lambda_1+\frac{1}{6}\Big(1-\frac{1}{N_{c1}}\Big)\Big(-39+10x_{f1}+\frac{128}{N_{c1}}\Big)h_1\lambda_1^2\\
&+96\Big(\frac{2}{N_{c1}^2}-\frac{3}{N_{c1}^3}\Big)h_1 f_1 \lambda_1+6\Big(\frac{1}{N_{c1}^2}-\frac{4}{N_{c1}^3}\Big)f_1 \lambda_1^2\\
&-\frac{1}{12}\Big(1-\frac{4}{N_{c1}}\Big)\Big(-3+2 x_{f1}+\frac{19}{N_{c1}}\Big)\lambda_1^3,\\
\beta_{f_1}^{\text{2-loop}}=&-192\Big(2+\frac{3}{N_{c1}}\Big)h_1^3-32\Big(5+\frac{10}{N_{c1}}+\frac{87}{N_{c1}^2}\Big)h_1^2 f_1-1408\Big(\frac{2}{N_{c1}^2}+\frac{1}{N_{c1}^3}\Big)h_1 f_1^2\\
&-192\Big(\frac{3}{N_{c1}^2}+\frac{14}{N_{c1}^4}\Big)f_1^3+24\Big(4-\frac{1}{N_{c1}}\Big)h_1^2\lambda_1+32\Big(4-\frac{2}{N_{c1}}+\frac{1}{N_{c1}^2}\Big)h_1 f_1\lambda_1\\
&+9\Big(1-\frac{1}{N_{c1}}\Big) h_1 \lambda_1^2+32\Big(1-\frac{1}{N_{c1}}+\frac{5}{N_{c1}^2}-\frac{5}{N_{c1}^3}\Big)f_1^2\lambda_1\\
&+\frac{1}{6}\Big(-21+10 x_{f1}+\frac{149-10 x_{f1}}{N_{c1}}-\frac{20}{N_{c1}^2}\Big)f_1 \lambda_1^2+\frac{1}{4}\Big(3-2 x_{f1}-\frac{19}{N_{c1}}\Big)\lambda_1^3\\
&-256\Big(\frac{2}{N_{c2}^2}+\frac{1}{N_{c1}N_{c2}^2}\Big)h_1 \zeta^2-64\Big(\frac{5}{N_{c1}^2}+\frac{4}{N_{c2}^2}+\frac{8}{N_{c1}^2N_{c2}^2}\Big)f_1\zeta^2+32\Big(1-\frac{1}{N_{c2}}\Big)\lambda_{2}\zeta^2.\\
\end{split}
\end{equation}
The 2-loop corrections to the beta functions of $h_2$ and $f_2$ can be obtained by $(1\leftrightarrow 2)$ exchange of the indices in the above expressions.

%%%%%%%%%%%%%%%%%%%%%%%%%%%%%%%%%%%%%%%%%%%%%%%%%%%%%%%%%%%%%%%%%%%%%%%%

\newpage

\subsubsection{Evaluation of $\beta_\zeta^{\text{2-loop}}$}

Let us now determine the 2-loop corrections to the beta function of $\zeta$. For this we will have to consider such corrections to the beta function of the coupling   $\lambda_{a_1i_1,b_1j_1,c_2k_2,d_2l_2}$. These corrections are given by 
\begin{equation}
\begin{split}
&(4\pi)^4\beta_{a_1i_1,b_1j_1,c_2k_2,d_2l_2}^{\text{2-loop}}\\
=&\frac{1}{2}\sum_k\Lambda^2(k)\lambda_{a_1i_1,b_1j_1,c_2k_2,d_2l_2}-\overline{\Lambda}^3_{a_1i_1,b_1j_1,c_2k_2,d_2l_2}\\
&+\sum_{i=1}^2\sum_{\beta=1}^2 g_i^2\Big(2(\overline{\Lambda}^{2S})_{a_1i_1,b_1j_1,c_2k_2,d_2l_2}^{i\beta}-6(\Lambda^{2g})_{a_1i_1,b_1j_1,c_2k_2,d_2l_2}^{i\beta}\Big)\\
&-\sum_{i=1}^2\sum_{\beta=1}^2 g_i^4\Big[\frac{35}{3}C_2^{i\beta}(G)-\frac{5}{3}S_2^{i\beta}(F)-\frac{11}{12}S_2^{i\beta}(S)\Big](\Lambda^{S})_{a_1i_1,b_1j_1,c_2k_2,d_2l_2}^{i\beta}\\
&+\sum_{i,j=1}^2\sum_{\beta,\gamma=1}^2g_{i}^2g_{j}^2\Bigg\{\frac{3}{2}(\Lambda^{SS})_{a_1i_1,b_1j_1,c_2k_2,d_2l_2}^{i\beta,j\gamma}+\frac{5}{2}(A^{\lambda})_{a_1i_1,b_1j_1,c_2k_2,d_2l_2}^{i\beta,j\gamma}+\frac{1}{2}(\overline{A}^{\lambda})_{a_1i_1,b_1j_1,c_2k_2,d_2l_2}^{i\beta,j\gamma}\Bigg\}\\
&+\sum_{i,j=1}^2\sum_{\beta,\gamma=1}^2 g_i^4g_j^2\Bigg\{\Big[\frac{161}{6}C_2^{i\beta}(G)-\frac{16}{3}S_2^{i\beta}(F)-\frac{7}{3}S_2^{i\beta}(S)\Big](A)_{a_1i_1,b_1j_1,c_2k_2,d_2l_2}^{i\beta,j\gamma}\Bigg\}\\
&-\frac{15}{2}\sum_{i,j,r=1}^2\sum_{\beta,\gamma,\rho=1}^2 g_i^2g_j^2g_r^2(A^S)_{a_1i_1,b_1j_1,c_2k_2,d_2l_2}^{i\beta,j\gamma,r\rho}+27 \sum_{i=1}^2\sum_{\beta=1}^2 g_i^6(A^g)_{a_1i_1,b_1j_1,c_2k_2,d_2l_2}^{i\beta}.
\end{split}
\end{equation}
Here
\begin{equation}
\begin{split}
&\Lambda^2(k)=\frac{1}{6}\lambda_{k,eu,fv,gw}\lambda_{k,eu,fv,gw},\\
&\overline{\Lambda}^3_{a_1i_1,b_1j_1,c_2k_2,d_2l_2}=\frac{1}{4}\sum_{\text{perms}}\lambda_{a_1 i_1,b_1j_1,eu,fv}\lambda_{c_2k_2,eu,gw,ht}\lambda_{d_2l_2,fv,gw,ht},\\
&(\overline{\Lambda}^{2S})_{a_1i_1,b_1j_1,c_2k_2,d_2l_2}^{i\beta}=\frac{1}{8}\sum_{\text{perms}}\Big(C_2^{i\beta}(S)\Big)_{fv,gw}\lambda_{a_1i_1,b_1j_1,eu,fv}\lambda_{c_2k_2,d_2l_2,eu,gw},\\
&(\Lambda^{2g})_{a_1i_1,b_1j_1,c_2k_2,d_2l_2}^{i\beta}=\frac{1}{8}\sum_{\text{perms}}\lambda_{a_1i_1,b_1j_1,eu,fv}\lambda_{c_2k_2,d_2l_2,gw,ht}\Big(T^{i\beta}_A(S)\Big)_{eu,gw}\Big(T^{i\beta}_A(S)\Big)_{fv,ht},\\
&(\Lambda^{S})_{a_1i_1,b_1j_1,c_2k_2,d_2l_2}^{i\beta}=2\Big(C_2^{i\beta}(S_1)+C_2^{i\beta}(S_2)\Big)\lambda_{a_1i_1,b_1j_1,c_2k_2,d_2l_2},\\
&(\Lambda^{SS})_{a_1i_1,b_1j_1,c_2k_2,d_2l_2}^{i\beta,j\gamma}=2\Big(C_2^{i\beta}(S_1)C_2^{j\gamma}(S_1)+C_2^{i\beta}(S_2)C_2^{j\gamma}(S_2)\Big)\lambda_{a_1i_1,b_1j_1,c_2k_2,d_2l_2},\\
&(A^{\lambda})_{a_1i_1,b_1j_1,c_2k_2,d_2l_2}^{i\beta,j\gamma}=\frac{1}{4}\sum_{\text{perms}}\lambda_{a_1i_1,b_1j_1,eu,fv}\{T_A^{i\beta}(S),T_B^{j\gamma}(S)\}_{eu,fv}\{T_A^{i\beta}(S),T_B^{j\gamma}(S)\}_{c_2k_2,d_2l_2},\\
&(\overline{A}^{\lambda})_{a_1i_1,b_1j_1,c_2k_2,d_2l_2}^{i\beta,j\gamma}=\frac{1}{4}\sum_{\text{perms}}\lambda_{a_1i_1,b_1j_1,eu,fv}\{T_A^{i\beta}(S),T_B^{j\gamma}(S)\}_{c_2k_2,eu}\{T_A^{i\beta}(S),T_B^{j\gamma}(S)\}_{d_2l_2,fv},\\
&(A^S)_{a_1i_1,b_1j_1,c_2k_2,d_2l_2}^{i\beta,j\gamma,r\rho}=2\Big(C_2^{r\rho}(S_1)+C_2^{r\rho}(S_2)\Big)(A)_{a_1i_1,b_1j_1,c_2k_2,d_2l_2}^{i\beta,j\gamma},\\
&(A^g)_{a_1i_1,b_1j_1,c_2k_2,d_2l_2}^{i\beta}=\frac{1}{8}(f_{i\beta})^{ACE}(f_{i\beta})^{BDE}\sum_{\text{perms}}\{T_A^{i\beta}(S),T_B^{i\beta}(S)\}_{a_1i_1,b_1j_1}\{T_C^{i\beta}(S),T_D^{i\beta}(S)\}_{c_2k_2,d_2l_2}.
\end{split}
\end{equation}
The contributions of the terms in the above expressions to the 2-loop corrections in the beta function of $\zeta$ are given in table \ref{tab:zeta-2loop}.

%%%%%%%%%%%%%%%%%%%%%%%%%%%%%%%%%%%%%%%%%%

\begin{table}[H]
\centering
\caption{Contribution of different terms to $\beta_{\zeta}^{\text{2-loop}}$.} 
\scalebox{0.8}{
\begin{tabular}{|c|c|}
\hline
 Contributing term & Contribution to $\beta_{\zeta}^{\text{2-loop}}$ \\
  \hline 
  $\frac{1}{2}\sum \limits_k\Lambda^2(k)\lambda_{a_1i_1,b_1j_1,c_2k_2,d_2l_2}$ & $\zeta\sum \limits_{i=1}^2\Big[16\Big(1+\frac{2}{N_{ci}}+\frac{3}{N_{ci}^2}\Big)h_i^2+32\Big(\frac{1}{N_{ci}^2}+\frac{2}{N_{ci}^4}\Big)f_i^2$\\
   & $+64\Big(\frac{2}{N_{ci}^2}+\frac{1}{N_{ci}^3}\Big)h_i f_i+ \frac{32}{N_{ci}^2}\zeta^2\Big]$\\
  \hline
  & $-32\zeta\Bigg[\sum\limits_{i=1}^2\Bigg\{3\Big(1+\frac{2}{N_{ci}}+\frac{3}{N_{ci}^2}\Big)h_i^2+6\Big(\frac{1}{N_{ci}^2}+\frac{2}{N_{ci}^4}\Big)f_i^2$\\
$-\overline{\Lambda}^3_{a_1i_1,b_1j_1,c_2k_2,d_2l_2}$ & $+12\Big(\frac{2}{N_{ci}^2}+\frac{1}{N_{ci}^3}\Big)h_i f_i +12\Big(2+\frac{1}{N_{ci}}\Big)\frac{\zeta h_i}{N_{c1}N_{c2}}$\\
 & $+12\Big(1+\frac{2}{N_{ci}^2}\Big)\frac{\zeta f_i}{N_{c1}N_{c2}}+\frac{2}{N_{ci}^2}\zeta^2\Bigg\}+\frac{16}{N_{c1}^2 N_{c2}^2}\zeta^2\Bigg]$\\
  \hline
  $2\sum \limits_{i=1}^2\sum \limits_{\beta=1}^2 g_i^2(\overline{\Lambda}^{2S})_{a_1i_1,b_1j_1,c_2k_2,d_2l_2}^{i\beta}$ & $ 8\zeta\sum\limits_{i=1}^2\lambda_i\Big(1-\frac{1}{N_{ci}}\Big)\Bigg\{\Big(2+\frac{1}{N_{ci}}\Big)h_i+\Big(1+\frac{2}{N_{ci}^2}\Big)f_i+\frac{2}{N_{c1}N_{c2}}\zeta\Bigg\}$\\
  \hline
  $-6\sum \limits_{i=1}^2\sum \limits_{\beta=1}^2 g_i^2(\Lambda^{2g})_{a_1i_1,b_1j_1,c_2k_2,d_2l_2}^{i\beta}$ & $24\zeta\sum\limits_{i=1}^2\lambda_i \Big(1-\frac{1}{N_{ci}}\Big)\Bigg\{\Big(2+\frac{1}{N_{ci}}\Big)h_i+\Big(1+\frac{2}{N_{ci}^2}\Big)f_i\Bigg\}$\\
  \hline
  $-\sum \limits_{i=1}^2\sum \limits_{\beta=1}^2 g_i^4\Big[\frac{35}{3}C_2^{i\beta}(G)-\frac{5}{3}S_2^{i\beta}(F)-\frac{11}{12}S_2^{i\beta}(S)\Big]$ & $-\zeta\sum\limits_{i=1}^2\Bigg\{\frac{\lambda_i^2}{24}\Big(129 -20 x_{fi}-\frac{280}{N_{ci}}\Big)\Big(1-\frac{1}{N_{ci}}\Big)\Bigg\}$ \\
  $(\Lambda^{S})_{a_1i_1,b_1j_1,c_2k_2,d_2l_2}^{i\beta}$ & \\
  \hline
  $\frac{3}{2}\sum \limits_{i,j=1}^2\sum \limits_{\beta,\gamma=1}^2g_{i}^2g_{j}^2 (\Lambda^{SS})_{a_1i_1,b_1j_1,c_2k_2,d_2l_2}^{i\beta,j\gamma}$ & $\frac{3}{4}\zeta \sum\limits_{i=1}^2\lambda_i^2 \Big(1-\frac{1}{N_{ci}}\Big)^2$\\
  \hline
    $\frac{5}{2}\sum \limits_{i,j=1}^2\sum \limits_{\beta,\gamma=1}^2g_{i}^2g_{j}^2 (A^{\lambda})_{a_1i_1,b_1j_1,c_2k_2,d_2l_2}^{i\beta,j\gamma}$ &  $\frac{5}{2}\zeta \sum\limits_{i=1}^2 \lambda_i^2 \Big(1-\frac{1}{N_{ci}}\Big)$\\
  \hline
    $\frac{1}{2}\sum \limits_{i,j=1}^2\sum \limits_{\beta,\gamma=1}^2g_{i}^2g_{j}^2 (\overline{A}^{\lambda})_{a_1i_1,b_1j_1,c_2k_2,d_2l_2}^{i\beta,j\gamma}$ & $\frac{1}{8}\zeta\sum\limits_{i=1}^2\Big(3-\frac{5}{N_{ci}}+\frac{2}{N_{ci}^2}\Big)\lambda_i^2$\\
  \hline
  $\sum \limits_{i,j=1}^2\sum \limits_{\beta,\gamma=1}^2 g_i^4g_j^2\Big[\frac{161}{6}C_2^{i\beta}(G)-\frac{16}{3}S_2^{i\beta}(F)-\frac{7}{3}S_2^{i\beta}(S)\Big]$ & $0$\\
  $(A)_{a_1i_1,b_1j_1,c_2k_2,d_2l_2}^{i\beta,j\gamma}$ & \\
  \hline
  $-\frac{15}{2}\sum \limits_{i,j,r=1}^2\sum \limits_{\beta,\gamma,\rho=1}^2 g_i^2g_j^2g_r^2(A^S)_{a_1i_1,b_1j_1,c_2k_2,d_2l_2}^{i\beta,j\gamma,r\rho}$ & $0$\\
  \hline
  $27 \sum \limits_{i=1}^2\sum \limits_{\beta=1}^2 g_i^6(A^g)_{a_1i_1,b_1j_1,c_2k_2,d_2l_2}^{i\beta}$ & $0$\\
  \hline
\end{tabular}
}
\label{tab:zeta-2loop}
\end{table}
Adding these contributions we get
\begin{equation}
\begin{split}
\beta_{\zeta}^{\text{2-loop}}=&\zeta\Bigg[\sum_{i=1}^2\Bigg\{-80\Big(1+\frac{2}{N_{ci}}+\frac{3}{N_{ci}^2}\Big)h_i^2-160\Big(\frac{1}{N_{ci}^2}+\frac{2}{N_{ci}^4}\Big)f_i^2-320\Big(\frac{2}{N_{ci}^2}+\frac{1}{N_{ci}^3}\Big)h_i f_i\\
&\qquad\qquad+\frac{1}{12}\Big(1-\frac{1}{N_{ci}}\Big)\Big(-21+10 x_{fi}+\frac{128}{N_{ci}}\Big)\lambda_i^2+32\Big(1-\frac{1}{N_{ci}}\Big)\Big(2+\frac{1}{N_{ci}}\Big)\lambda_i h_i\\
&\qquad\qquad+32\Big(1-\frac{1}{N_{ci}}\Big)\Big(1+\frac{2}{N_{ci}^2}\Big)\lambda_i f_i-384\Big(\frac{2}{N_{ci}N_{ci^\prime}}+\frac{1}{N_{ci}^2 N_{ci^\prime}}\Big)\zeta h_i\\
&\qquad\qquad-384\Big(\frac{1}{N_{ci}N_{ci^\prime}}+\frac{2}{N_{ci}^3 N_{ci^\prime}}\Big)\zeta f_i+16\Big(\frac{1}{N_{ci}N_{ci^\prime}}-\frac{1}{N_{ci}^2 N_{ci^\prime}}\Big)\zeta \lambda_i\Bigg\}\\
&\qquad-32\Big(\frac{1}{N_{c1}^2}+\frac{1}{N_{c2}^2}+\frac{16}{N_{c1}^2 N_{c2}^2}\Big)\zeta^2\Bigg],
\end{split}
\end{equation}
where $i^{\prime}$ is the complement of $i$, i.e., for $i=1$, $i^\prime=2$, and for $i=2$, $i^\prime=1$.

%%%%%%%%%%%%%%%%%%%%%%%%%%%%%%%%%%%%%%%%%%
%%%%%%%%%%%%%%%%%%%%%%%%%%%%%%%%%%%%%%%%%%%%%%%%%%%%%%%%%%%%%%%%%%%%%%%%

\section{Constraints on the fixed points in the large N limit}
\label{app: fixed point constraints}
In this appendix, we will discuss some constraints on the fixed points of the RG flow of the couplings in the real double bifundamental model. For this, we will restrict our attention to just the 2-loop planar beta functions of the gauge couplings and the 1-loop planar beta functions of the quartic couplings. In the planar limit $(N_{c1},N_{c2}\rightarrow\infty)$,  the 1-loop beta functions of the quartic couplings have the following forms:
\begin{equation}
\begin{split}
&\beta_{h_i}^{\text{1-loop}}=16 h_i^2-6 h_i\lambda_i+\frac{3}{16}\lambda_i^2\ ,\\
&\beta_{f_i}^{\text{1-loop}}=8 f_i^2+32 f_ih_i-6 f_i\lambda_i+24h_i^2+\frac{9}{16}\lambda_i^2+8\zeta^2\ ,\\
&\beta_{\zeta}^{\text{1-loop}}=\zeta\Bigg[8 f_1+8 f_2+16 h_1+16 h_2-3\lambda_1-3\lambda_2\Bigg]\ .
\end{split}
\end{equation}
Here $\lambda_1$ and $\lambda_2$ are fixed by demanding $\beta_{\lambda_1}=\beta_{\lambda_2}=0$ which leads to the following nontrivial solutions:
\begin{equation}
\begin{split}
\lambda_i=\frac{21-4x_{fi}}{-27+13x_{fi}}.
\end{split}
\label{lambda fixed point expression}
\end{equation}
 
In what follows, we will demonstrate the following constraints on unitary fixed points\footnote{By a unitary fixed point we mean a fixed point where the couplings are real. The reality of the couplings is a necessary condition for the unitarity of the theory.} of the above beta functions.
\begin{itemize}
\item \textbf{Constraint 1:} When $x_{f1}\neq x_{f2}$, there is no unitary fixed point where the two sectors are coupled, i.e., $\zeta\neq0$.
\item \textbf{Constraint 2:} When $x_{f1}= x_{f2}$, at any unitary fixed point with $\zeta\neq0$, we must have $h_1=h_2=\frac{3-\sqrt{6}}{16}\lambda$, where $\lambda$ is the common value of the gauge couplings in the two sectors.
\end{itemize} 

\subsection{Proof of constraint 1}
When $x_{f1}\neq x_{f2}$, from equation \eqref{lambda fixed point expression} we have  $\lambda_1\neq \lambda_2$. In this case, by demanding that $\beta_{h_i}^{\text{1-loop}}=0$, we get 
\begin{equation}
 h_i=\Big(\frac{3+\sigma_i\sqrt{6}}{16}\Big)\lambda_i,
 \label{hi expression}
\end{equation}
where $\sigma_1$ and $\sigma_2$ can be $1$ or $-1$. Therefore,
\begin{equation}
\begin{split}
 h_p\equiv\frac{h_1+h_2}{2}=\frac{3(\lambda_1+\lambda_2)+(\sigma_1\lambda_1+\sigma_2\lambda_2)\sqrt{6}}{32},\\
  h_m\equiv\frac{h_1-h_2}{2}=\frac{3(\lambda_1-\lambda_2)+(\sigma_1\lambda_1-\sigma_2\lambda_2)\sqrt{6}}{32}.
  \end{split} 
 \label{hp-hm expression}
\end{equation}
One can define similar linear combinations for the double trace couplings to simplify the analyisis:
\begin{equation}
f_p\equiv\frac{f_1+f_2}{2}, f_m\equiv\frac{f_1-f_2}{2}. 
\end{equation}
The 1-loop beta functions of these couplings (along with $\zeta$) are as follows:
\begin{equation}
\begin{split}
\beta_{f_p}^{\text{1-loop}}=&f_m\Big[32 h_m-3(\lambda_1-\lambda_2)\Big]+8 \Big[\zeta^2+(f_p+ h_p)(f_p+ 3h_p)\Big]+ 8f_m^2-3 f_p(\lambda_1+\lambda_2)\\
&+24 h_m^2+\frac{9}{32}(\lambda_1^2+\lambda_2^2),\\
\beta_{f_m}^{\text{1-loop}}=& f_m\Big[16 f_p+32h_p-3(\lambda_1+\lambda_2)\Big] +32 f_p h_m-3 f_p(\lambda_1-\lambda_2)+48 h_p h_m+\frac{9}{32}(\lambda_1^2-\lambda_2^2),\\
\beta_{\zeta}^{\text{1-loop}}=&\zeta\Big[16 f_p+32h_p-3(\lambda_1+\lambda_2)\Big].
\end{split}
\end{equation}
By setting $\beta_{\zeta}^{\text{1-loop}}=0$ and demanding that $\zeta\neq 0$, we get
\begin{equation}
\begin{split}
&f_p=-2h_p+\frac{3}{16}(\lambda_1+\lambda_2).
\end{split}
\label{first equation fp}
\end{equation}
Similarly, setting $\beta_{f_m}^{\text{1-loop}}=0$, we get
\begin{equation}
\begin{split}
&f_p=\frac{-48 h_p h_m-\frac{9}{32}(\lambda_1^2-\lambda_2^2)}{32  h_m-3 (\lambda_1-\lambda_2)}\\
\end{split}
\label{second equation fp}
\end{equation}
Here, we have assumed that 
\begin{equation}
\begin{split}
32  h_m-3 (\lambda_1-\lambda_2)\neq 0.
\end{split}
\end{equation}
From the value of $h_m$ given in equation \eqref{hp-hm expression}, we can see that this is equivalent to demanding 
\begin{equation}
\begin{split}
\sigma_1\lambda_1\neq\sigma_2\lambda_2.
\end{split}
\end{equation}
When $\sigma_1=\sigma_2$, this holds true trivially because $x_{f1}\neq x_{f2}$. When $\sigma_1=-\sigma_2$, this is true because otherwise $\lambda_1$ and $\lambda_2$ would have opposite signs. This is not admissible because $\lambda_i$ is related to the gauge coupling by the relation
\begin{equation}
\begin{split}
\lambda_i\equiv \frac{ N_{ci}g_i^2}{(4\pi)^2}\ ,
\end{split}
\end{equation}
and hence both $\lambda_1$ and $\lambda_2$ must be positive. \footnote{Here the reality of $g_i$ is the crucial assumption which is a necessary condition for the unitarity of the theory.} Therefore, we can trust equation \eqref{second equation fp}. 

Now, combining \eqref{first equation fp} with \eqref{second equation fp}, we get
\begin{equation}
16h_p h_m-6h_m(\lambda_1+\lambda_2)-6h_p(\lambda_1-\lambda_2)+\frac{9}{32}(\lambda_1^2-\lambda_2^2)=0.
\end{equation}
Substituting the values of $h_p$ and $h_m$ (given in \eqref{hp-hm expression}) into the above equation, we get
\begin{equation}
\begin{split}
 & (13+6\sqrt{6}\sigma_1)\lambda_1^2=(13+6\sqrt{6}\sigma_2)\lambda_2^2.
\end{split}
\end{equation}
When $\sigma_1=\sigma_2$, we get $\lambda_1^2=\lambda_2^2\implies \lambda_1=\lambda_2$ since we have already shown that both $\lambda_1$ and $\lambda_2$ must be positive.  However, this cannot be true for $x_{f1}\neq x_{f2}$.

The other possibility is that $\sigma_1=-\sigma_2$. For instance, consider the case $\sigma_1=-\sigma_2=1$. Then we get 
\begin{equation}
\begin{split}
 & (13+6\sqrt{6})\lambda_1^2=(13-6\sqrt{6})\lambda_2^2.
\end{split}
\label{sigm1=-sigm2=1}
\end{equation}
Now the coefficient $(13+6\sqrt{6})>0$, whereas $(13-6\sqrt{6})\approx-1.69694<0$. Then the signs of $\lambda_1^2$ and $\lambda_2^2$  are opposite. However, this is not consistent with the reality of $\lambda_1$ and $\lambda_2$. Therefore, equation \eqref{sigm1=-sigm2=1} cannot be satisfied. Similarly, we can rule out the existence of any unitary fixed point with $\sigma_1=-\sigma_2=-1$. Thus, from the above analysis, we can conclude that when $x_{f1}\neq x_{f2}$, there is no unitary fixed point of the 1-loop beta functions with $\zeta\neq 0$.

\subsection{Proof of constraint 2 }
When $x_{f1}=x_{f2}=x_f$, from equation \eqref{lambda fixed point expression}, we have  $\lambda_1=\lambda_2=\lambda\equiv\frac{21-4x_{f}}{-27+13x_{f}}$. As before the couplings for the single trace interactions are given by 
\begin{equation}
 h_i=\Big(\frac{3+\sigma_i\sqrt{6}}{16}\Big)\lambda.
\end{equation}
\begin{equation}
\begin{split}
 h_p=\frac{6+(\sigma_1+\sigma_2)\sqrt{6}}{32}\lambda,\ h_m=\frac{(\sigma_1-\sigma_2)\sqrt{6}}{32}\lambda.
  \end{split} 
\end{equation}
The 1-loop beta functions of the couplings corresponding to the double trace interactions simplify in this case as follows:
\begin{equation}
\begin{split}
\beta_{f_p}^{\text{1-loop}}=&\ 32 f_m h_m+8 \Big[\zeta^2+(f_p+ h_p)(f_p+ 3h_p)\Big]+ 8f_m^2-6 f_p\lambda+24 h_m^2+\frac{9}{16}\lambda^2,\\
\beta_{f_m}^{\text{1-loop}}=& f_m\Big[16 f_p+32h_p-6\lambda\Big] +32 f_p h_m+48 h_p h_m,\\
\beta_{\zeta}^{\text{1-loop}}=&\zeta\Big[16 f_p+32h_p-6\lambda\Big].
\end{split}
\end{equation}
Let us focus on the last two beta functions given above. Setting them equal to zero and searching for fixed points with $\zeta\neq 0$, we get
\begin{equation}
\begin{split}
16 f_p=-32h_p+6\lambda,\ 32 f_p h_m+48 h_p h_m=0.
\end{split}
\end{equation}
Substituting  the value of $f_p$  obtained from the first equation into the second one, we get
\begin{equation}
\begin{split}
(12 \lambda -16 h_p) h_m=0\implies \Bigg(\frac{18-(\sigma_1+\sigma_2)\sqrt{6}}{2}\Bigg)(\sigma_1-\sigma_2)=0 .
\end{split}
\end{equation}
The only way in which the above equation can be satisfied is if $\sigma_1=\sigma_2=\sigma$. In this case, we have
\begin{equation}
\begin{split}
 h_p=\frac{3+\sigma\sqrt{6}}{16}\lambda,\ h_m=0,\  f_p=-\frac{\sigma\sqrt{6}}{8}\lambda.
  \end{split} 
\end{equation}
Now, setting $\beta_{f_p}^{\text{1-loop}}=0$, we get
\begin{equation}
\begin{split}
& 8(\zeta^2+f_m^2)=-\frac{1}{32}\Bigg[18\sigma\sqrt{6}+39\Bigg]\lambda^2.
\end{split}
\end{equation}
Note that the reality of the couplings $f_m$ and $\zeta$ leads to the LHS of the above equation being non-negative. The RHS of the same equation is, however,  manifestly negative if $\sigma=1$. Thus, the only admissible unitary fixed point is the one where $\sigma=-1$. Therefore, we can conclude that when $x_{f_1}=x_{f_2}$, at any unitary fixed point where the two sectors are coupled, the couplings corresponding to the single trace interactions have the following values:
\begin{equation}
 h_1=h_2=\Big(\frac{3-\sqrt{6}}{16}\Big)\lambda.
\end{equation}
 
%%%%%%%%%%%%%%%%%%%%%%%%%%%%%%%%%%%%%%%%%%%%%%%%%%%%%%%%%%%%%%%%%%%%%%%%
%%%%%%%%%%%%%%%%%%%%%%%%%%%%%%%%%%%%%%%%%%

%%%%%%%%%%%%%%%%%%%%%%%%%%%%%%%%%%%%%%%%%%%%%%%%%%%%%%%%%%%%%%%%%%%%%%%%
%%%%%%%%%%%%%%%%%%%%%%%%%%%%%%%%%%%%%%%%%%%%%%%%%%%%%%%%%%%%%%%%%%%%%%%%

\section{Minima of the thermal effective potential}
\label{app: minimum of  potential}
In this appendix, we will investigate the minima of the thermal effective potential of the scalar fields in the real double bifundamental model. When both the thermal masses (squared), $m_{\text{th},1}^2$ and $m_{\text{th},2}^2$, are positive, one can trivially conclude that the minimum lies at the origin of the field space. On the other hand, if either $m_{\text{th},1}^2$ or $m_{\text{th},2}^2$ is negative, the minima would lie away from the origin, and the baryon symmetry would be broken. To analyze the location of the minima in such a situation, we would restrict our attention to the fixed points discussed in section \ref{subsec: rdb symmetry breaking analysis} for which $r\equiv \frac{N_{c2}}{N_{c1}}<1$, and $m_{\text{th},1}^2>0$ while  $m_{\text{th},2}^2<0$.  

First, let us employ gauge transformations to bring the matrices of the scalar fields to the following diagonal forms:
\beq  
\Phi_i=\text{diag}\{\phi_{i1},\cdots,\phi_{iN_{ci}}\}.
\eeq
The thermal effective potential (up to leading order in $\lambda$) for such a configuration  is
\beq
V_{\text{eff}}=&\frac{1}{2}\sum_{i=1}^2\sum_{a_i=1}^{N_{ci}}m_{\text{th},i}^2(\phi_{ia_i})^2+\sum_{i=1}^2\sum_{a_i=1}^{N_{ci}}\tilde h_i(\phi_{ia_i})^4\\
&+\sum_{i=1}^2\sum_{a_i=1}^{N_{ci}}\sum_{b_i=1}^{N_{ci}}\tilde f_i(\phi_{ia_i})^2(\phi_{ib_i})^2+2\tilde \zeta\sum_{a_1=1}^{N_{c1}}\sum_{b_2=1}^{N_{c2}}(\phi_{1a_1})^2(\phi_{2b_2})^2.
\eeq
One can determine the saddle points of this potential by setting its partial derivatives with respect to all the scalar fields equal to zero as shown below:
\beq \label{eq:ext}
\frac{\partial V_{\text{eff}}}{\partial \phi_{ia_i}}=&2\phi_{ia_i}\Bigg[\frac{1}{2}m_{\text{th},i}^2+2\tilde h_i(\phi_{ia_i})^2+2\tilde f_i\sum_{b_i=1}^{N_{ci}}(\phi_{ib_i})^2+2\tilde \zeta\sum_{b_{i^\prime}=1}^{N_{ci^\prime}}(\phi_{i^\prime b_{i^\prime}})^2\Bigg]=0,
\eeq
where $i^\prime$ denotes the complement of $i$. The above equation has the following possible solutions:
\beq
\phi_{ia_i}=0\ \ \text{or}\ \  \frac{1}{2}m_{\text{th},i}^2+2\tilde h_i(\phi_{ia_i})^2+2\tilde f_i\sum_{b_i=1}^{N_{ci}}(\phi_{ib_i})^2+2\tilde \zeta\sum_{b_{i^\prime}=1}^{N_{ci^\prime}}(\phi_{i^\prime b_{i^\prime}})^2=0.
\eeq
Let us consider a saddle where $n_i$ of the $N_{ci}$ diagonal entries of $\Phi_i$ are nonzero. In \ref{imaginary values of fields}, we will show that for $n_1\neq 0$, the solution corresponds to negative values of $(\phi_{1a_1})^2$ which is in conflict with the reality  of $\phi_{1a_1}$. So such a saddle point cannot correspond to a minimum of the potential. Later in \ref{true minimum of potential}, we will argue that the minima actually correspond to $n_1=0,\ n_2=N_{c2}$.

\subsection{Saddle points with $n_1\neq 0$ correspond to imaginary field configurations}
\label{imaginary values of fields}

In this subsection we will prove that all the saddle points with $n_1\neq0$ correspond to imaginary values of $\phi_{1a_1}$. For this, let us consider such a saddle where  $\phi_{a_i}\neq 0$ for the first $n_i$ values of $a_i$. The nonzero components of the scalar fields satisfy the following equations:
\beq
 \frac{1}{2}m_{\text{th},1}^2+2\tilde h_1(\phi_{1a_1})^2+2\tilde f_1\sum_{b_1=1}^{n_1}(\phi_{1b_1})^2+2\tilde \zeta\sum_{b_2=1}^{n_2}(\phi_{2 b_2})^2=0 \ \ \forall\ \ a_1\in\{1,\cdots,n_1\},\\
  \frac{1}{2}m_{\text{th},2}^2+2\tilde h_2(\phi_{2a_2})^2+2\tilde f_2\sum_{b_2=1}^{n_2}(\phi_{2b_2})^2+2\tilde \zeta\sum_{b_1=1}^{n_1}(\phi_{1 b_1})^2=0 \ \ \forall\ \ a_2\in\{1,\cdots,n_2\}.\\
\eeq
Solving these equations, we get the following values of the squares of the fields after substituting the original couplings by the corresponding 't Hooft couplings: 
\beq
(\phi_{1a_1})^2=-\frac{N_{c1}^2}{64\pi^2}\Bigg[\frac{(h_2 N_{c2}+f_2 n_2)m_{\text{th},1}^2- r \zeta  n_2 m_{\text{th},2}^2}{( h_1 N_{c1}+ f_1 n_1)( h_2 N_{c2}+ f_2 n_2)- \zeta^2 n_1  n_2}\Bigg] \ \ \forall\ \ a_1\in\{1,\cdots,n_1\},\\
(\phi_{2a_2})^2=-\frac{N_{c2}^2}{64\pi^2}\Bigg[\frac{( h_1 N_{c1}+ f_1 n_1)m_{\text{th},2}^2- \frac{\zeta}{r}  n_1 m_{\text{th},1}^2}{( h_1 N_{c1}+ f_1 n_1)( h_2 N_{c2}+ f_2 n_2)- \zeta^2 n_1  n_2}\Bigg] \ \ \forall\ \ a_2\in\{1,\cdots,n_2\}.
\label{equations for a general saddle}
\eeq
To prove that the nonzero values of $(\phi_{1a_1})^2$ are negative, we will show that both the numerator and the denominator of the quantity within the brackets in the first line of \eqref{equations for a general saddle} are positive. First, let us consider the denominator:
\beq
( h_1 N_{c1}+ f_1 n_1)( h_2 N_{c2}+ f_2 n_2)- \zeta^2 n_1  n_2=& h_1 h_2 N_{c1} N_{c2}+h_1 f_2 N_{c1} n_2+h_2 f_1 N_{c2} n_1\\
&+ (f_1 f_2- \zeta^2) n_1  n_2.
\eeq
Note that in section \ref{subsubsec rdb:potential stability}, we have already shown that $h_i>0$, $f_i>0$ and $(f_1 f_2-\zeta^2)>0$ for all points on the fixed circle. Therefore, all the terms in the above expression are positive-definite which ensures the positivity of the denominator. Now, let us look at the numerator: 
\begin{equation}
\begin{split}
&(h_2 N_{c2}+f_2 n_2)m_{\text{th},1}^2- r \zeta  n_2 m_{\text{th},2}^2\\
&=h(N_{c2}-n_2)+\frac{\pi^2}{48} \beta_{\text{th}}^{-2} n_2\Bigg[3(25-2\sqrt{6})\lambda^2-16 (f_m+r \zeta)   \Big(9-\sqrt{6}\Big)\lambda\Bigg]\\
\end{split}
\end{equation}
Here we have substituted the thermal masses by their values given in \eqref{rdb: fixed point thermal masses}. Moreover we have used the values of the different couplings and the equation of the fixed circle:
\beq
h_1=h_2=h=\Big(\frac{3-\sqrt{6}}{16}\Big)\lambda,\ f_p\equiv\frac{f_1+f_2}{2}=\frac{\sqrt{6}}{8}\lambda, \ \zeta^2+f_m^2= \Big(\frac{18\sqrt{6}-39}{256}\Big)\lambda^2,
\eeq
where $f_m\equiv \frac{f_1-f_2}{2}$. The equation of the fixed circle and the fact that $r<1$ further impose the condition  $(f_m+r\zeta)<|f_m|+|\zeta|<\frac{\sqrt{18\sqrt{6}-39}}{8}\lambda$. This inequality, together with the positivity of $h$ and the fact that $n_2\leq N_{c2}$, ensures that the numerator is positive as shown below:
\begin{equation}
\begin{split}
(h_2 N_{c2}+f_2 n_2)m_{\text{th},1}^2- r \zeta  n_2 m_{\text{th},2}^2 & > \frac{\pi^2}{48} \lambda^2\beta_{\text{th}}^{-2} n_2\Bigg[3(25-2\sqrt{6})-2\sqrt{18\sqrt{6}-39}\Big(9-\sqrt{6}\Big)\Bigg]\\
&\approx 6.32 \lambda^2\beta_{\text{th}}^{-2} n_2>0.\\
\end{split}
\end{equation}
Therefore, we can conclude  that
\beq
 (\phi_{1a_1})^2<0 \ \ \forall\ \ a_1\in \{1,\cdots, n_1\},
\eeq
which implies that $\phi_{1a_1}$ is imaginary for these values of $a_1$. This rules out the possibility of any saddle point with $n_1\neq 0$ being a minimum of the potential.

\subsection{The minima correspond to the saddle points with $n_1=0,n_2=N_{c2}$}
\label{true minimum of potential}
Let us now consider the saddle points with $n_1=0$. For these saddle points, $\phi_{1a_1}=0$  for all  $a_1\in\{1,\cdots, N_{c1}\}$, whereas the nonzero values of $\phi_{2a_2}$ are as follows:
\beq
(\phi_{2a_2})^2=-\frac{N_{c2}^2}{64\pi^2}\frac{m_{\text{th},2}^2}{h_2 N_{c2}+ f_2 n_2} \ \ \forall\ \ a_2\in\{1,\cdots,n_2\}.
\label{solution n1=0,n2}
\eeq
For these saddle points to be candidates for the minima of the potential, the above values of $(\phi_{2a_2})^2$ have to be positive. This can be verified by noting that for the fixed points under consideration, we have $m_{\text{th},2}^2<0$, $h_2>0$ and $f_2>0$. The values of potential at these saddle points are
\beq
V_{\text{eff}}\Big|_{n_1=0,n_2}
=&-\frac{N_{c2}^2}{256\pi^2}\Bigg[\frac{(m_{\text{th},2}^2)^2}{h_2 (\frac{N_{c2}}{n_2})+ f_2}\Bigg].\\
\eeq
Due to the positivity of $h_2$ at the fixed points, we can see from the above expression that the minima of the potential correspond to $n_2=N_{c2}$. Substituting $n_2$ by $N_{c2}$ in \eqref{solution n1=0,n2}, we get
\beq
(\phi_{2a_2})^2=-\frac{N_{c2}}{64\pi^2}\frac{m_{\text{th},2}^2}{h_2 + f_2 } \ \ \forall\ \ a_2\in\{1,\cdots,N_{c2}\}.
\label{solution for minima}
\eeq
Let us note here that there are $2^{N_{c2}}$ different field configurations that satisfy the above equation. These configurations are as follows:
\beq
\Phi_1=0,\ \Phi_2
=\Big(-\frac{N_{c2}}{64\pi^2}\frac{m_{\text{th},2}^2}{h_2+f_2}\Big)^{\frac{1}{2}}\text{diag}\{\sigma_1,\cdots,\sigma_{N_{c2}}\},
\eeq
where $\sigma_1,\cdots,\sigma_{N_{c2}}$ can be $1$ or $-1$. However, many of these configurations are related to each other by $SO(N_{c2})\times SO(N_{c2})$ gauge transformations. It can be shown that all these configurations can be categorized into two equivalence classes. The configurations in each class are related to each other by gauge transformations. One can go from one class to the other by flipping the sign of just one of the diagonal entries. Thus, we can choose one representative from each of these classes as follows:
\beq
\Phi_1=0,\ \Phi_2
=\Big(-\frac{N_{c2}}{64\pi^2}\frac{m_{\text{th},2}^2}{h_2+f_2}\Big)^{\frac{1}{2}}\text{diag}\{\pm1,1,\cdots,1\},
\eeq
and treat these as two distinct minima unrelated by gauge transformations. Since we are working in a perturbative regime where the 't Hooft couplings are small, the thermal expectation values of the field $\Phi_1$ and $\Phi_2$ would correspond to these minima at leading order in perturbation theory.

\section{Large N scaling of planar diagrams}
\label{app: scaling of diagrams}
In this appendix we will discuss the scaling of different planar diagrams in the  double bifundamental models. For simplicity, we will restrict our attention to the complex double bifundamental model. This would enable us to use  't Hooft's large $N$ expansion in terms of oriented surfaces. The $\mathfrak{s}\mathfrak{o}(N_{ci})$ valued gauge connections in the real double bifundamental model generate both orientable and non-orientable diagrams. However, the large N scaling of the planar diagrams are identical in both the double bifundamental models.

In the following analysis, we will take the two ranks $N_{c1}$ and $N_{c2}$ to be of comparable magnitudes, say $O(N)$, and then work in the limit $N\rightarrow\infty$. In this  limit, we will look at how different planar diagrams scale with $N$. In the process, we will demonstrate an important feature of these diagrams, viz., any double trace vertex in such a diagram links two otherwise disconnected subdiagrams. As discussed in section \ref{sec: survival of fixed circle}, this feature  plays a crucial role in proving the survival of the fixed circles in these models under all loop corrections at the planar limit. Throughout  this appendix, we  will be using  't Hooft's double line notation to represent the  diagrams. As we are interested only in how different diagrams scale with $N$ where $N$ is the order of magnitude of the ranks in both the sectors, we will not distinguish between the fields in the two sectors. Thus, unlike section \ref{sec: survival of fixed circle}, we will not use different colors to show propagators and vertices belonging to different sectors.

\subsection{Scaling of bubble diagrams}
Let us begin our analysis by considering bubble diagrams, i.e., diagrams which have no external legs. For convenience, we will temporarily rescale all the fields such that their propagators scale as $O(1/N)$. This would not affect the scaling of the bubble diagrams due to the absence of any external leg in such diagrams. Under the above rescaling of the fields, the single trace and double trace couplings scale as $O(N)$ and $O(1)$ respectively. 

Now, let us first consider a  bubble diagram with only single trace vertices. Each vertex in such a diagram contributes a factor of $N$, while each propagator (an edge in the double line notation) contributes a factor of $\frac{1}{N}$. Moreover, each color loop (a face in the double line notation) in the diagram contributes a factor of $N$. Therefore, the overall scaling of the diagram is given by
\beq
N^{V-E+F}=N^{2-2\tilde g}
\eeq
where $V, E$ and $F$ are the number of vertices, edges and faces in the diagram respectively. $\tilde g$ is the genus number corresponding to the diagram \footnote{Departing from the usual convention, we put a tilde over $g$ to avoid any confusion with the gauge couplings.}. For a planar diagram, we have $\tilde g=0$. Hence such a diagram scales as $N^2$.

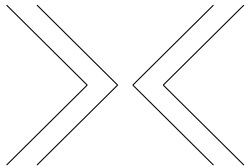
\begin{figure}[H]
\centering
\begin{tikzpicture}
  \begin{feynman}
    \vertex (a);
    \vertex [above right=1.5cm of a] (b);
     \vertex [below right=1.5cm of a] (c);
       \vertex [left=0.4cm of a] (d);
        \vertex [left =0.4cm of b] (e);
     \vertex [left=0.4cm of c] (f);
     \vertex [left=0.2cm of d] (d1);
      \vertex [left=0.15cm of d] (o);
     \vertex [left=0.4cm of d1] (a1);
    \vertex [above left=1.5cm of a1] (b1);
     \vertex [below left=1.5cm of a1] (c1);
        \vertex [right=0.4cm of b1] (e1);
     \vertex [right =0.4cm of c1] (f1);
    \diagram* {
       (c)-- (a),
 (a)--(b), 
      (f)-- (d),
 (d)--(e),
    (c1)-- (a1),
 (a1)--(b1), 
      (f1)-- (d1),
 (d1)--(e1), 
%d1[label={[font=\large]$~~f$}]
    };
  \end{feynman}
\end{tikzpicture}
\caption{Structure of a double trace vertex}
\label{fig:doubletrace}
\end{figure}

Now let us extend our discussion to connected bubble diagrams with double trace vertices. To be specific, let us consider a diagram with $m$ double trace vertices.  Each such double trace vertex has the structure shown in figure \ref{fig:doubletrace}.
To understand how this diagram scales with $N$, let us employ the following trick. Let us remove all the double trace vertices from the diagram and  join the pairs of legs with identical colors attached to each such vertex. This would leave us with a set of disconnected  bubble diagrams which consist only of single trace vertices and propagators.  Let the number of such disconnected pieces be $(m^\prime+1)$. Here, $m^{\prime}\leq m$ as the removal of each double trace vertex can lead to at most two disconnected pieces. Now, as we saw earlier, each of the disconnected pieces, comprising only of single trace vertices and propagators, scales as $N^{2-2\tilde g_i}$ where $\tilde g_i$ is the genus number corresponding to that piece. Therefore, the overall scaling of the disconnected diagram obtained by removing the double trace vertices is 
\beq
N^{\sum_{i=1}^{m^\prime+1}(2-2\tilde g_i)}=N^{2(m^\prime+1)-2\sum_{i=1}^{m^\prime+1}\tilde g_i}.
\label{scaling of disconnected bubbles}
\eeq
 From this we can now estimate the scaling of the original diagram with the double trace vertices. For this, we just need to recall that in our present convention, each double trace coupling scales as $O(1)$. Therefore, the removal of such vertices from the diagram, by itself, does not lead to any change in the scaling of the diagram. However, while removing  a double trace vertex, we are also joining two pairs of propagators in the diagram. This leads to a reduction in the overall number of propagators by $2m$. Since each such propagator scales as $1/N$ in our convention, the above reduction in their number leads to an enhancement by a factor of $N^{2m}$. Therefore, comparing with the overall scaling of the disconnected diagram as given in \eqref{scaling of disconnected bubbles}, we find that the original diagram scales as
\beq
N^{2(m^\prime+1)-2\sum_{i=1}^{m^\prime+1}\tilde g_i-2m}=N^{2-2(m-m^\prime)-2\sum_{i=1}^{m^\prime+1}\tilde g_i}.
\label{scaling of bubbles with double trace vertices}
\eeq
From the above expression, we can see that the leading contributions come from diagrams where $m^\prime=m$ and $\tilde g_i=0$ for all the disconnected pieces. These are precisely the planar diagrams with double trace vertices. Each double trace vertex in such a diagram connects two disconnected planar subdiagrams. As we can see from \eqref{scaling of bubbles with double trace vertices}, these planar diagrams scale as $N^2$, which is identical to the scaling of the planar diagrams with only single trace vertices. Thus, our choice of scaling of the different couplings leads to a consistent large $N$ scaling of all planar bubble diagrams. 

Next, we will determine the scaling of different diagrams that contribute to the wave function and vertex renormalizations considered in section \ref{sec: survival of fixed circle}. Henceforth, we will revert back to our previous convention of unrescaled fields where the corresponding propagators are $O(1)$.

\subsection{Scaling of diagrams with 2 external legs}

\begin{figure}[H]
\centering
\begin{tikzpicture}
\draw[] (-3,0.25) -- (-1.2,0.25);
\draw[] (-3,-0.25) -- (-1.2,-0.25);
\draw[] (3,0.25) -- (1.2,0.25);
\draw[] (3,-0.25) -- (1.2,-0.25);
 \draw (-1.2,0.25) arc (180:0:1.2);
  \draw (-1.2,-0.25) arc (-180:0:1.2);
    \draw (-1.2,-0.25) arc (-180:0:1.2);
\draw[ pattern=north east lines, pattern color=gray!30] (0,0) circle (1);
\end{tikzpicture}
\caption{Planar connected diagram with 2 external legs: Here, as well as in all subsequent diagrams, a shaded blob represents a planar connected component.}
\label{fig:diagram for wfn. renorm.}
\end{figure}
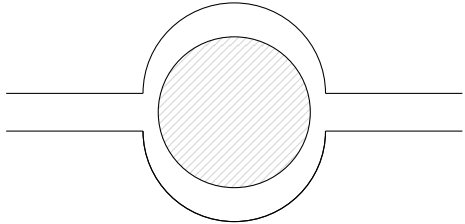

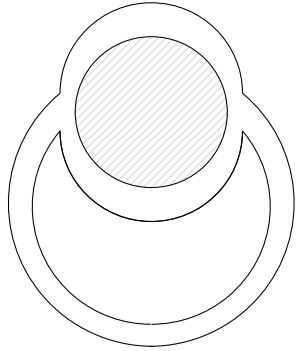
\begin{figure}[H]
\centering
\begin{tikzpicture}
\draw (-1.2,0.25) arc (180:0:1.2);
\draw (-1.2,-0.25) arc (-180:0:1.2);
\draw (-1.2,-0.25) arc (-180:0:1.2);
\draw[ pattern=north east lines, pattern color=gray!30] (0,0) circle (1);
\draw(-1.2,-0.25) arc (140:270:1.56);
\draw(1.2,-0.25) arc (40:-90:1.56);
\draw(-1.2,0.25) arc (130:270:1.9);
\draw(1.2,0.25) arc (50:-90:1.9);
\end{tikzpicture}
\caption{Bubble diagram obtained by joining the two external legs in figure \ref{fig:diagram for wfn. renorm.}.}
\label{fig: bubble diagram for wfn. renorm.}
\end{figure}

Let us first consider planar connected diagrams that contribute to the wave function renormalizations of the scalar fields. Such a diagram has 2 external legs as show in figure \ref{fig:diagram for wfn. renorm.}.
We can estimate the scaling of this diagram by first joining the two external legs and then summing  over the color indices in this propagator. This leads to an enhancement by a factor of $N^2$ due to the introduction of two new color loops. The resulting diagram is a planar bubble as shown in figure \ref{fig: bubble diagram for wfn. renorm.}. As we have already argued, such a planar bubble scales as $N^2$. Therefore, taking into account the above-mentioned enhancement, we can conclude that the original diagram in figure \ref{fig:diagram for wfn. renorm.} scales as $N^0$.

\subsection{Scaling of diagrams with 4 external legs}
Now, let us look at planar connected diagrams with 4 external legs of the scalar fields. We can have two different types of such diagrams: 
\begin{enumerate}
\item Diagrams which contribute to the renormalization of the single trace vertices,
\item Diagrams which contribute to the renormalization of the double trace vertices.
\end{enumerate}
We will derive the scaling of both these types of diagrams below.
\subsubsection{Diagrams corresponding to vertex renormalizations of single trace couplings}
\begin{figure}[H]
\centering
\begin{tikzpicture}
\draw[] (-3,2.25) -- (-1.2,0.25);
\draw[] (-3,-2.25) -- (-1.2,-0.25);
\draw[] (3,2.25) -- (1.2,0.25);
\draw[] (3,-2.25) -- (1.2,-0.25);
 \draw (-1.2,0.25) arc (180:0:1.2);
  \draw (-1.2,-0.25) arc (-180:0:1.2);
    \draw (-1.2,-0.25) arc (-180:0:1.2);
\draw[ pattern=north east lines, pattern color=gray!30] (0,0) circle (1);
\draw[] (-3.2,2) -- (-1.4,0);
\draw[] (-3.2,-2) -- (-1.4,0);
\draw[] (3.2,2) -- (1.4,0);
\draw[] (3.2,-2) -- (1.4,0);
\end{tikzpicture}
\caption{Planar connected diagram contributing to renormalization of a single trace vertex.}
\label{fig:diagram for single trace vertex renorm.}
\end{figure}
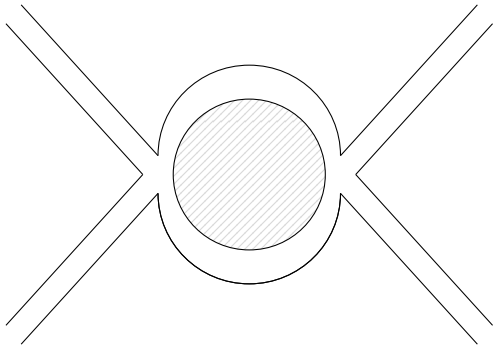

\begin{figure}[H]
\centering
\begin{tikzpicture}
\newcommand \x {3.5};
\draw (-1.2,0.25) arc (180:0:1.2);
\draw (-1.2,-0.25) arc (-180:0:1.2);
\draw (-1.2,-0.25) arc (-180:0:1.2);
\draw[ pattern=north east lines, pattern color=gray!30] (0,0) circle (1);
\draw (-1.2,0.25-\x) arc (180:0:1.2);
\draw (-1.2,-0.25-\x) arc (-180:0:1.2);
\draw (-1.2,-0.25-\x) arc (-180:0:1.2);
\draw[ pattern=north east lines, pattern color=gray!30] (0,0-\x) circle (1);
   
\draw (-1.2,-0.25) arc (110:250:\x/2-0.15);
\draw (-1.3,0) arc (116:244:\x/2+0.23);
\draw (-1.3,0) arc (70:290:\x/2+0.15);
\draw (-1.2,0.25) arc (70:290:\x/2+0.4);

\draw (1.2,-0.25) arc (70:-70:\x/2-0.15);
\draw (1.3,0) arc (64:-64:\x/2+0.23);
\draw (1.3,0) arc (110:-110:\x/2+0.15);
\draw (1.2,0.25) arc (110:-110:\x/2+0.4);

\end{tikzpicture}
\caption{Bubble diagram obtained by joining the corresponding external legs of two copies of the diagram in figure \ref{fig:diagram for single trace vertex renorm.}.}
\label{fig:bubble diagram for single trace vertex renorm.}
\end{figure}
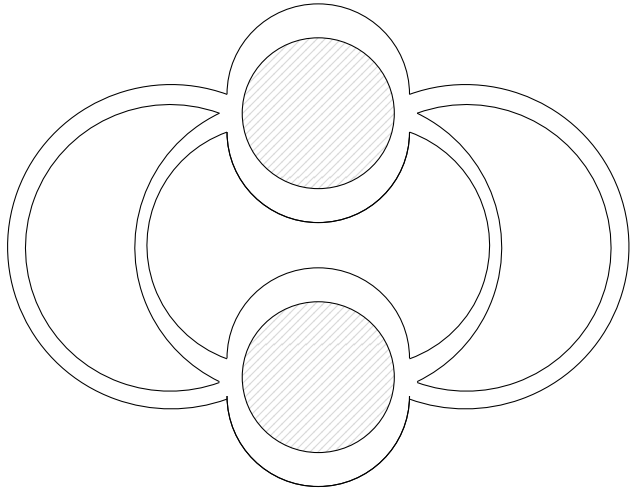

Any planar diagram which contributes to the renormalization of a single trace vertex has the structure shown in figure \ref{fig:diagram for single trace vertex renorm.}.
Suppose such a diagram scales as $N^x$. We can estimate the value of $x$ by constructing a bubble diagram via the following procedure:
Take two identical copies of the diagram in figure \ref{fig:diagram for single trace vertex renorm.} and join the external legs with identical colors in these two copies as shown in figure \ref{fig:bubble diagram for single trace vertex renorm.}. While joining these legs, sum over corresponding color indices. 
This procedure introduces 4 new color loops which leads to an enhancement by a factor of $N^4$. The planar bubble diagram at the end of the procedure scales as $N^2$. Therefore, we have the following equation:
\beq
2x+4=2\implies x=-1.
\eeq
Therefore, all planar connected diagrams of the kind shown in figure \ref{fig:diagram for single trace vertex renorm.} scale as $N^{-1}$. 

\subsubsection{Diagrams corresponding to vertex renormalizations of double trace couplings}
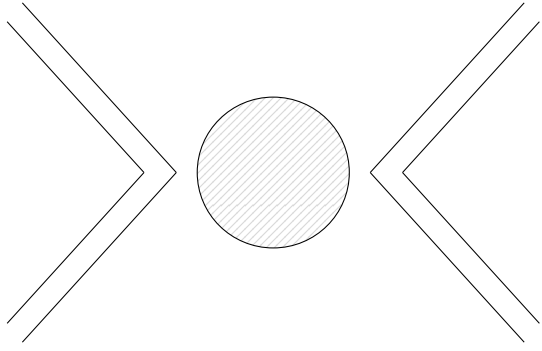
\begin{figure}[H]
\centering
\begin{tikzpicture}
\newcommand \x{0.3}
\draw[] (-3-\x,2.25) -- (-0.975-\x,0);
\draw[] (-3-\x,-2.25) -- (-0.975-\x,0);
\draw[] (3+\x,2.25) -- (0.975+\x,0);
\draw[] (3+\x,-2.25) -- (0.975+\x,0);
\draw[ pattern=north east lines, pattern color=gray!30] (0,0) circle (1);
\draw[] (-3.2-\x,2) -- (-1.4-\x,0);
\draw[] (-3.2-\x,-2) -- (-1.4-\x,0);
\draw[] (3.2+\x,2) -- (1.4+\x,0);
\draw[] (3.2+\x,-2) -- (1.4+\x,0);
\end{tikzpicture}
\caption{Planar connected diagram contributing to renormalization of a double trace vertex.}
\label{fig:diagram for double trace vertex renorm.}
\end{figure}

\begin{figure}[H]
\centering
\begin{tikzpicture}
\newcommand \x{0.3}
\newcommand \y{1}

\draw[ pattern=north east lines, pattern color=gray!30] (0,0) circle (1);

\draw (-0.975-\x,0) arc (0:180:\y);

\draw (-0.975-\x,0) arc (0:-180:\y);

\draw (-0.975-\x-0.2,0) arc (0:180:\y-0.2);

\draw (-0.975-\x-0.2,0) arc (0:-180:\y-0.2);

\draw (0.975+\x,0) arc (180:0:\y);

\draw (0.975+\x,0) arc (-180:0:\y);

\draw (0.975+\x+0.2,0) arc (180:0:\y-0.2);

\draw (0.975+\x+0.2,0) arc (-180:0:\y-0.2);

\end{tikzpicture}
\caption{Bubble diagram obtained by joining the external legs with identical colors in figure \ref{fig:diagram for double trace vertex renorm.}.}
\label{fig:bubble diagram for double trace vertex renorm.}
\end{figure}
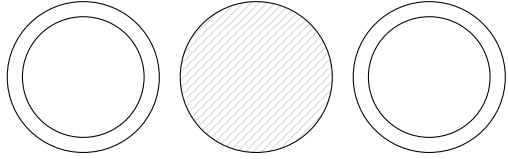

Planar diagrams which contribute to the renormalization of a double trace vertex are of the form shown in figure \ref{fig:diagram for double trace vertex renorm.}.\footnote{Here let us remark that given the characterization of a planar diagram as a cell decomposition of a sphere (i.e., a compactified plane) in terms of a ribbon graph where the external lines correspond to open intervals in the boundaries of the faces, the only requirement for these external lines to represent a double trace vertex is that for each pair of them with identical colors, the two lines must lie on the same boundary of a common face. This implies that in addition to the diagrams shown in figure \ref{fig:diagram for double trace vertex renorm.}, there can be  diagrams where the two pairs of external lines do not share a common face. In particular, one of the pairs of external lines in such a diagram may not be continuously connected to the infinity of the plane. An example of this would be a ribbon graph version of the diagram shown in figure \ref{fig: diagram with single trace couplings contributing to Zf1}.  Despite this subtle  difference, the large N scaling of such diagrams would be the same as the scaling of the diagrams shown in figure \ref{fig:diagram for double trace vertex renorm.} since the following arguments are equally applicable to them.}
As in the case of the diagrams with 2 external legs, we can estimate the scaling of the diagram in figure \ref{fig:diagram for double trace vertex renorm.} by joining the pairs of external legs with identical colors and then summing over the color indices. As a result, we obtain the bubble diagram  shown in figure  \ref{fig:bubble diagram for double trace vertex renorm.}.
The joining of the external legs introduces 4 new color loops leading to an enhancement by a factor of $N^4$. As we have already argued, the bubble diagram in figure \ref{fig:bubble diagram for double trace vertex renorm.} scales as $N^2$. Therefore, taking into account the above-mentioned enhancement, we can conclude that the original diagram in figure \ref{fig:diagram for double trace vertex renorm.} scales as $N^{-2}$.

\subsection{Comments}
From the above analysis, we can draw the following two conclusions about  connected planar diagrams with 2 and 4 external legs:
\begin{enumerate}
\item All  such diagrams with a particular set of external legs scale identically with $N$.
\item Since we have shown that these diagrams can be augmented by definite procedures to construct planar bubble diagrams, the result that double trace vertices link otherwise disconnected planar subdiagrams can be extended to these diagrams as well.
\end{enumerate}
In section \ref{sec: survival of fixed circle}, we have repeatedly used these two features of the connected planar diagrams  to determine the structure of the planar beta functions of the different couplings.

 \section{General expressions for the planar beta functions of the double trace couplings in the double bifundamental models}
\label{app: double trace beta fns.} 
In this appendix, we will derive some general expressions for the planar beta functions of the double trace couplings ($f_i$ and $\zeta$) in terms of the corresponding wave function and vertex renormalizations. For this, we will restrict our attention to the subspace where the planar beta functions of the single trace couplings ($\lambda_i$ and $h_i$) vanish. As we have shown in section \ref{sec: survival of fixed circle}, these  beta functions ($\beta_{\lambda_i}$ and $\beta_{h_i}$) are independent of the double trace couplings. Hence, their roots can be determined independently. There is a discrete set of such roots out of which only one corresponds to unitary fixed points where the two sectors are coupled\footnote{Here, let us remind the reader that we need to have $x_{f1}=x_{f2}$ for the existence of this root.}. At this root, the single trace couplings in the two sectors are equal, i.e., we have 
 \beq
 \lambda_1=\lambda_2=\lambda_0\ ,\ h_1=h_2=h_0.
 \eeq
 We will now derive the expressions for the beta functions of the double trace couplings on the subspace defined by the above equation.  
 
To derive these expressions, we will work in  dimensional regularization and the $\overline{\text{MS}}$ scheme. In this scheme, the renormalized double trace couplings are related to the corresponding bare couplings as follows:
 \beq
f_i^B =\tilde\mu^\varepsilon Z_{f_i}   Z_{\Phi_i}^{-2} f_i \ ,\ \zeta^B =\tilde\mu^\varepsilon Z_{\zeta}   Z_{\Phi_1}^{-1}Z_{\Phi_2}^{-1} \zeta.
\eeq
Here $\tilde \mu$ is an energy scale which is related to the renormalization scale $\mu$ in the $\overline{\text{MS}}$ scheme by $\tilde \mu^2=\mu^2 \frac{e^{\gamma_E}}{4\pi}$ where $\gamma_E$ is  the Euler-Mascheroni constant. $Z_{\Phi_i}$ is the wave function renormalization of the field $\Phi_i$, and $Z_{f_i}$ and $Z_{\zeta}$ are the vertex renormalizations of the respective couplings. The superscript $B$ indicates the bare couplings. Since these bare couplings are independent of $\mu$, they drop out once we differentiate the above equations with respect to $\mu$, and we get
\begin{equation}
\begin{split}
& \beta_{f_i}=-\varepsilon f_i-f_i\Big(\sum_{j=1}^2 \beta_{\lambda_j}\frac{\partial}{\partial \lambda_j}+\sum_{j=1}^2 \beta_{h_j}\frac{\partial}{\partial h_j}+\sum_{A^\prime} \beta_{\kappa_{A^\prime}}\frac{\partial}{\partial \kappa_{A^\prime}}\Big)\ln\Big[Z_{f_i}   Z_{\Phi_i}^{-2} \Big],\\
&\beta_{\zeta}=-\varepsilon \zeta-\zeta\Big(\sum_{j=1}^2 \beta_{\lambda_j}\frac{\partial}{\partial \lambda_j}+\sum_{j=1}^2 \beta_{h_j}\frac{\partial}{\partial h_j}+\sum_{A^\prime} \beta_{\kappa_{A^\prime}}\frac{\partial}{\partial \kappa_{A^\prime}}\Big)\ln\Big[ Z_{\zeta}   Z_{\Phi_1}^{-1}Z_{\Phi_2}^{-1} \Big],\\
\end{split}
\label{double trace coupling beta fn expression eq1}
\end{equation}
where   $\kappa_{A^\prime}$ runs over the double trace couplings in the model.
Recall that we are working in the subspace where the planar beta functions $\beta_{\lambda_i}$ and $\beta_{h_i}$ vanish in the $\varepsilon\rightarrow 0$ limit. Therefore, for small nonzero values of $\varepsilon$, these beta functions are $\beta_{\lambda_i}=-\varepsilon \lambda_0$, $\beta_{h_i}=-\varepsilon h_0$ on the subspace of our interest. In the neighborhood of this subspace, it is convenient to switch to the following basis for the single trace couplings:
 \beq
 \lambda_{p,m}\equiv\frac{\lambda_1\pm\lambda_2}{2},\ h_{p,m}\equiv\frac{h_1\pm h_2}{2}.
 \eeq
In terms of this basis, the operator $\Big(\sum\limits_{j=1}^2 \beta_{\lambda_j}\frac{\partial}{\partial \lambda_j}+\sum\limits_{j=1}^2 \beta_{h_j}\frac{\partial}{\partial h_j}\Big)$ on the above-mentioned subspace is as follows
\begin{equation}
\begin{split}
\sum_{j=1}^2 \beta_{\lambda_j}\frac{\partial}{\partial \lambda_j}+\sum_{j=1}^2 \beta_{h_j}\frac{\partial}{\partial h_j}
&=-\varepsilon\lambda_0\frac{\partial}{\partial \lambda_p}
-\varepsilon h_0\frac{\partial}{\partial h_p}.
\end{split}
\end{equation}
Thus, the equations in \eqref{double trace coupling beta fn expression eq1} can be rewritten as
\begin{equation}
\begin{split}
& \beta_{f_i}=-\varepsilon f_i-f_i\Big(-\varepsilon\lambda_0 \frac{\partial}{\partial \lambda_p}-\varepsilon h_0\frac{\partial}{\partial h_p}+\sum_{A^\prime} \beta_{\kappa_{A^\prime}}\frac{\partial}{\partial \kappa_{A^\prime}}\Big)\ln\Big[Z_{f_i}   Z_{\Phi_i}^{-2} \Big],\\
&\beta_{\zeta}=-\varepsilon \zeta-\zeta\Big(-\varepsilon\lambda_0 \frac{\partial}{\partial \lambda_p}-\varepsilon h_0\frac{\partial}{\partial h_p}+\sum_{A^\prime} \beta_{\kappa_{A^\prime}}\frac{\partial}{\partial \kappa_{A^\prime}}\Big)\ln\Big[ Z_{\zeta}   Z_{\Phi_1}^{-1}Z_{\Phi_2}^{-1} \Big].\\
\end{split}
\label{double trace coupling beta fn expression eq2}
\end{equation}

Now,  we can expand the wave function and vertex renormalizations in powers of $\frac{1}{\varepsilon}$ as follows:
\beq
&Z_{f_i}=1+\sum_{n=1}^\infty\frac{Z_{f_i}^{(n)}}{\varepsilon^n},\ Z_{\zeta}=1+\sum_{n=1}^\infty\frac{Z_{\zeta}^{(n)}}{\varepsilon^n},\ Z_{\Phi_i}=1+\sum_{n=1}^\infty\frac{Z_{\Phi_i}^{(n)}}{\varepsilon^n}. 
\eeq
To satisfy the equations in \eqref{double trace coupling beta fn expression eq2} at each order in the $\frac{1}{\varepsilon}$-expansion, we need the beta functions to have the following forms:
\beq
\beta_{f_i}=\varepsilon\beta_{f_i}^{(1)}+\beta_{f_i}^{(0)},\ \beta_{\zeta}=\varepsilon\beta_{\zeta}^{(1)}+\beta_{\zeta}^{(0)}.\\
\eeq
Solving the $O(\varepsilon)$ and $O(1)$ terms in \eqref{double trace coupling beta fn expression eq2} we get
\begin{equation}
\begin{split}
\beta_{f_i}^{(1)}=- f_i,\ \beta_{\zeta}^{(1)}=- \zeta,
\end{split}
\end{equation}
\begin{equation}
\begin{split}
& \beta_{f_i}^{(0)}=f_i\Big(\lambda_0 \frac{\partial}{\partial \lambda_p}+ h_0\frac{\partial}{\partial h_p}+\sum_{A^\prime} \kappa_{A^\prime}\frac{\partial}{\partial \kappa_{A^\prime}}\Big)\Bigg[ Z_{f_i}^{(1)} -2 Z_{\Phi_i}^{(1)}\Bigg],\\
&\beta_{\zeta}^{(0)}= \zeta\Big(\lambda_0 \frac{\partial}{\partial \lambda_p}+ h_0\frac{\partial}{\partial h_p}+\sum_{A^\prime}\kappa_{A^\prime}\frac{\partial}{\partial \kappa_{A^\prime}}\Big)\Bigg[  Z_{\zeta}^{(1)} -  Z_{\Phi_1}^{(1)}- Z_{\Phi_2}^{(1)} \Bigg].\\
\end{split}
\end{equation}

Finally, taking the $\varepsilon\rightarrow 0$ limit, we get the following expressions of the planar beta functions of the double trace couplings on the subspace $\lambda_1=\lambda_2=\lambda_0,\ h_1=h_2=h_0$:
\begin{equation}
\begin{split}
& \beta_{f_i}=f_i\sum_{A} \kappa_{A}\frac{\partial}{\partial \kappa_{A}}\Bigg[ Z_{f_i}^{(1)} -2 Z_{\Phi_i}^{(1)}\Bigg],\\
&\beta_{\zeta}= \zeta\sum_{A}\kappa_{A}\frac{\partial}{\partial \kappa_{A}}\Bigg[  Z_{\zeta}^{(1)} -  Z_{\Phi_1}^{(1)}- Z_{\Phi_2}^{(1)} \Bigg],\\
\end{split}
\label{beta fn of double trace couplings: final expressions}
\end{equation}
where $\kappa_A$ runs over the values $\lambda_p, h_p, f_1, f_2$ and $\zeta$. Note that the expressions in \eqref{beta fn of double trace couplings: final expressions} do not involve differentiation with respect to $\lambda_m$ and $h_m$. Therefore, while evaluating these beta functions, we can restrict our attention to the wave function and vertex renormalizations in a region where $\lambda_m=h_m=0$, or equivalently, $\lambda_1=\lambda_2=\lambda$ and $h_1=h_2=h$ with $(\lambda, h)$ lying in the neighborhood of $(\lambda_0,h_0)$. As discussed in section \ref{sec:proof}, this plays an important role in proving the survival of the fixed circle under all loop corrections in the planar limit.

%%%%%%%%%%%%%%%%%%%%%%%%%%%%%%%%%%%%%%%%%%
%%%%%%%%%%%%%%%%%%%%%%%%%%%%%%%%%%%%%%%%%%
%%%%%%%%%%%%%%%%%%%%%%%%%%%%%%%%%%%%%%%%%%

\section{Finite $N$ corrections}
\label{app: finite N corrections}

In this appendix we will explore the fate of the fixed circle in the real double bifundamental model when the corrections due to finiteness of $N_{ci}$ are taken into account. We will assume that $N_{c1}$ and $N_{c2}$ are of comparable magnitudes (say, $O(N)$, where $N$ is a large number). As we have already seen in appendix \ref{app: fixed point constraints}, the values of $x_{f1}$ and $x_{f2}$ must be the same (say, $x_{0f}$) at the leading order in $\frac{1}{N}$ for the existence of unitary fixed points where the two sectors are coupled to each other. While considering  finite N corrections to such a fixed point, we will allow for a difference between $x_{f1}$ and $x_{f2}$ at subleading orders in the $\frac{1}{N}$ expansion. So, we will take
 \begin{equation}
\begin{split}
x_{fi}=\sum_{n=0}^\infty  x_{fi}^{(n)}.
\end{split}
\end{equation}
where $x_{fi}^{(n)}$ is the $O(\frac{1}{N^n})$ term in the expansion of $x_{fi}$. Here, $x_{f1}^{(0)}=x_{f2}^{(0)}=x_{0f}=\frac{21}{4}-\epsilon$ with $0<\epsilon\ll1$. The common value ($\lambda_0$) of the gauge couplings in the two sectors for the fixed points  at the planar limit is determined by this small parameter $\epsilon$. This relation can be used to express $x_{0f}$ as a perturbative expansion in $\lambda_0$.

Therefore, we see that there are two small parameters in the problem: $\frac{1}{N}$ and $\lambda_0$. We will work in a regime where $\frac{1}{N}\ll \lambda_0$. We will denote the $n^{\text{th}}$ order term in the $1/N$-expansion of a coupling $g$ which lies at a fixed point of the beta functions by $g^{(n)}$. For each value of $n$, all the couplings  have perturbative expansions in $\lambda_0$. We will determine the first few terms in such perturbative expansions for the $O(1/N)$ and $O(1/N^2)$ corrections to the couplings. 

For later convenience, we find it useful to define the following quantities:
 \begin{equation}
\begin{split}
&\lambda_{p}^{(n)}=\frac{\lambda_1^{(n)}+\lambda_2^{(n)}}{2},\ \lambda_{m}^{(n)}=\frac{\lambda_1^{(n)}-\lambda_2^{(n)}}{2},\\
h_{p}^{(n)}=\frac{h_1^{(n)}+h_2^{(n)}}{2},&\ h_{m}^{(n)}=\frac{h_1^{(n)}-h_2^{(n)}}{2},\
 f_{p}^{(n)}=\frac{f_1^{(n)}+f_2^{(n)}}{2},\ f_{m}^{(n)}=\frac{f_1^{(n)}-f_2^{(n)}}{2}.\\
\end{split}
\end{equation}
The $\lambda_0$-expansion of these quantities for $n=0$ are as follows:
\begin{equation}
\begin{split}
&\lambda_p^{(0)}\equiv\lambda_0,\ \lambda_m^{(0)}=0,\\ &\ h_m^{(0)}=0,\ h_p^{(0)}\equiv h_0=\Big(\frac{3-\sqrt{6}}{16}\Big)\lambda_0+\Bigg[\frac{93\sqrt{6}-201+8(7-5\sqrt{6})x_{0f}}{768\sqrt{6}}\Bigg]\lambda_0^2+O(\lambda_0^3)\\
&\qquad\qquad\qquad\qquad\quad=\Big(\frac{3-\sqrt{6}}{16}\Big)\lambda_0-\Bigg(\frac{234-31\sqrt{6}}{1536}\Bigg)\lambda_0^2+O(\lambda_0^3),\\
&f_m^{(0)}\equiv f_{0m},\ \zeta^{(0)}\equiv \zeta_0,\ f_p^{(0)}\equiv f_{0p}=\frac{\sqrt{6}}{8}\lambda_0-\Bigg[\frac{2160+339\sqrt{6}+56\sqrt{6}x_{0f}}{2304}\Bigg]\lambda_0^2+O(\lambda_0^3)\\
&\qquad\qquad\qquad\qquad\qquad\qquad\qquad\quad=\frac{\sqrt{6}}{8}\lambda_0-\Bigg(\frac{720+211\sqrt{6}}{768}\Bigg)\lambda_0^2+O(\lambda_0^3).
\end{split}
\end{equation}
Here we have substituted $x_{0f}$ in the expression of $h_0$ and $f_{0p}$ by its leading order value in the $\lambda_0$ expansion:
\beq
x_{0f}=\frac{21}{4}+O(\lambda_0). 
\eeq
Note that the quantities $f_{0m}$, $\zeta_0$ are  constrained to lie on a circle which is  given by
 \begin{equation}
\begin{split}
(\zeta_0^2+f_{0m}^2)&=\Big(\frac{18\sqrt{6}-39}{256}\Big)\lambda_0^2-\Bigg[\frac{-5784+5607\sqrt{6}+8(51\sqrt{6}-172)x_{0f}}{12288}\Bigg]\lambda_0^3+O(\lambda_0^4)\\
&=\Big(\frac{18\sqrt{6}-39}{256}\Big)\lambda_0^2-\Bigg[\frac{2583\sqrt{6}-4336}{4096}\Bigg]\lambda_0^3+O(\lambda_0^4).
\end{split}
\label{fixed line  zeroth order}
\end{equation}
We will now check whether the circle of fixed points satisfying the above constraints survives when the finite N corrections are taken into account. In particular, we will focus on possible fixed points where the two sectors are coupled, i.e., $\zeta\neq 0$.

\subsection{Finite $N$ corrections to the gauge couplings}
First, let us look at the finite N corrections to the values of the gauge couplings. From the form of the planar beta functions of the gauge couplings, we see that  $\lambda_0$ satisfies an equation of the following form at any fixed point:
\begin{equation}
\begin{split}
-\lambda_0^2\Bigg(\frac{21-4x_{0f}}{6}\Bigg)-\lambda_0^3\Bigg(\frac{27-13x_{0f}}{6}\Bigg)+A_0 \lambda_0^4+O(\lambda_0^5)=0.
\end{split}
\end{equation}
Here, we have included a contribution of the 3-loop diagrams to the planar beta functions of the gauge couplings since it is necessary for determining the $O(1/N)$ and $O(1/N^2)$ terms in the values of the different couplings at the fixed points.\footnote{There may be terms involving the coupling $h_0$ in such 3-loop contributions to the planar beta functions. As we have already seen, the coupling $h_0=(\frac{3-\sqrt{6}}{16})\lambda_0+O(\lambda_0)$ at any fixed point. One can substitute this relation between $h_0$ and $\lambda_0$ to express the quartic terms in the planar beta function of $\lambda_0$ as $A_0\lambda_0^4$.} Although we have not evaluated the  coefficient $A_0$, we will see that most of the important conclusions in this appendix do not rely on its exact value. 

The value of $\lambda_0$ in the above expression is determined by $x_{0f}$. One can invert this relation to express $x_{0f}$ as follows:
\begin{equation}
\begin{split}
x_{0f}=\frac{21}{4}-\frac{165}{16}\lambda_0+\Big(\frac{2145-96 A_0}{64}\Big)\lambda_0^2+O(\lambda_0^3).
\end{split}
\end{equation}

Now, we will determine the values of the $O(1/N)$ and $O(1/N^2)$ corrections to the gauge couplings. As we will see, these corrections take the form of perturbative expansions in $\lambda_0$ where the leading order terms are independent of $\lambda_0$. For the $O(1/N^2)$ contributions to the couplings, we will discuss only these leading order terms. However, for the $O(1/N)$ contributions, we will need to take into account the first subleading terms in the $\lambda_0$-expansion as they play a role in determining the leading order terms in the $O(1/N^2)$ corrections. Moreover, we will allow the quantities $x_{fi}^{(n)}$ to have a perturbative expansion in $\lambda_0$ as shown below:
\beq
x_{fi}^{(n)}=(x_{fi}^{(n)})_0+\lambda_0 (x_{fi}^{(n)})_1+O(\lambda_0^2).
\eeq
We will check if the coefficients in this expansion can be fine-tuned such that the degeneracy in the fixed points survives at a given order.

Having provided a general outline of our strategy, let us now turn to the analysis of the $O(1/N)$ and $O(1/N^2)$ corrections to the gauge couplings at the fixed points.

\subsubsection{$O(1/N)$ terms in the gauge couplings}

To study the $O(1/N)$ corrections to the gauge couplings, we will assume that the leading order term in $\lambda_i^{(1)}$ is independent of $\lambda_0$. We will soon provide a consistency check for this assumption.  Keeping this assumption in mind, one can get the following perturbative  expansion for the beta function of $\lambda_i^{(1)}$:
\begin{equation}
\begin{split}
\beta_{\lambda_i^{(1)}}=&-\lambda_0\lambda_i^{(1)}\Bigg(\frac{21-4x_{0f}}{3}\Bigg)+\lambda_0^2\Bigg(\frac{2 x_{fi}^{(1)}}{3}+\frac{22}{3N_{ci}}\Bigg)-\lambda_0^2\lambda_i^{(1)}\Bigg(\frac{27-13x_{0f}}{2} \Bigg)\\
&+\lambda_0^3\Bigg(\frac{13  x_{fi}^{(1)}}{6} +\frac{29}{24 N_{ci}}\Bigg)+4A_0 \lambda_0^3 \lambda_i^{(1)}+O(\lambda_0^4/N).
\end{split}
\end{equation}
While writing the second last term in the above expansion, we have substituted $x_{0f}$ by $\frac{21}{4}$ as all higher order terms in $x_{0f}$ lead to contributions at $O(\lambda_0^4/N)$ or a higher order in $\beta_{\lambda_i^{(1)}}$. 

Now, setting $\beta_{\lambda_i^{(1)}}=0$, we get the following value of $\lambda_i^{(1)}$ at the fixed point:
\begin{equation}
\begin{split}
\lambda_i^{(1)}=&-\frac{16\Big(11+N_{ci}(x_{fi}^{(1)})_0\Big)}{165 N_{ci}}\\
&+\Bigg[\frac{-55\Bigg(601+8 N_{ci}\Big(13 (x_{fi}^{(1)})_0+2(x_{fi}^{(1)})_1\Big)\Bigg)+256 A_0\Big(11+N_{ci} (x_{fi}^{(1)})_0\Big)}{9075 N_{ci}}\Bigg]\lambda_0\\
&+O(\lambda_0^2/N).
\end{split}
\end{equation}
Note that the leading order term in $\lambda_i^{(1)}$ is independent of $\lambda_0$ as mentioned earlier. We will see that this feature is shared by the leading order terms in the $O(1/N)$ and $O(1/N^2)$ corrections to all the couplings.

\subsubsection{$O(1/N^2)$ terms in the gauge couplings}
The beta functions of the $O(1/N^2)$ correction to the gauge couplings can be similarly expanded as follows:
\begin{equation}
\begin{split}
\beta_{\lambda_i^{(2)}}=&-\Big(2\lambda_0\lambda_i^{(2)}+\lambda_i^{(1)2}\Big)\Bigg(\frac{21-4x_{0f}}{6}\Bigg)+\frac{4}{3}\lambda_0\lambda_i^{(1)}\Bigg(x_{fi}^{(1)}+\frac{11}{N_{ci}}\Bigg)+\frac{2}{3}\lambda_0^2 x_{fi}^{(2)}\\
&-\Big(\lambda_0^2 \lambda_i^{(2)}+\lambda_0\lambda_i^{(1)2}\Big)\Bigg(\frac{27-13x_{0f}}{2}\Bigg)+\lambda_0^2\lambda_i^{(1)}\Bigg(\frac{13x_{fi}^{(1)}}{2} +\frac{29}{8N_{ci}} \Bigg)\\
&+6A_0\lambda_0^2\lambda_i^{(1)2}+O(\lambda_0^3/N^2).
\end{split}
\end{equation}
At the fixed point, we have
\begin{equation}
\begin{split}
\lambda_i^{(2)}=&-\frac{16}{1497375 N_{ci}^2}\Bigg[128A_0\Big(11+N_{ci}(x_{fi}^{(1)})_0\Big)^2\\
&\qquad\qquad\qquad\qquad-55\Bigg(319+601 N_{ci} (x_{fi}^{(1)})_0+N_{ci}^2\Big(52 (x_{fi}^{(1)})_0^2-165 x_{fi}^{(2)}\Big)\Bigg)\Bigg]\\
&+O(\lambda_0/N^2).
\end{split}
\end{equation}

From the expressions of $\lambda_i^{(1)}$ and $\lambda_i^{(2)}$ one can  extract the expressions of $\lambda_{p,m}^{(1)}\equiv \frac{\lambda_1^{(1)}\pm\lambda_2^{(1)}}{2}$ and $\lambda_{p,m}^{(2)}\equiv \frac{\lambda_1^{(2)}\pm\lambda_2^{(2)}}{2}$. These will be useful in determining the finite $N$ corrections to the quartic couplings as we will discuss below.
\subsection{Finite $N$ corrections to the quartic couplings}
The analysis of the finite $N$ corrections to the quartic couplings  at the fixed points is quite similar to the one for the gauge couplings presented above.  We will first discuss the $O(1/N)$ corrections and show that by tuning $x_{fi}^{(1)}$ and $x_{f2}^{(1)}$ appropriately, the degeneracy in the fixed points can be preserved up to the first subleading order in $\lambda_0$. We will then look at the $O(1/N^2)$ corrections. There we will again see that the degeneracy in the fixed points can be preserved up to the leading order in $\lambda_0$ by appropriately tuning the values of $x_{f1}^{(2)}$ and $x_{f2}^{(2)}$.
 
\subsubsection{$O(1/N)$  terms in the quartic couplings}
The beta functions of the $O(1/N)$ corrections to the quartic couplings have the following forms:

 \begin{equation}
\begin{split}
\beta_{h_p^{(1)}}=&(32 h_0-6 \lambda_0) h_p^{(1)}-\Bigg((6 h_0-\frac{3}{8}\lambda_0) \lambda_p^{(1)}-(16 h_0^2+3 h_0 \lambda_0-\frac{3\lambda_0^2}{8})\Big(\frac{1}{N_{c1}}+\frac{1}{N_{c2}}\Big)\Bigg)\\
&+\Big(-288 h_0^2+80\lambda_0 h_0+\frac{9}{4}\lambda_0^2\Big)h_p^{(1)} +\Big(40 h_0^2+\frac{9}{2} h_0\lambda_0-\frac{15}{8}\lambda_0^2\Big)\lambda_p^{(1)}+O(\lambda_0^3/N),
\end{split}
\end{equation}

\begin{equation}
\begin{split}
\beta_{h_m^{(1)}}
=&(32 h_0-6 \lambda_0) h_m^{(1)}-\Bigg((6 h_0-\frac{3}{8}\lambda_0) \lambda_m^{(1)}-(16 h_0^2+3 h_0 \lambda_0-\frac{3\lambda_0^2}{8})\Big(\frac{1}{N_{c1}}-\frac{1}{N_{c2}}\Big)\Bigg)\\
&+\Big(-288 h_0^2+80\lambda_0 h_0+\frac{9}{4}\lambda_0^2\Big)h_m^{(1)} +\Big(40 h_0^2+\frac{9}{2} h_0\lambda_0-\frac{15}{8}\lambda_0^2\Big)\lambda_m^{(1)}+O(\lambda_0^3/N),
\end{split}
\end{equation}

\begin{equation}
\begin{split}
\beta_{\zeta^{(1)}}=& \zeta_0\Bigg[16 f_p^{(1)}+32 h_p^{(1)}-6\lambda_p^{(1)}+(8h_0+3\lambda_0)\Big(\frac{1}{N_{c1}}+\frac{1}{N_{c2}}\Big)\\
&\quad+\Big(128\lambda_0-320 h_0\Big) h_p^{(1)}+\Big(64f_{0p}+128 h_0+\frac{21}{2}\lambda_0\Big) \lambda_p^{(1)}+64\lambda_0 f_p^{(1)}+64 f_{0m}\lambda_m^{(1)}\Bigg]\\
&+O(\lambda_0^3/N),
\end{split}
\end{equation}

\begin{equation}
\begin{split}
\beta_{f_p^{(1)}}=& 16(1+4\lambda_0) (\zeta_0\zeta^{(1)}+f_{0m} f_m^{(1)})+
(48 h_0+32 f_{0p})  h_p^{(1)}+
\Big(\frac{9\lambda_0}{8}-6 f_{0p}\Big)\lambda_p^{(1)}\\
&+ (16 f_{0p}+32h_{0}- 6\lambda_0) f_p^{(1)}+  f_{0m} (32 h_m^{(1)}-6 \lambda_m^{(1)})\\
&+(8h_0+3\lambda_0)\Big(\frac{f_{0p}+f_{0m}}{N_{c1}}+\frac{f_{0p}-f_{0m}}{N_{c2}}\Big)+32 (f_{0m}^2+\zeta_0^2)\lambda_p^{(1)}\\
&+\Big(-1152 h_0^2-320  h_0 f_{0p}+192\lambda_0 h_0+9\lambda_0^2+128\lambda_0 f_{0p}\Big) h_p^{(1)}\\
&+\Big(96 h_0^2+18 h_0\lambda_0 -\frac{45}{8}\lambda_0^2+128 h_0 f_{0p}+\frac{21}{2}\lambda_0 f_{0p}+32 f_{0p}^2\Big)\lambda_p^{(1)}\\
&+\Big(-160 h_0^2+128\lambda_0 h_0+\frac{21}{4}\lambda_0^2+ 64\lambda_0 f_{0p}\Big) f_p^{(1)} \\
&+f_{0m}\Bigg(-320  h_0 h_m^{(1)}+128\lambda_0 h_m^{(1)}+128 h_0 \lambda_m^{(1)}+64 f_{0p} \lambda_m^{(1)}+\frac{21}{2}\lambda_0 \lambda_m^{(1)}\Bigg)+O(\lambda_0^3/N),
 \end{split}
\end{equation}

\begin{equation}
\begin{split}
\beta_{f_m^{(1)}}=& (48 h_0+32 f_{0p})  h_m^{(1)}+
\Big(\frac{9\lambda_0}{8}-6 f_{0p}\Big)\lambda_m^{(1)}+ (16 f_{0p}+32h_{0}- 6\lambda_0) f_m^{(1)}\\
&+  f_{0m} (16 f_p^{(1)}+32 h_p^{(1)}-6 \lambda_p^{(1)})+(8h_0+3\lambda_0)\Big(\frac{f_{0p}+f_{0m}}{N_{c1}}-\frac{f_{0p}-f_{0m}}{N_{c2}}\Big)\\
&+\Big(-1152 h_0^2-320 h_0 f_{0p}+192\lambda_0 h_0+128\lambda_0 f_{0p}+9\lambda_0^2 \Big) h_m^{(1)}\\
&+\Bigg(96 h_0^2+128 h_0  f_{0p}+18 h_0\lambda_0+\frac{21}{2}\lambda_0 f_{0p}-\frac{45}{8}\lambda_0^2+32\Big(f_{0p}^2+f_{0m}^2-\zeta_0^2\Big)\Bigg)\lambda_m^{(1)}\\
&+ f_{0m}\Big(-320 h_0 h_p^{(1)} +128\lambda_0  h_p^{(1)} +128 h_0\lambda_p^{(1)}+64 f_{0p}\lambda_p^{(1)}+64\lambda_0 f_p^{(1)}+\frac{21}{2}\lambda_0 \lambda_p^{(1)}\Big)\\
&+\Big(-160 h_0^2+128\lambda_0 h_0+ 64\lambda_0 f_{0p} +\frac{21}{4}\lambda_0^2\Big) f_m^{(1)}+O(\lambda_0^3/N).
 \end{split}
\end{equation}
In the above expression of $\beta_\zeta^{(1)}$, we have ignored terms of the form $\zeta^{(1)}\Big[\cdots\Big]$ where the dots stand for a coefficient which is not suppressed by 1/N.  This coefficient is identical to the coefficient of $\zeta_0$ in $\beta_{\zeta_0}$, and hence is set equal to zero at the fixed point. 

Now, setting  $\beta_{h_p^{(1)}}$ and $\beta_{h_m^{(1)}}$ to zero, we get the following expressions for the $O(1/N)$ corrections to $h_p$ and $h_m$:

\begin{equation}
\begin{split}
h_p^{(1)}=&\frac{(-3+\sqrt{6})}{330 N_{c1}N_{c2}}\Bigg[11(N_{c1}+N_{c2})+N_{c1}N_{c2}\Big((x_{f1}^{(1)})_0+(x_{f2}^{(1)})_0\Big)\Bigg]\\
&+\frac{\lambda_0}{871200 N_{c1}N_{c2}}\Bigg[\Big(55(-7290+2947\sqrt{6})-8448(-3+\sqrt{6})A_0\Big)(N_{c1}+N_{c2})\\
&\qquad\qquad\quad+N_{c1}N_{c2}\Bigg\{\Big(55(-702+281\sqrt{6})-768(-3+\sqrt{6})A_0\Big)\Big((x_{f1}^{(1)})_0+(x_{f2}^{(1)})_0\Big)\\
&\qquad\qquad\qquad\qquad\qquad+2640(-3+\sqrt{6})\Big((x_{f1}^{(1)})_1+(x_{f2}^{(1)})_1\Big)\Bigg\}\Bigg]+O(\lambda_0^2/N),
 \end{split}
\end{equation}

\begin{equation}
\begin{split}
h_m^{(1)}=&\frac{(-3+\sqrt{6})}{330 N_{c1}N_{c2}}\Bigg[-11(N_{c1}-N_{c2})+N_{c1}N_{c2}\Big((x_{f1}^{(1)})_0-(x_{f2}^{(1)})_0\Big)\Bigg]\\
&+\frac{\lambda_0}{871200 N_{c1}N_{c2}}\Bigg[-\Big(55(-7290+2947\sqrt{6})-8448(-3+\sqrt{6})A_0\Big)(N_{c1}-N_{c2})\\
&\qquad\qquad\quad+N_{c1}N_{c2}\Bigg\{\Big(55(-702+281\sqrt{6})-768(-3+\sqrt{6})A_0\Big)\Big((x_{f1}^{(1)})_0-(x_{f2}^{(1)})_0\Big)\\
&\qquad\qquad\qquad\qquad\qquad+2640(-3+\sqrt{6})\Big((x_{f1}^{(1)})_1-(x_{f2}^{(1)})_1\Big)\Bigg\}\Bigg]+O(\lambda_0^2/N),
 \end{split}
\end{equation}

Substituting these values of $h_p^{(1)}$ and $h_m^{(1)}$ in the expression of $\beta_{\zeta^{(1)}}$ and then setting $\beta_{\zeta^{(1)}}=0$, we get the following value of $f_p^{(1)}$:
\begin{equation}
\begin{split}
f_p^{(1)}=&-\frac{2}{55\sqrt{6}N_{c1}N_{c2}}\Bigg[11(N_{c1}+N_{c2})+N_{c1}N_{c2}\Big((x_{f1}^{(1)})_0+(x_{f2}^{(1)})_0\Big)\Bigg]\\
&+\frac{\lambda_0}{871200 N_{c1}N_{c2}}\Bigg[11(101475-7195\sqrt{6}+1536\sqrt{6}A_0)(N_{c1}+N_{c2})\\
&\qquad\qquad +N_{c1}N_{c2}\Bigg\{\Big(110(720-101\sqrt{6})+1536\sqrt{6}A_0\Big)\Big((x_{f1}^{(1)})_0+(x_{f2}^{(1)})_0\Big)\\
&\qquad\qquad\qquad\qquad-5280\sqrt{6}\Big((x_{f1}^{(1)})_1+(x_{f2}^{(1)})_1\Big)\Bigg\}\Bigg]\\
&+\frac{32f_{0m}}{165N_{c1}N_{c2}}\Bigg[-11(N_{c1}-N_{c2})+N_{c1}N_{c2}\Big((x_{f1}^{(1)})_0-(x_{f2}^{(1)})_0\Big)\Bigg]+O(\lambda_0^2/N).
 \end{split}
\end{equation}

Now substituting the values of $h_p^{(1)}$, $h_m^{(1)}$  and $f_p^{(1)}$  in the expression of $\beta_{f_p^{(1)}}$, and then setting $\beta_{f_p^{(1)}}=0$, we get
\begin{equation}
\begin{split}
&2(1+4\lambda_0)\Big(\zeta_0\zeta^{(1)}+f_{0m}f_{0m}^{(1)}\Big)\\
&=\frac{(13-6\sqrt{6})\lambda_0}{880 N_{c1}N_{c2}}\Bigg[11(N_{c1}+N_{c2})+N_{c1} N_{c2}\Big((x_{f1}^{(1)})_0+(x_{f2}^{(1)})_0\Big)\Bigg]\\
&\quad+\frac{4f_{0m}}{55\sqrt{6}N_{c1}N_{c2}}\Bigg[11(N_{c1}-N_{c2})-N_{c1}N_{c2}\Big((x_{f1}^{(1)})_0-(x_{f2}^{(1)})_0\Big)\Bigg]\\
%%%%%%%%%%%%%%%%%%%%%%%%%%%%%%%%%%%%%%%%%%
&\quad+\frac{\lambda_0^2}{1548800 N_{c1}N_{c2}}\Bigg[11\Big(5(8793\sqrt{6}-5516)+512(6\sqrt{6}-13)A_0\Big)(N_{c1}+N_{c2})\\
&\qquad\qquad\quad+N_{c1}N_{c2}\Bigg\{\Big(55(32+567\sqrt{6})+512(6\sqrt{6}-13)A_0\Big)\Big((x_{f1}^{(1)})_0+(x_{f2}^{(1)})_0\Big)\\
&\qquad\qquad\qquad\qquad\qquad+55(416-192\sqrt{6})\Bigg\}\Big((x_{f1}^{(1)})_1+(x_{f2}^{(1)})_1\Big)\Bigg]\\
&\quad+\frac{f_{0m}\lambda_0}{435600 N_{c1}N_{c2}}\Bigg[11(101475-28315\sqrt{6}+1536\sqrt{6}A_0)(N_{c2}-N_{c1})\\
&\qquad\qquad\qquad\qquad +N_{c1}N_{c2}\Bigg\{(79200-32230\sqrt{6}+1536\sqrt{6}A_0)\Big((x_{f1}^{(1)})_0-(x_{f2}^{(1)})_0\Big)\\
&\qquad\qquad\qquad\qquad\qquad\qquad\quad-5280\sqrt{6}\Big((x_{f1}^{(1)})_1-(x_{f2}^{(1)})_1\Big)\Bigg\}\Bigg]+O(\lambda_0^3/N).
 \end{split}
\end{equation}
 The above expression gives the $O(1/N)$ correction to the location of the fixed points along the radial direction of fixed circle in the $N\rightarrow\infty$ limit.
 
 Finally, setting $\beta_{f_m^{(1)}}=0$, we get the following two constraints on the values of $(x_{f_i}^{(1)})_0$ and $(x_{f_i}^{(1)})_1$:
\begin{equation}
\begin{split}
&(x_{f_1}^{(1)})_0-(x_{f_2}^{(1)})_0=11\Big(\frac{1}{N_{c2}}-\frac{1}{N_{c1}}\Big),\\
&(x_{f_1}^{(1)})_1-(x_{f_2}^{(1)})_1=-\frac{9}{376}(437-55\sqrt{6})\Big(\frac{1}{N_{c2}}-\frac{1}{N_{c1}}\Big).
\end{split}
\end{equation}
The important point to note here is that after substituting the expressions of all the quantities evaluated earlier, the coefficients of $f_{0m}$ and $f_m^{(1)}$ in $\beta_{f_m^{(1)}}$ vanish and one is left with a quantity that vanishes under the above constraints. This means that when the values of $(x_{f_i}^{(1)})_0$ and $(x_{f_i}^{(1)})_1$ are tuned to satisfy these constraints, no additional condition is imposed on the coupling $f_m$ and the degeneracy in the fixed points survives up to $O(\lambda_0/N)$ corrections to the couplings. Now, substituting these relations between  $x_{f_1}^{(1)}$ and $x_{f_2}^{(1)}$ in the expressions of $f_p^{(1)}$ and $2(1+4\lambda_0)\Big(\zeta_0\zeta^{(1)}+f_{0m}f_m^{(1)}\Big)$, we get 

\begin{equation}
\begin{split}
f_p^{(1)}=&-\frac{4}{55\sqrt{6} N_{c2}}\Big(11+N_{c2}(x_{f2}^{(1)})_0\Big)\\
&+\frac{\lambda_0}{40946400 N_{c1} N_{c2}}\\
&\qquad\Bigg[81675(165-7\sqrt{6})N_{c2}+11\Big(5(1662705-124871\sqrt{6})+144384\sqrt{6}A_0\Big)N_{c1}\\
&\qquad+188N_{c1}N_{c2}\Bigg\{(39600-5555\sqrt{6}+768\sqrt{6}A_0)(x_{f2}^{(1)})_0-2640\sqrt{6}(x_{f2}^{(1)})_1\Bigg\}\Bigg]\\
&+O(\lambda_0^2/N),
\end{split}
\end{equation}

\begin{equation}
\begin{split}
&2(1+4\lambda_0)\Big(\zeta_0\zeta^{(1)}+f_{0m}f_{0m}^{(1)}\Big)\\
&=\frac{(13-6\sqrt{6})\lambda_0}{440 N_{c2}}\Big(11+N_{c2} (x_{f2}^{(1)})_0\Big)\\
&\quad+\frac{\lambda_0^2}{774400 N_{c2}}\Bigg[11\Big(5(8793\sqrt{6}-5516)+512(6\sqrt{6}-13)A_0\Big) \\
&\qquad\qquad\qquad\quad+\Big(55(567\sqrt{6}+32)+512(6\sqrt{6}-13)A_0\Big)N_{c2}(x_{f2}^{(1)})_0\\
&\qquad\qquad\qquad\quad-1760(6\sqrt{6}-13)N_{c2}(x_{f2}^{(1)})_1\Bigg]-\frac{3(165-7\sqrt{6}) f_{0m}\lambda_0}{752 N_{c1} N_{c2}}(N_{c1}-N_{c2})\\
&\quad+O(\lambda_0^3/N).
 \end{split}
\end{equation}
The fact that $h_p^{(1)}$, $h_m^{(1)}$ and $f_p^{(1)}$ are independent of $f_{0m}$ implies that even after taking the $O(1/N)$ corrections into account, the closed curve of fixed points still lies on a plane. When $N_{c1}\neq N_{c2}$, the dependence of the quantity $2(1+4\lambda_0)(f_{0m} f_m^{(1)} +\zeta_0\zeta^{(1)})$ on $f_{0m}$  indicates a deformation of the shape of the closed curve away from its circular form in the planar limit. When the two ranks are equal, i.e., $N_{c1}=N_{c2}$,  the quantity $2(1+4\lambda_0)(f_{0m} f_m^{(1)} +\zeta_0\zeta^{(1)})$ is independent of $f_{0m}$ which means that  the closed curve of fixed points remains a circle up to this order.

To explore whether the above features survive at higher orders in $\lambda_0$, one needs to evaluate the contributions of the higher loop diagrams  to the beta functions. For now, we will assume that the degeneracy in the fixed points survives at $O(1/N)$ up to all orders in $\lambda_0$ and check whether the leading order contributions at $O(1/N^2)$ lift the degeneracy.

\subsubsection{$O(1/N^2)$  terms in the quartic couplings}

The beta functions of the $O(1/N^2)$ corrections to the quartic couplings have the following forms:
\begin{equation}
\begin{split}
\beta_{h_p^{(2)}}=& (32 h_0-6  \lambda_0 ) h_p^{(2)}+\Big(\frac{3}{8}\lambda_0-6 h_0\Big)\lambda_p^{(2)}+16 (h_p^{(1)2}+h_m^{(1)2})+\frac{3}{16}(\lambda_p^{(1)2}+\lambda_m^{(1)2})\\
&-6(\lambda_p^{(1)} h_p^{(1)}+\lambda_m^{(1)} h_m^{(1)})+(3 \lambda_0+32 h_0)\Big(\frac{h_p^{(1)}+h_m^{(1)}}{N_{c1}}+\frac{h_p^{(1)}-h_m^{(1)}}{N_{c2}}\Big)\\
&+(3 h_0 -\frac{3 \lambda_0}{4})\Big(\frac{\lambda_p^{(1)}+\lambda_m^{(1)}}{N_{c1}}+\frac{\lambda_p^{(1)}-\lambda_m^{(1)}}{N_{c2}}\Big)+(40 \lambda_0-288 h_0) \Big(h_p^{(1)2}+h_m^{(1)2}\Big)\\
&+\Big(\frac{9}{4}h_0-\frac{15}{8}\lambda_0\Big)\Big(\lambda_p^{(1)2}+\lambda_m^{(1)2}\Big)+\Big(\frac{9}{2}\lambda_0+80 h_0\Big) \Big(\lambda_p^{(1)} h_p^{(1)}+\lambda_m^{(1)} h_m^{(1)}\Big)+O(\lambda_0^2/N^2),
\end{split}
\end{equation}

\begin{equation}
\begin{split}
\beta_{h_m^{(2)}}=& (32 h_0-6  \lambda_0 ) h_m^{(2)}+\Big(\frac{3}{8}\lambda_0-6 h_0\Big)\lambda_m^{(2)}+32 h_p^{(1)}h_m^{(1)}+\frac{3}{8}\lambda_p^{(1)}\lambda_m^{(1)}\\
&-6\Big(\lambda_p^{(1)} h_m^{(1)}+\lambda_m^{(1)} h_p^{(1)}\Big)+(3 \lambda_0+32 h_0)\Big(\frac{h_p^{(1)}+h_m^{(1)}}{N_{c1}}-\frac{h_p^{(1)}-h_m^{(1)}}{N_{c2}}\Big)\\
&+(3 h_0 -\frac{3 \lambda_0}{4})\Big(\frac{\lambda_p^{(1)}+\lambda_m^{(1)}}{N_{c1}}-\frac{\lambda_p^{(1)}-\lambda_m^{(1)}}{N_{c2}}\Big)+(80 \lambda_0-576 h_0) h_p^{(1)}h_m^{(1)}\\
&+\Big(\frac{9}{2}h_0-\frac{15}{4}\lambda_0\Big)\lambda_p^{(1)}\lambda_m^{(1)}+\Big(\frac{9}{2}\lambda_0+80 h_0\Big) \Big(\lambda_p^{(1)} h_m^{(1)}+\lambda_m^{(1)} h_p^{(1)}\Big)+O(\lambda_0^2/N^2),
\end{split}
\end{equation}

\begin{equation}
\begin{split}
\beta_{\zeta^{(2)}}=& \zeta_0\Bigg[16 f_p^{(2)}+32 h_p^{(2)}-6\lambda_p^{(2)}\\
&\quad+\Big(\frac{8h_p^{(1)}+3\lambda_p^{(1)}+8h_m^{(1)}+3\lambda_m^{(1)}}{N_{c1}}+\frac{8h_p^{(1)}+3\lambda_p^{(1)}-8h_m^{(1)}-3\lambda_m^{(1)}}{N_{c2}}\Big)\\
&\quad-160\Big(h_p^{(1)2}+h_m^{(1)2}\Big)+\frac{21}{4}\Big(\lambda_p^{(1)2}+\lambda_m^{(1)2}\Big)+128 \Big(\lambda_p^{(1)} h_p^{(1)}+\lambda_m^{(1)} h_m^{(1)}\Big)\\
&\quad+64\Big(\lambda_p^{(1)}f_p^{(1)}+\lambda_m^{(1)}f_m^{(1)}\Big)\Bigg]+O(\lambda_0^2/N^2),\\
\end{split}
\end{equation}

\begin{equation}
\begin{split}
\beta_{f_p^{(2)}}=&  16 \Big(\zeta_0\zeta^{(2)}+f_{0m}f_m^{(2)}\Big)+8(1+4\lambda_0)\Big(f_m^{(1)2}+\zeta^{(1)2}\Big)+64 \lambda_p^{(1)}\Big(\zeta_0\zeta^{(1)}+f_{0m} f_m^{(1)}\Big)\\
&+\Big(16 f_{0p}+32 h_0-6\lambda_0\Big) f_p^{(2)}+\Big(32 f_{0p}+48 h_0\Big) h_p^{(2)}+\Big(\frac{9}{8}\lambda_0-6 f_{0p}\Big)\lambda_p^{(2)}\\
&+24 \Big(h_p^{(1)2}+h_m^{(1)2}\Big)+\frac{9}{16}\Big(\lambda_p^{(1)2}+\lambda_m^{(1)2}\Big)+ f_{0m}\Big(32 h_m^{(2)}-6 \lambda_m^{(2)}\Big)\\
&+ \Big(32 h_p^{(1)}-6\lambda_p^{(1)}\Big) f_p^{(1)}+ \Big(32 h_m^{(1)}-6\lambda_m^{(1)}\Big) f_m^{(1)}+8 f_p^{(1)2}\\
&+\Big( \frac{(8 h_0+3 \lambda_0)(f_p^{(1)}+f_m^{(1)})+ (f_{0p}+f_{0m}) (8 h_p^{(1)}+3\lambda_p^{(1)}+8 h_m^{(1)}+3\lambda_m^{(1)})}{N_{c1}}\\
&\qquad+\frac{(8 h_0+3 \lambda_0)(f_p^{(1)}-f_m^{(1)})+ (f_{0p}-f_{0m}) (8 h_p^{(1)}+3\lambda_p^{(1)}-8 h_m^{(1)}-3\lambda_m^{(1)})}{N_{c2}}\Big)\\
&+\Big(9 h_0-\frac{45}{8}\lambda_0+\frac{21}{4} f_{0p}\Big)\Big(\lambda_p^{(1)2}+\lambda_m^{(1)2}\Big)+(96\lambda_0-1152 h_0-160 f_{0p}) \Big(h_p^{(1)2}+h_m^{(1)2}\Big)\\
&+2(96 h_0+9\lambda_0+64 f_{0p})\Big(\lambda_p^{(1)} h_p^{(1)}+\lambda_m^{(1)} h_m^{(1)}\Big)+64(2\lambda_0-5 h_0)\Big( h_p^{(1)} f_p^{(1)}+h_m^{(1)} f_m^{(1)}\Big)\\
&+2\Bigg(64 h_0+\frac{21}{4}\lambda_0+32 f_{0p}\Bigg)\Big( \lambda_p^{(1)}  f_p^{(1)}+ \lambda_m^{(1)}  f_m^{(1)}\Big)+32\lambda_0 f_p^{(1)2} \\
&+f_{0m}\Big(-320 h_p^{(1)} h_m^{(1)}+64\lambda_m^{(1)} f_p^{(1)}+128\lambda_p^{(1)} h_m^{(1)}+128 \lambda_m^{(1)} h_p^{(1)}+\frac{21}{2} \lambda_p^{(1)}\lambda_m^{(1)}\Big)\\
&+O(\lambda_0^2/N^2),
 \end{split}
\end{equation}

\begin{equation}
\begin{split}
\beta_{f_m^{(2)}}=&  \Big(16 f_{0p}+32 h_0-6\lambda_0\Big) f_m^{(2)}+ f_{0m}\Big(16 f_p^{(2)}+32 h_p^{(2)}-6 \lambda_p^{(2)}\Big)\\
&+\Big(32 f_{0p}+48 h_0\Big) h_m^{(2)}+\Big(\frac{9}{8}\lambda_0-6 f_{0p}\Big)\lambda_m^{(2)}+\Big(16 f_p^{(1)}+32h_p^{(1)}-6\lambda_p^{(1)} \Big) f_m^{(1)}\\
&+\Big(32 f_p^{(1)}+48 h_p^{(1)}\Big)h_m^{(1)}+\Big(\frac{9}{8}\lambda_p^{(1)}-6 f_p^{(1)}\Big)\lambda_m^{(1)}\\
&+\Big( \frac{(8 h_0+3 \lambda_0)(f_p^{(1)}+f_m^{(1)})+ (f_{0p}+f_{0m}) (8 h_p^{(1)}+3\lambda_p^{(1)}+8 h_m^{(1)}+3\lambda_m^{(1)})}{N_{c1}}\\
&\qquad-\frac{(8 h_0+3 \lambda_0)(f_p^{(1)}-f_m^{(1)})+ (f_{0p}-f_{0m}) (8 h_p^{(1)}+3\lambda_p^{(1)}-8 h_m^{(1)}-3\lambda_m^{(1)})}{N_{c2}}\Big)\\
&+4(48\lambda_0-576 h_0-80 f_{0p}) h_p^{(1)}h_m^{(1)}+\Big(18 h_0-\frac{45}{4}\lambda_0+\frac{21}{2}f_{0p}\Big)\lambda_p^{(1)}\lambda_m^{(1)}\\
&+2(64\lambda_0-160 h_0)\Big( h_p^{(1)} f_m^{(1)}+h_m^{(1)} f_p^{(1)}\Big)\\
&+2(96 h_0+9\lambda_0+64 f_{0p})\Big(\lambda_p^{(1)} h_m^{(1)}+\lambda_m^{(1)} h_p^{(1)}\Big)\\
&+2\Bigg(64 h_0+\frac{21}{4}\lambda_0+32 f_{0p}\Bigg)\Big( \lambda_p^{(1)}  f_m^{(1)}+\lambda_m^{(1)}  f_p^{(1)}\Big)\\
&+64\lambda_0 f_p^{(1)}f_m^{(1)}+64f_{0m}\Big(f_p^{(1)}\lambda_p^{(1)}+f_m^{(1)}\lambda_m^{(1)}\Big)-64\zeta_0\zeta^{(1)}\lambda_m^{(1)}\\
&+ f_{0m}\Bigg[128\Big( \lambda_p^{(1)} h_p^{(1)}+ \lambda_m^{(1)} h_m^{(1)}\Big) -160 \Big( h_p^{(1)2}+ h_m^{(1)2}\Big)+\frac{21}{4} \Big(\lambda_p^{(1)2}+\lambda_m^{(1)2}\Big)\Bigg]\\
&+O(\lambda_0^2/N^2).
 \end{split}
\end{equation}
In the above expression of $\beta_{\zeta^{(2)}}$, we have ignored terms of the forms $\zeta^{(2)}\Big[\cdots\Big]$ and $\zeta^{(1)}\Big[\cdots\Big]$ because the dots correspond to coefficients of $\zeta_0$ in $\beta_{\zeta_0}$ and $\beta_{\zeta^{(1)}}$ respectively, and these coefficients have already been set to zero at the fixed point.

As earlier, by setting $\beta_{h_p^{(2)}}$, $\beta_{h_m^{(2)}}$ and $\beta_{\zeta^{(2)}}$ to zero, one can determine the values of $h_p^{(2)}$, $h_m^{(2)}$ and $f_p^{(2)}$. Then by setting $\beta_{f_p^{(2)}}$ to zero, one can find  the $O(1/N^2)$ correction to the quantity $(1+4\lambda_p)(\zeta^2+f_m^2)$. We will give the forms of these corrections shortly. But first let us discuss the constraint following from setting $\beta_{f_m^{(2)}}$ to zero. Just like the $O(1/N)$ corrections, the coefficients of $f_{0m}$, $f_m^{(1)}$ and $f_m^{(2)}$ in $\beta_{f_m^{(2)}}$ vanish, and no additional condition is imposed on $f_m$ up to the leading order term in the $\lambda_0$-expansion when $\beta_{f_m^{(2)}}=0$. In fact, this equation leads to a relation between the quantities $x_{f1}^{(2)}$ and $x_{f2}^{(2)}$ which is as follows:
\beq
x_{f1}^{(2)}=x_{f2}^{(2)}-\frac{6(71\sqrt{6}-163)(N_{c2}-N_{c1})}{55(6\sqrt{6}-13)N_{c1}N_{c2}^2}\Big(11+N_{c2}(x_{f2}^{(1)})_0\Big)+O(\lambda_0^2/N^2).
\eeq
Thus, when this relation is satisfied, the degeneracy in the fixed points survives at this order. Now, imposing the above relation as well as  the relation between $x_{f1}^{(1)}$ and $x_{f2}^{(1)}$ obtained earlier, we get
\beq
h_p^{(2)}=&\frac{1}{1497375(36-13\sqrt{6})N_{c1}N_{c2}^2}\\
&\Bigg[\Big(4235(8821-3606\sqrt{6})+46464(25\sqrt{6}-62)A_0\Big)N_{c1}\\
&+27225(32-13\sqrt{6})N_{c2}\Big(11+N_{c2}(x_{f2}^{(1)})_0\Big)\\
&+\Big(55(148262-60519\sqrt{6})+8448(25\sqrt{6}-62)A_0\Big)N_{c1}N_{c2}(x_{f2}^{(1)})_0\\
&+N_{c1}N_{c2}^2\Bigg\{\Big(55(7865-3207\sqrt{6})+384(25\sqrt{6}-62)A_0\Big)(x_{f2}^{(1)})_0^2\\
&\qquad\qquad\quad+27225(25\sqrt{6}-62)x_{f2}^{(2)}\Bigg\}\Bigg]+O(\lambda_0/N^2),
\eeq

\beq
h_m^{(2)}=&\frac{(13\sqrt{6}-32)(N_{c1}-N_{c2})}{55(36-13\sqrt{6})N_{c1}N_{c2}^2}\Bigg[11+N_{c2}(x_{f2}^{(1)})_0\Bigg]+O(\lambda_0/N^2),
\eeq

\begin{equation}
\begin{split}
f_p^{(2)}
&=\frac{(165-7\sqrt{6})(N_{c1}-N_{c2})}{5170 N_{c1}N_{c2}^2}\Bigg[11+N_{c2}(x_{f2}^{(1)})_0\Bigg]\\
&\quad-\frac{\sqrt{6}}{8984250 N_{c2}^2}\Bigg[\Bigg(11+N_{c2}(x_{f2}^{(1)})_0\Bigg)\Bigg\{1536A_0\Big(11+N_{c2}(x_{f2}^{(1)})_0\Big)\\
&\qquad\qquad\qquad\qquad+55\Big(-656+4125\sqrt{6}+2(-101+120\sqrt{6})N_{c2}(x_{f2}^{(1)})_0\Big)\Bigg\}\\
&\qquad\qquad\qquad\qquad+108900 N_{c2}^2 x_{f2}^{(2)}\Bigg]+O(\lambda_0/N^2),
 \end{split}
\end{equation}

\begin{equation}
\begin{split}
&2\Big(\zeta_0\zeta^{(2)}+f_{0m}f_m^{(2)}\Big)+(1+4\lambda_0)\Big(f_m^{(1)2}+\zeta^{(1)2}\Big)+8 \lambda_p^{(1)}\Big(\zeta_0\zeta^{(1)}+f_{0m} f_m^{(1)}\Big)\\
&=\frac{6\sqrt{6}-13}{9075 N_{c2}^2}\Big(11+N_{c2}(x_{f2}^{(1)})_0\Big)^2\\
&\quad+\frac{\lambda_0}{145200 N_{c2}^2}\Bigg[\Big((-2216+441\sqrt{6})+\frac{768}{55}(13-6\sqrt{6})A_0\Big)\Big(11+N_{c2}(x_{f2}^{(1)})_0\Big)^2\\
&\qquad\qquad\qquad\qquad+\Big(72(163-71\sqrt{6})-32(13-6\sqrt{6})N_{c2} (x_{f2}^{(1)})_1\Big)\Big(11+N_{c2}(x_{f2}^{(1)})_0\Big)\\
&\qquad\qquad\qquad\qquad+330(13-6\sqrt{6})N_{c2}^2 x_{f2}^{(2)}\Bigg]-\Bigg[\frac{3(165-7\sqrt{6})(N_{c1}-N_{c2})\lambda_0}{752 N_{c1}N_{c2}}\Bigg]f_m^{(1)}\\
&\quad+\Bigg[\frac{(165-7\sqrt{6})(N_{c1}-N_{c2})\Big(11+N_{c2}(x_{f2}^{(1)})_0\Big)}{2585 N_{c1}N_{c2}^2}\Bigg]f_{0m}+O(\lambda_0^2/N^2).
 \end{split}
\end{equation}
The last quantity is the $O(1/N^2)$ correction to $(1+4\lambda_p)(\zeta^2+f_{m}^2)$ up to the first subleading order in the $\lambda_0$-expansion. When $N_{c1}\neq N_{c2}$, this quantity has terms which are dependent on $f_{0m}$ and $f_{m}^{(1)}$. These terms  provide the correction to the shape of the closed curve of  fixed points at $O(1/N^2)$. Notice that when $N_{c1}=N_{c2}$, this quantity becomes independent of $f_{0m}$ and $f_m^{(1)}$ which indicates that the circular form of the closed curve is preserved up to this order.

To determine whether the degeneracy in the fixed points survives at higher orders in $\lambda_0$ as well as to evaluate the higher order corrections to the couplings at the fixed points, one needs to compute the contributions of the higher loop diagrams  to the beta functions. This lies beyond the scope of this paper.

\bibliographystyle{JHEP}
\bibliography{NonRe}

\providecommand{\href}[2]{#2}\begingroup\raggedright\begin{thebibliography}{10}

\bibitem{Chai:2020zgq}
N.~Chai, S.~Chaudhuri, C.~Choi, Z.~Komargodski, E.~Rabinovici and M.~Smolkin,
  \emph{{Thermal Order in Conformal Theories}},
  \href{https://doi.org/10.1103/PhysRevD.102.065014}{\emph{Phys. Rev. D}
  {\bfseries 102} (2020) 065014}
  [\href{https://arxiv.org/abs/2005.03676}{{\ttfamily 2005.03676}}].

\bibitem{Chai:2020onq}
N.~Chai, S.~Chaudhuri, C.~Choi, Z.~Komargodski, E.~Rabinovici and M.~Smolkin,
  \emph{{Symmetry Breaking at All Temperatures}},
  \href{https://doi.org/10.1103/PhysRevLett.125.131603}{\emph{Phys. Rev. Lett.}
  {\bfseries 125} (2020) 131603}.

\bibitem{Kirzhnits:1972iw}
D.~Kirzhnits, \emph{{Weinberg model in the hot universe}}, {\emph{JETP Lett.}
  {\bfseries 15} (1972) 529}.

\bibitem{Kirzhnits:1972ut}
D.~Kirzhnits and A.~D. Linde, \emph{{Macroscopic Consequences of the Weinberg
  Model}}, \href{https://doi.org/10.1016/0370-2693(72)90109-8}{\emph{Phys.
  Lett. B} {\bfseries 42} (1972) 471}.

\bibitem{Dolan:1973qd}
L.~Dolan and R.~Jackiw, \emph{{Symmetry Behavior at Finite Temperature}},
  \href{https://doi.org/10.1103/PhysRevD.9.3320}{\emph{Phys. Rev.} {\bfseries
  D9} (1974) 3320}.

\bibitem{Kirzhnits:1976ts}
D.~Kirzhnits and A.~D. Linde, \emph{{Symmetry Behavior in Gauge Theories}},
  \href{https://doi.org/10.1016/0003-4916(76)90279-7}{\emph{Annals Phys.}
  {\bfseries 101} (1976) 195}.

\bibitem{Klimenko:1988ng}
K.~Klimenko, \emph{{1/N expansion in the O(N) x O(N) scalar theory and the
  problem of symmetry restoration at high temperature}},
  \href{https://doi.org/10.1007/BF01016185}{\emph{Theor. Math. Phys.}
  {\bfseries 80} (1989) 929}.

\bibitem{Bimonte:1996cw}
G.~Bimonte and G.~Lozano, \emph{{Symmetry nonrestoration and inverse symmetry
  breaking on the lattice}},
  \href{https://doi.org/10.1016/S0370-2693(96)01230-0}{\emph{Phys. Lett. B}
  {\bfseries 388} (1996) 692}
  [\href{https://arxiv.org/abs/hep-th/9603201}{{\ttfamily hep-th/9603201}}].

\bibitem{Gavela:1998ux}
M.~B. Gavela, O.~Pene, N.~Rius and S.~Vargas-Castrillon, \emph{{The Fading of
  symmetry nonrestoration at finite temperature}},
  \href{https://doi.org/10.1103/PhysRevD.59.025008}{\emph{Phys. Rev.}
  {\bfseries D59} (1998) 025008}
  [\href{https://arxiv.org/abs/hep-ph/9801244}{{\ttfamily hep-ph/9801244}}].

\bibitem{Weinberg:1974hy}
S.~Weinberg, \emph{{Gauge and Global Symmetries at High Temperature}},
  \href{https://doi.org/10.1103/PhysRevD.9.3357}{\emph{Phys. Rev.} {\bfseries
  D9} (1974) 3357}.

\bibitem{Orloff:1996yn}
J.~Orloff, \emph{{The UV price for symmetry nonrestoration}},
  \href{https://doi.org/10.1016/S0370-2693(97)00552-2}{\emph{Phys. Lett.}
  {\bfseries B403} (1997) 309}
  [\href{https://arxiv.org/abs/hep-ph/9611398}{{\ttfamily hep-ph/9611398}}].

\bibitem{Bimonte:1999tw}
G.~Bimonte, D.~Iniguez, A.~Tarancon and C.~L. Ullod, \emph{{Inverse symmetry
  breaking on the lattice: An Accurate MC study}},
  \href{https://doi.org/10.1016/S0550-3213(99)00421-6}{\emph{Nucl. Phys.}
  {\bfseries B559} (1999) 103}
  [\href{https://arxiv.org/abs/hep-lat/9903027}{{\ttfamily hep-lat/9903027}}].

\bibitem{Pinto:1999pg}
M.~B. Pinto and R.~O. Ramos, \emph{{A Nonperturbative study of inverse symmetry
  breaking at high temperatures}},
  \href{https://doi.org/10.1103/PhysRevD.61.125016}{\emph{Phys. Rev.}
  {\bfseries D61} (2000) 125016}
  [\href{https://arxiv.org/abs/hep-ph/9912273}{{\ttfamily hep-ph/9912273}}].

\bibitem{Komargodski:2017dmc}
Z.~Komargodski, A.~Sharon, R.~Thorngren and X.~Zhou, \emph{{Comments on Abelian
  Higgs Models and Persistent Order}},
  \href{https://doi.org/10.21468/SciPostPhys.6.1.003}{\emph{SciPost Phys.}
  {\bfseries 6} (2019) 003} [\href{https://arxiv.org/abs/1705.04786}{{\ttfamily
  1705.04786}}].

\bibitem{Tanizaki:2017qhf}
Y.~Tanizaki, T.~Misumi and N.~Sakai, \emph{{Circle compactification and 't
  Hooft anomaly}}, \href{https://doi.org/10.1007/JHEP12(2017)056}{\emph{JHEP}
  {\bfseries 12} (2017) 056}
  [\href{https://arxiv.org/abs/1710.08923}{{\ttfamily 1710.08923}}].

\bibitem{Dunne:2018hog}
G.~V. Dunne, Y.~Tanizaki and M.~Ünsal, \emph{{Quantum Distillation of Hilbert
  Spaces, Semi-classics and Anomaly Matching}},
  \href{https://doi.org/10.1007/JHEP08(2018)068}{\emph{JHEP} {\bfseries 08}
  (2018) 068} [\href{https://arxiv.org/abs/1803.02430}{{\ttfamily
  1803.02430}}].

\bibitem{Wan:2019oax}
Z.~Wan and J.~Wang, \emph{{Higher Anomalies, Higher Symmetries, and Cobordisms
  III: QCD Matter Phases Anew}},
  \href{https://arxiv.org/abs/1912.13514}{{\ttfamily 1912.13514}}.

\bibitem{Hong:2000rk}
S.-I. Hong and J.~B. Kogut, \emph{{Symmetry nonrestoration in a Gross-Neveu
  model with random chemical potential}},
  \href{https://doi.org/10.1103/PhysRevD.63.085014}{\emph{Phys. Rev. D}
  {\bfseries 63} (2001) 085014}
  [\href{https://arxiv.org/abs/hep-th/0007216}{{\ttfamily hep-th/0007216}}].

\bibitem{Mohapatra:1979qt}
R.~N. Mohapatra and G.~Senjanovic, \emph{{Soft CP Violation at High
  Temperature}}, \href{https://doi.org/10.1103/PhysRevLett.42.1651}{\emph{Phys.
  Rev. Lett.} {\bfseries 42} (1979) 1651}.

\bibitem{Langacker:1980kd}
P.~Langacker and S.-Y. Pi, \emph{{Magnetic Monopoles in Grand Unified
  Theories}}, \href{https://doi.org/10.1103/PhysRevLett.45.1}{\emph{Phys. Rev.
  Lett.} {\bfseries 45} (1980) 1}.

\bibitem{Salomonson:1984rh}
P.~Salomonson, B.~S. Skagerstam and A.~Stern, \emph{{On the Primordial Monopole
  Problem in Grand Unified Theories}},
  \href{https://doi.org/10.1016/0370-2693(85)90843-3}{\emph{Phys. Lett.}
  {\bfseries 151B} (1985) 243}.

\bibitem{Dodelson:1989ii}
S.~Dodelson and L.~M. Widrow, \emph{{BARYON SYMMETRIC BARYOGENESIS}},
  \href{https://doi.org/10.1103/PhysRevLett.64.340}{\emph{Phys. Rev. Lett.}
  {\bfseries 64} (1990) 340}.

\bibitem{Dodelson:1991iv}
S.~Dodelson, B.~R. Greene and L.~M. Widrow, \emph{{Baryogenesis, dark matter
  and the width of the Z}},
  \href{https://doi.org/10.1016/0550-3213(92)90328-9}{\emph{Nucl. Phys.}
  {\bfseries B372} (1992) 467}.

\bibitem{Dvali:1995cj}
G.~R. Dvali, A.~Melfo and G.~Senjanovic, \emph{{Is There a monopole problem?}},
  \href{https://doi.org/10.1103/PhysRevLett.75.4559}{\emph{Phys. Rev. Lett.}
  {\bfseries 75} (1995) 4559}
  [\href{https://arxiv.org/abs/hep-ph/9507230}{{\ttfamily hep-ph/9507230}}].

\bibitem{Meade:2018saz}
P.~Meade and H.~Ramani, \emph{{Unrestored Electroweak Symmetry}},
  \href{https://doi.org/10.1103/PhysRevLett.122.041802}{\emph{Phys. Rev. Lett.}
  {\bfseries 122} (2019) 041802}
  [\href{https://arxiv.org/abs/1807.07578}{{\ttfamily 1807.07578}}].

\bibitem{Gubser:2008px}
S.~S. Gubser, \emph{{Breaking an Abelian gauge symmetry near a black hole
  horizon}}, \href{https://doi.org/10.1103/PhysRevD.78.065034}{\emph{Phys.
  Rev.} {\bfseries D78} (2008) 065034}
  [\href{https://arxiv.org/abs/0801.2977}{{\ttfamily 0801.2977}}].

\bibitem{Hartnoll:2008kx}
S.~A. Hartnoll, C.~P. Herzog and G.~T. Horowitz, \emph{{Holographic
  Superconductors}},
  \href{https://doi.org/10.1088/1126-6708/2008/12/015}{\emph{JHEP} {\bfseries
  12} (2008) 015} [\href{https://arxiv.org/abs/0810.1563}{{\ttfamily
  0810.1563}}].

\bibitem{Hartnoll:2008vx}
S.~A. Hartnoll, C.~P. Herzog and G.~T. Horowitz, \emph{{Building a Holographic
  Superconductor}},
  \href{https://doi.org/10.1103/PhysRevLett.101.031601}{\emph{Phys. Rev. Lett.}
  {\bfseries 101} (2008) 031601}
  [\href{https://arxiv.org/abs/0803.3295}{{\ttfamily 0803.3295}}].

\bibitem{Buchel:2009ge}
A.~Buchel and C.~Pagnutti, \emph{{Exotic Hairy Black Holes}},
  \href{https://doi.org/10.1016/j.nuclphysb.2009.08.017}{\emph{Nucl. Phys. B}
  {\bfseries 824} (2010) 85} [\href{https://arxiv.org/abs/0904.1716}{{\ttfamily
  0904.1716}}].

\bibitem{Donos:2011ut}
A.~Donos and J.~P. Gauntlett, \emph{{Superfluid black branes in
  AdS\_4\textbackslash{}times S\textasciicircum{}7}},
  \href{https://doi.org/10.1007/JHEP06(2011)053}{\emph{JHEP} {\bfseries 06}
  (2011) 053} [\href{https://arxiv.org/abs/1104.4478}{{\ttfamily 1104.4478}}].

\bibitem{Alberte:2017oqx}
L.~Alberte, M.~Ammon, A.~Jim\'enez-Alba, M.~Baggioli and O.~Pujol\`as,
  \emph{{Holographic Phonons}},
  \href{https://doi.org/10.1103/PhysRevLett.120.171602}{\emph{Phys. Rev. Lett.}
  {\bfseries 120} (2018) 171602}
  [\href{https://arxiv.org/abs/1711.03100}{{\ttfamily 1711.03100}}].

\bibitem{Gursoy:2018umf}
U.~G\"ursoy, E.~Kiritsis, F.~Nitti and L.~Silva~Pimenta, \emph{{Exotic
  holographic RG flows at finite temperature}},
  \href{https://doi.org/10.1007/JHEP10(2018)173}{\emph{JHEP} {\bfseries 10}
  (2018) 173} [\href{https://arxiv.org/abs/1805.01769}{{\ttfamily
  1805.01769}}].

\bibitem{Buchel:2018bzp}
A.~Buchel, \emph{{Klebanov-Strassler black hole}},
  \href{https://doi.org/10.1007/JHEP01(2019)207}{\emph{JHEP} {\bfseries 01}
  (2019) 207} [\href{https://arxiv.org/abs/1809.08484}{{\ttfamily
  1809.08484}}].

\bibitem{Buchel:2020thm}
A.~Buchel, \emph{{Thermal order in holographic CFTs and no-hair theorem
  violation in black branes}},
  \href{https://arxiv.org/abs/2005.07833}{{\ttfamily 2005.07833}}.

\bibitem{Buchel:2020xdk}
A.~Buchel, \emph{{SUGRA/Strings like to be bald}},
  \href{https://arxiv.org/abs/2007.09420}{{\ttfamily 2007.09420}}.

\bibitem{Buchel:2020jfs}
A.~Buchel, \emph{{The fate of the conformal order}},
  \href{https://arxiv.org/abs/2011.11509}{{\ttfamily 2011.11509}}.

\bibitem{Hogervorst:2015akt}
M.~Hogervorst, S.~Rychkov and B.~C. van Rees, \emph{{Unitarity violation at the
  Wilson-Fisher fixed point in 4-$\epsilon$ dimensions}},
  \href{https://doi.org/10.1103/PhysRevD.93.125025}{\emph{Phys. Rev. D}
  {\bfseries 93} (2016) 125025}
  [\href{https://arxiv.org/abs/1512.00013}{{\ttfamily 1512.00013}}].

\bibitem{Belavin:1974gu}
A.~Belavin and A.~Migdal, \emph{{Calculation of anomalous dimensions in
  non-abelian gauge field theories}}, {\emph{Pisma Zh. Eksp. Teor. Fiz.}
  {\bfseries 19} (1974) 317}.

\bibitem{PhysRevLett.33.244}
W.~E. Caswell, \emph{Asymptotic behavior of non-abelian gauge theories to
  two-loop order},
  \href{https://doi.org/10.1103/PhysRevLett.33.244}{\emph{Phys. Rev. Lett.}
  {\bfseries 33} (1974) 244}.

\bibitem{Banks:1981nn}
T.~Banks and A.~Zaks, \emph{{On the Phase Structure of Vector-Like Gauge
  Theories with Massless Fermions}},
  \href{https://doi.org/10.1016/0550-3213(82)90035-9}{\emph{Nucl. Phys.}
  {\bfseries B196} (1982) 189}.

\bibitem{Veneziano:1976wm}
G.~Veneziano, \emph{{Some Aspects of a Unified Approach to Gauge, Dual and
  Gribov Theories}},
  \href{https://doi.org/10.1016/0550-3213(76)90412-0}{\emph{Nucl. Phys. B}
  {\bfseries 117} (1976) 519}.

\bibitem{Senaha:2020mop}
E.~Senaha, \emph{{Symmetry Restoration and Breaking at Finite Temperature: An
  Introductory Review}},
  \href{https://doi.org/10.3390/sym12050733}{\emph{Symmetry} {\bfseries 12}
  (2020) 733}.

\bibitem{tHooft:1974pnl}
G.~'t~Hooft, \emph{{A Two-Dimensional Model for Mesons}},
  \href{https://doi.org/10.1016/0550-3213(74)90088-1}{\emph{Nucl. Phys. B}
  {\bfseries 75} (1974) 461}.

\bibitem{Bardeen:1983rv}
W.~A. Bardeen, M.~Moshe and M.~Bander, \emph{{Spontaneous Breaking of Scale
  Invariance and the Ultraviolet Fixed Point in O($N$) Symmetric
  $(\bar{\phi}^6_3$ in Three-Dimensions) Theory}},
  \href{https://doi.org/10.1103/PhysRevLett.52.1188}{\emph{Phys. Rev. Lett.}
  {\bfseries 52} (1984) 1188}.

\bibitem{Aharony:2011jz}
O.~Aharony, G.~Gur-Ari and R.~Yacoby, \emph{{d=3 Bosonic Vector Models Coupled
  to Chern-Simons Gauge Theories}},
  \href{https://doi.org/10.1007/JHEP03(2012)037}{\emph{JHEP} {\bfseries 03}
  (2012) 037} [\href{https://arxiv.org/abs/1110.4382}{{\ttfamily 1110.4382}}].

\bibitem{Ryttov:2012ur}
T.~A. Ryttov and R.~Shrock, \emph{{Scheme Transformations in the Vicinity of an
  Infrared Fixed Point}},
  \href{https://doi.org/10.1103/PhysRevD.86.065032}{\emph{Phys. Rev. D}
  {\bfseries 86} (2012) 065032}
  [\href{https://arxiv.org/abs/1206.2366}{{\ttfamily 1206.2366}}].

\bibitem{Ryttov:2012nt}
T.~A. Ryttov and R.~Shrock, \emph{{An Analysis of Scheme Transformations in the
  Vicinity of an Infrared Fixed Point}},
  \href{https://doi.org/10.1103/PhysRevD.86.085005}{\emph{Phys. Rev. D}
  {\bfseries 86} (2012) 085005}
  [\href{https://arxiv.org/abs/1206.6895}{{\ttfamily 1206.6895}}].

\bibitem{Shrock:2014zca}
R.~Shrock, \emph{{Question of an ultraviolet zero of the beta function of the
  $\lambda (\vec{\phi}^2)^2_4$ theory}},
  \href{https://doi.org/10.1103/PhysRevD.90.065023}{\emph{Phys. Rev. D}
  {\bfseries 90} (2014) 065023}
  [\href{https://arxiv.org/abs/1408.3141}{{\ttfamily 1408.3141}}].

\bibitem{Kiritsis:2008at}
E.~Kiritsis and V.~Niarchos, \emph{{Interacting String Multi-verses and
  Holographic Instabilities of Massive Gravity}},
  \href{https://doi.org/10.1016/j.nuclphysb.2008.12.010}{\emph{Nucl. Phys. B}
  {\bfseries 812} (2009) 488}
  [\href{https://arxiv.org/abs/0808.3410}{{\ttfamily 0808.3410}}].

\bibitem{Bashmakov:2017rko}
V.~Bashmakov, M.~Bertolini and H.~Raj, \emph{{On non-supersymmetric conformal
  manifolds: field theory and holography}},
  \href{https://doi.org/10.1007/JHEP11(2017)167}{\emph{JHEP} {\bfseries 11}
  (2017) 167} [\href{https://arxiv.org/abs/1709.01749}{{\ttfamily
  1709.01749}}].

\bibitem{Harlow:2018tng}
D.~Harlow and H.~Ooguri, \emph{{Symmetries in quantum field theory and quantum
  gravity}},  \href{https://arxiv.org/abs/1810.05338}{{\ttfamily 1810.05338}}.

\bibitem{Elitzur:1975im}
S.~Elitzur, \emph{{Impossibility of Spontaneously Breaking Local Symmetries}},
  \href{https://doi.org/10.1103/PhysRevD.12.3978}{\emph{Phys. Rev. D}
  {\bfseries 12} (1975) 3978}.

\bibitem{Banks:1979fi}
T.~Banks and E.~Rabinovici, \emph{{Finite Temperature Behavior of the Lattice
  Abelian Higgs Model}},
  \href{https://doi.org/10.1016/0550-3213(79)90064-6}{\emph{Nucl. Phys. B}
  {\bfseries 160} (1979) 349}.

\bibitem{Fradkin:1978dv}
E.~H. Fradkin and S.~H. Shenker, \emph{{Phase Diagrams of Lattice Gauge
  Theories with Higgs Fields}},
  \href{https://doi.org/10.1103/PhysRevD.19.3682}{\emph{Phys. Rev. D}
  {\bfseries 19} (1979) 3682}.

\bibitem{Ding:2015ona}
H.-T. Ding, F.~Karsch and S.~Mukherjee, \emph{{Thermodynamics of
  strong-interaction matter from Lattice QCD}},
  \href{https://doi.org/10.1142/S0218301315300076}{\emph{Int. J. Mod. Phys. E}
  {\bfseries 24} (2015) 1530007}
  [\href{https://arxiv.org/abs/1504.05274}{{\ttfamily 1504.05274}}].

\bibitem{Witten:1998zw}
E.~Witten, \emph{{Anti-de Sitter space, thermal phase transition, and
  confinement in gauge theories}},
  \href{https://doi.org/10.4310/ATMP.1998.v2.n3.a3}{\emph{Adv. Theor. Math.
  Phys.} {\bfseries 2} (1998) 505}
  [\href{https://arxiv.org/abs/hep-th/9803131}{{\ttfamily hep-th/9803131}}].

\bibitem{Benini:2019dfy}
F.~Benini, C.~Iossa and M.~Serone, \emph{{Conformality Loss, Walking, and 4D
  Complex Conformal Field Theories at Weak Coupling}},
  \href{https://doi.org/10.1103/PhysRevLett.124.051602}{\emph{Phys. Rev. Lett.}
  {\bfseries 124} (2020) 051602}
  [\href{https://arxiv.org/abs/1908.04325}{{\ttfamily 1908.04325}}].

\bibitem{Hansen:2017pwe}
F.~F. Hansen, T.~Janowski, K.~Lang\ae{}ble, R.~B. Mann, F.~Sannino, T.~G.
  Steele et~al., \emph{{Phase structure of complete asymptotically free
  SU($N_c$) theories with quarks and scalar quarks}},
  \href{https://doi.org/10.1103/PhysRevD.97.065014}{\emph{Phys. Rev. D}
  {\bfseries 97} (2018) 065014}
  [\href{https://arxiv.org/abs/1706.06402}{{\ttfamily 1706.06402}}].

\bibitem{Choi:2018ohn}
C.~Choi, M.~Ro\v{c}ek and A.~Sharon, \emph{{Dualities and Phases of $3D N=1$
  SQCD}}, \href{https://doi.org/10.1007/JHEP10(2018)105}{\emph{JHEP} {\bfseries
  10} (2018) 105} [\href{https://arxiv.org/abs/1808.02184}{{\ttfamily
  1808.02184}}].

\bibitem{Aitken:2019mtq}
K.~Aitken, A.~Baumgartner, C.~Choi and A.~Karch, \emph{{Generalization of
  QCD$_{3}$ symmetry-breaking and flavored quiver dualities}},
  \href{https://doi.org/10.1007/JHEP02(2020)060}{\emph{JHEP} {\bfseries 02}
  (2020) 060} [\href{https://arxiv.org/abs/1906.08785}{{\ttfamily
  1906.08785}}].

\bibitem{Rabinovici:1987tf}
E.~Rabinovici, B.~Saering and W.~A. Bardeen, \emph{{Critical Surfaces and Flat
  Directions in a Finite Theory}},
  \href{https://doi.org/10.1103/PhysRevD.36.562}{\emph{Phys. Rev.} {\bfseries
  D36} (1987) 562}.

\bibitem{Karananas:2019fox}
G.~K. Karananas, V.~Kazakov and M.~Shaposhnikov, \emph{{Spontaneous Conformal
  Symmetry Breaking in Fishnet CFT}},
  \href{https://doi.org/10.1016/j.physletb.2020.135922}{\emph{Phys. Lett. B}
  {\bfseries 811} (2020) 135922}
  [\href{https://arxiv.org/abs/1908.04302}{{\ttfamily 1908.04302}}].

\bibitem{Kachru:1998ys}
S.~Kachru and E.~Silverstein, \emph{{4-D conformal theories and strings on
  orbifolds}}, \href{https://doi.org/10.1103/PhysRevLett.80.4855}{\emph{Phys.
  Rev. Lett.} {\bfseries 80} (1998) 4855}
  [\href{https://arxiv.org/abs/hep-th/9802183}{{\ttfamily hep-th/9802183}}].

\bibitem{Lawrence:1998ja}
A.~E. Lawrence, N.~Nekrasov and C.~Vafa, \emph{{On conformal field theories in
  four-dimensions}},
  \href{https://doi.org/10.1016/S0550-3213(98)00495-7}{\emph{Nucl. Phys.}
  {\bfseries B533} (1998) 199}
  [\href{https://arxiv.org/abs/hep-th/9803015}{{\ttfamily hep-th/9803015}}].

\bibitem{Bershadsky:1998cb}
M.~Bershadsky and A.~Johansen, \emph{{Large N limit of orbifold field
  theories}}, \href{https://doi.org/10.1016/S0550-3213(98)00526-4}{\emph{Nucl.
  Phys.} {\bfseries B536} (1998) 141}
  [\href{https://arxiv.org/abs/hep-th/9803249}{{\ttfamily hep-th/9803249}}].

\bibitem{Schmaltz:1998bg}
M.~Schmaltz, \emph{{Duality of nonsupersymmetric large N gauge theories}},
  \href{https://doi.org/10.1103/PhysRevD.59.105018}{\emph{Phys.\ Rev.\ D}
  {\bfseries 59} (1999) 105018}
  [\href{https://arxiv.org/abs/hep-th/9805218}{{\ttfamily hep-th/9805218}}].

\bibitem{Dymarsky:2005nc}
A.~Dymarsky, I.~Klebanov and R.~Roiban, \emph{{Perturbative gauge theory and
  closed string tachyons}},
  \href{https://doi.org/10.1088/1126-6708/2005/11/038}{\emph{JHEP} {\bfseries
  11} (2005) 038} [\href{https://arxiv.org/abs/hep-th/0509132}{{\ttfamily
  hep-th/0509132}}].

\bibitem{Dymarsky:2005uh}
A.~Dymarsky, I.~Klebanov and R.~Roiban, \emph{{Perturbative search for fixed
  lines in large N gauge theories}},
  \href{https://doi.org/10.1088/1126-6708/2005/08/011}{\emph{JHEP} {\bfseries
  08} (2005) 011} [\href{https://arxiv.org/abs/hep-th/0505099}{{\ttfamily
  hep-th/0505099}}].

\bibitem{Pomoni:2008de}
E.~Pomoni and L.~Rastelli, \emph{{Large N Field Theory and AdS Tachyons}},
  \href{https://doi.org/10.1088/1126-6708/2009/04/020}{\emph{JHEP} {\bfseries
  04} (2009) 020} [\href{https://arxiv.org/abs/0805.2261}{{\ttfamily
  0805.2261}}].

\bibitem{Cherman:2010jj}
A.~Cherman, M.~Hanada and D.~Robles-Llana, \emph{{Orbifold equivalence and the
  sign problem at finite baryon density}},
  \href{https://doi.org/10.1103/PhysRevLett.106.091603}{\emph{Phys. Rev. Lett.}
  {\bfseries 106} (2011) 091603}
  [\href{https://arxiv.org/abs/1009.1623}{{\ttfamily 1009.1623}}].

\bibitem{Hanada:2011ju}
M.~Hanada and N.~Yamamoto, \emph{{Universality of Phases in QCD and QCD-like
  Theories}}, \href{https://doi.org/10.1007/JHEP02(2012)138}{\emph{JHEP}
  {\bfseries 02} (2012) 138} [\href{https://arxiv.org/abs/1103.5480}{{\ttfamily
  1103.5480}}].

\bibitem{Dunne:2016nmc}
G.~V. Dunne and M.~\"Unsal, \emph{{New Nonperturbative Methods in Quantum Field
  Theory: From Large-N Orbifold Equivalence to Bions and Resurgence}},
  \href{https://doi.org/10.1146/annurev-nucl-102115-044755}{\emph{Ann. Rev.
  Nucl. Part. Sci.} {\bfseries 66} (2016) 245}
  [\href{https://arxiv.org/abs/1601.03414}{{\ttfamily 1601.03414}}].

\bibitem{Aitken:2019shs}
K.~Aitken, C.~Choi and A.~Karch, \emph{{New and Old Fermionic Dualities from 3d
  Bosonization}}, \href{https://doi.org/10.1007/JHEP01(2020)035}{\emph{JHEP}
  {\bfseries 01} (2020) 035}
  [\href{https://arxiv.org/abs/1909.04036}{{\ttfamily 1909.04036}}].

\bibitem{Jepsen:2020czw}
C.~B. Jepsen, I.~R. Klebanov and F.~K. Popov, \emph{{RG Limit Cycles and
  $``$Spooky$"$ Fixed Points in Perturbative QFT}},
  \href{https://arxiv.org/abs/2010.15133}{{\ttfamily 2010.15133}}.

\bibitem{Kovtun:2003hr}
P.~Kovtun, M.~Unsal and L.~G. Yaffe, \emph{{Nonperturbative equivalences among
  large N(c) gauge theories with adjoint and bifundamental matter fields}},
  \href{https://doi.org/10.1088/1126-6708/2003/12/034}{\emph{JHEP} {\bfseries
  12} (2003) 034} [\href{https://arxiv.org/abs/hep-th/0311098}{{\ttfamily
  hep-th/0311098}}].

\bibitem{Kovtun:2004bz}
P.~Kovtun, M.~Unsal and L.~G. Yaffe, \emph{{Necessary and sufficient conditions
  for non-perturbative equivalences of large N(c) orbifold gauge theories}},
  \href{https://doi.org/10.1088/1126-6708/2005/07/008}{\emph{JHEP} {\bfseries
  07} (2005) 008} [\href{https://arxiv.org/abs/hep-th/0411177}{{\ttfamily
  hep-th/0411177}}].

\bibitem{Armoni:2004uu}
A.~Armoni, M.~Shifman and G.~Veneziano, \emph{{From superYang-Mills theory to
  QCD: Planar equivalence and its implications}},
  \href{https://arxiv.org/abs/hep-th/0403071}{{\ttfamily hep-th/0403071}}.

\bibitem{Unsal:2006pj}
M.~Unsal and L.~G. Yaffe, \emph{{(In)validity of large N orientifold
  equivalence}}, \href{https://doi.org/10.1103/PhysRevD.74.105019}{\emph{Phys.
  Rev. D} {\bfseries 74} (2006) 105019}
  [\href{https://arxiv.org/abs/hep-th/0608180}{{\ttfamily hep-th/0608180}}].

\bibitem{kleinert2001critical}
H.~Kleinert and V.~Schulte-Frohlinde, \emph{Critical Properties of
  $\Phi^4$-theories}. World Scientific, 2001.

\bibitem{OBrien:1984hvc}
K.~O'Brien and J.~Zuber, \emph{{Strong Coupling Expansion of Large $N$ \{QCD\}
  and Surfaces}},
  \href{https://doi.org/10.1016/0550-3213(85)90549-8}{\emph{Nucl. Phys. B}
  {\bfseries 253} (1985) 621}.

\bibitem{Linde:1980ts}
A.~D. Linde, \emph{{Infrared Problem in Thermodynamics of the Yang-Mills Gas}},
  \href{https://doi.org/10.1016/0370-2693(80)90769-8}{\emph{Phys. Lett. B}
  {\bfseries 96} (1980) 289}.

\bibitem{Gross:1980br}
D.~J. Gross, R.~D. Pisarski and L.~G. Yaffe, \emph{{QCD and Instantons at
  Finite Temperature}},
  \href{https://doi.org/10.1103/RevModPhys.53.43}{\emph{Rev. Mod. Phys.}
  {\bfseries 53} (1981) 43}.

\bibitem{Witten:1998xy}
E.~Witten, \emph{{Baryons and branes in anti-de Sitter space}},
  \href{https://doi.org/10.1088/1126-6708/1998/07/006}{\emph{JHEP} {\bfseries
  07} (1998) 006} [\href{https://arxiv.org/abs/hep-th/9805112}{{\ttfamily
  hep-th/9805112}}].

\bibitem{Gukov:1998kn}
S.~Gukov, M.~Rangamani and E.~Witten, \emph{{Dibaryons, strings and branes in
  AdS orbifold models}},
  \href{https://doi.org/10.1088/1126-6708/1998/12/025}{\emph{JHEP} {\bfseries
  12} (1998) 025} [\href{https://arxiv.org/abs/hep-th/9811048}{{\ttfamily
  hep-th/9811048}}].

\bibitem{Berenstein:2002ke}
D.~Berenstein, C.~P. Herzog and I.~R. Klebanov, \emph{{Baryon spectra and AdS
  /CFT correspondence}},
  \href{https://doi.org/10.1088/1126-6708/2002/06/047}{\emph{JHEP} {\bfseries
  06} (2002) 047} [\href{https://arxiv.org/abs/hep-th/0202150}{{\ttfamily
  hep-th/0202150}}].

\bibitem{Klebanov:2007us}
I.~R. Klebanov and A.~Murugan, \emph{{Gauge/Gravity Duality and Warped Resolved
  Conifold}}, \href{https://doi.org/10.1088/1126-6708/2007/03/042}{\emph{JHEP}
  {\bfseries 03} (2007) 042}
  [\href{https://arxiv.org/abs/hep-th/0701064}{{\ttfamily hep-th/0701064}}].

\bibitem{Martelli:2007mk}
D.~Martelli and J.~Sparks, \emph{{Baryonic branches and resolutions of
  Ricci-flat Kahler cones}},
  \href{https://doi.org/10.1088/1126-6708/2008/04/067}{\emph{JHEP} {\bfseries
  04} (2008) 067} [\href{https://arxiv.org/abs/0709.2894}{{\ttfamily
  0709.2894}}].

\bibitem{Martelli:2008cm}
D.~Martelli and J.~Sparks, \emph{{Symmetry-breaking vacua and baryon
  condensates in AdS/CFT}},
  \href{https://doi.org/10.1103/PhysRevD.79.065009}{\emph{Phys. Rev. D}
  {\bfseries 79} (2009) 065009}
  [\href{https://arxiv.org/abs/0804.3999}{{\ttfamily 0804.3999}}].

\bibitem{Chai:2020hnu}
N.~Chai, E.~Rabinovici, R.~Sinha and M.~Smolkin, \emph{{The bi-conical vector
  model at $1/N$}},  \href{https://arxiv.org/abs/2011.06003}{{\ttfamily
  2011.06003}}.

\bibitem{Machacek:1983tz}
M.~E. Machacek and M.~T. Vaughn, \emph{{Two Loop Renormalization Group
  Equations in a General Quantum Field Theory. 1. Wave Function
  Renormalization}},
  \href{https://doi.org/10.1016/0550-3213(83)90610-7}{\emph{Nucl. Phys.}
  {\bfseries B222} (1983) 83}.

\bibitem{Machacek:1984zw}
M.~E. Machacek and M.~T. Vaughn, \emph{{Two Loop Renormalization Group
  Equations in a General Quantum Field Theory. 3. Scalar Quartic Couplings}},
  \href{https://doi.org/10.1016/0550-3213(85)90040-9}{\emph{Nucl. Phys.}
  {\bfseries B249} (1985) 70}.

\bibitem{Luo:2002ti}
M.-x. Luo, H.-w. Wang and Y.~Xiao, \emph{{Two loop renormalization group
  equations in general gauge field theories}},
  \href{https://doi.org/10.1103/PhysRevD.67.065019}{\emph{Phys. Rev. D}
  {\bfseries 67} (2003) 065019}
  [\href{https://arxiv.org/abs/hep-ph/0211440}{{\ttfamily hep-ph/0211440}}].

\end{thebibliography}\endgroup
	
\end{document}